\documentclass[a4paper,11pt,twoside]{report}

\pdfoutput=1

\usepackage[UKenglish]{babel}   

\usepackage[T1]{fontenc}        
\usepackage{mathptmx}           
\usepackage{helvet}             
\usepackage{courier}            

\usepackage{amsmath,amssymb}  
\usepackage{microtype}        
\usepackage{pbox}             
\usepackage{textcomp}
\usepackage{bm}                 

\usepackage{float}
\usepackage{subcaption}       
\usepackage{graphicx}

\usepackage{fancyhdr}         
\usepackage{titlesec}         
\titleformat{\chapter}[hang]{\LARGE\bfseries}{\thechapter\hspace{26pt}}{0pt}{\LARGE\bfseries}
\titlespacing*{\chapter}{0pt}{3.5\baselineskip}{1.8\baselineskip}

\usepackage[font=small,labelfont=bf,labelsep=period]{caption}  

\usepackage[top=30mm, bottom=30mm, inner=30mm, outer=30mm, bindingoffset=0mm, marginparsep=3mm, marginparwidth=20mm, heightrounded]{geometry}
\newlength{\logobox}
\usepackage{color}
\usepackage{soul}
\usepackage{multirow}

\usepackage{esvect}

\usepackage{siunitx} 

\usepackage[dvipsnames]{xcolor}
\usepackage{longtable}

\usepackage[numbers, square, comma, sort&compress]{natbib}

\usepackage[colorlinks=true,allcolors=blue,bookmarks=true,bookmarksnumbered=true]{hyperref}

\hypersetup{                    
pdfauthor = {The NEXTLASERS consortium},
pdftitle = {Guidelines for designs for ultrastable laser with 1E-17 fractional frequency instability},
pdfsubject = {Ultrastable Lasers},
pdfkeywords = {ultrastable lasers, optical resonators, spectral hole burning},
pdfstartview = {FitV},
}

\newcommand{\authorlist}[1]{#1 \vspace{0.5\baselineskip}}

\newcommand{\affil}[2]{\noindent {\footnotesize $^{#1}$ \emph{#2} \par}}
\newcommand{\corr}[1]{\noindent {\footnotesize $^\dag$ e-mail: \href{mailto:#1}{#1} \par}}

\newcommand{\chapstart}{\vspace{1.5\baselineskip} \noindent}

\newcommand{\ptb}{2}
\newcommand{\inrim}{4}

\newcommand{\op}{5}
\newcommand{\rise}{{10}}
\newcommand{\vtt}{6}

\newcommand{\hhud}{9}
\newcommand{\links}{{11}}
\newcommand{\ulund}{7}
\newcommand{\tubs}{8}
\newcommand{\umk}{3}
\newcommand{\femto}{1}

\newcommand{\PTBaff}{Physikalisch-Technische Bundesanstalt, Bundesallee 100, 38116 Braunschweig, Germany}

\newcommand{\INRIMaff}{Istituto Nazionale di Ricerca Metrologica, Strada delle Cacce 91, 10135 Torino, Italy}

\newcommand{\VTTaff}{VTT Technical Research Centre of Finland Ltd, National Metrology Institute VTT MIKES, P.O.\ Box 1000, FI-02044 VTT, Finland}

\newcommand{\OPaff}{LNE-SYRTE, Observatoire de Paris, Universit\'e PSL, CNRS, Sorbonne Universit\'e, 61 Avenue de l'Observatoire, 75014 Paris, France}

\newcommand{\UMKaff}{Institute of Physics, Faculty of Physics, Astronomy and Informatics, Nicolaus Copernicus University, Grudzi\c{a}dzka 5, PL-87-100 Toru\'n, Poland}

\newcommand{\HHUDaff}{Institut f\"ur Experimentalphysik, Heinrich-Heine-Universit\"at D\"usseldorf, 40225 D\"usseldorf, Germany}

\newcommand{\LINKSaff}{Fondazione LINKS, 10135 Torino, Italy}

\newcommand{\FEMTOaff}{Universit\'e de Franche-Comt\'e, SUPMICROTECH, CNRS, Institut FEMTO-ST, F-25000 Besan\c{c}on, France}

\newcommand{\ULUNDaff}{Department of Physics, Lund University, PO Box 118, SE-22100 Lund, Sweden}

\newcommand{\TUBSaff}{Technical University of Braunschweig, Institute for Semiconductor Technology, Hans-Sommer-Str. 66, Braunschweig, Germany}

\newcommand{\RISEaff}{Measurement Science and Technology, RISE Research Institutes of Sweden, SE-501 15 Borås, Sweden}

\begin{document}

\graphicspath{{Title/figures}}

\enlargethispage{1\baselineskip}
\thispagestyle{empty}
\vspace*{-2\baselineskip}
\pbox{0.12\textwidth}{\vspace{-5mm}\hspace{-5mm}\href{https://www.ptb.de/empir2021/nextlasers}{\includegraphics[width=0.42\textwidth]{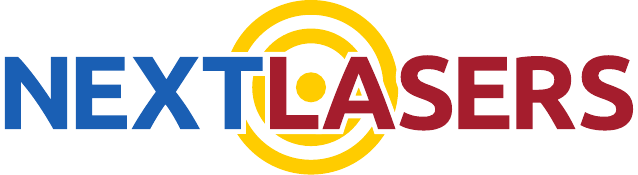}}}
\hfill
\pbox{0.5\textwidth}{\vspace{-6mm}\hspace{1mm}\href{http://www.empir-online.eu/}{\includegraphics[width=0.5\textwidth]{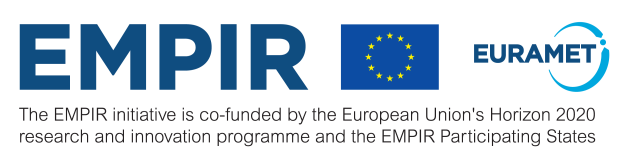}}}

\vspace*{3.5\baselineskip}

\begin{flushleft}
{\huge \textbf{Guidelines for designs for ultrastable laser with $\mathbf{10^{-17}}$ fractional frequency instability}}\\  
\pdfbookmark[0]{Guidelines for designs for ultrastable laser with 1E-17 fractional frequency instability}{title}
\vspace*{1.25\baselineskip}

\noindent
Joannès Barbarat$^\femto$,
Erik Benkler$^\ptb$,
Marcin Bober$^\umk$,
Cecilia Clivati$^\inrim$,
Johannes Dickmann$^\tubs$,
Bess Fang$^\op$,
Christophe Fluhr$^\femto$,
Thomas Fordell$^\vtt$, 
Jonathan Gillot$^\femto$,
Vincent Giordano$^\femto$, 
David Gustavsson$^\ulund$,
Kalle Hanhij\"arvi$^\vtt$, 
Michael Hartman$^\op$, 
Sofia Herbers$^\ptb$, 
Angelina Jaros$^\ptb$, 
Jan Kawohl$^\ptb$, 
Yann Kersalé$^\femto$, 
Stefanie Kroker$^\tubs$, 
Chang Jian Kwong$^\hhud$,
Clément Lacroûte$^\femto$,
Rodolphe Le Targat$^\op$,
Thomas Legero$^\ptb$, 
Marcus Lindén$^\rise$, 
Thomas Lindvall$^\vtt$,
J\'er\^ome Lodewyck$^\op$,
Chun Yu Ma$^\ptb$, 
Rémi Meyer$^\femto$, 
Piotr Morzy\'nski$^\umk$,
Jacques Millo$^\femto$,
Mateusz Naro\.znik$^\umk$, 
Daniele Nicolodi$^\ptb$,
Alessia Penza$^\inrim$, 
Benjamin Pointard$^\op$,
Lars Rippe$^\ulund$,  
Matias Risaro$^\inrim$, 
Pierre Roset$^\femto$,
Steffen Sauer$^\tubs$,
Paolo Savio$^\links$,
Stephan Schiller$^\hhud$,
Liam Shelling Neto$^\tubs$,
Uwe Sterr$^\ptb$,
Omid Vartehparvar$^\umk$,
Victor Vogt$^\hhud$, 
Nico Wagner$^\tubs$,
Anders E. Wallin$^\vtt$,  
Eugen Wiens$^\hhud$, 
Jialiang Yu$^\ptb$, 
Micha\l{} Zawada$^\umk$, 
and 
Martin Zelan$^\rise$ \\

\vspace*{0.7\baselineskip}

\noindent
{Edited by Uwe Sterr$^\ptb$}\\

\vspace*{0.7\baselineskip}

\affil{\femto}{\FEMTOaff}
\affil{\ptb}{\PTBaff}
\affil{\umk}{\UMKaff}
\affil{\inrim}{\INRIMaff}
\affil{\op}{\OPaff}
\affil{\vtt}{\VTTaff}
\affil{\ulund}{\ULUNDaff}
\affil{\tubs}{\TUBSaff}
\affil{\hhud}{\HHUDaff}
\affil{\rise}{\RISEaff}
\affil{\links}{\LINKSaff}


\end{flushleft}


\noindent
\begin{tabular}{c @{\hspace{0.055\textwidth}} c @{\hspace{0.045\textwidth}} c  @{\hspace{0.045\textwidth}} c  @{\hspace{0.045\textwidth}} c}
\centering
\vspace{3mm}
\pbox{\logobox}{\href{http://www.ptb.de/cms/en.html}{\includegraphics[width=0.22\textwidth]{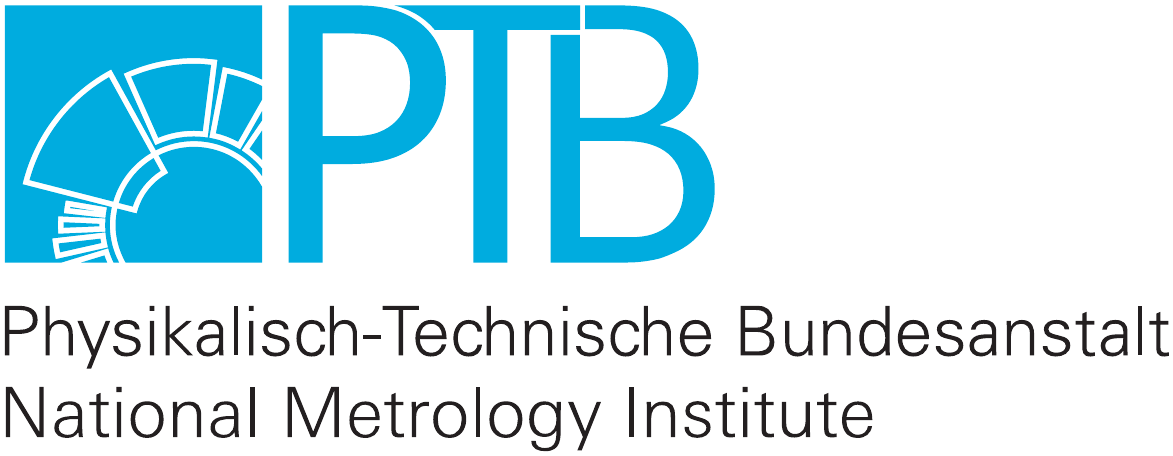}}} &
\pbox{\logobox}{\vspace{-2mm}\href{https://www.inrim.eu/}{\includegraphics[width=0.2\textwidth]{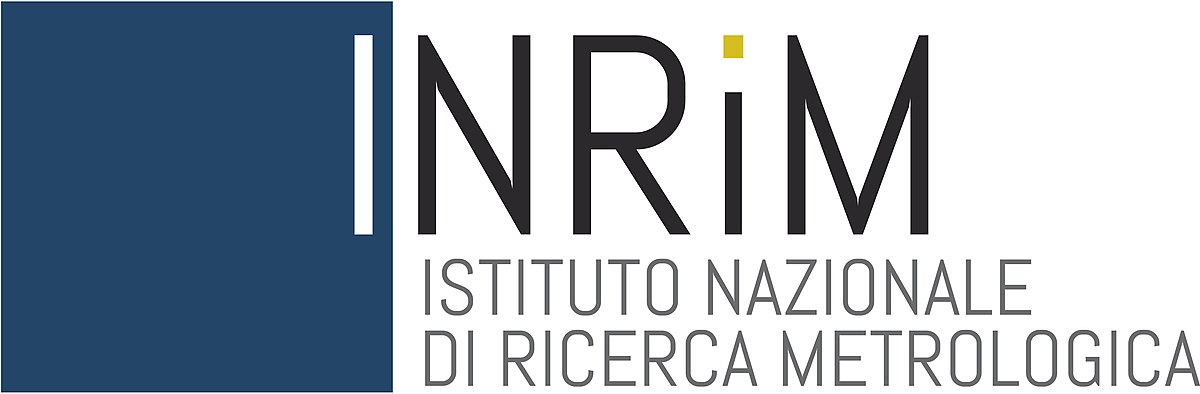}}} &
\pbox{\logobox}{\vspace{-2mm}\href{https://www.lne.fr/en}{\includegraphics[width=0.2\textwidth]{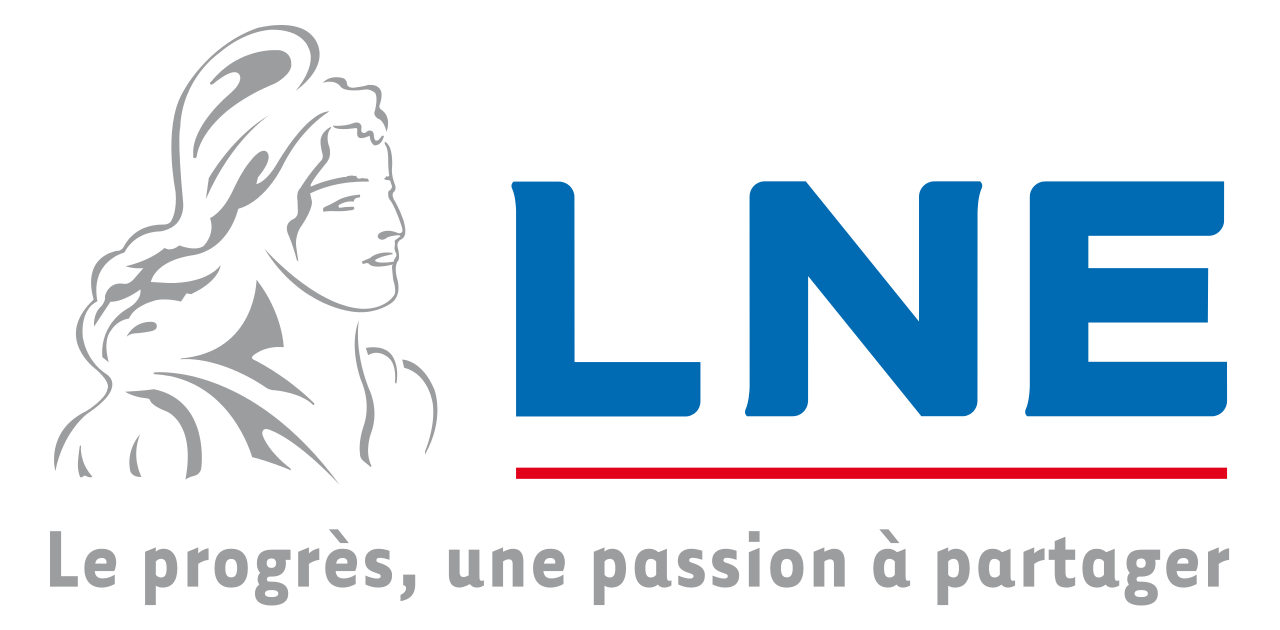}}} &
\pbox{\logobox}{\href{https://www.obspm.fr/?lang=en}{\includegraphics[width=0.22\textwidth]{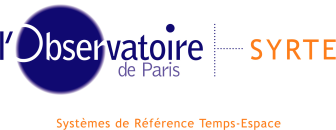}}} 
\vspace{-1mm} \\
\pbox{\logobox}{\href{https://www.umk.pl/en/}{\includegraphics[width=0.28\textwidth]{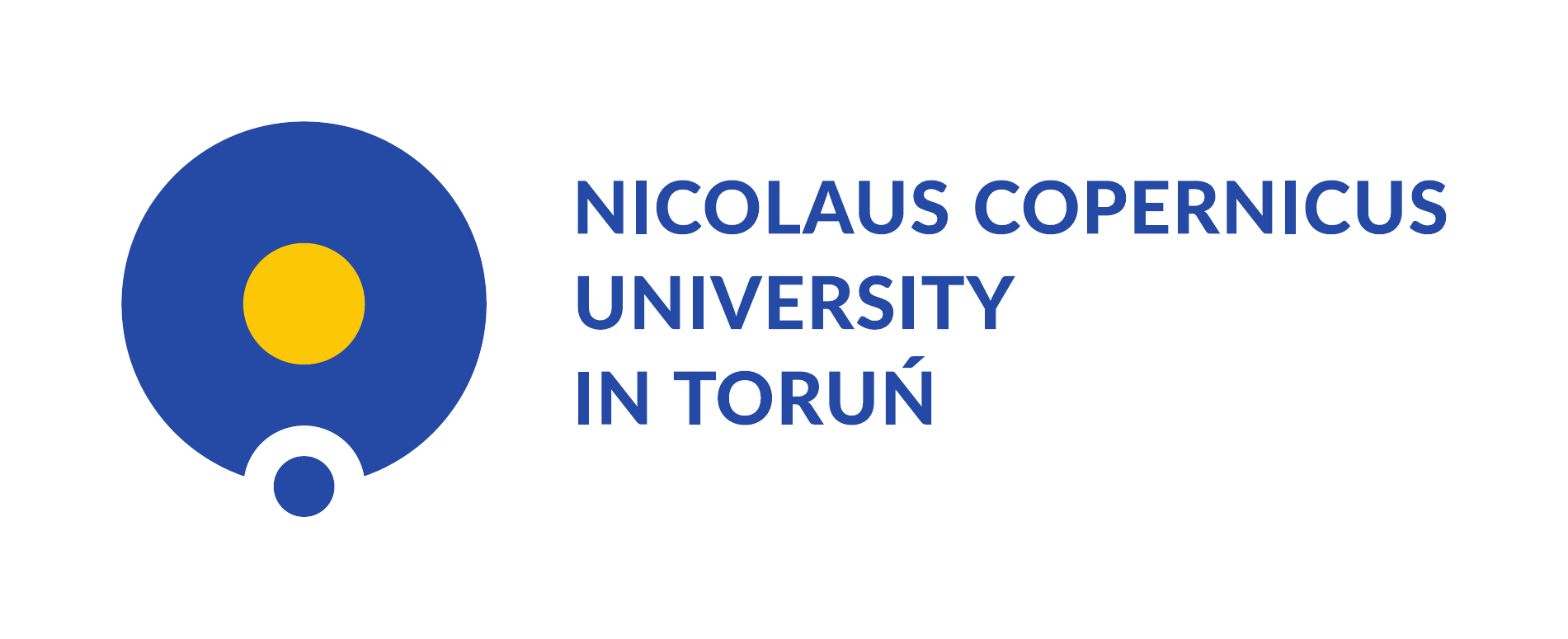}}} &
\pbox{\logobox}{\href{https://linksfoundation.com/en/}{\includegraphics[width=0.2\textwidth]{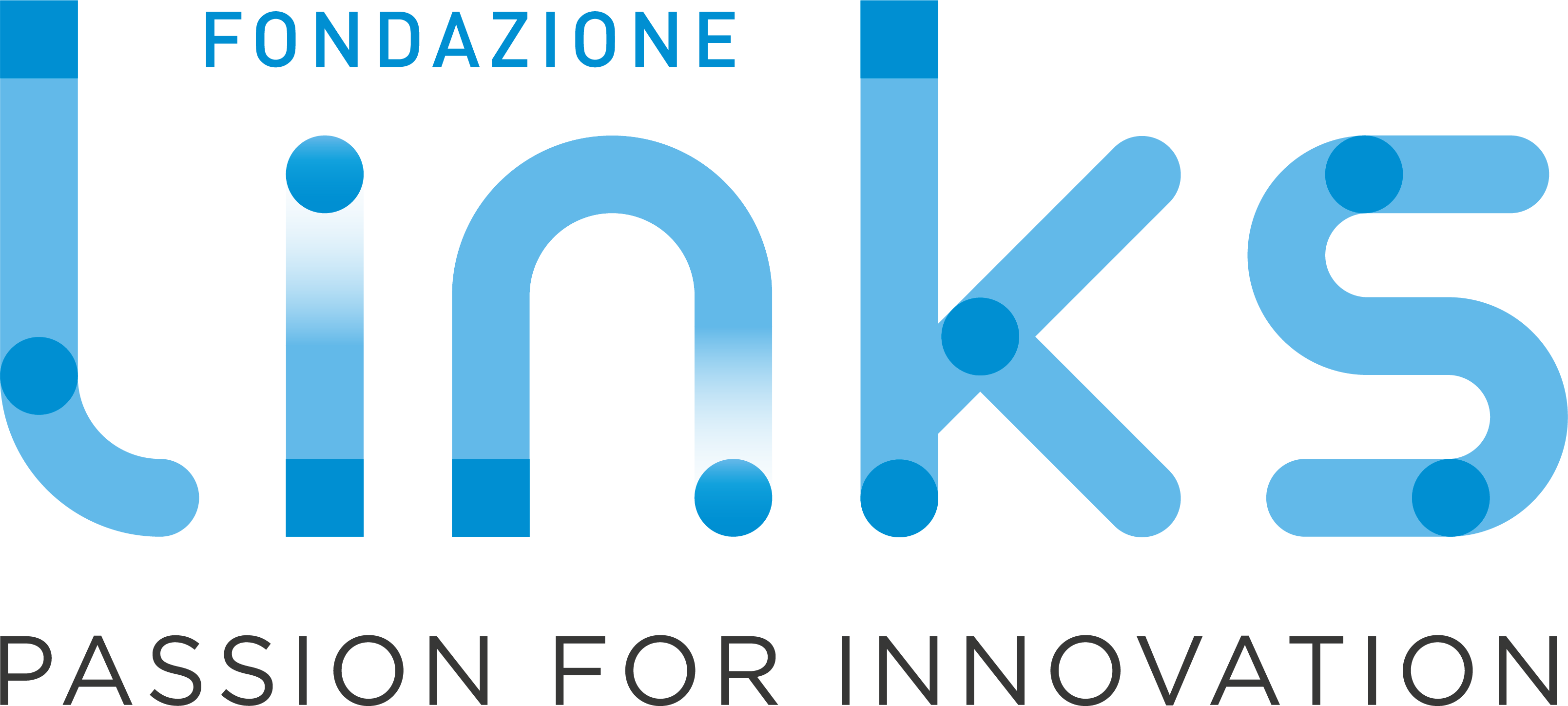}}} &  
\pbox{\logobox}{\href{http://www.cnrs.fr/index.php/en}{\includegraphics[width=0.12\textwidth]{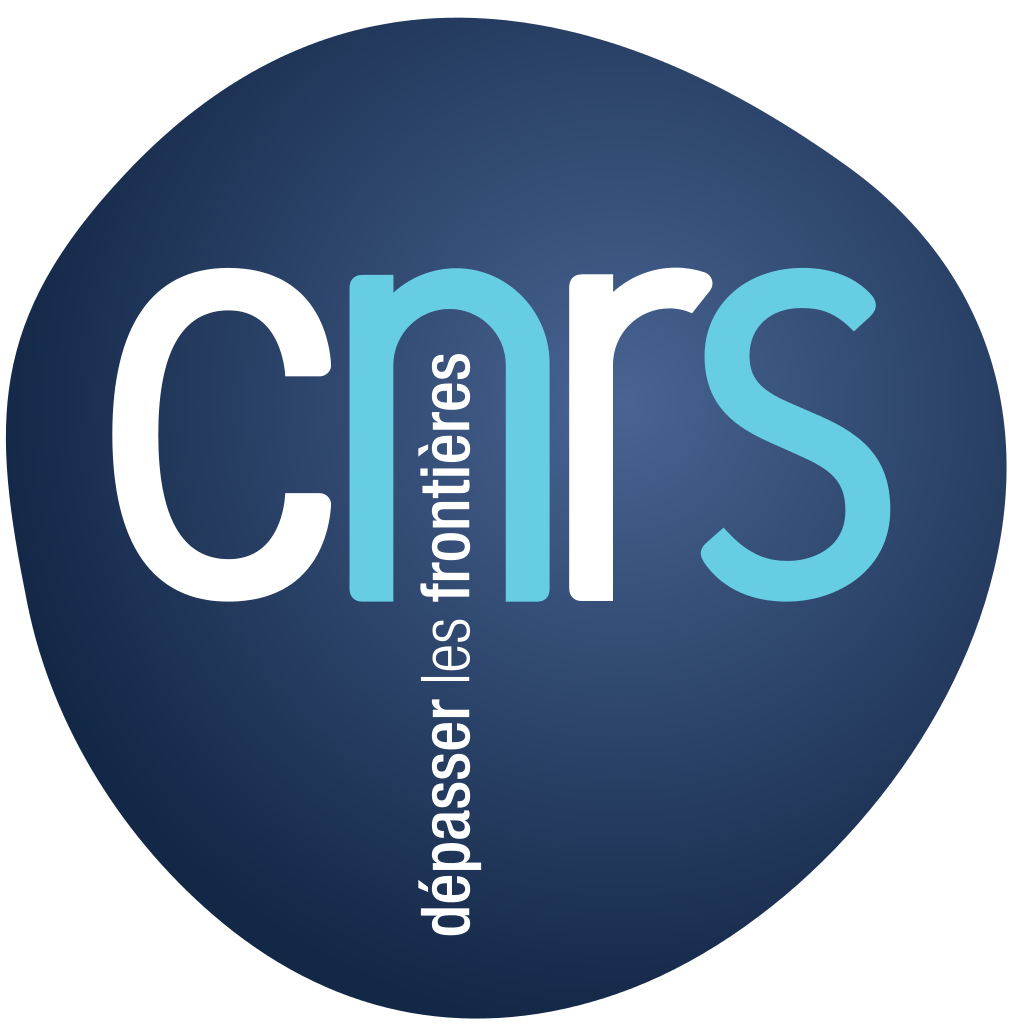}}} &
\pbox{\logobox}{\href{https://www.femto-st.fr/en}{\includegraphics[width=0.16\textwidth]{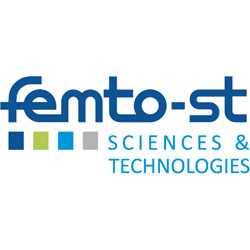}}} 
\vspace{-3mm} \\
\pbox{\logobox}{\href{https://www.hhu.de/en/}{\includegraphics[width=0.30\textwidth]{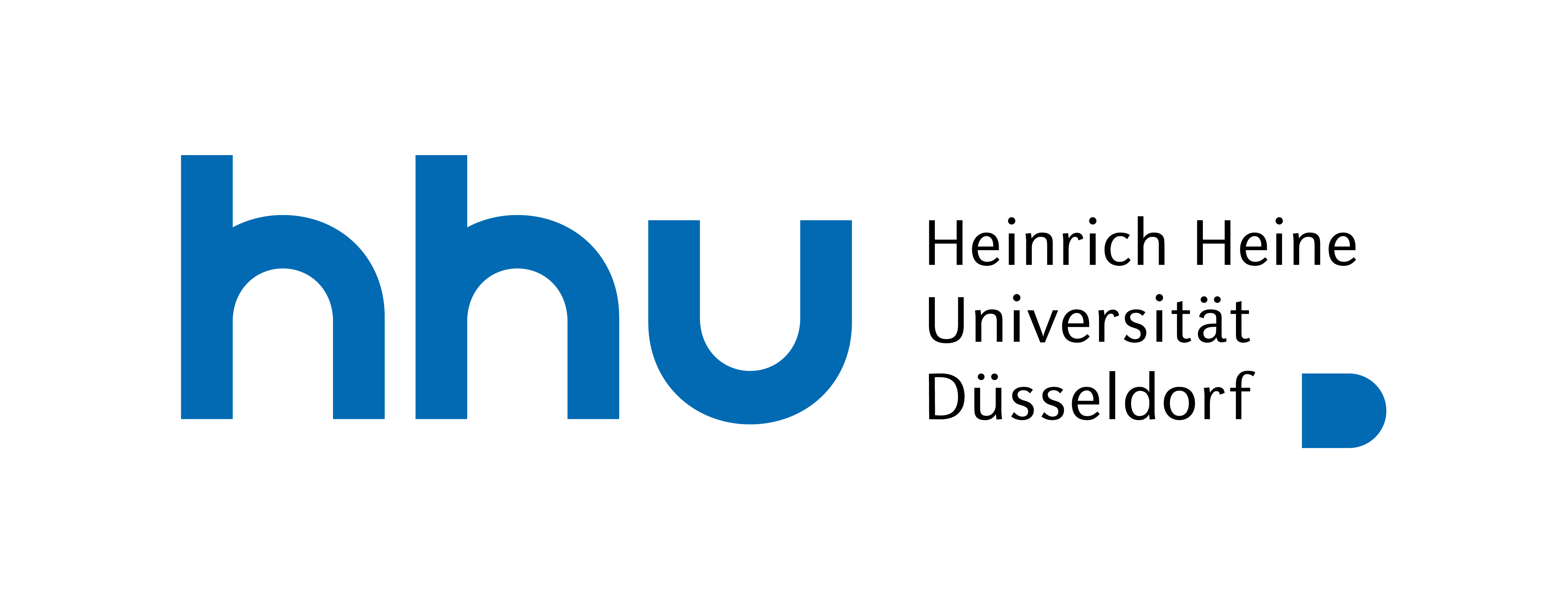}}} &
\pbox{\logobox}{\href{https://www.ri.se/en}{\includegraphics[width=0.075\textwidth]{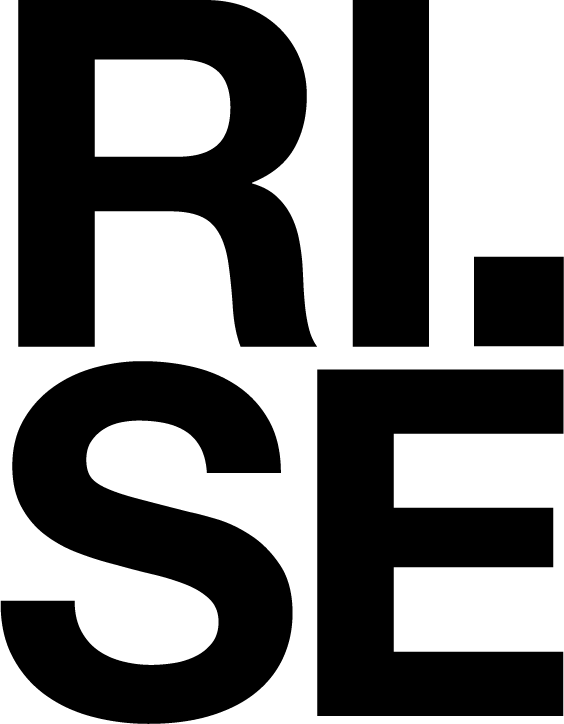}}} &
\pbox{\logobox}{\href{https://www.vttresearch.com/en}{\includegraphics[width=0.22\textwidth]{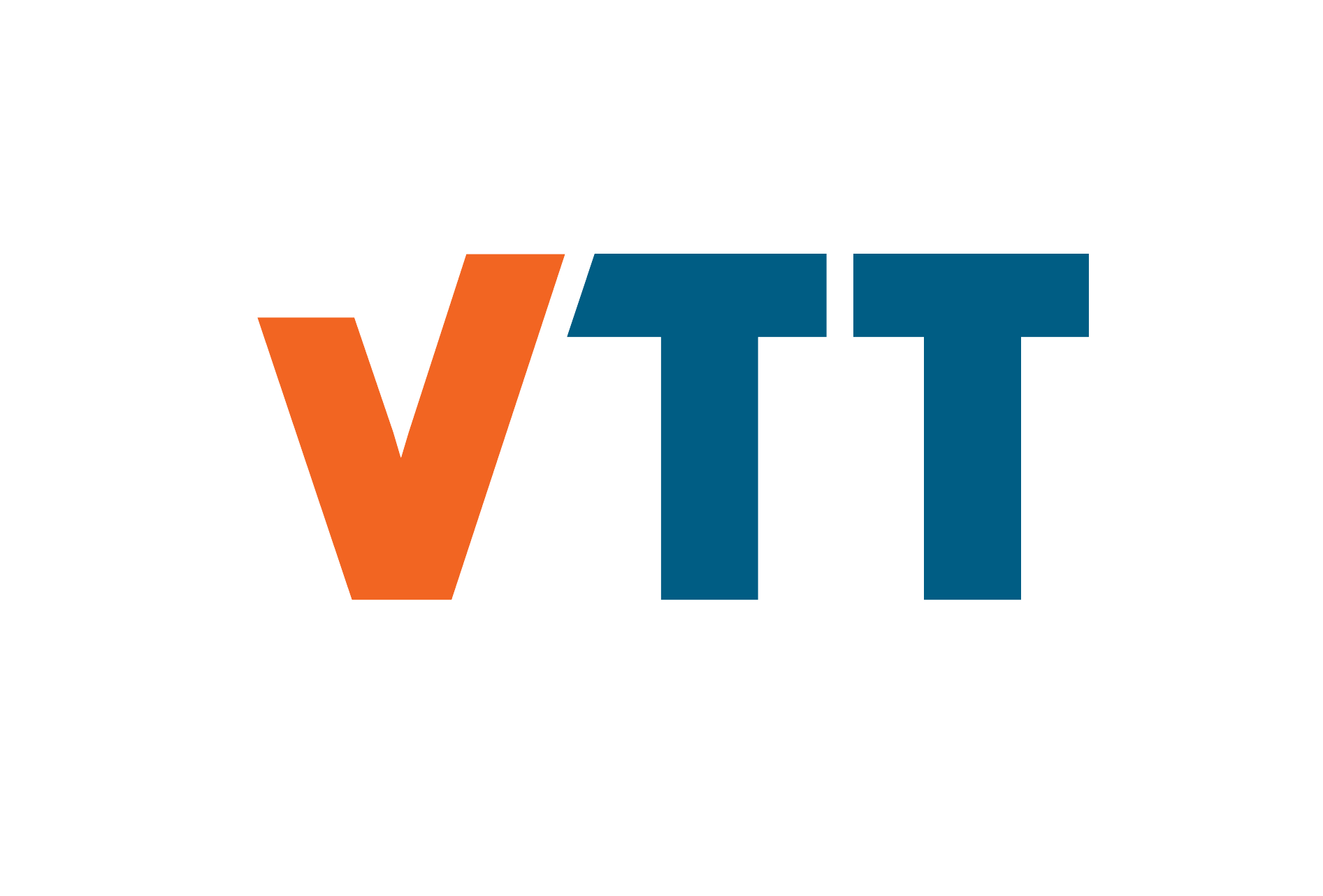}}} & 
\pbox{\logobox}{\href{https://www.tu-braunschweig.de}{\includegraphics[width=0.22\textwidth]{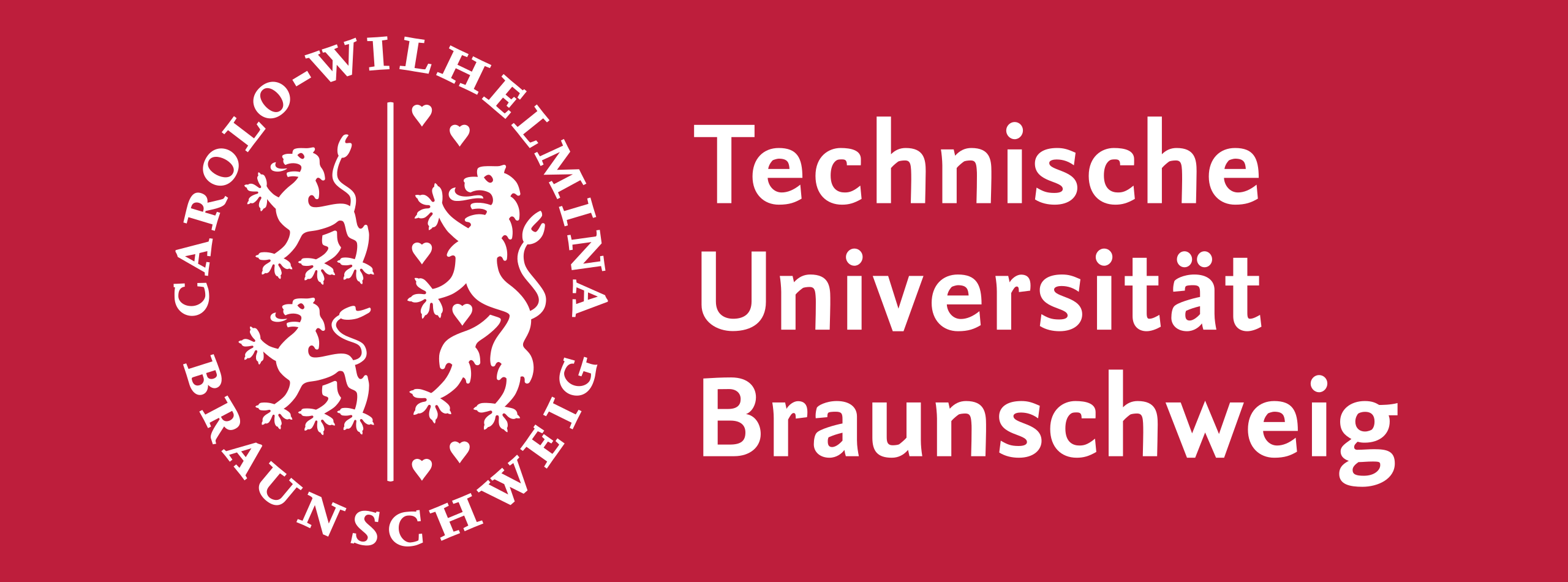}}}
\vspace{1mm} \\
\pbox{\logobox}{\href{https://www.lunduniversity.lu.se}{\includegraphics[width=0.36\textwidth]{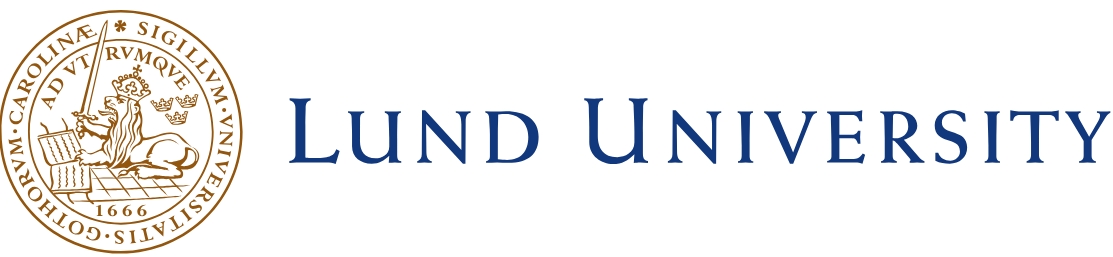}}}
&
&
&
\\\\

\end{tabular}

\clearpage

\thispagestyle{plain}
\vspace*{5\baselineskip}

Guidelines for designs for ultrastable laser with $10^{-17}$ fractional frequency instability \par
{\href{https://www.ptb.de/empir2021/nextlasers}{https://www.ptb.de/empir2021/nextlasers}}\par
2024\par
\vspace{0.5\baselineskip}
Edited by Uwe~Sterr\par
e-mail: \href{mailto:uwe.sterr@ptb.de}{uwe.sterr@ptb.de}\par

\vfill

\noindent
This work was carried out within the EMPIR project 20FUN08 NEXTLASERS. This project has received funding from the EMPIR programme co-financed by the European Union's Horizon 2020 research and innovation programme and the EMPIR Participating States. 
\clearpage

\cleardoublepage
\phantomsection
\addcontentsline{toc}{chapter}{Contents}
\tableofcontents

\chapter*{Preface}
\addcontentsline{toc}{chapter}{Preface}

Lasers with long coherence time and narrow linewidth are an essential tool for quantum sensors and clocks.
Ultrastable cavities and laser systems are now commercially available with fractional frequency instabilities in the mid $10^{-16}$ range.  
Ultrastable frequencies can now be distributed both locally and over long distances \cite{sch22a}.
With femtosecond combs, lasers at different wavelengths can be compared, allowing frequency stability transfer from one wavelength to another \cite{nic14,ben19, li22a} with added fluctuations well below $10^{-17}$ and  microwaves generation with exceptionally low phase noise \cite{xie17}. 
This enables interrogation times of optical clocks longer than a second \cite{ori18}.
Cavity-stabilized lasers have become very reliable and can even replace masers as flywheels to bridge downtimes of  optical clocks to create a fully optical and continuous timescale \cite{mil19}.

For both in-field or in-space use of transportable clocks, many transportable cavity designs and setups have been developed, the best reaching $1.6 \times 10^{-16}$ instability \cite{her22}.  
Laser instabilities as low as $7 \times 10^{-17}$ and $4 \times 10^{-17}$ have been reported for room temperature and cryogenic cavities respectively \cite{sch22a,mat17a}. 

This document aims to provide technical guidance for researchers starting in the field of ultrastable lasers. 
The target audience includes national metrology institutes (NMIs) wanting to set up optical clocks (or subsystems thereof) and PhD students and postdocs entering the field. Another potential audience is academic groups with experience in atomic physics and atom or ion trapping, but with less experience of time and frequency metrology and optical clock requirements.

These guidelines have arisen from the scope of the EMPIR project ``Next generation ultrastable lasers'' (\href{https://www.ptb.de/empir2021/nextlasers}{NEXTLASERS}). Examples are given from European partners, even though similar work is carried out all over the world.

NEXTLASER aimed to push the development of ultrastable lasers, based both on ultrastable optcal resonators and on spectral holes that are burnt in special crystals. It also covers the transfer of frequency stability from the ultrastable laser to other wavelength, required for interrogation of optical clocks. 
As one of the project's deliverables, this document fulfills objective 6 of the NEXTLASERS project to facilitate the takeup of the technology by sharing the knowledge acquired by the project consortium, and by providing practical guidance on each of these aspects in the following chapters.

\vspace{2\baselineskip}
\noindent
July 2024\\


\noindent\begin{minipage}[t]{0.45\linewidth}
Uwe Sterr\\
NEXTLASERS coordinator
\end{minipage}
\begin{minipage}[t]{0.5\linewidth}
Thomas Legero\\
NEXTLASERS work package  'Creating Impact'
\end{minipage}

\clearpage

\chapter*{Acronyms}
\addcontentsline{toc}{chapter}{Acronyms}

\begin{centering}

\renewcommand{\arraystretch}{1.2}
    \begin{longtable}{l|l}
\endhead
\endlastfoot

AC & Alternate Current\\
\hline
ADC & Analog to Digital Converter\\
\hline
ADEV & Allan Deviation\\
\hline
AM & Amplitude Modulation\\
\hline
AOM & Acousto-Optic Modulator\\
\hline
AR & Anti Reflection\\
\hline
ARTIQ & Advanced Real-Time Infrastructure for Quantum physics\\
\hline
AVI & Anti-Vibration Isolation\\
\hline
BS & Beam Splitter\\
\hline
CCD & Charge Coupled Device\\
\hline
CNRS & Centre National de la Recherche Scientifique\\
\hline
CTE & Coefficient of Thermal Expansion\\
\hline
DAC & Digital to Analog Converter\\
\hline
DC & Direct Current\\
\hline
DDS & Direct Digital Synthesis\\
\hline
EMPIR & European Metrology Programme for Innovation and Research \\
\hline
EOM & Electro-Optic Modulator\\
\hline
FEM & Finite Element Method \\
\hline
FF & Feed Forward \\
\hline
FPGA & Field-Programmable Gate Array\\
\hline
FS & Fused Silica \\
\hline
FSR & Free Spectral Range\\
\hline
FWHM & Full Width at Half Maximum\\
\hline
HDL & Hardware Description Language\\
\hline
HHUD & Heinrich Heine University Düsseldorf\\
\hline
HNLF & Highly Nonlinear Fiber\\
\hline
HOM & Higher Order Modes\\
\hline
I/Q & In-Phase/Quadrature\\
\hline
IIR & Infinite Impulse Response (filter)\\
\hline
LED & Light Emitting Diode\\
\hline
LN2 & Liquid Nitrogen\\
\hline
LNE & Laboratoire National de M\'etrologie et d'Essais\\
\hline
MADEV & Modified Allan Deviation\\
\hline
MIMO & Multiple Input / Multiple Output\\
\hline
MX & Mixing Plate\\
\hline
NEXCERA\textsuperscript{\textregistered} & low thermal expansion ceramic material (trademark of Krosaki Harima Corporation)\\
\hline
NHNM & New High (seismic) Noise Model\\
\hline
NLNM & New Low (seismic) Noise Model\\
\hline
NMI & National Metrology Institute \\
\hline
OBSPARIS & Observatoire de Paris\\
\hline
PBS & Polarizing Beam Splitter\\
\hline
PD & Photodiode\\
\hline
PDH & Pound-Drever-Hall (laser stabilization)\\
\hline
PID & Proportional Integral Differential\\
\hline
PM & Phase Modulation\\
\hline
PPLN & Periodically Poled Lithium Niobate\\
\hline
PSD & Power Spectral Density\\
\hline
PTB & Physikalisch-Technische Bundesanstalt\\
\hline
RAM & Residual Amplitude Modulation\\
\hline
RCWA & Rigorous Coupled Wave Analysis\\
\hline
RF & Radio Frequency\\
\hline
RIN & Relative Intensity Noise\\
\hline
RMS & Root Mean Squared\\
\hline
ROC & Radius of Curvature\\
\hline
SCS & Silicon Cryogenic Cavity\\
\hline
SHB & Spectral Hole Burning\\
\hline
SHG & Second Harmonic Generation\\
\hline
SNR & Signal to Noise Ratio\\
\hline
SYRTE &Syst\`eme de R\'ef\'erences Temps-Espace\\
\hline
SoC & System on Chip\\
\hline
TEM & Transverse Electro-Magnetic (mode)\\
\hline
THA & Tracking and Hold Amplifier\\
\hline
TUBS & Technical University Braunschweig\\
\hline
ULE\textsuperscript{\textregistered} & Ultra-Low Expansion (glass) (trademark of Corning) \\
\hline
UMK & Uniwersytet Mikołaja Kopernika w Toruniu \\
\hline
USRP & Universal Software-defined Radio Peripheral \\
\hline
VTT & Technical Research Centre of Finland Ltd\\
\hline
YSO & Yttrium Orthosilicate (Y$_2$SiO$_5$)\\
\hline
ZCTE & Zero of Coefficient of Thermal Expansion\\

\end{longtable}

\end{centering}
\clearpage

\graphicspath{{D1_low_thermal_noise/Figures}}
\chapter{Low thermal noise cavity designs}
\authorlist{%
Marcin Bober$^4$, 
Johannes Dickmann$^2$, 
Thomas Fordell$^5$,
Stefanie Kroker$^2$, 
Chang Jian Kwong$^1$, 
Thomas Legero$^3$, 
Chun Yu Ma$^3$, 
Mateusz Naro\.znik$^4$, 
Liam Shelling Neto$^2$, 
Steffen Sauer$^2$, 
Stephan Schiller$^1$,
Uwe Sterr$^3$,
Omid Vartehparvar$^4$, 
Victor Vogt$^1$, 
Nico Wagner$^2$, 
Eugen Wiens$^{1,\dag}$, 
Jialiang Yu$^3$, 
Micha\l{} Zawada$^4$
}

\affil{1}{\HHUDaff}
\affil{2}{\TUBSaff}
\affil{3}{\PTBaff}
\affil{4}{\UMKaff}
\affil{5}{\VTTaff}

\corr{eugen.wiens@hhu.de}

\chapstart

This chapter presents the results of investigations aimed toward the reduction of the thermal noise limit in state-of-the-art reference systems based on lasers stabilized to optical Fabry-Perot resonators. 

The reduction of thermal noise contribution of mirror coatings was studied on crystalline AlGaAs/GaAs coatings with low thermal noise and on newly developed meta-surface coated mirrors. 
NEXCERA ceramic as a material for cavity spacers was investigated by setting up and characterizing two first-generation NEXCERA cavities for their vibration sensitivities, long-term drift, and thermal noise. Additionally, a measurement of NEXCERA loss angle was determined. 
An optical and optomechanical study was carried out to determine a viable configuration for short room temperature cavities with large spot sizes on the cavity mirrors to reduce thermal noise.

\section{Introduction}
\label{sec:D1_intro} 

State-of-the-art laser instabilities as low as $7 \times 10^{-17}$ and $4 \times 10^{-17}$ have been reported for room-temperature and cryogenic cavities, respectively \cite{sch22a,mat17a}. 
Similar to large gravitational-wave detectors \cite{har06b, gra20a}, the performance of these best sub-meter sized ultrastable optical resonators is currently limited by the Brownian fluctuations of the usually employed dielectric mirror coatings \cite{ked23,yu23a}.
Brownian thermal noise is related to the mechanical loss by internal friction \cite{lev98} through the fluctuation-dissipation theorem \cite{cal51,kub66}. 
It can be in principle reduced by employing optical coatings with lower mechanical loss, using spacers and mirror substrates made of suitable materials with low loss, or by increasing the spot size of the laser beam using curved mirrors with a large radius of curvature. 

The general design of ultrastable lasers and the techniques to narrow the linewidth has been described elsewhere in great detail \cite{abd19, boy24}. 
This chapter describes ideas and work in progress toward the reduction of the thermal noise limit in state-of-the-art reference systems based on lasers stabilized to optical Fabry-Perot resonators. 
The reduction of thermal noise contribution of mirror coatings was studied by implementing novel crystalline AlGaAs/GaAs coatings with low thermal noise and investigating meta-surface coated mirrors. 

Ceramic material (NEXCERA) as a cavity spacer promises very small frequency drifts from aging of the material \cite{ito17}. 
The mechanical loss angle of the material was determined, as well as the characterization of two first-generation ceramic cavities for their vibration sensitivities, long-term drift and thermal noise.
Additionally, optical and optomechanical aspects for a viable configuration for short room-temperature cavities with large spot size on the cavity mirrors to reduce thermal noise are presented.

\section{Crystalline AlGaAs/GaAs coatings}
\label{sec:D1_S1}

Crystalline Al$_{0.92}$Ga$_{0.08}$As/GaAs Bragg reflectors \cite{col13,col23} have appeared as a promising approach for reducing coating thermal noise. 
These monocrystalline multilayers exhibit lower mechanical loss inferred from mechanical ring-down  \cite{col12, pen19} than conventional dielectric coatings (loss angle $\phi \approx 4\times 10^{-4}$ \cite{yam06,rob21}). 
Their mechanical loss, which is about a factor of ten lower than conventional coatings, can reduce the Brownian noise, expressed in Allan deviation $\sigma_y$, by a factor of 3 \cite{col13}. 
However, recently unexpected novel noise types exceeding the Brownian thermal noise level were observed at cryogenic temperatures of 124~K, 16~K and 4~K \cite{ked23,yu23a}. 

Here we discuss the performance of low-loss crystalline AlGaAs/GaAs coatings on silicon and fused silica mirror substrates at cryogenic (124~K) and room (297 K) temperatures, and on the influence of intracavity and external light on the birefringence of these mirrors.

In the next subsection we will first introduce the two different Fabry-Perot cavities contacted with crystalline coatings. 
Using two lasers locked to the same cavity allows for efficient and systematic optimization of the setup and characterization of the mirror noise.

\subsection{Experimental setup}

A 21~cm long Si cavity (Si5) and 48~cm long ULE cavity operated at a wavelength of 1.5~$\mu$m were set up at PTB to investigate the crystalline AlGaAs coating performance at 124 and 297~K respectively. 
Table \ref{tab:setups} summarizes their properties.
The Si cavity employs a monocrystalline silicon spacer with AlGaAs coatings bonded to silicon mirror substrates \cite{ked23,yu23a}. 
The cavity temperature is stabilized to the zero crossing of the coefficient of thermal expansion (CTE) of Si at 124~K.  
Based on our previous design \cite{hae15a}, the room temperature resonator employs a 48~cm long ULE spacer and a pair of fused silica substrates coated with AlGaAs coatings. 
ULE rings are attached to the back of the mirrors to compensate for CTE mismatch \cite{leg10}, leading to a measured zero crossing of the CTE at 297~K. 

In these setups, two lasers can be locked independently from opposite sides of the cavity \cite{yu23a} to study the coating-specific noises. 
An optical heterodyne beat signal between the two lasers is detected from the transmitted and reflected beam at one side of the cavity, forming a common-path geometry to suppress the influence of external path length fluctuations to the Fabry-Perot eigenfrequency. 
By locking to adjacent modes of the same cavity, influences from temperature and pressure fluctuations, vibrations, and photo-thermal contributions that lead to common optical length changes can be largely suppressed in the observed frequency difference.

The frequency of each cavity can be compared to another cryogenic 21~cm long Si cavity (Si2) equipped with conventional SiO$_2$/Ta$_2$O$_5$ coatings, which shows a fractional frequency instability of $4 \times 10^{-17}$ \cite{mat17a}.

In both cavities, the static birefringence (${n_\mathrm{biref}}\approx 4\times 10^{-4}$) of the crystalline coatings leads to a splitting of the resonances into two eigenmodes with orthogonal linear polarization as stated in Table \ref{tab:setups}. 
Besides the different eigenfrequencies, we also observe in both cavities a difference between the finesse of the two eigenmodes of 0.5 to 2\%.

\begin{center}
	\begin{table}[h]
		\centering
        \renewcommand{\arraystretch}{1.2}
		\caption{\label{tab:setups}
		Summary of the cryogenic Si cavity and the room temperature ULE cavity with AlGaAs coatings at PTB. Mirror configuration is plane concave (pl/cc) or concave - concave (cc/cc) with the indicated radius of curvature (ROC) and the beam radius on the curved (Si cavity) or on the plane mirror (ULE cavity) is indicated. 
		}
		
		\begin{tabular}{c r r c c c c}
			\hline
		    Cavity & Length & Temperature & Mirror          &   FSR   & Birefringent & Beam  \\
			         &        &             & configuration   &         & splitting    & radius\\
			\hline
			  Si5    & 21 cm  &  124 K      & cc/cc (2 m ROC) & 710 MHz & 200 kHz      & 482 $\mu$m\\
		    ULE    & 48 cm  &  297 K      & pl/cc (1 m ROC) & 310 MHz & 104 kHz      & 495 $\mu$m\\ 
			\hline
		\end{tabular}
	\end{table}
\end{center}

\subsection{Noise characterization of the Si cavity at 124~K}

\begin{figure}[ht]
	\begin{subfigure}[b]{0.50\textwidth}
		\includegraphics[width=18pc]{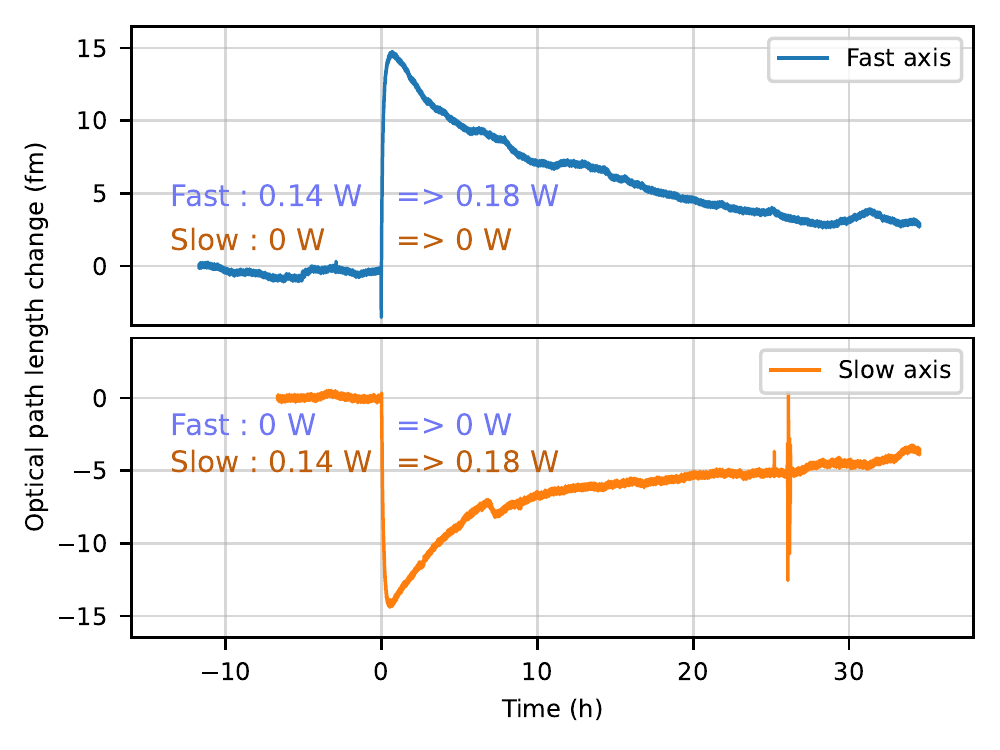}
	\end{subfigure}
	\begin{subfigure}[b]{0.50\textwidth}
		\includegraphics[width=18pc]{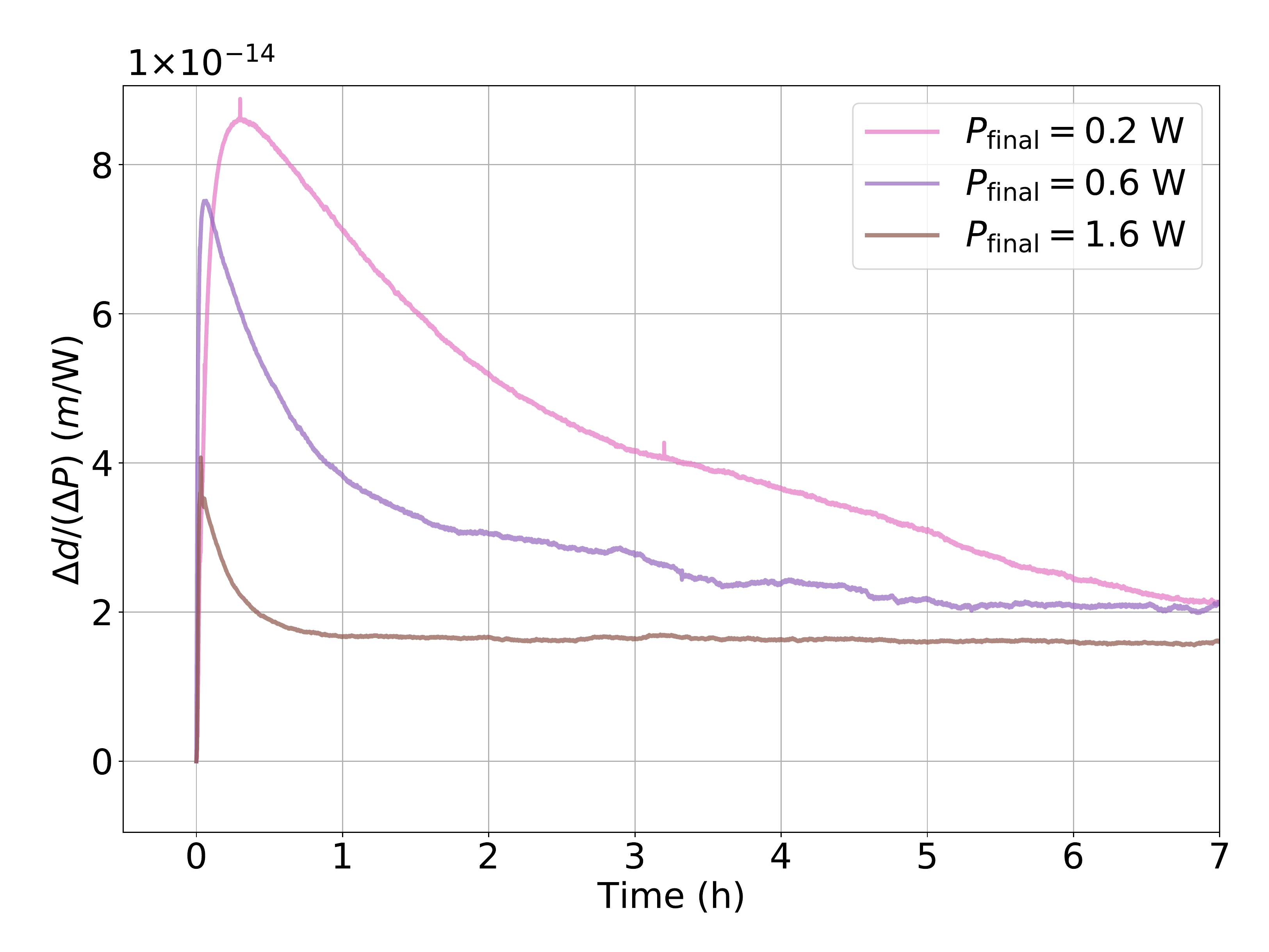}
	\end{subfigure}
	\caption{
	Left: Single polarization transient response of the 21 cm cavity to a step change of intracavity power measured in succession. 
	The optical path length changes of the fast (blue) and the slow (orange) axes are opposite. Right: Normalized response of the optical path length $\Delta d$ (slow axis) to a step $\Delta P$ in intracavity optical power. Both amplitude and time constant show a strong dependence on the final intracavity power $P_\mathrm{final}$.	}
	\label{fig:transient_a}
\end{figure}

The study on AlGaAs coatings in Si5 has revealed an unexpected dependence of the coating birefringence on intracavity power \cite{yu23a}. 
This effect is beyond the photo-thermo-optic effect observed in dielectric coatings \cite{far12} arising from the temperature change caused by the laser power absorbed in the coatings. 
The change in the optical path length is opposite for the two polarization eigenmodes at 124~K, as shown in the left panel of Fig.\ \ref{fig:transient_a}.  
Therefore, the effect is called "the photo-birefringent effect". 
The temporal response of the effect is also dependent on intracavity power, as shown in the right panel of Fig.~\ref{fig:transient_a}. 
At higher final optical power both sensitivity  and time constant are smaller. 
As the Si cavity is operated at the zero-CTE temperature of both spacer and mirror substrates, the photo thermal optic effect from these constituents is negligible. 
Nevertheless, stabilizing the laser power is critical as power fluctuations will lead to fluctuations of the birefringence and hence results in additional frequency noise of a laser locked to a polarization eigenmode. 

\begin{figure}[ht]
	\begin{subfigure}[b]{0.50\textwidth}
		\includegraphics[width=19pc]{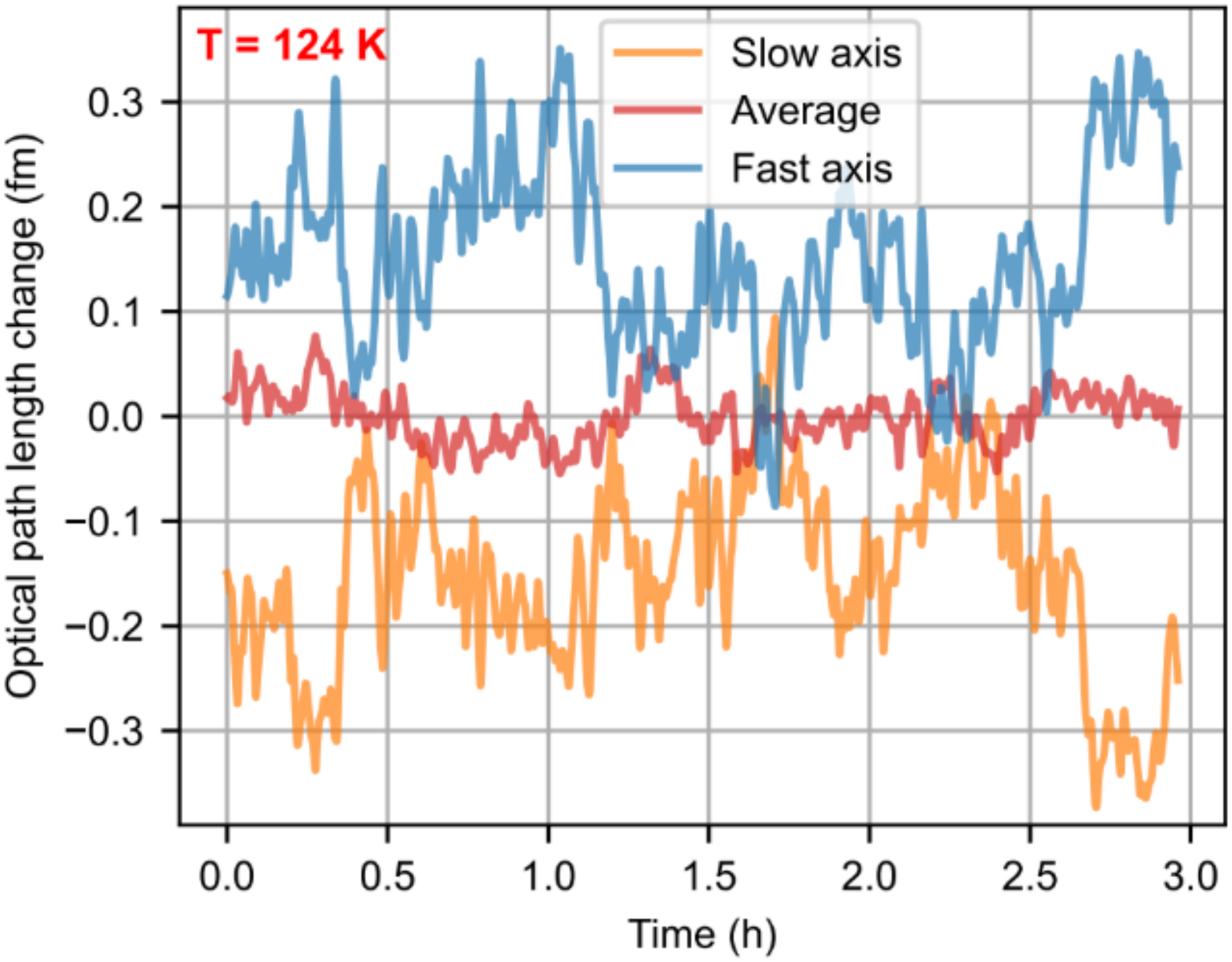}
	\end{subfigure}
	\begin{subfigure}[b]{0.50\textwidth}
		\includegraphics[width=19pc]{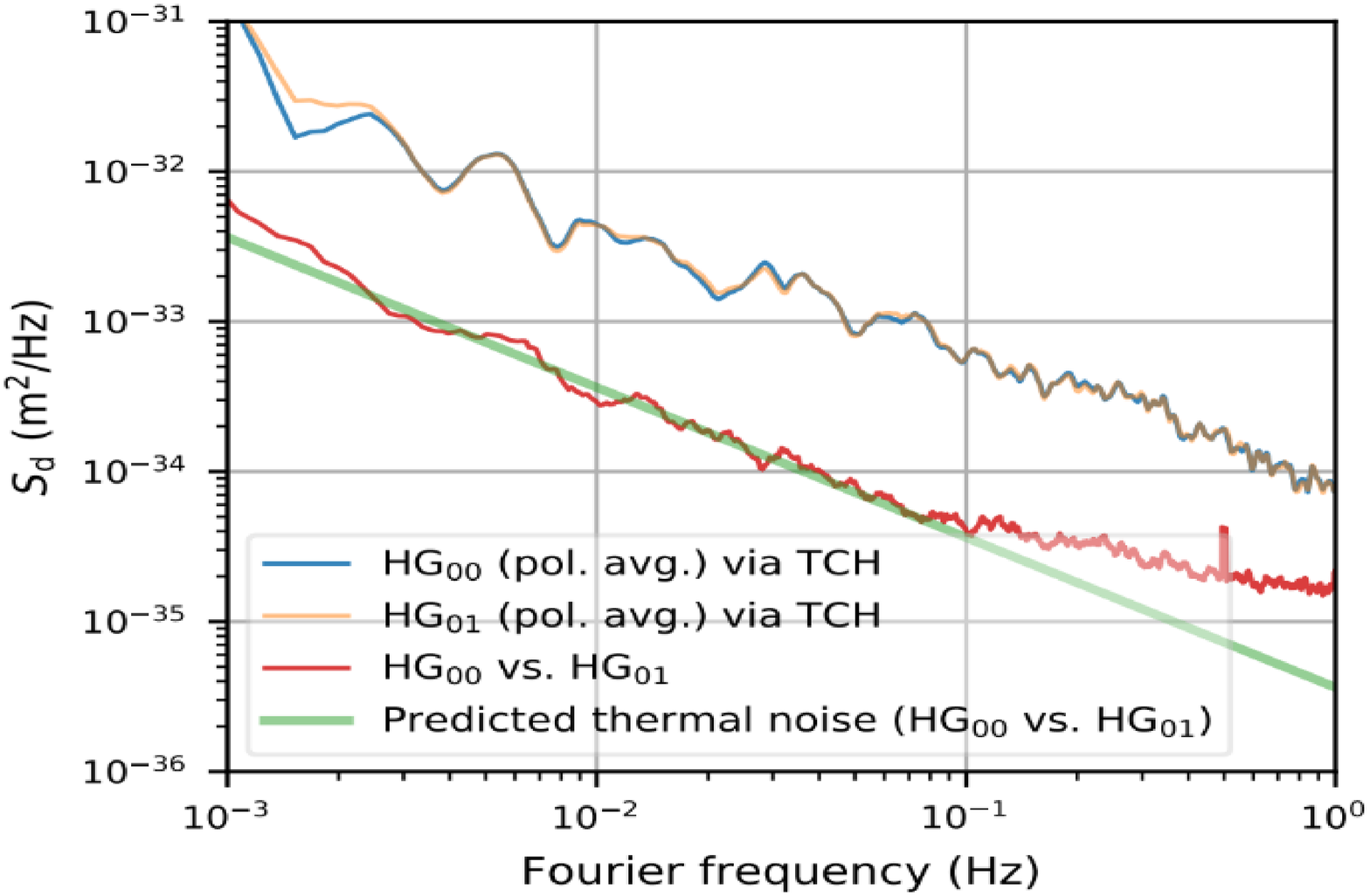}
	\end{subfigure}
	\caption{
		Left: Optical path length fluctuations of the 21 cm cavity measured with the beat signal between Si2 and the two
		polarization eigenmodes with intracavity power of 1.3 W in the fast and 0.7 W in the slow axis (blue, orange). Right: Spectral power densities of the individual displacement noise of HG\textsubscript{00} (blue) and HG\textsubscript{01} mode (orange) determined by three-cornered-hat (TCH) method, and of their difference (red). The estimated differential Brownian noise between the two modes is shown in green.	}
	\label{fig:birefSi5}
\end{figure}

After suppressing all technical noises including the photo-birefringent noise due to laser power fluctuations, we observe a characteristic noise from the coatings while measuring the stability against Si2. 
This noise type shows a strong anti-correlated frequency fluctuations between the two polarization eigenmodes as shown in the left panel of Fig.\ \ref{fig:birefSi5}. 
As this intrinsic birefringent noise of the polarization eigenmodes is anti correlated , the laser stability could be further improved by locking to the average of the two eigenmodes, using the dual frequency locking scheme \cite{yu23a}. 
The remaining noise level after polarization averaging still exhibits an excess noise beyond the expected Brownian thermal noise of the coatings \cite{yu23a,ked23}. 
The power spectral density of this excess noise was found to be independent of laser-power and shows an approximate $1/f$ behaviour, similar to Brownian thermal noise. 
After careful examination of the spatial correlation of this residual noise, it was found that its correlation length is much larger than the beam diameter, ruling out Brownian noise as the source, which is why we call it "global excess noise". \cite{yu23a}. 
By comparing the noise of different transverse modes, we can give an upper limit of the local noise that includes the Brownian noise. We could confirm, that the mechanical loss factor of the AlGaAs coating at 124~K is at at most equal to or lower than its room-temperature value.
To provide more understanding of the noise mechanism,
as a next step, the semiconductor properties of crystalline coatings were further investigated at PTB with a ULE cavity with fused silica substrates that enable the study of these effects at room temperature as well their dependence on illumination with light in a broad range of wavelengths. 

\subsection{Photo-induced effects at room temperature}
   
\begin{figure}[ht]
	\begin{subfigure}[b]{0.50\textwidth}
		\includegraphics[width=18pc]{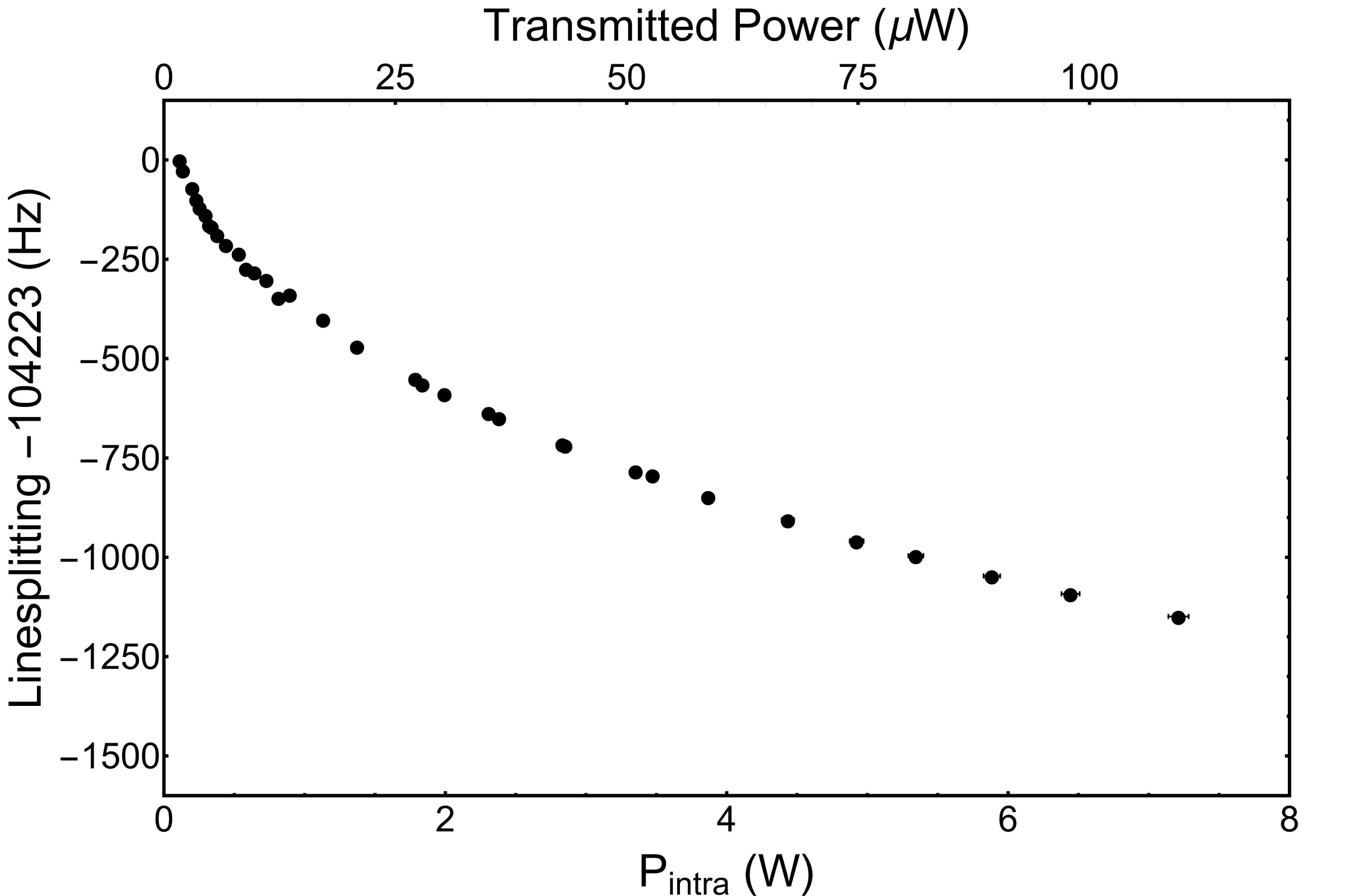}
	\end{subfigure}
	\begin{subfigure}[b]{0.50\textwidth}
		\includegraphics[width=18pc]{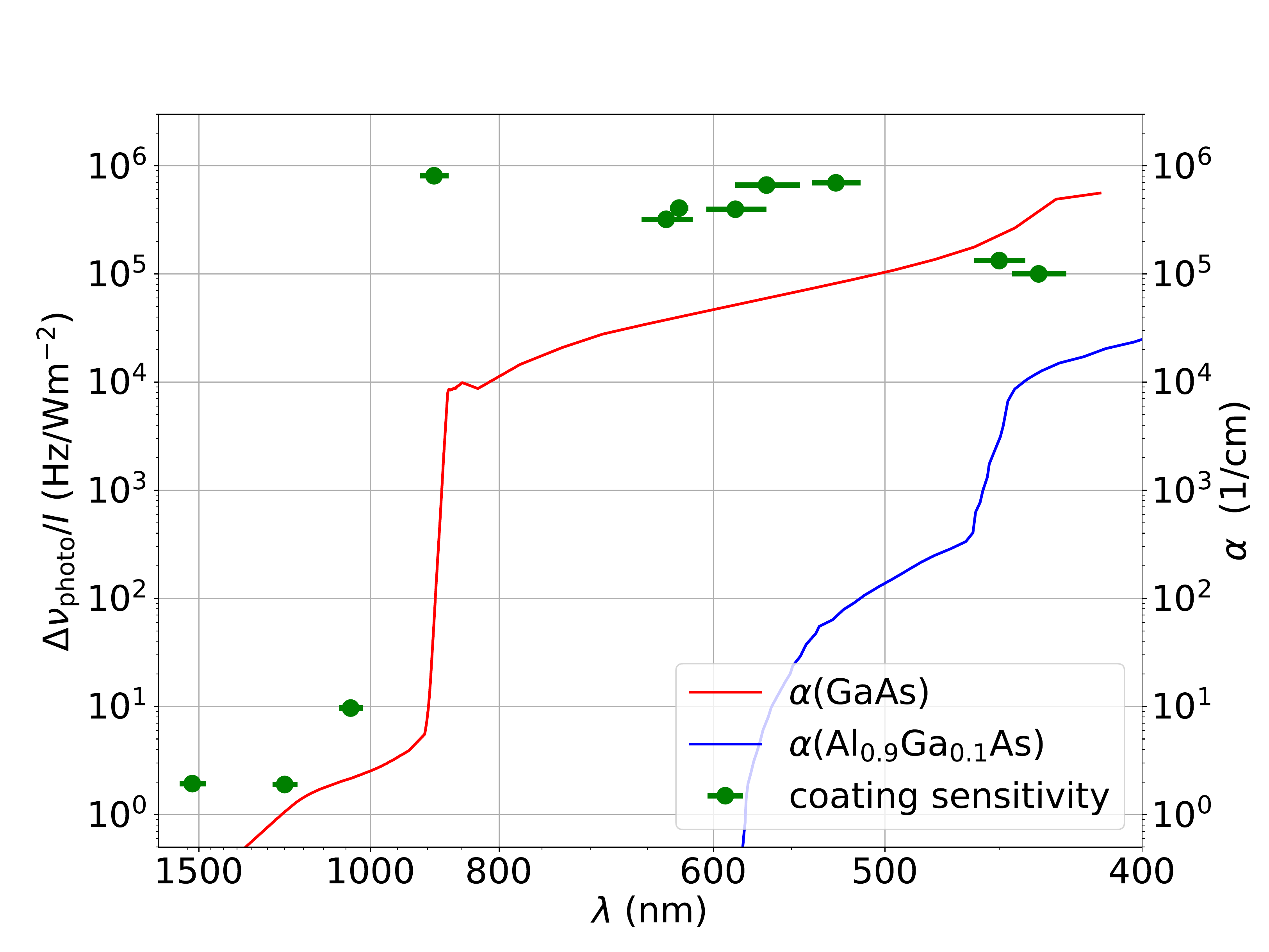}
	\end{subfigure}
	\caption{
		Left: Birefringent splitting as a function of intracavity laser intensity (left) for the ULE cavity at PTB after reaching steady state. The corresponding power transmitted by the cavity is also indicated on the top horizontal axis. 
		Right: Wavelength dependent sensitivity $\Delta \nu_\mathrm{photo}/I$ of photo-modified birefringence measured with narrowband LEDs. 
		The bandwidths of the LEDs are indicated by the horizontal bars.
		The uncertainty of the sensitivity is less than the marker size.
		The absorption coefficients $\alpha$ of GaAs (red) and  Al$_{0.9}$Ga$_{0.1}$As (blue) in cm$^{-1}$ are included for comparison \cite{mon76,stu62,hut08}.
	}
	\label{fig:intensity}
\end{figure}

Similar to the coatings at 124~K, we observed a photo-modified birefringence in the 48 cm ULE cavity \cite{ma24a}. Fig.\ \ref{fig:intensity} (left) shows the line splitting as a function of the intracavity laser intensity on the mirror coatings measured with the ULE cavity. 
The range of intensity investigated here is expanded to a transmitted power of 1.7~\textmu W to 108~\textmu W to study the linearity of this effect. 
Increasing intracavity intensity reduces the line splitting for the room temperature cavity, in contrast to the results from our Si cavity, as well as in recent studies at room temperature \cite{zhu23, kra23a}.
This is possibly related to the fact that all setups besides our 297~K setup use ``flipped''  AlGaAs coatings \cite{col16}, where the mirror top surface is ($\overline{1}00$)-oriented, different from our ($100$) orientation \cite{per24}. 
The line splitting as a function of intensity of the room temperature cavity at PTB is highly nonlinear, possibly indicating a saturation at high intensity. 
At the highest intracavity intensity level studied, the static mirror birefringence, ${n_\mathrm{biref}}$ can be modified by at most 1\%. 

As the fused silica mirror substrates are transparent to visible and near infrared light, the AlGaAs coatings of the PTB room temperature cavity are illuminated from outside the vacuum chamber with narrow-band LEDs with wavelengths from 450 nm to 1.5~\textmu m, covering  the bandgap of GaAs. 
The intensity $I$ of the LEDs at the mirror coatings was LEDs determined, accounting for divergence and the loss from the windows. 
A strong nonlinearity of the photo modified birefringence on the intensity of the LED light was also observed.  
The right panel of Fig.\ \ref{fig:intensity} displays the largest observed shift $\Delta \nu_\mathrm{photo}$ after turning on the LED at each wavelength, normalized to the intensity $I$ at the mirror.
The absorption profiles of GaAs and AlGaAs are shown for comparison \cite{mon76,stu62,hut08}. 
The sensitivity drops by five orders of magnitude across the GaAs bandgap between 890 nm to 1050 nm, as for longer wavelengths no direct photoabsorption can occur in pure GaAs. 
For photon energies above the GaAs bandgap a slight decrease in the sensitivity towards shorter wavelengths, which might be due to the reduced penetration depth at shorter wavelength.
E.g. Blue light is already absorbed in the first GaAs layer in comparison to light at 890 nm that has an absorption length beyond ten layers of GaAs. 

Practically, this would mean that for ULE cavities with fused silica substrates one has to take special care of the shielding of external stray light to the cavity as only nW of visible light can cause frequency change in the sub Hz region.

\begin{figure}[ht]
	\begin{subfigure}[b]{0.50\textwidth}
		\includegraphics[width=\textwidth]{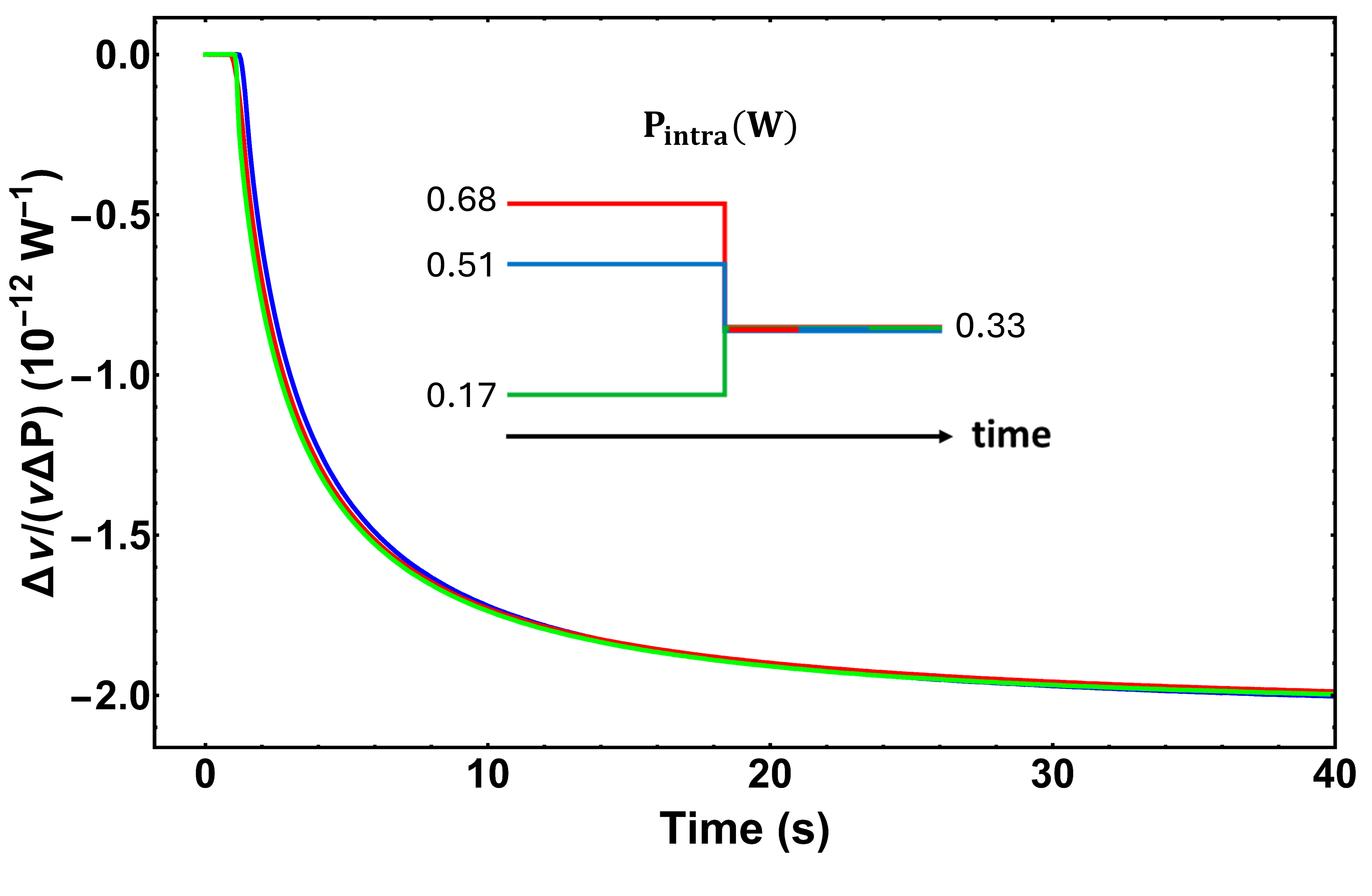}
	\end{subfigure}
	\begin{subfigure}[b]{0.50\textwidth}
		\includegraphics[width=\textwidth]{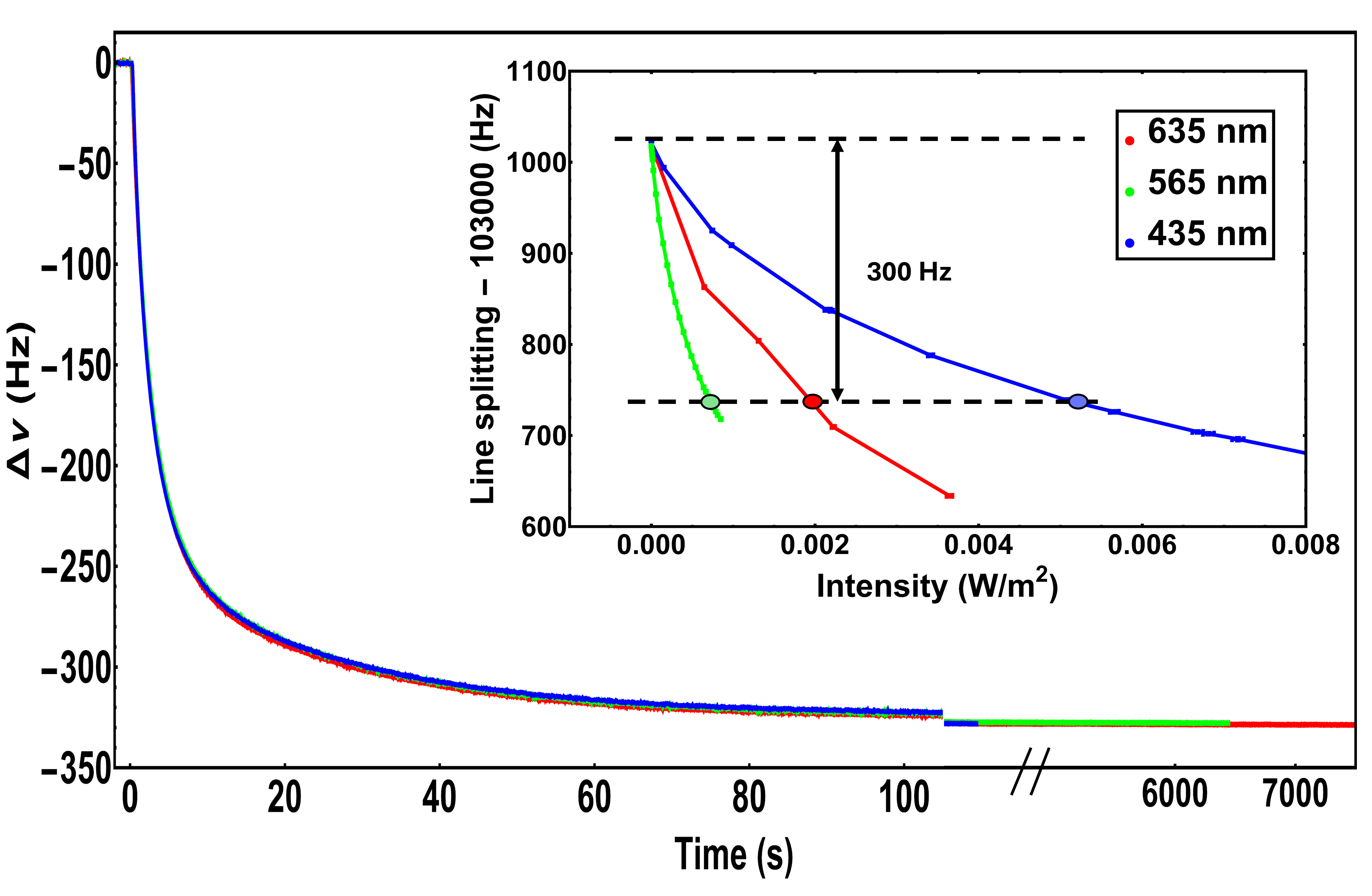}
	\end{subfigure}\hspace{2pc}%
	\caption{
	Left: Temporal response of the birefringent splitting by intracavity light for different step sizes of intensities to the same final value. 
    The colors of the curves correspond to the line colors of the steps in the inset.
	Right: Response on switching on LEDs of different wavelengths. 
	The intensity (determined from the current) is chosen to get the same $\Delta \nu_\mathrm{photo}$ of 300 Hz (see inset). 
	}
	\label{fig:temporal}
\end{figure}

The temporal behavior of the modified birefringence by intracavity light at room temperature was also investigated to understand the dynamics. The observed dynamics is solely dependent on the final intracavity power, as shown in the left panel of Fig.\ \ref{fig:temporal}. 
Three independent step responses of the normalized frequency change per intensity were studied while keeping the final intensity constant. 
Due to the nonlinear sensitivity of line splitting with intensity (as seen in Fig.\ \ref{fig:intensity}), the sensitivities were further scaled (with scaling factors of 0.8, 1 and 1.09) to check if the step responses can be overlapped in time.
The good overlap here indicates that the dynamical behavior is independent on the initial condition and the step size of intensity level. 
Similarly, the temporal response of the photo modified birefringence to external light is studied by switching on LEDs at different wavelengths (Fig.\ \ref{fig:temporal} right).
The intensities of the LEDs were adjusted to result in the same birefringent frequency change $\Delta \nu_\mathrm{photo}$ = 300 Hz. 
We observe the same temporal behavior, irrespective of the LED wavelength.
The observation of the wavelength dependence of the modified birefringence and the associated dynamics bring us to a simple model at PTB that photo-excitation of free charge carriers is likely to be the primary effect upon illumination. 
The carriers that are created in the first few~\textmu m thick absorption region on the back of the mirror coating stack migrate in the GaAs/AlGaAs heterostructure and finally create an electric field at the top GaAs layer. 
To modify the birefringence through the electro-optic effect by the observed amount of 1300 Hz, an electric field perpendicular to the coating surface of 155 kV/m, corresponding to a voltage of 17 mV over one GaAs layer would be required. In future work, the coating associated noise contributions with and without external light will be investigated to determine the utmost performance of AlGaAs coatings at room temperature. 

\section{Meta-surface coated mirrors}

\begin{figure*}[htb]
    \centering
    \includegraphics[width=0.7\linewidth]{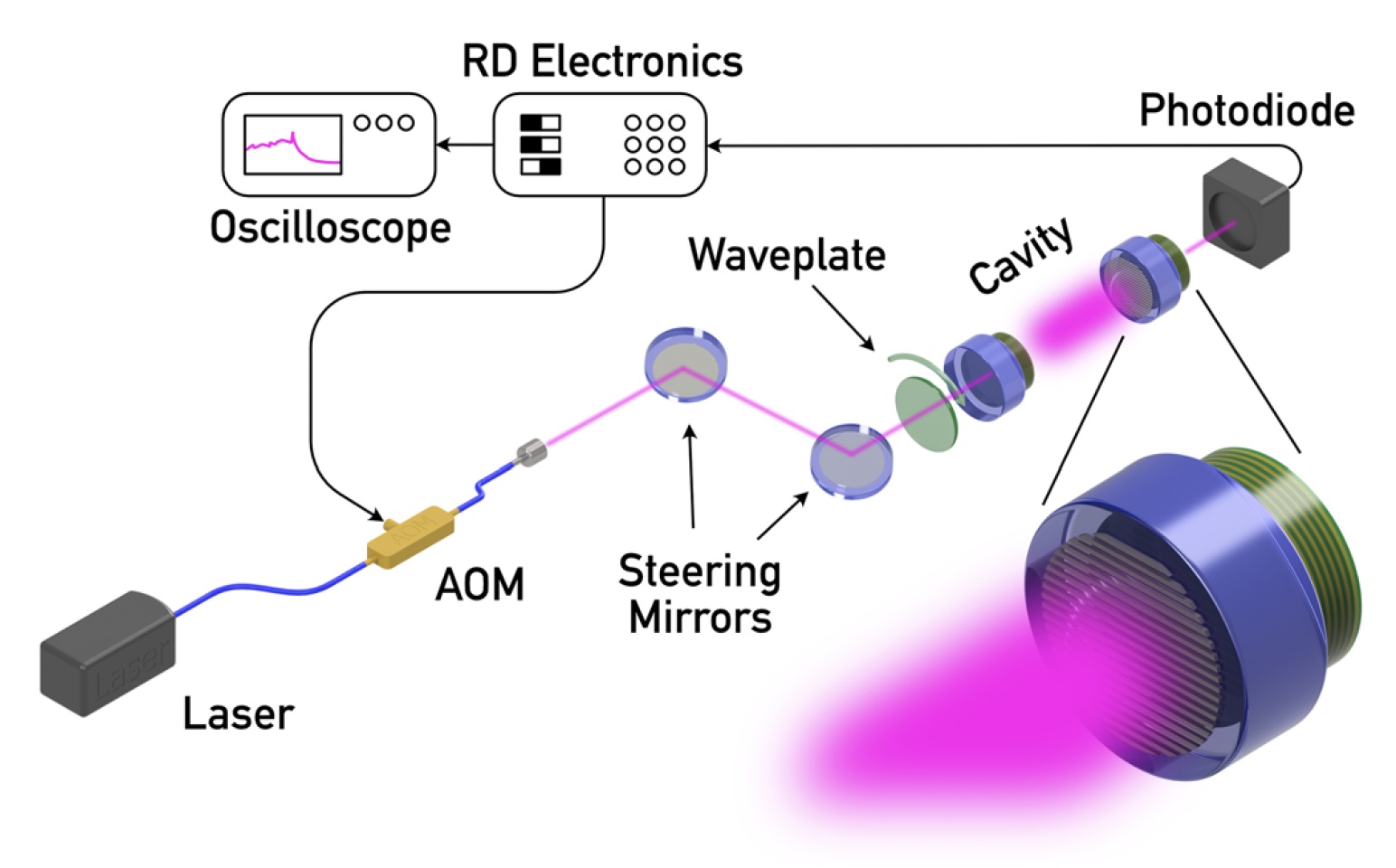}
    \caption{Schematic of the cavity ringdown setup. Here, the cavity is formed of a conventional mirror (left) and a meta-etalon consisiting of a meta-mirror at the front side of the second mirror and multilayer mirror on the backside (right).}
    \label{fig:CRD}
\end{figure*}

\begin{figure*}[tb]
    \centering
    \includegraphics[width=0.65\linewidth]{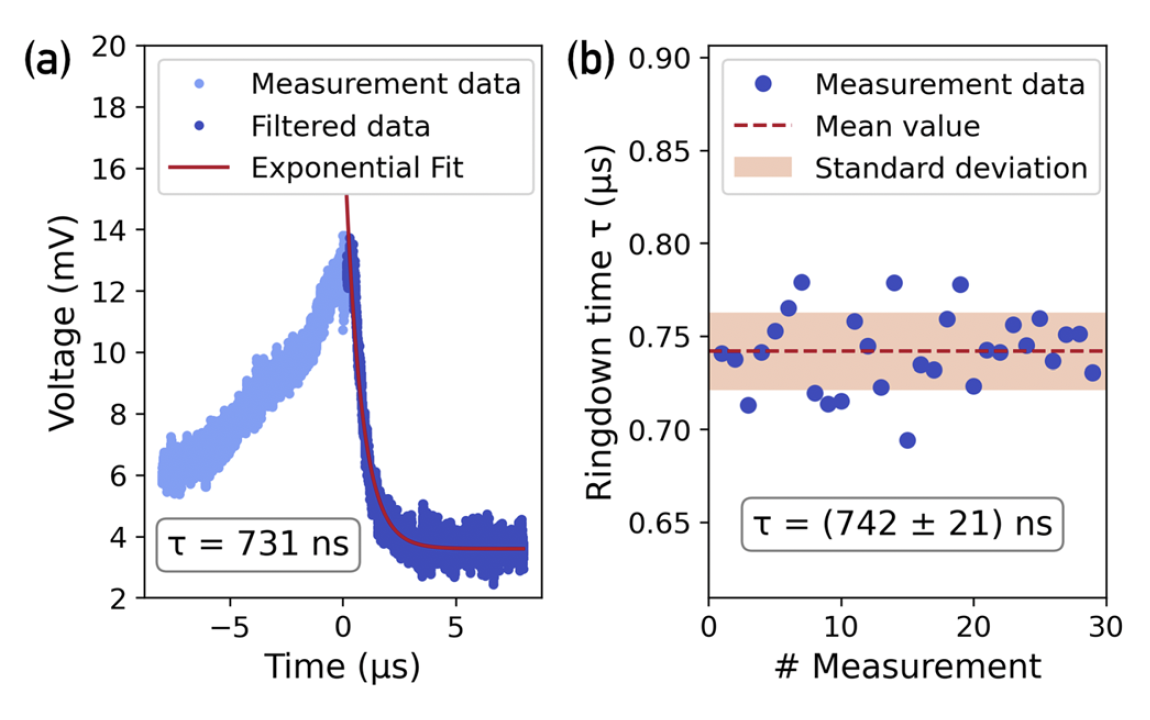}
    \caption{Exemplary measurement results of the cavity ringdown. (a) Single ringdown measurement. (b) Ringdown times of 29 measurements including mean value and standard deviation.}
    \label{fig:CRDmeas}
\end{figure*}

Meta-surfaces made of low-loss optical materials are an alternative pathway for overcoming current limitations in thermomechanical noise of the mirrors \cite{DICKMANN20182275}. TUBS with the support of PTB and VTT investigated low noise meta-mirrors and compound meta-etalons with photospectrometry and cavity ringdown spectroscopy. The latter may provide the possibility to enhance the accessible mirror reflectivity. For these investigations, the cavity ringdown setup shown in Fig.~\ref{fig:CRD} was used. In the setup, the laser light (Keysight
Laser 81606A, tunable) entering the cavity is switched off by an acousto-optic modulator (SFO4504-T-M080-0.5C8J-3-F2P) at a threshold voltage, which is measured by a DC photodiode (Thorlabs PDA20CS2) behind the cavity. The transmitted light behind the cavity is also directed to a CCD camera (Xenics Bobcat-320) via a beam splitter so that the resonant mode is filmed simultaneously. 
This ensures that only the desired mode is excited—in our case the TEM$_{00}$ mode. 
The setup can measure a minimum reflectivity of a 6 cm cavity at 1550 nm of 99.6 $\%$. Fig.~\ref{fig:CRDmeas} shows the results of the ringdown measurements of the meta-etalon. The evaluation of all ring-down events results in a ringdown time of $(742\pm21)\, \mathrm{ns}$ and a finesse of $(11655\pm330)$. Compared to standalone meta-mirrors \cite{Brueckner2010} this corresponds to an increase of the finesse by a factor of about ten. Preliminary investigations indicate that scattering due to material defects and side wall roughness accounts causes a major fraction of the optical losses. In future work they need to be mitigated by optimized fabrication processes.

\begin{figure*}[tb]
    \centering
    \includegraphics[width=0.4\linewidth]{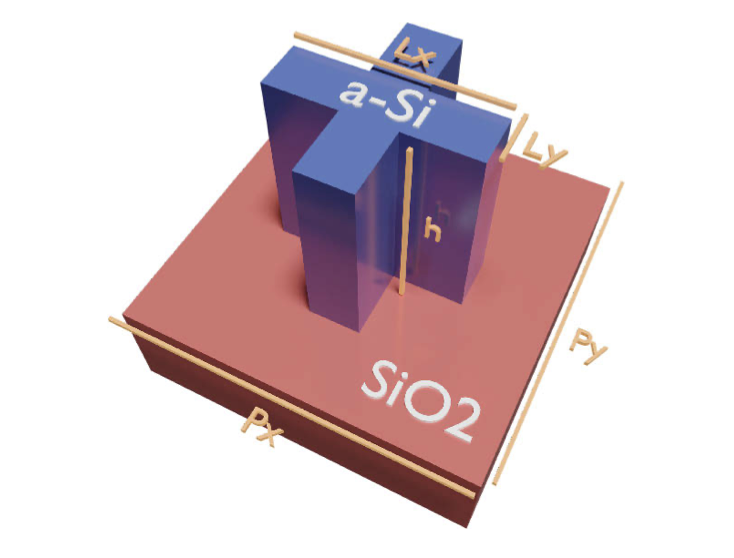}
    \caption{Schematic of the cross-like meta-atom with variable lateral height and width (Lx, Ly). Their structural height and period in both lateral directions are fixed to be 550 nm and 820 nm, respectively.}
    \label{fig:cross}
\end{figure*}

In addition to planar mirrors, we also investigated strategies for realizing focusing meta-mirrors. So, far such devices have not been demonstrated for reflectivities close to unity as required for ultra-stable laser cavities. Reaching high reflectivities and have control on the phase requires a careful design. 
To this end, we developed a tandem neural network framework to render a focusing meta-mirror with high mean and maximum reflectivity of $R_\mathrm{mean}=99.993~\%$  and 
$R_\mathrm{max}=99.9998~\%$, respectively, and a minimal phase mismatch of $\delta\varphi=0.016~\%$ that is comparable to state-of-art dielectric mirrors. 
As first topology, we utilized cross-shaped meta-atoms as shown in Fig.~\ref{fig:cross} to benchmark the network with conventional rigorous coupled wave analysis (RCWA) simulations. 

\begin{figure*}[tb]
    \centering
    \includegraphics[width=0.8\linewidth]{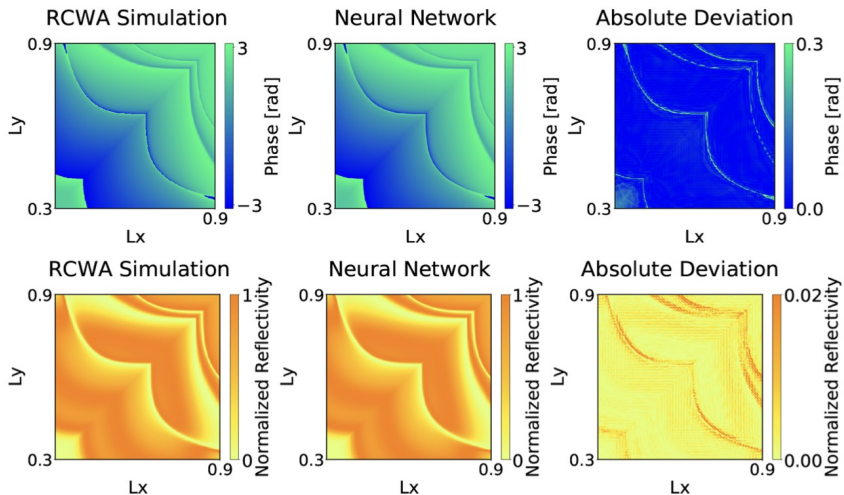}
    \caption{Left: Ground truth training data from the RCWA simulation. Middle: Forward predictions of the trained neural networks (for reflectivity and phase) on the ground truth data. Right: Absolute deviation of both, phase and reflectivity predictions, showing excellent performance of the neural networks.}
    \label{fig:benchmarking}
\end{figure*}
\begin{figure*}[tb]
    \centering
    \includegraphics[width=0.5\linewidth]{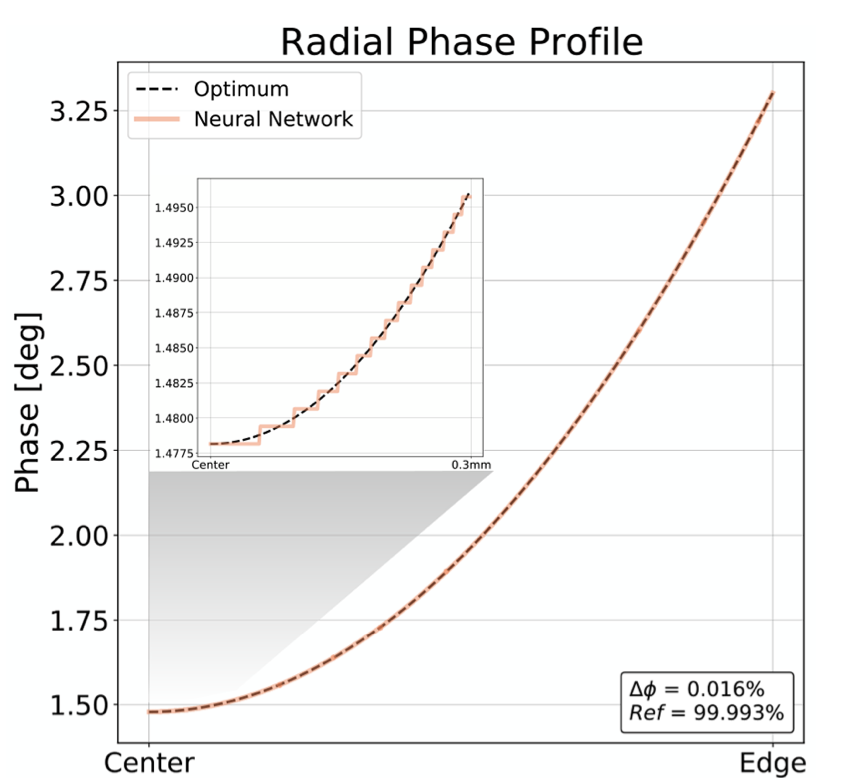}
    \caption{Radial phase profile of the metasurface with excellent phase agreement and high reflectivity.}
    \label{fig:focusing}
\end{figure*}

Fig.~\ref{fig:benchmarking} illustrates the deviation of the neural network simulations from RCWA in dependence of the lateral widths of the cross showing an excellent agreement encouraging also more complex topologies. 
Finally, the tandem neural network was used to design a focusing mirror consisting of the cross-shaped meta-atoms. Fig.~\ref{fig:focusing} displays the phase profile with an overall design reflectivity of $R=99.993\%$. 
First structures have been fabricated and are currently being characterized.

\section[NEXCERA spacer]{NEXCERA ceramic as spacer material for optical resonators}
\label{sec:D1_S2}
NEXCERA ceramic, developed by Krosaki Harima in Japan, is an ultra-low expansion cordierite ($\mathrm{2MgO-2Al_2O_3-5SiO_2}$) based  ceramic material. The NEXCERA ceramic has a nominal zero coefficient of thermal expansion at room temperature, i.e. $\left| \alpha \right| < 0.03 \times 10^{-6} \mathrm{K}^{-1}$ at 23~°C. Apart from ultra-low expansion properties, the mechanical properties of NEXCERA ceramic, with its larger ratio of Young's modulus to density, offers the possibility of lower acceleration sensitivity compared to ultra-low expansion (ULE) glass. Furthermore, a preliminary characterisation of an optical cavity made of N117B NEXCERA material has found a long-term linear drift of 4.9~mHz/s \cite{ito17}, which is another potential advantage compared to ULE cavity.

Some of the relevant properties of NEXCERA ceramic are shown in Tab.~\ref{tab:NEXCERA_Properties}.
\begin{table}[h!]
\centering
\renewcommand{\arraystretch}{1.2}
    \begin{tabular} { |l|c|c|c|c|c| }
        \hline
            & N113B & N117B & N118C & N119C & ULE\\
        \hline
            Bulk density ($\mathrm{g/cm^{3}}$) & 2.5 & 2.55 & 2.58 & 2.5 & 2.21 \\
        \hline
            Young's modulus (GPa) & 130 & 140 & 140 & 130 & 68\\
        \hline
            Poisson ratio ($\mathrm{GPa/(g/cm^3)}$) & 0.30 & 0.31 & 0.31 & 0.31 & 0.17\\
        \hline
            Thermal expansion coefficient & & & & & \\
            at 23°C ($\times 10^{-6}/$°C) & < 0.03 & < 0.03 & < 0.05 & < 0.05 & $\left(0 \pm 0.03 \right)$\\
        \hline
    \end{tabular}
    \caption{Properties of NEXCERA ceramic \cite{KrosakiHarimaWebsite} and ULE glass.}
     \label{tab:NEXCERA_Properties}
\end{table}

\subsection{Mechanical loss measurements of NEXCERA \label{sec:mech_loss}}

The mechanical loss is a critical parameter for the thermomechanical noise of spacer and mirror materials in ultra-stable laser cavities. To  minimize excess suspension (or clamping) losses we employed the Gentle Nodal Suspension (GeNS) technique developed by Cesarini \textit{et al.}~\cite{Cesarini2009}. In this technique, the sample is suspended on a sphere minimizing the contact area of the sample and the suspension. In our configuration, using a 4.5 mm steel sphere with a quality class of G3 according to DIN 5401 standards, the contact area radius of the disks (diameter 50~mm, thickness 0.5~mm) is approximately 10 $\mu$m. 

The mechanical vibration excited by an AC electric field through a comb structure is detected as a spatial oscillation of the beam of a laser diode on a quadrant detector. The electric field is controlled by a custom-built high-voltage amplifier. Positioned approximately 2~mm  above the sample, the structure applies an oscillating voltage up to 800~V to excite the eigenmodes of the disc.
The mechanical loss of 3 disks of raw material has been measured at room temperature. All mechanical losses are at or below 3$\times 10^{-5}$. The losses of the different samples do not show systematic deviations. It is known from literature that the mechanical loss increases with the increasing surface-to-volume ratio. This means that ``geometrically small'' samples (such as thin disks) will have larger mechanical losses than bigger samples, for example, in geometries as used for typical spacers in laser cavities. That is why the determined minimum loss of $1.894(41)\times 10^{-5}$ of the NEXCERA N117B disks is very promising as novel low-noise material~\cite{Wagner2024_nexcera}. The determined values can be considered as a worst-case scenario for the thermo-mechanical noise performance of typical spacer geometries.

\subsection{First-generation NEXCERA cavities}
\label{sec:first_gen_NEX}

At HHUD a 30~cm long cavity made of the N117B NEXCERA ceramic has been developed for the frequency stabilization of a 698~nm laser. Another cavity with a length of 12~cm made of N118C NEXCERA material was developed for 1.5~$\mu$m laser stabilization.  

We characterized the frequency instabilities and other properties of these cavities integrated with respective laser systems to investigate the potential of NEXCERA ceramic as a spacer material for ultrastable cavities.

\subsubsection{N117B NEXCERA cavity}
\label{sec:N117B_cavity}
\begin{figure}[ht!]
  \centering
    \includegraphics[width=0.75\linewidth]{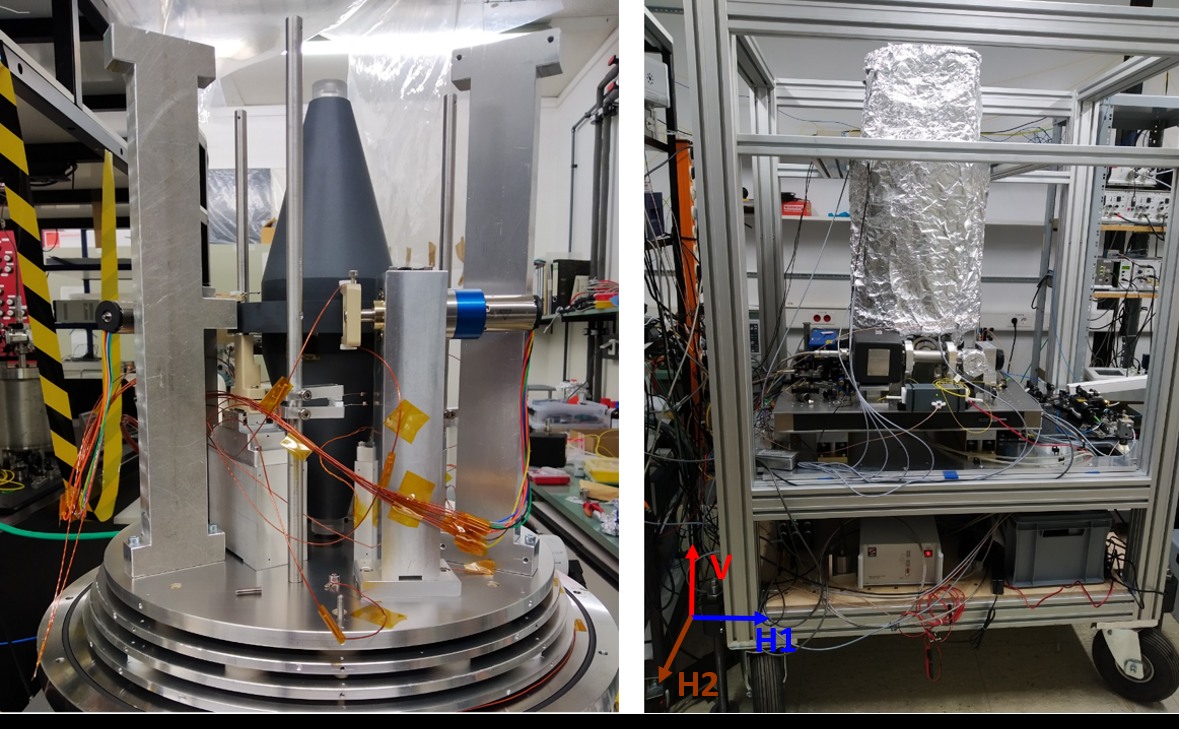}
    \caption{The transportable 30~cm N117B NEXCERA vertical optical resonator system. (Left) Optical resonator with support after removal of the outer vacuum shield and the two thermal insulation shields. (Right) The closed vacuum chamber containing the optical resonator sitting on the active vibration isolation (AVI) system inside the transportable trolley.}
    \label{fig:N117B-NEXCERAResonator}
\end{figure}

The N117B NEXCERA cavity consists of the NEXCERA spacer and a pair of ULE mirrors with a finesse of $2\times10^5$. 
The cavity is operated inside its dedicated vacuum chamber that is installed within a home-built trolley to allow for transportation of the cavity system. The cavity system is shown in Fig.~\ref{fig:N117B-NEXCERAResonator}.

To characterize the acceleration sensitivity of the NEXCERA cavity, we swept a phase-locked 698 nm laser across the resonance of the cavity while recording the cavity-transmitted laser power. Simultaneously, we used the active vibration isolation as a shaker by driving it with a sinusoidal signal at frequencies of 5~Hz, 7~Hz, 10~Hz, and 15~Hz, and measured the excited vibrations on the cavity breadboard with a 731A accelerometer amplified by a P31 amplifier from Wilcoxon Sensing Technologies.

\begin{figure}[ht!]
  \centering
    \includegraphics[width=0.75\linewidth]{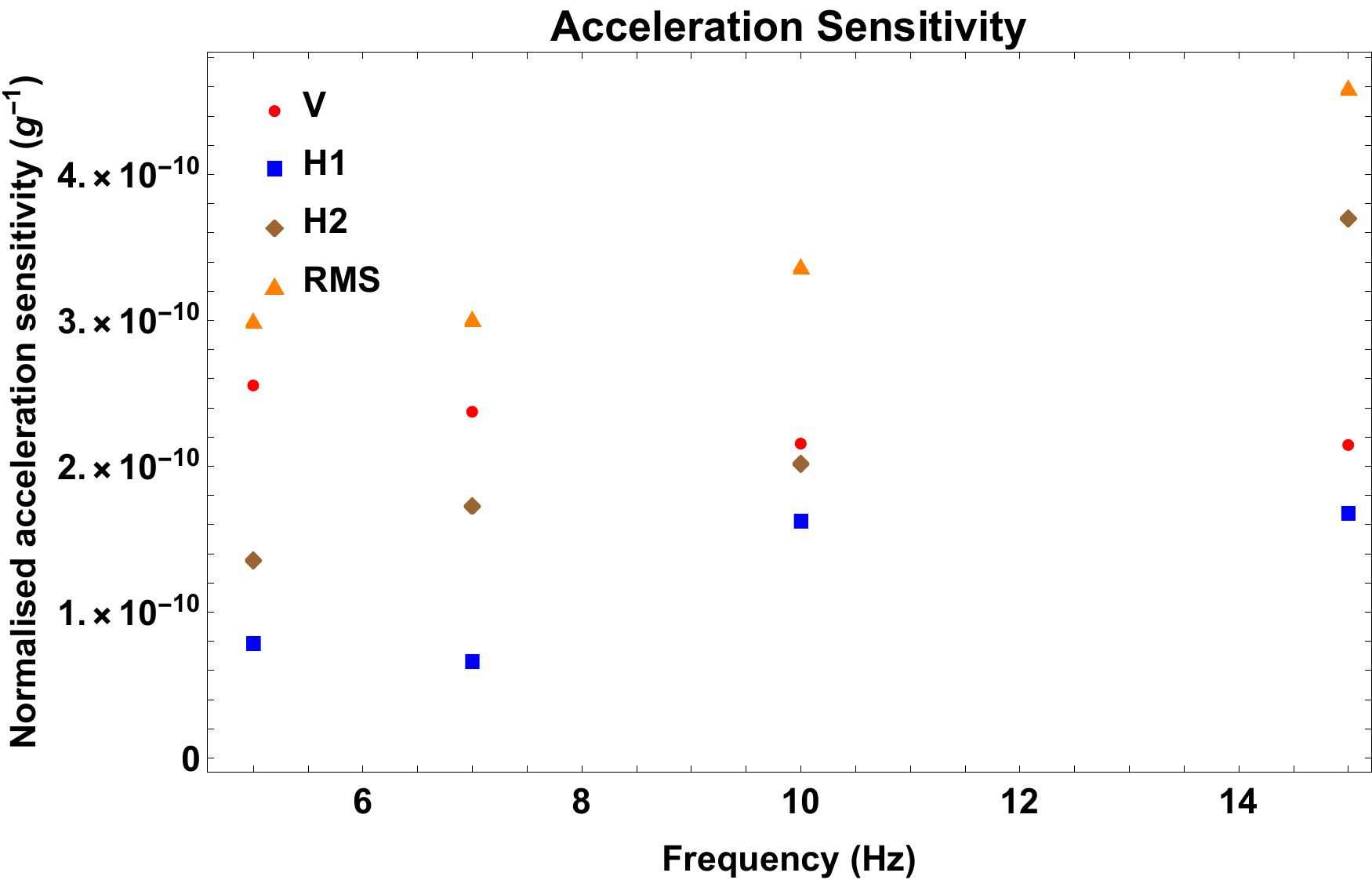}
    \caption{Acceleration sensitivities of the N117B NEXCERA cavity along the three spatial axes and the RMS acceleration sensitivity.}
    \label{fig:N117B-AccelerationSensitivity}
\end{figure}

The acceleration sensitivities for all three directions as well as the total acceleration sensitivity calculated as the root-mean-squared (RMS) value of the three sensitivities are plotted in Fig.~\ref{fig:N117B-AccelerationSensitivity}. The directions of horizontal vibrations, H1 and H2, are defined with respect to the AVI mounts (see Fig.~\ref{fig:N117B-NEXCERAResonator}). The acceleration sensitivities of the cavity in all three directions are in the range of low to mid $10^{-10} g^{-1}$ for all frequencies, which are comparable to state-of-the-art room temperature optical cavity~\cite{hae15a}. These acceleration sensitivities are obtained by placing the Viton ball mounts at its nominal vibration-insensitive support points. The acceleration sensitivities of the cavity can potentially be reduced to the level of $10^{-11} g^{-1}$ by optimizing the positions of the Viton balls that hold the optical cavity.

\begin{figure}[t!]
  \centering
    \includegraphics[width=0.75\linewidth]{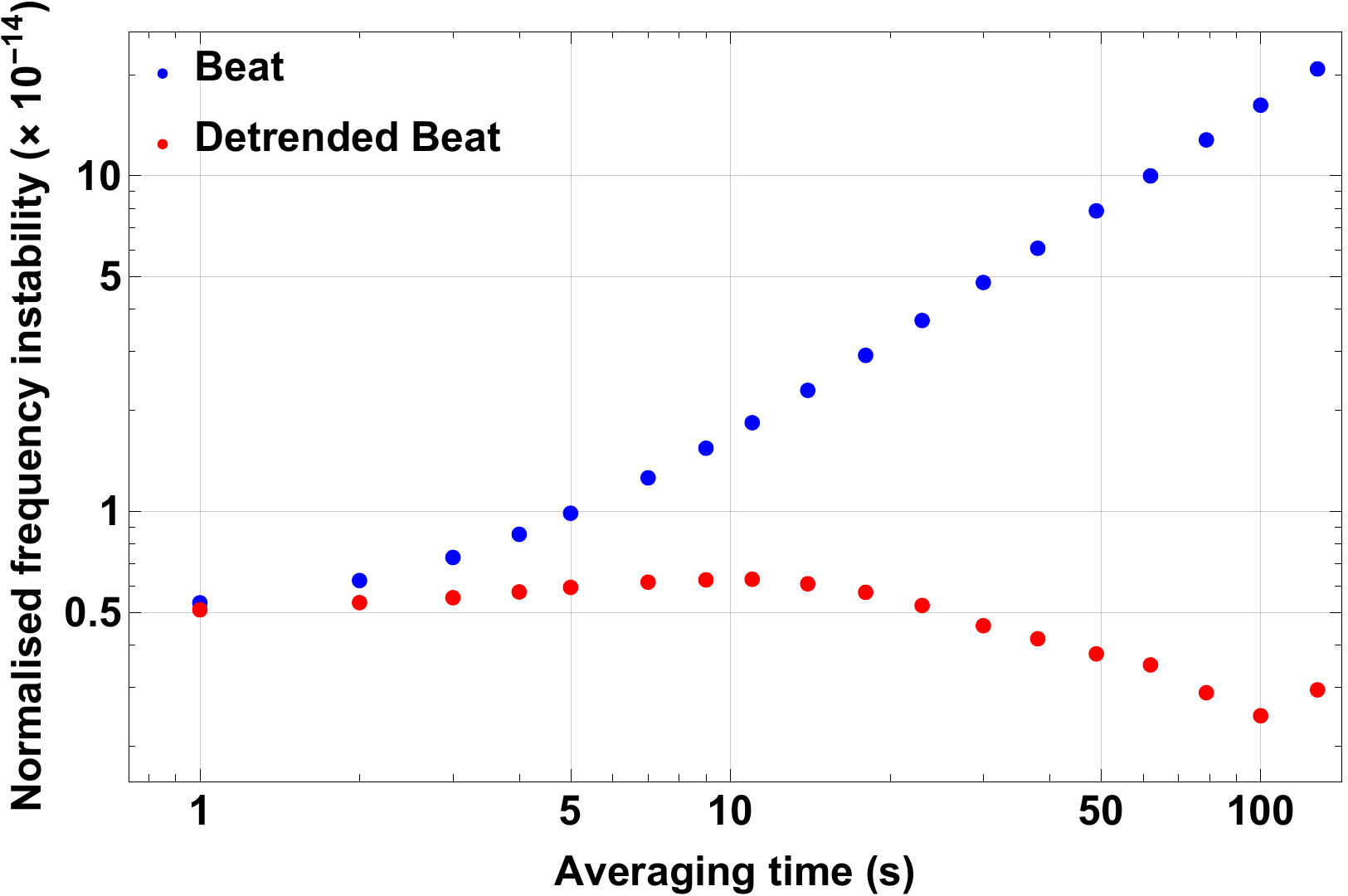}
    \caption{Fractional 698 nm laser frequency instability locked to the 30 cm N117B NEXCERA cavity.}
    \label{fig:N117B-AllanDeviation}
\end{figure}

A Toptica DLPro 698~nm laser is locked to the NEXCERA cavity via the Pound-Drever-Hall (PDH) scheme. An initial characterization of the laser frequency stability was done using a frequency comb phase-locked to a 1064~nm reference laser with a frequency instability of $3 \times 10^{-15}$. The fractional laser frequency instability was found to be $5.1\times 10^{-15}$ at 1~s averaging time that averages down to a minimum of $2.5\times 10^{-15}$ at 100~s averaging time. Fig.~\ref{fig:N117B-AllanDeviation} shows the results of the characterization of the laser frequency instability. The observed beat note instability is limited by the instability of the reference laser. Assuming a mechanical loss of $3\times10^{-5}$ measured in Sec.~\ref{sec:mech_loss} and using the evaluation method from Numata \textit{et al.}\cite{Numata2004}, the thermal noise of the N117B 30~cm NEXCERA cavity is evaluated to be $2\times10^{-16}$. This suggests that an improvement of a factor of 10 may be possible for the N117B cavity.

The temperature of zero coefficient thermal expansion (ZCTE) is found to lie within the range of 17.6 to 17.7 \si{\celsius}.

\subsubsection{N118C NEXCERA cavity}

This cavity comprises two ultra-high finesse $\left(7\times10^5\right)$ fused silica mirrors, optimized for a wavelength of 1.5\,$\mu$m, which have been optically contacted to a 12\,cm long NEXCERA spacer that includes ULE compensation rings \cite{Schenkel2024}. 
With a mechanical loss of $3\times10^{-5}$ for the spacer material, this configuration has a predicted thermal noise floor of approximately $3\times10^{-16}$.
The resonator assembly is oriented vertically within a vacuum chamber, that is actively stabilized to maintain a constant temperature. 
Previous work on this cavity spacer with a different mirror configuration has been shown in \cite{Kwong_2018}, including ZCTE and drift rate characterization.

To lock the laser frequency to the resonator, the PDH technique is utilized. Additionally, several noise reduction strategies are employed, including active vibration isolation and power stabilization, to minimize disturbances.  Measurements determined the ZCTE temperature to be 17.8$^\circ$C.

\begin{figure}[t!]
  \centering
    \includegraphics[width=\linewidth]{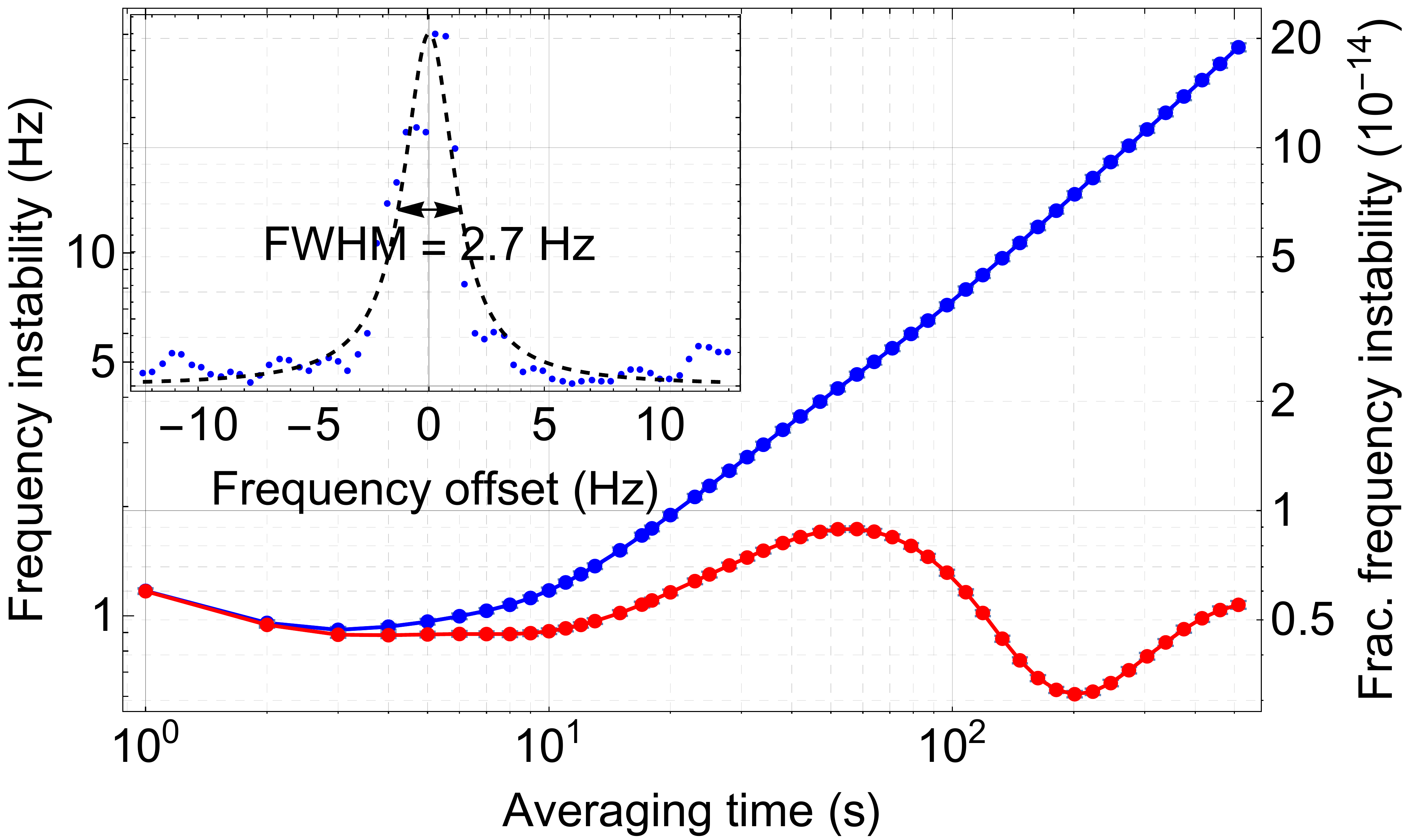}
    \caption{Performance of the short NEXCERA (N118C) resonator: The blue data represents the measurement of the beat-note with another ultra stable laser. The red curve shows the Allan deviation after subtracting a linear drift of the raw data. }
    \label{fig:N118C-Performance}
\end{figure}

Linewidth characterization was performed using beat note analysis with a 1064 nm ULE-stabilized reference laser and a frequency comb, while the Allan deviation was evaluated through a direct beat measurement with another 1.5 µm ULE-stabilized laser.
Representative data is shown in Fig.\,\ref{fig:N118C-Performance}. The Allan deviation remains below $1\times10^{-14}$ for averaging times ranging from 1 second to 500 seconds, reaching a minimum of $3\times10^{-15}$ at an averaging time of 200 seconds. Detrending of resonator drifts was performed for this analysis. This level provides another upper limit for the thermal noise; the actual level is limited by the reference laser instability. The linewidth of the beat-note is approximately 3\,Hz.  Long-term stability assessment revealed that the resonator drift is less than 3\,mHz per second over 110 days. This is an outstanding value.

\section{Compact cavities with large spot size on mirror }

\begin{figure}[hbt]
    \centering
    \includegraphics[width=0.5\linewidth]{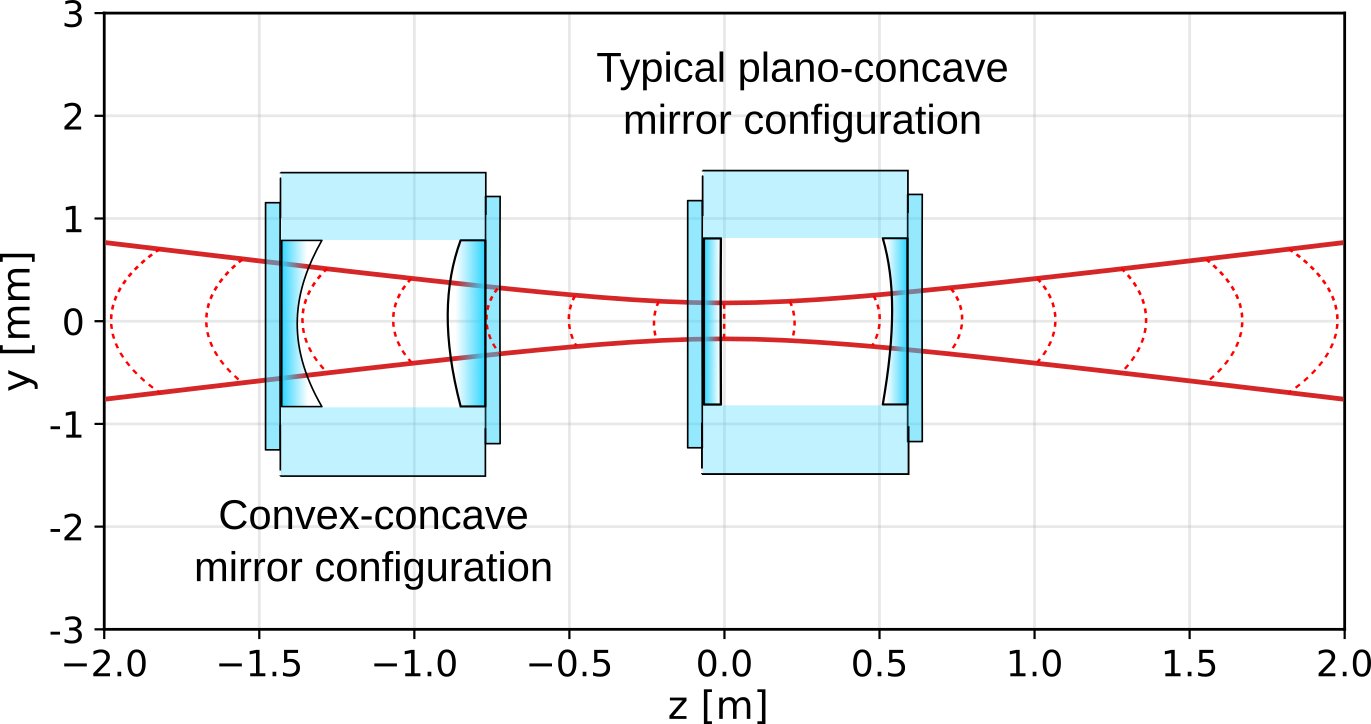}
    \caption{Conceptual scheme illustrating the Gaussian beams and their wavefront curvatures (dashed lines), with a typical plano-concave setup and proposed convex-concave mirror configuration.}
    \label{fig:UMK_Schematic_Gaussian_beam}
\end{figure}

Based on the fluctuation-dissipation theorem, Brownian thermal noise depends on temperature $T$, mechanical loss $\phi$, and the laser beam spot radius $w$ on the coating's surface $S_{x,ct}^{Br} \sim \phi T/w^{2}$. 
To reduce Brownian noise, one can look for substrates and coatings with a lower $\phi$ or lower the operational temperature (as demonstrated with SCS and sapphire cavities \cite{Zhang2017, Schiller2004}). 
We propose an alternative method for reducing thermal noise by increasing the beam spot on mirrors using a convex-concave mirror configuration. 
We demonstrate the feasibility and advantages of using a cavity with a radius of curvature (ROC) ranging from 20 m to -25 m for 10 cm and 30 cm long cavity spacers. 
This approach significantly reduces both the coatings' and substrate's thermal noises as well as thermoelastic and thermo-optic noises, compared to the usual ROCs and plano-concave design. Fig.\ \ref{fig:UMK_Schematic_Gaussian_beam} shows the conceptual scheme of placing the convex-concave configuration in comparison to the typical plano-concave with respect to the Gaussian beam and its wavefront curvature. 
In Fig.~\ref{fig:UMK_Schematic_Gaussian_beam}, the increase in the beam spot size on the mirror is visible. 
Unlike plano-concave or concave-concave configuration waist is placed outside the cavity.

\subsection{Coatings and substrates thermal noise reduction }

\begin{figure}[hbt]
    \centering
    \includegraphics[width = 0.45\linewidth ]{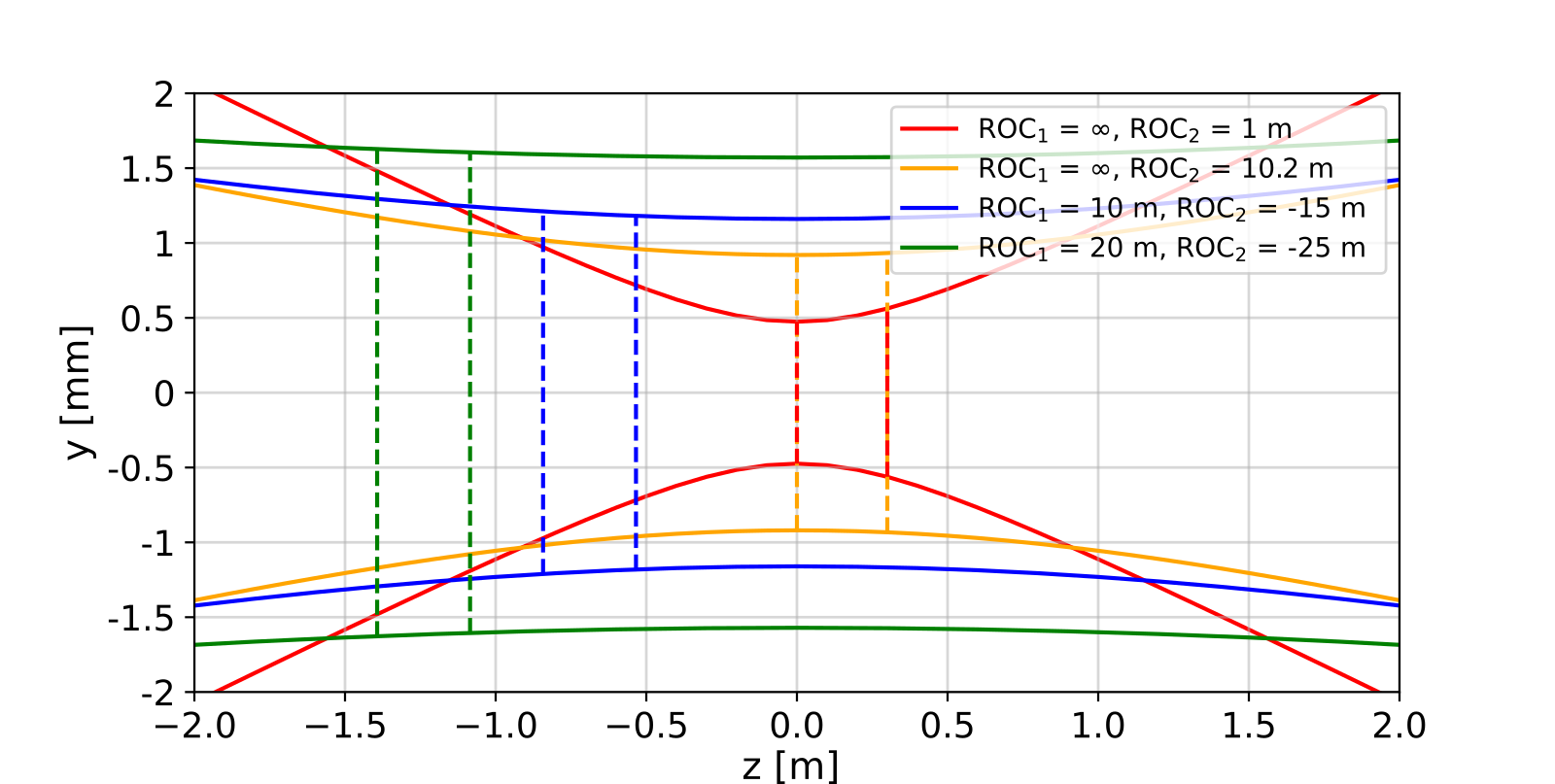}
    \caption{Mode-matched Gaussian beams together with the four investigated stable and near-stable configurations of the spacer length and mirrors' ROC. The dashed line indicates the location of the cavity mirrors.}
    \label{fig:UMK_Beamsize_cavity}
\end{figure}

Brownian thermal noise from coatings and substrates is the dominant noise source in an ultra-stable cavity, both at room and cryogenic temperatures. 
Tab. \ref{tab:ROC_waist_noises} illustrates the numerical improvement in the PSD of the coatings $S_{y,ct}$ and substrates $S_{y,sb}$ in the concave-convex ROC compared to the typical configuration $S_{y, ct}^{(\infty, 1)}$ and $S_{y, sb}^{(\infty, 1)}$. 
We observe an order of magnitude improvement between $S_{y, sb}^{(20, -25)}$ and $S_{y, sb}^{(\infty, 1)}$. 
Additionally, Fig.\ \ref{fig:UMK_Beamsize_cavity} depicts the mode-matched mirror positions for a given radius of curvature (ROC) for four configurations. 
In all cases, the beam waist is at a distance of 0.

\begin{table}[hbt]
    \caption{Several configurations of the ROCs for the 30~cm~long cavity, including plano-concave, concave-concave and convex-concave mirrors. The mirror's curvature and spacer's length determine the fundamental optical stability parameter $g_{1}g_{2}$, along with laser beam spot size on the mirror $w_{1}$, $w_{2}$, which determines the thermal noise floor. Thermal noise of the coatings (ct) and substrates (sb) is depicted as a ratio between the R$_{m1} = \infty$, R$_{m2} = 1$~m and the particular setup to show the effect of the beam size enhancement on the fundamental thermal noise.}
    \centering
    \renewcommand{\arraystretch}{1.1}
    \begin{tabular}{c c c c c c}
    R$_{m1}$, R$_{m2}$ [m] & g$_{1}$g$_{2}$ & $w_{1}$, $w_{2}$ [mm] &S$_{y, ct}^{(R_{m1}, R_{m2})}$/S$_{y,ct}^{(\infty, 1)}$ & S$_{y, sb}^{(R_{m1}, R_{m2})}$/S$_{x,sb}^{(\infty, 1)}$&  \\
        \hline \hline
        $\infty$, 1 & 0.7000 & 0.474, 0.567 &  1.0 &  1.0  \\
        $\infty$, 5.0 & 0.9400 & 0.764, 0.787 & 0.440 &  0.666   \\
        $\infty$, 10.2 & 0.9705 & 0.920, 0.934 & 0.308 & 0.557  \\
        10.2, 10.2 & 0.9421 & 0.782, 0.782 & 0.433 & 0.660  \\
        10, -15 & 0.9894 & 1.211, 1.181 &  0.185 & 0.432  \\
        10, -20 & 0.9845 & 1.100, 1.076 &   0.223 & 0.475  \\
        12.5, -25 &  0.9877 & 1.163, 1.142 &  0.199 &  0.448  \\
        20, -25 & 0.9968 &1.627, 1.605 &  0.101 & 0.320  \\
         \hline
    \end{tabular}
    \label{tab:ROC_waist_noises}
\end{table}

\subsection{Stability conditions}
\subsubsection{Optical stability near-unstable convex-concave configuration}

The reduction of thermal noises by increasing the mirror's laser beam spot size has already been proposed for use in future gravitational wave detection~\cite{Wang2018} in the form of a near-unstable solution using a plano-concave cavity. 
In this work, we propose a convex-concave design for the cavity mirrors to increase the beam spot size further.  
For a convex-concave configuration to achieve the same spot size as a plano-concave setup, less extreme mirror curvatures are needed.
For a 30 cm long cavity with ROC$_{1} = 10$ m and ROC$_{2} = -15$~m, we can obtain beam sizes on the mirrors of $w_{1} = 1.211$ mm and $w_{2} = 1.181$ mm, with $g_{1}g_{2} = 0.9894$. 
To achieve similar spot sizes with a plano-concave configuration, ROC$_{1} = \infty$, and ROC$_{2} = -28$ m are needed, and $g_{1}g_{2} = 0.9892$. 
Furthermore, to achieve a comparable spot size with a convex-concave setup using ROC$_{1} = 20$ m and ROC$_{2} = -25$ m  with $g_{1}g_{2} = 0.9946$, a plano-concave setup would need ROC$_{1} = \infty$ and ROC$_{2} = 50$ m, results in slightly more stable optical condition with $g_{1}g_{2} = 0.9940$.

\begin{figure}[hbt]
    \centering
    \includegraphics[width=0.45\linewidth]{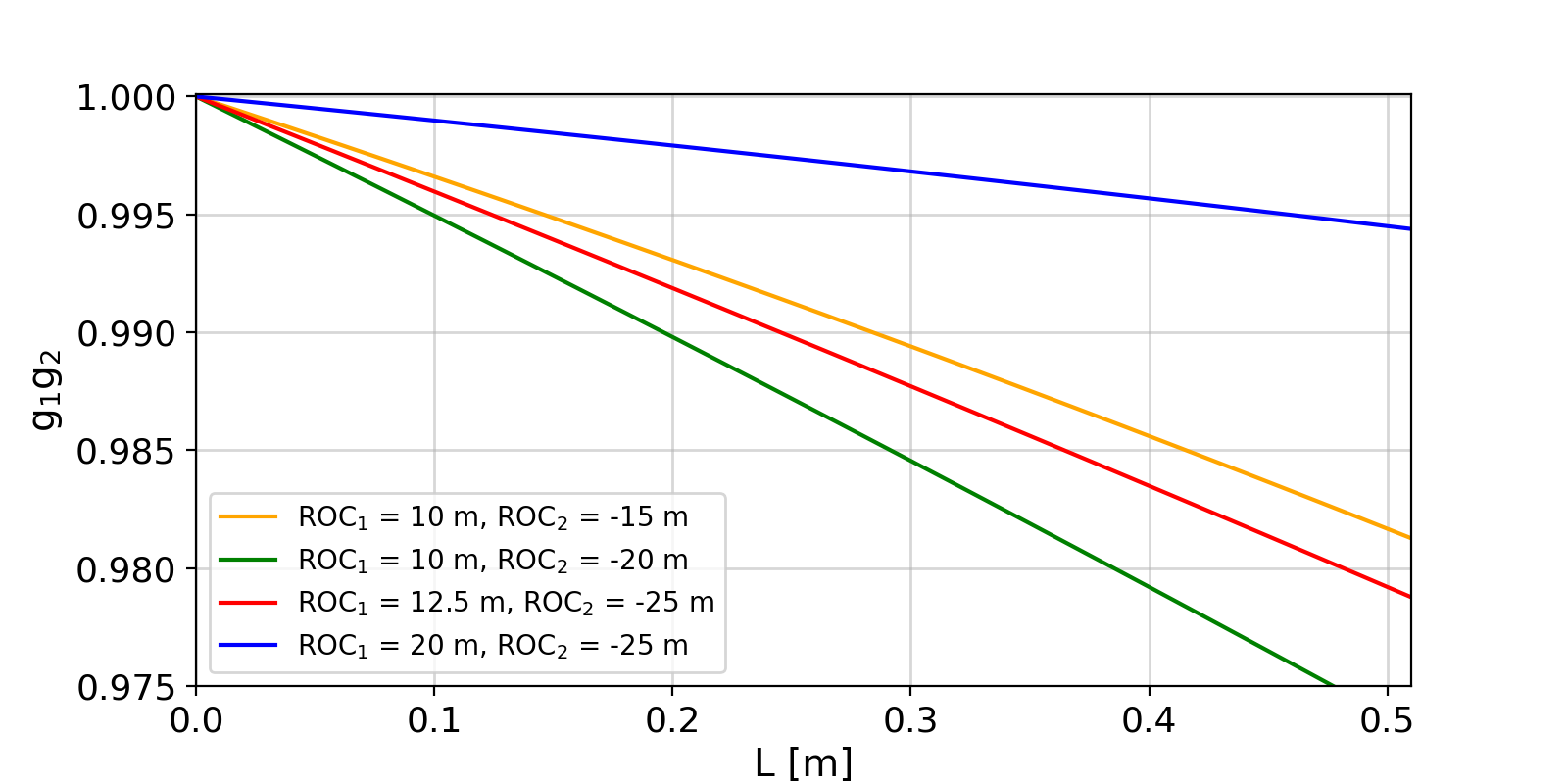}
    \caption{Stability factor $g_{1}g_{2}$ as a function of the spacer length for four configurations of mirror curvature.}
    \label{fig:UMK_Stability_contour}
\end{figure}

\begin{figure}[hbt]
    \centering
    \includegraphics[width=0.45\linewidth]{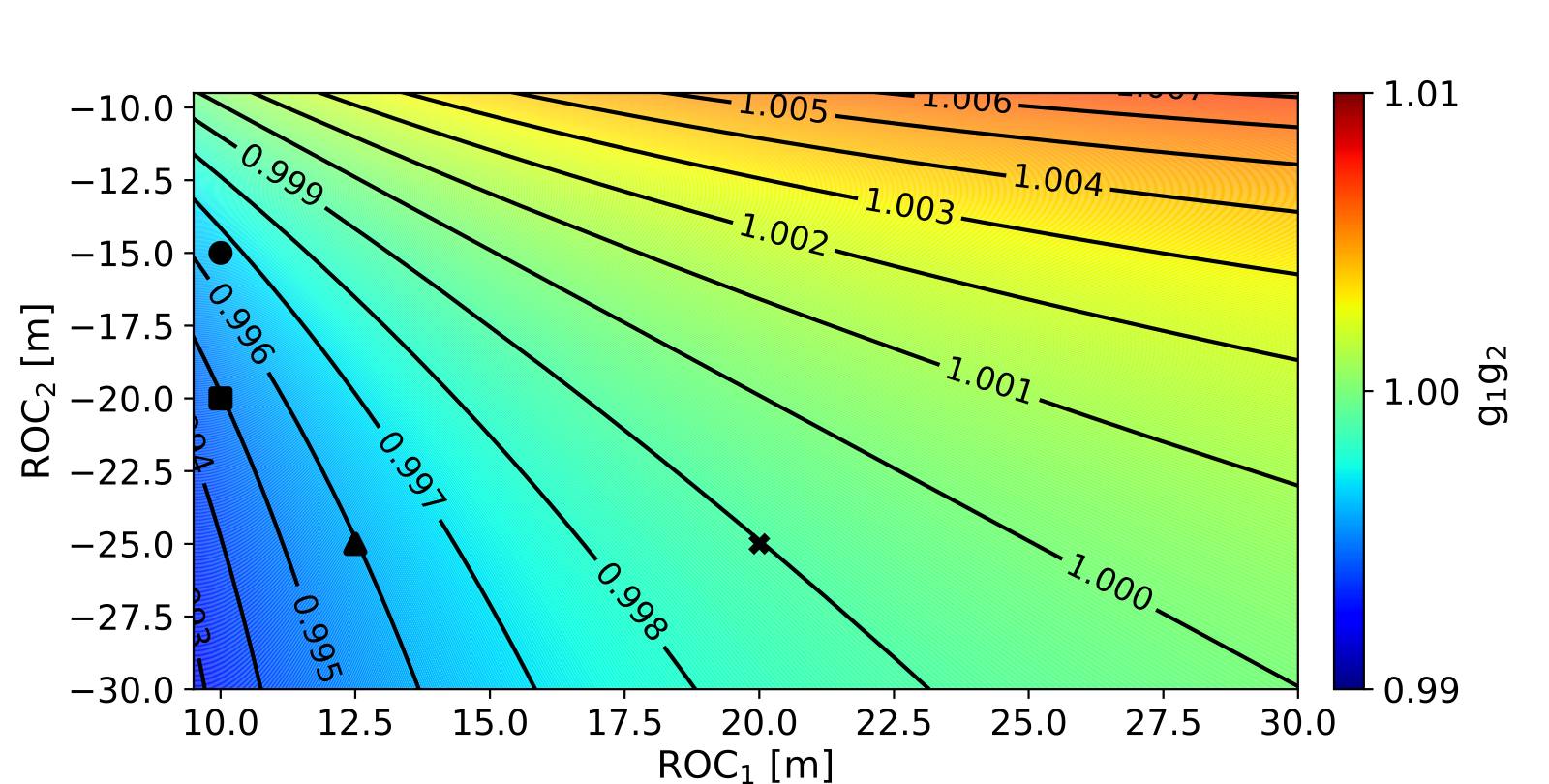}
    \includegraphics[width = 0.45\linewidth]{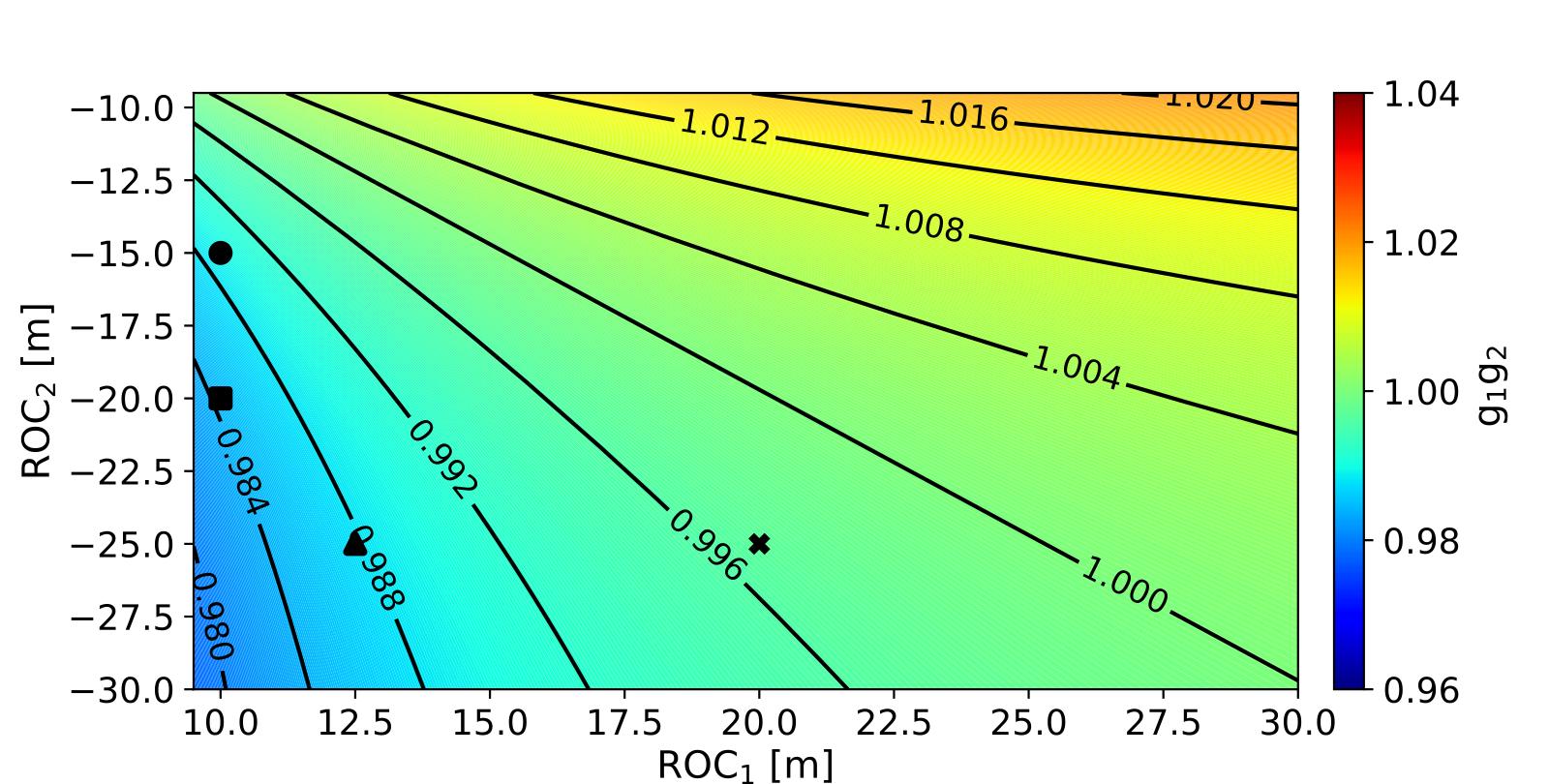}\quad
    \includegraphics[width = 0.45\linewidth]{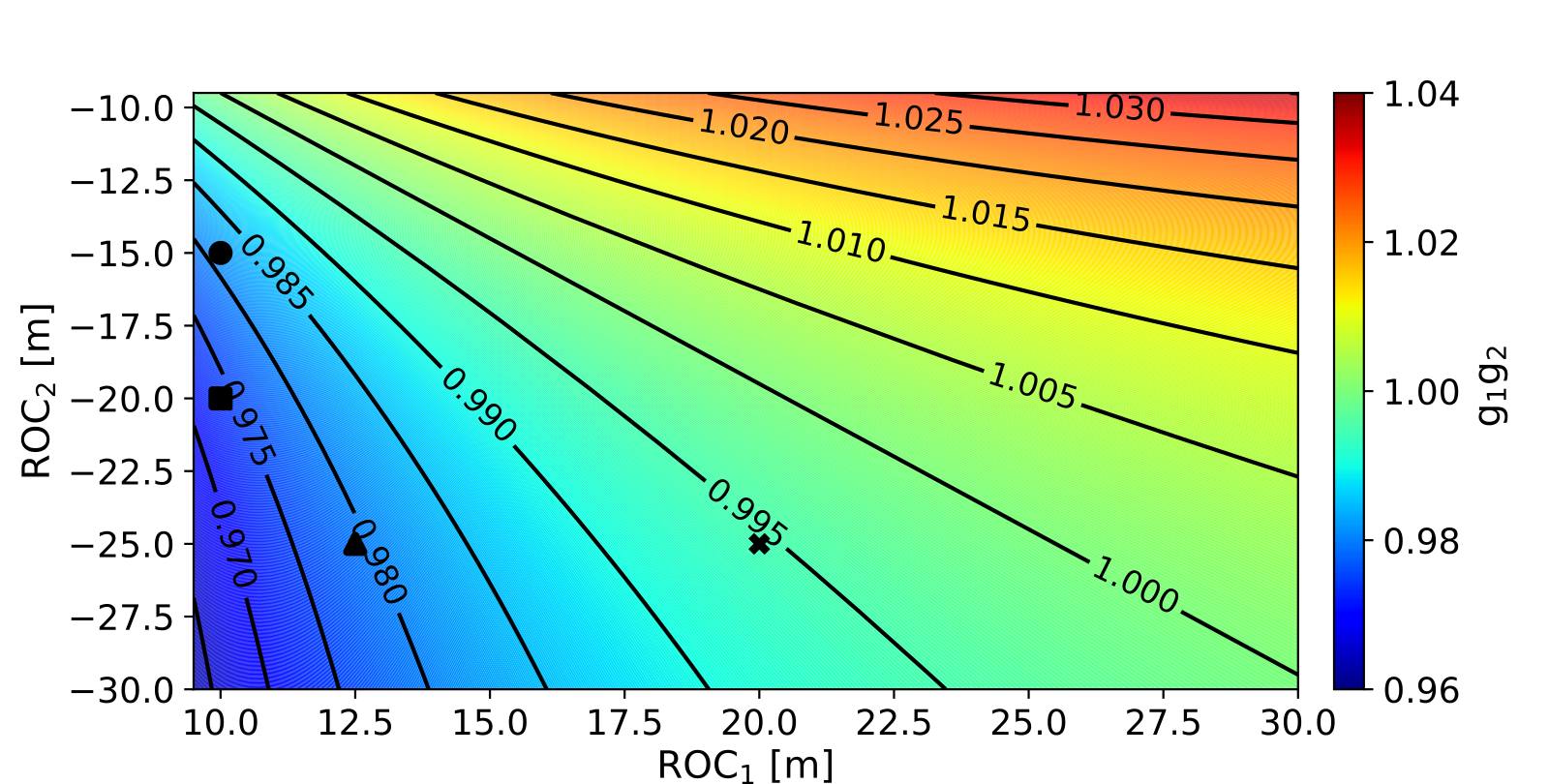}\quad
    \caption{ Stability maps for the near-unstable convex-concave mirrors' cavity configuration. Four possible ROCs pairs are marked by black points for 0.1~m (Top), 0.3~m (Middle) and 0.5~m (Bottom) long spacer.}
    \label{fig:UMK_Stability_maps_g1g2}
\end{figure}

Fig.\ \ref{fig:UMK_Stability_contour} illustrates how the stability factor $g_{1}g_{2}$ varies with changes in the spacer's length for four specified configurations of the mirror's curvature. 
It is observed that increasing the length for ROC$_{1} = 10$~m, and ROC$_{2} = -20$~m, is most beneficial in terms of enhancing stability among the presented cases. 
However, an increase in length may also increase sensitivity to external vibrations. Fig.\ \ref{fig:UMK_Stability_maps_g1g2} shows stability maps for three spacer lengths: 0.1~m, 0.3~m, and 0.5~m, with markers denoting the chosen ROC, also presented in Fig. \ref{fig:UMK_Stability_contour}. These maps and calculations of the beam spot on the mirrors indicate whether increasing the ROC is beneficial for cavity overall stability.

\subsubsection{Geometrical stability}

The stability condition ($g_1g_2$) is not the only factor that matters in our computations. Mechanical deformation of a resonant cavity adds another crucial factor to take into account in actual laser systems~\cite{boy24}.
Deformation can change and affect the optical path length of the cavity which is related to the vibration sensitivity of the optical resonator.

\begin{figure}[hbt]
\centering
\includegraphics[width=0.3\columnwidth]{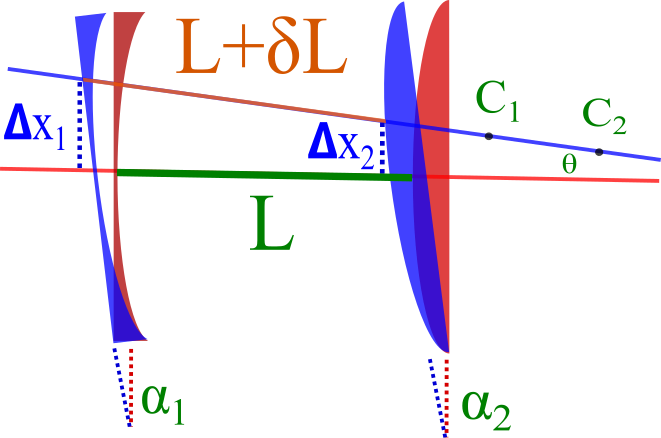}
\caption{Misalignment and mode-shifts on each mirror. Red and blue colours show the original position and misaligned position, respectively.}\label{fig:UMK_misalignment_scheme}
\end{figure}

In the case of mirror tilting, the resonator axis rotates by an angle $\Theta_i$ and the centres of the field intensity patterns shift if the mirror $M_i$ is tilted by an angle $\alpha_i$ (See Fig.~\ref{fig:UMK_misalignment_scheme}) the modified cavity length will be $L + \delta L_{tilt}$. The relations can be obtained with a geometric study. From \cite{Hauck1980, Siegman1986, Alvarez2019} we have

\begin{equation}
\Delta x_1 = \frac{g_2}{1-g_1g_2} \times L\alpha_1 + \frac{1}{1-g_1g_2} \times L \alpha_2, \label{eq:deltax_1}
\end{equation}
\begin{equation}
\Delta x_2 = \frac{1}{1-g_1g_2} \times L\alpha_1 + \frac{g_1}{1-g_1g_2} \times L \alpha_2, \label{eq:deltax_2}
\end{equation}
\begin{equation}
\Theta = \frac{(1-g_2)\alpha_1-(1-g_1)\alpha_2}{1-g_1g_2}. \label{eq:delta Theta}
\end{equation}

Finally, 
\begin{equation}
\delta L_{\text{tilt}} = \frac{\Delta x_2 - \Delta x_1}{\sin(\Theta)}- L. \label{eq:deltaL_1}
\end{equation}

Manufacturers can attach mirrors with an accuracy of $2.91 \times 10^{-4}\ \text{radians}$ (60 arcsecs). 
Using this precision, we can assess the effect of the rotation of the mirrors on the cavity length. 
For example, among the cases in Tab.\ \ref{UMK_tab:angle}, a cavity with $R_2 = 10$ m and $R_2 = -20$ m has less length change ($3.589 \times 10^{-8}$ m). Although the plane-concave configuration has better results of the required length accuracy, the convex-concave configuration has a beam spot size that is approximately twice as large. 

\begin{table}[hbt]
\caption{The theoretical calculations of $2.91 \times 10^{-4}\ \text{radians}$ (60 arcsec) tilt of each mirror result in a change in length and result in mode displacement. The required length accuracy for different configurations with 30 cm length shows the limited length tolerance to prevent overlapping the higher order modes (HOM's) with fundamental modes. }
  \centering
  \renewcommand{\arraystretch}{1.2}
  \begin{tabular}{c c c c c c c c c}
    \hline
    $R_1$, $R_2$ [m]& $\Delta x_1$ [m]   & $\Delta x_2$ [m]& $\delta L_{tilt}$ [m] &Required length accuracy [m] \\
    \hline\hline
    $\infty$, 1 &$5.818 \times 10^{-4}$ & $4.945 \times 10^{-4}$ & $4.231 \times 10^{-9}$ & $1.271 \times 10^{-2}$  \\
    $\infty$, 5.0 & $2.909 \times 10^{-3}$ & $2.822 \times 10^{-3}$ & $4.231 \times 10^{-9}$ &  $4.827 \times 10^{-3}$ \\
    $\infty$, 10.2 & $5.847 \times 10^{-3}$ & $5.934 \times 10^{-3}$ & $4.231 \times 10^{-9}$ &  $3.325 \times 10^{-3}$\\
    10, -15 & $1.663 \times 10^{-2}$ & $1.622 \times 10^{-2}$ & $9.413 \times 10^{-8}$ & $2.096 \times 10^{-3}$ \\
    10, -20 &$1.138 \times 10^{-2}$ & $1.113 \times 10^{-2}$ & $3.589 \times 10^{-8}$ & $2.465 \times 10^{-3}$\\
    12.5, -25 & $1.429 \times 10^{-2}$ & $1.403 \times 10^{-2}$ & $3.631 \times 10^{-8}$ & $2.182 \times 10^{-3}$ \\
    20, -25 &  $5.522 \times 10^{-2}$ & $5.447 \times 10^{-2}$ & $3.050 \times 10^{-7}$ &$1.144 \times 10^{-3}$ \\
    \hline
  \end{tabular}
  
  \label{UMK_tab:angle}
\end{table}

\subsubsection{Higher-order modes overlapping}

The consideration of higher-order modes (HOM) is of crucial importance, since imperfection in mode matching results in the excitation of HOM when using stabilising techniques such as the Pound-Drever-Hall (PDH) technique. 
This coupling defect causes higher-order modes to have resonance frequencies different from the fundamental mode. 
Neglecting these modes may affect the accuracy of the error signal obtained in the PDH technique, introducing potential instabilities in the laser frequency and influencing overall stability assessments~\cite{Amairi2013}. 
HOM produce theirs own small error signals which adds a small offset to the main error signal of the TEM$_{00}$ mode. 
And if HOM depend on the coupling and vibrations, then also the offset change in time, resulting in variation of the lock point.

To calculate the  higher order modes we used Hermit-Gauss modes which is a family of modes characterised by two numbers, $m$ and $n$, denoting the modes' extensions in transverse directions.

\begin{figure}[hbt]
\centering\includegraphics[width=0.45\columnwidth]{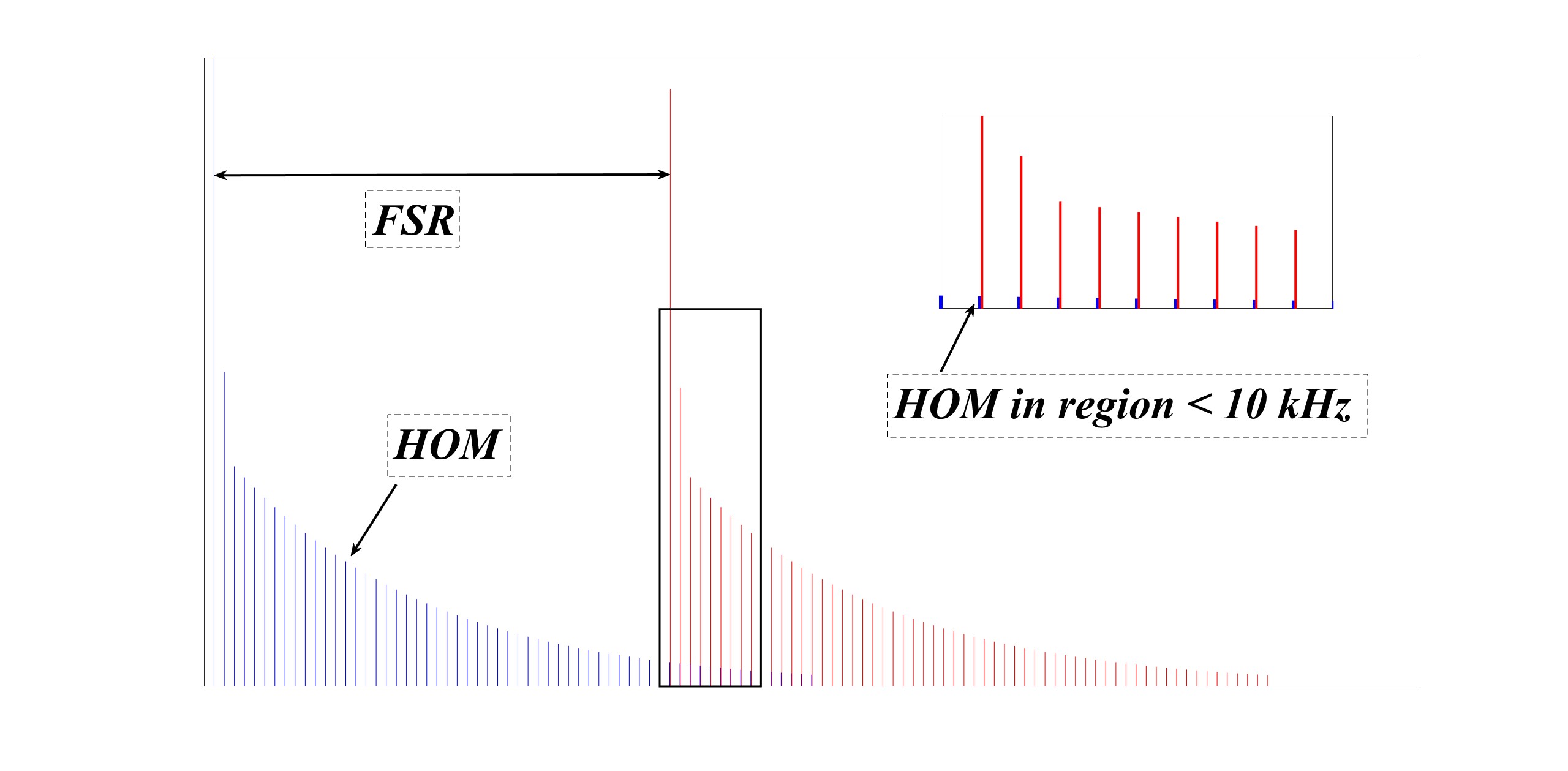}
\caption{Example of two longitudinal modes with TEM${}_{mnq}$ HOMs in an optical cavity. Blue and red lines are sequences of longitudinal modes separated by FSR, $\Delta\nu_{FSR} = 500 $MHz.}\label{UMK_freq_lines}
\end{figure}

\begin{figure}[hbt]
\centering
{\includegraphics[width=0.9\linewidth]{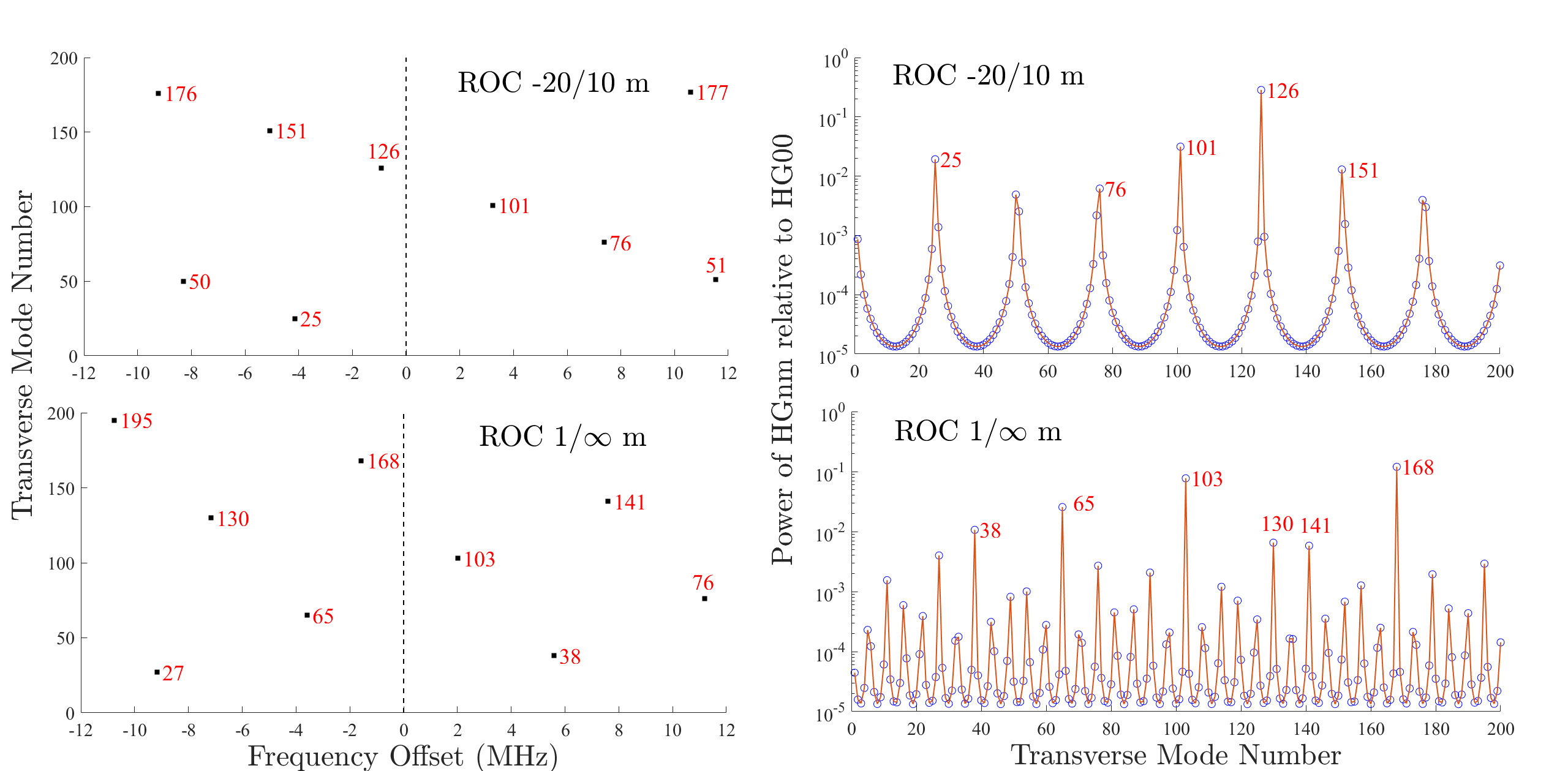}}
\caption{(left) HOM resonance frequency offsets for cavities (assuming range of $12$ MHz) with L $= 30$ cm for ROC $10,-20$ m (top) and ROC $1,\infty$ m (bottom). The red labels indicate the transverse mode number (m+n). (right) The transmitted power of the first $200$ HG modes relative to the power of the TEM${}_{00}$.}
\label{fig:UMK_TEM}
\end{figure}

Using 
 \begin{equation}
\frac{\Delta\nu_{mnp}}{\Delta\nu_{FSR}} = p + \frac{(m+n+1)}{\pi}\cos^{-1}(\sqrt{g_1g_2}), \label{eq:UMK_higher_mode}
\end{equation}
\noindent
makes it straightforward to calculate the resonance mode frequencies, as the integer $p$ indicates the longitudinal mode. The FSR range of the optical resonator separates the resonance frequencies of corresponding fixed $p$, $m$, and $n$. According to Eq. \eqref{eq:UMK_higher_mode}, the resonance peaks' linewidth and finesse are greatly dependent on the choice of mirror geometry, particularly the ratio of cavity length to mirror radius. It is crucial to place higher-order mode resonances apart from the fundamental mode and ensure they don't overlap with fundamental modes to create high-finesse cavities for laser locking. For example,  as illustrated in Fig.~\ref{UMK_freq_lines}, higher-order modes can shift close to the fundamental modes by misalignment and problematic mode matching. To ensure optimal performance of the cavity, an important constraint was introduced to avoid combinations of RoCs with higher-order modes close to the fundamental mode, with an arbitrarily chosen safety threshold of 12 MHz. The results show a region where higher-order modes are confined to the $12$ MHz range, in proximity to the TEM${}_{00}$ mode. For our analysis, we opted for specific radii of curvature—specifically, 10 and -20 meters, as outlined in Fig~\ref{fig:UMK_TEM}.

\subsection{Convex mirror substrate}

\begin{figure}[hbt]
\centering\includegraphics[width=0.35\columnwidth]{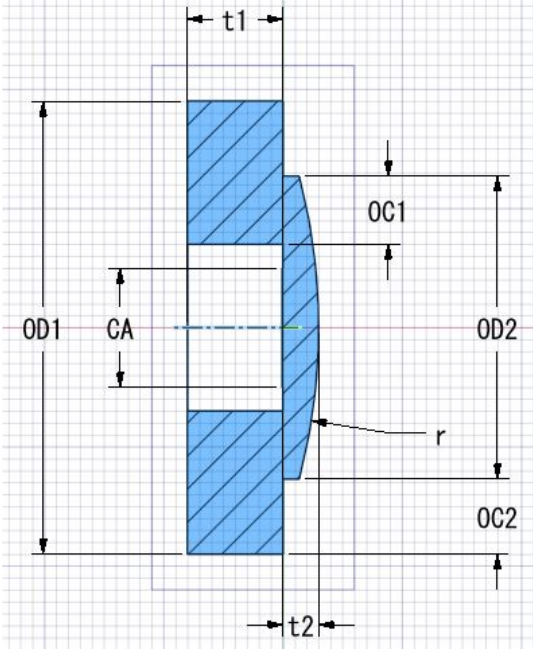}
\caption{Design of a convex mirror substrate that can be optically contacted to a cavity  spacer}\label{fig:UMK_mirrorcc}
\end{figure}

The required convex mirror poses
a serious technological obstacle that must be overcome - the optical contact between such a mirror and spacer. 
An ingenious solution was proposed by OptoSigma, presented in Fig.~\ref{fig:UMK_mirrorcc}, which is composed of a mirror and an additional ring that connect both ring and spacer. 
FEM simulations made by UMK shows that  such a configuration should be stable and keep the zero-crossing point in reasonable zone.

\section{Conclusion and Prospects}

In this chapter, the noise performance of crystalline AlGaAs coatings with expected low thermal noise floor in optical resonators is investigated. Novel noise types like photo birefringent noise and intrinsic birefringent noise that are related to the birefringence of the coatings are realized. The former can be suppressed by stabilizing laser power while the latter can be removed by using advanced locking schemes to average out the noise from the two polarizations. The results from the monocrystalline Si cavity operated at PTB confirm the coating thermal noise level at 124 K is at most equal to or lower than the theoretical valve based on mechanical loss values determined at room temperature. Yet, owing to an unexpected excess noise with a correlation length larger than the mode diameter, the ultimate noise in the coating length, expressed from the laser frequency locked to the cavity length, is substantially higher and eventually similar to the otherwise identical silicon cavity with dielectric coatings. As a result, we continued the investigation on a room-temperature cavity focusing on the understanding of the mechanisms that create the novel noises. We confirm the modified birefringence by light (and hence photo birefringent noise) is related to the photoabsorption inside the semiconductor coatings that is strongly wavelength dependent as photoexcitation above the bandgap is allowed. This in turn means that special care of the environment has to be taken such that no stray light from the windows would degrade the laser stability. The ultimate performance of the coatings at room temperature is to be determined with and without light beyond the GaAs bandgap. Our findings may also be of crucial importance for the next-generation crystalline coatings aiming at the thermal noise level.

A numerical simulation of the thermal noise of mirror coatings to identify the optimum metasurface designs based on the required optical properties (reflectivity, radius of curvature, polarization dependence, scatter, and absorption loss) was conducted. Based on the result low-noise meta-mirrors and compound meta-etalons were fabricated and investigated with photospectrometry and cavity ringdown spectroscopy. These preliminary investigations indicate that scattering due to material defects and side wall roughness accounts for the major fraction of the optical losses. 
In future work, we will study the influence of different fabrication steps and their respective parameters on optical losses and optimize them. This involves, for example etching hard masks made of different materials as well as different etching techniques.  In addition, for focusing metasurfaces the interaction of neighboring meta-atoms has to be studied to identify the arrangement of meta-atoms that optimizes high reflectivity and the desired phase profile on macroscopic areas containing several tens or hundreds of thousands of meta-atoms.

A reference cavity system based on a NEXCERA ceramic spacer with 30 cm length was developed during the project. It is equipped with conventional dielectric mirrors. 
The optimization of the system could not be fully completed during the project; at this stage, the initial characterization of its frequency instability yielded a level approximately one order higher than the thermal noise level computed from the measured mechanical loss. 
However, the measured instability is limited by the instability of the available reference laser. The acceleration sensitivity 
was measured and found to be at a reasonable level of several parts in $10^{-10}/g$. 
Very favourable is the extremely small long-term drift found for another NEXCERA cavity made out of a similar material. 

To make the final decision on the suitability of the NEXCERA ceramic for top-performance optical cavities, further measurements of the frequency stability against an appropriate optical reference need to be done. 

To further reduce the thermal noise limit we simulated numerically optical cavities with a laser beam diameter of several mm. 
To implement such beam sizes mirror substrates with a radius of curvature in the range of 20-25 m for 10-30 cm long cavity spacers are needed which can be produced by optically contacting separate parts.

\clearpage

\graphicspath{{D2_SHB/Figures}}

\chapter[Rare-earth ion-based frequency references]{Rare-earth ion-based frequency references in single pass and in highly dispersive cavities}

\authorlist{%
Yann Le Coq$^1$, 
Johannes Dickmann$^2$, 
Bess Fang$^{1,\dag}$, 
David Gustavsson$^3$, 
Michael Hartman$^1$, 
Stefanie Kroker$^2$, 
Thomas Legero$^4$,
Xiuji Lin$^1$, 
Marcus Lindén$^5$, 
Lars Rippe$^3$, 
Nico Wagner$^2$, 
Martin Zelan$^{5}$
}

\affil{1}{\OPaff}
\affil{2}{\TUBSaff}
\affil{3}{\ULUNDaff}
\affil{4}{\PTBaff}
\affil{5}{\RISEaff}

\corr{bess.fang@obspm.fr} 

\chapstart
This chapter summarizes various influences on laser frequency stabilisation based on spectral hole burning in rare-earth doped crystals.
Section \ref{sec:D2_S1} presents investigations of the thermal noise of rare-earth ion-based frequency references based on spectral hole burning (SHB) technique. 
In section \ref{sec:D2_S2} environmental influences which can impact the laser frequency stability are identified. 
Thereafter, section \ref{sec:D2_S3} concerns detection schemes for rare-earth ion-based frequency references.
Finally, section \ref{sec:D2_S4} presents the prospects of a highly dispersive Eu$^{3+}$:Y$_2$SiO$_5$ (Eu:YSO) slow light crystal cavity.

\section{Introduction}

Stabilizing a laser on a spectral pattern photo-imprinted by spectral hole burning (SHB) in a rare-earth doped material is a promising alternative to advanced high-finesse Fabry–Perot cavity stabilized lasers. With an appropriate choice of the dopant ion and crystalline matrix, narrow optical transition can be found. In the case of europium doped yttrium orthosilicate Eu$^{3+}$:Y$_2$SiO$_5$ (Eu:YSO), the homogeneous line width estimated from photon echo measurements indicates a width of $\approx 100$\,Hz for the optical transition of $^7{\rm F}_0 \rightarrow {}^5{\rm D}_0$ transition at 580\,nm, and the typical time scale of the redistribution of the population in the ground-state hyperfine manifold is of the order of 10\,h around 4\,K, allowing for engineering of persistent spectral structures for high performance frequency locking. We present here our investigations of various aspects of this technique in order to evaluate its metrological relevance, including several physical noise sources, ultra-low noise detection methods, as well as the use of the dispersive properties in an cavity configuration.  

\section[Thermal noise in SHB]{Thermal noise of rare-earth ion-based frequency references based on SHB technique}
\label{sec:D2_S1}

One of the interesting aspects of using spectral hole structures as a frequency reference is the purported property of having a low thermal noise limit~\cite{thorpe2011frequency}. Yet, neither material properties such as the loss angle at cryogenic temperatures nor detailed analysis of the thermal noise processes can be found in the literature. Both aspects have been considered here.

\begin{figure}[htb]
    \centering
    \includegraphics[width=8cm]{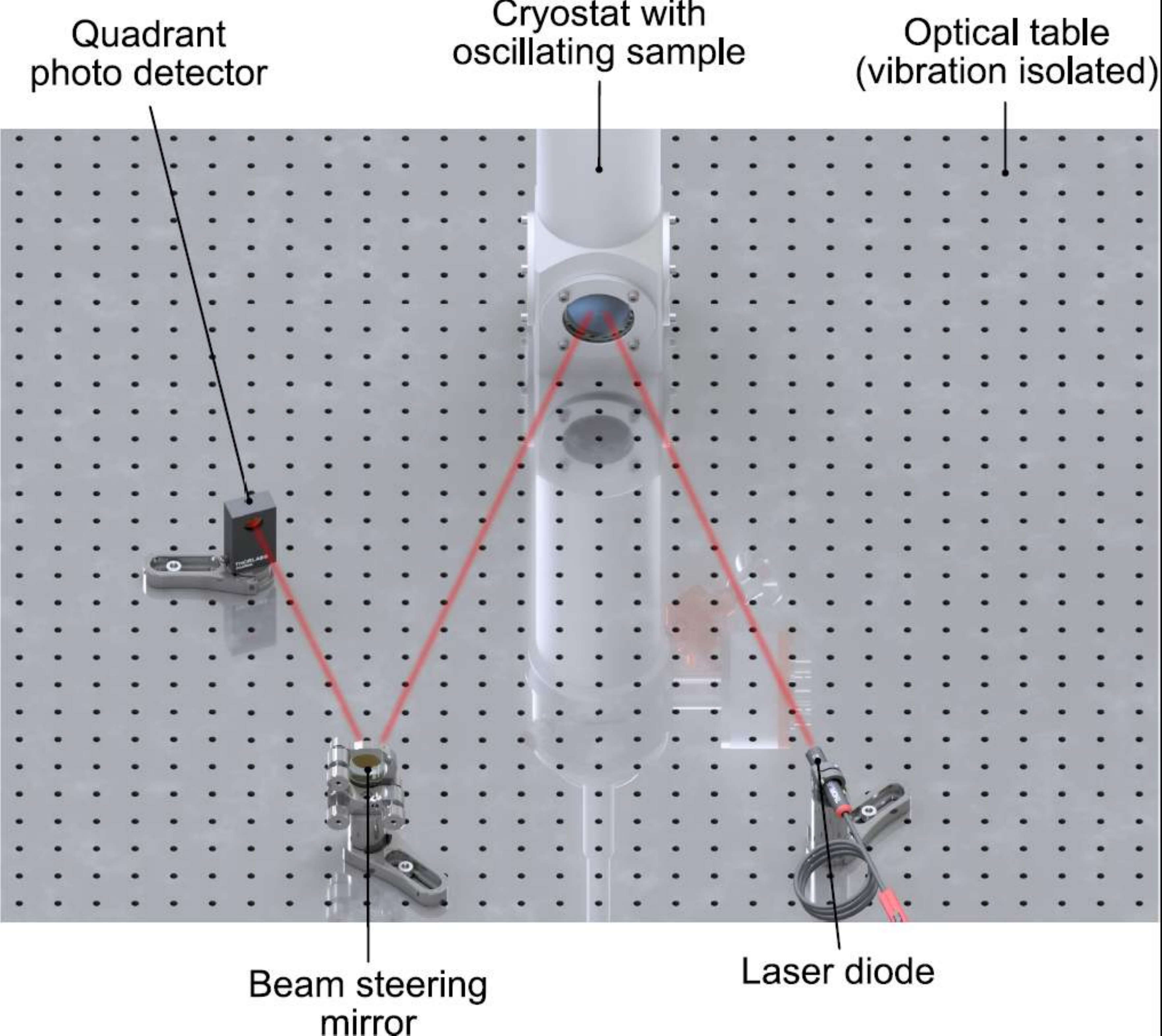}
    \caption{Schematic of the experimental setup for mechanical loss measurements. The laser beam is reflected off the oscillating sample within the cryostat. The motion of the sample is monitored through a quadrant photodetector.}
    \label{fig:explossYSO}
\end{figure}

Since the material of choice is Eu:YSO, a few macroscopic thin plates, with a thickness ranging from sub mm to a few mm and the other linear dimensions of the order of 1 cm, have been prepared by mechanical polishing. They are then incorporated in the experimental setup shown in Figure~\ref{fig:explossYSO}. The mechanical excitation is realized by a piezo actuator. For the readout of the mechanical vibration, the oscillator is illuminated by a laser diode. The motion of the reflected beam is detected with a quadrant diode. As soon as the sample has been excited to a mechanical resonance, the excitation is switched off and the mechanical decay is recorded. Due to material restrictions, the dimensions of the oscillator could not be chosen freely to avoid an overlap of its resonance frequencies with the sample holder. To avoid that additional damping through the clamping block occurs due to overlapping mechanical resonances, we checked the resonance frequencies of the system oscillator - sample holder by using the finite element simulation tool COMSOL Multiphysics. The sample resonances sufficiently isolated from holder resonances were chosen for temperature dependent measurements of the mechanical loss. The results of these measurements indicate a transition from $ \phi=1.2\times10^{-2}$ at room temperature to $\phi=(2.9\pm0.1)\times 10^{-4}$ at $T=52$ K~\cite{Wagner2024_yso}. These values should however be considered as upper limits of the bulk material properties since other loss mechanism, including but not limited to clamping losses cannot be isolated from the measurements, and the limited variation of aspect ratio does not permit reliable extrapolation as proposed in~\cite{Nawrodt_2013}.  

Theoretically, thermal fluctuations can couple to the frequency of the Eu$^{3+}$ ions via the fluctuations of the position of the ions, as well as the fluctuations of the local force and strain experienced by the ions \cite{har24}. 
Whereas the former turns out to be a second-order effect, the latter is relevant, based on previous characterization of the sensitivity to uniaxial strains~\cite{galland2020mechanical}. 
We derived, starting from the Fluctuation Dissipation Theorem~\cite{Callen1985}, the power spectral density of the thermomechanical component of fractional frequency noise of a spectral hole in terms of the complex elastic modulus, $E=E_0(1+i\phi)$, and the volume, $V$, of the crystal,
\begin{equation}
    S_{\frac{\delta{\nu}}{\nu_0}} (\omega) = \frac{4k_\mathrm{B}T}{\omega}\frac{k^2_{\mathrm{D}_m\mathrm{S}_n}}{\nu^2_0}\frac{E_0\phi}{V},
\end{equation}
where $k_{\mathrm{D}m\mathrm{S}n}$ is the stress/frequency-strain modulus measured along dielectric axis $\mathrm{D}_m$ when probing spectroscopic site $\mathrm{S}_n$~\cite{galland2020mechanical}, $k_{\rm B}$ is the Boltzmann's constant, $T$ is the absolute temperature, and $\nu_0$ is the optical frequency of the transition in Eu$^{3+}$ under consideration. Putting in the numerical values of the material and geometric properties of our crystal, and using the measured $\phi=(2.9\pm0.1){\times}10^{-4}$ at $T=52\,\mathrm{K}$, we estimate the thermal noise to be below $10^{-18}$ at 100\,mK for site 1, which is the metrologically relevant crystal substitution site.

\section[Environmental influences]{Identification of the environmental influences which can impact the laser frequency stability}
\label{sec:D2_S2}

\begin{figure}
    \centering
    \includegraphics[width=0.8\linewidth, trim=4.2cm 5cm 2cm 4cm, clip]{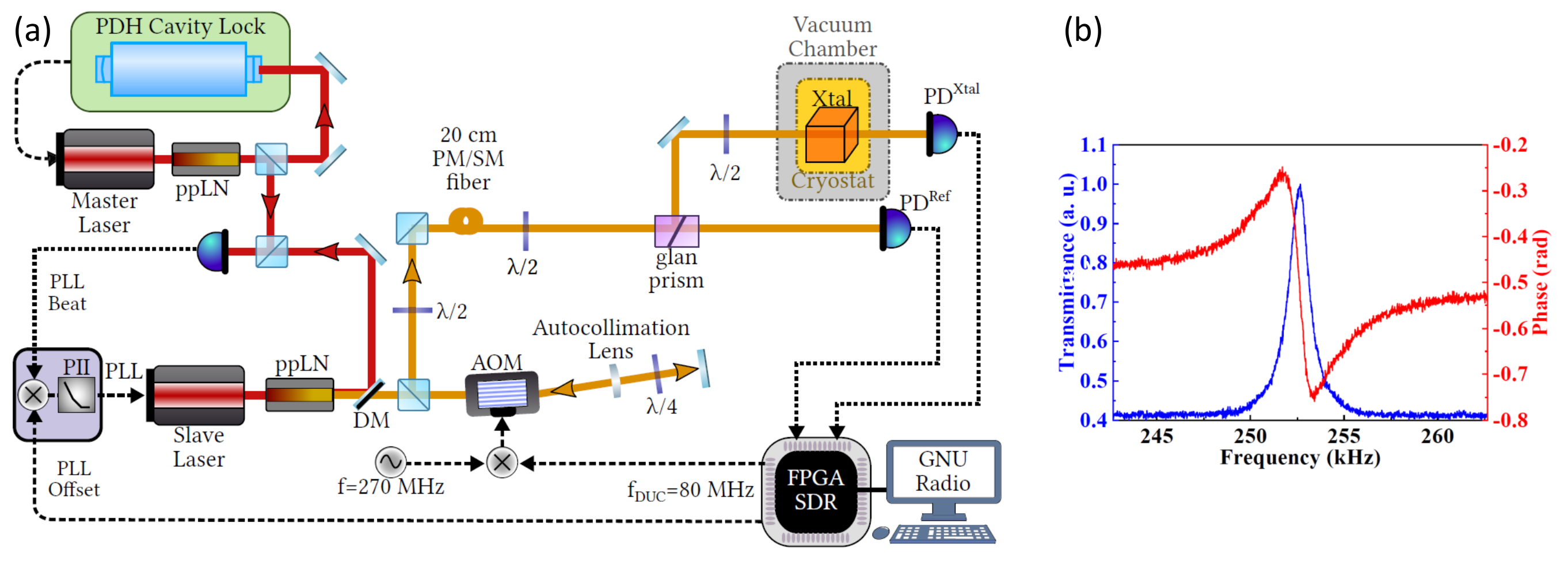}
    \caption{
		(a) A schematic of the experimental setup of the heterodyne interferometry. 
		Drawing reproduced from~\cite{lin2023multi-mode}. 
		(b) Typical transmittance and dephasing of a narrow spectral hole. }     
    \label{fig:SHB}   
\end{figure}

Heterodyne interferometry is used to carry out spectroscopic measurements in a single-pass configuration through the crystal. The experimental setup is shown in Fig.~\ref{fig:SHB}(a). By the Kramers-Kronig relation, a narrow transmission spectrum of a spectral hole induces a linear dephasing of a nearly resonant laser field, with the steepest slope located at the center frequency of the spectral hole, see Fig.~\ref{fig:SHB}(b). We optically beat down the probe signal to the radio frequency (RF) regime, and measure the amplitude and the phase against that collected on a reference path, so as to obtain both the transmittance as a probe of the spectral profile and the dephasing that serves as an error signal for the frequency servo loop. 

The measurement of the DC Stark effect is detailed in~\cite{zhang2020precision}. 
Simply speaking, we track the absolute frequency of the spectral holes by referencing against the primary frequency standards at the LNE-SYRTE laboratory via an optical frequency comb. 
Due to the symmetry of the YSO crystal, a spectral hole essentially splits into two with positive and negative frequency shift in the presence of a homogeneous electric field. 
By analysing both branches for residual asymmetries, and targeting a laser fractional frequency instability at the $10^{-17}$, we are able to infer an upper limit for electric field fluctuations to 2.3~V\,m$^{-1}$ RMS, which is experimentally within reach. 

In the case of a magnetic field, previously published results~\cite{tho13} have provided sufficient information to establish bounds of the magnetic environment of the crystal. 
For an asymmetric quadratic Zeeman coefficient of 1~Hz\,G$^{-2}$, the same reasoning as above requires an upper bound of 720~G magnetic field instability to achieve a fractional frequency instability of $10^{-17}$ or below. 
Alternatively, typical magnetic field instabilities in our facility (the Foucault laboratories in the center of the Paris city) is about $10^{-4}$~G or below, corresponding to an induced fractional frequency instability of below $2\times 10^{-23}$, making magnetic noise irrelevant to our set up for now. 

Concerning the impact of the temperature of the crystal $T$, based on the two-phonon Raman model, one expects that the shift of the resonance frequency of rare-earth dopants $f_{\rm shift} = \nu-\nu_0$ follows a scaling law of $f_{\rm shift} \propto T^4$~\cite{konz2003temperature}, which was later measured down to 1.5\,K~\cite{tho13,Oswald2018} without the presence of a buffer gas. 
We have observed that at the sensitivity to temperature fluctuations near 4~K largely limit the performance achieved in ~\cite{galland2020double}. 
In the presence of an external isotropic pressure in thermal equilibrium with the crystal, however, the thermal fluctuation induced pressure change would counteract the thermal expansion and contraction of the crystal itself, giving the possibility to cancel to the first order the temperature induced frequency shift of the spectral holes for a well chosen pressure-temperature couple. 
Similarly to earlier work in ~\cite{thorpe2011frequency}, we investigated these behaviours at temperatures achievable by our pulse-tube cryocooler. 
Whereas we were able to verify the cancellation of the temperature sensitivity in an environment that is coupled to an isotropic pressure~\cite{zhang2023first}, the technical challenges in realizing a hermetic gas cell that is robust against temperature cycling is tremendous and requires technical expertise in cryogenics and mechanical engineering as well as dedicated efforts. 
Furthermore, this approach is only valid for timescales long-enough where thermal equilibrium is reached. An alternative consisting of further decreasing the crystal temperature using a dilution refrigerator has been explored and detailed in Section~\ref{ch:cryostats}.

It is worth mentioning an earlier study, also relevant for the implementation of an ultra-stable oscillator based on the SHB technique, is the sensitivity to mechanical strains, or equivalently to vibrations. We employed a cryo-compatible 3-axis translation stage in order to place or remove small masses (aluminium plates) on the crystal, while probing the transmission spectrum in the vicinity of a previously photo-imprinted spectral hole~\cite{galland2020mechanical, zhang2020inhomogeneous}. 
By characterizing the linear frequency shift, we established the intrinsic acceleration sensitivity to range from $1.5\times 10^{-11}$~(m\,s$^{-2}$)$^{-1}$ 
down to $1.3\times 10^{-12}$~(m\,s$^{-2}$)$^{-1}$ (depending on the direction of the strain and the crystal substitution site), 
giving an upper bound, for a target laser fractional frequency instability of $10^{-17}$ at 1~s, to the tolerated vibration noise to be below $10^{-5}$~m\,s$^{-2}/\sqrt{\rm Hz}$, which is manageable even in the presence of a pulse-tube cryocooler that allows for continuous operation without the need of regular replacement of cryogen.

\section[Detection schemes]{Detection schemes for rare-earth ion-based frequency references}
\label{sec:D2_S3}

The considerations of the choice of detection scheme is also based on the setup mentioned in Section~\ref{sec:D2_S2}, using a heterodyne interferometry. 
In order to be insensitive to differential noise sources between the two measurement paths, monitor modes can be generated and placed in a broad spectral hole to measure and their contribution subtracted from the error signal. 
Multimode generation also constitutes a means to spread the probing optical power into several narrow spectral holes in order to increase the signal to noise ratio while keeping overburning (continuous optical pumping from the probe laser field) in check.  
Note that this strategy was previously explored in~\cite{PhysRevA.104.063111} using analog generation and detection system.

The technological challenge in multimode generation and detection is overcome by using all digital signal processing, with python-based GNURadio software package coupled with a universal software-defined radio peripheral (USRP), Ettus X310 in our case, that offers phase coherent RF multiple inputs and outputs. 
Multiple frequency modes in time domain can be generated using inverse fast Fourier transform of the desired spectrum centered on zero frequency. 
The USRP uses digital up conversion to shift the spectrum to a suitable RF signal to drive an acousto-optic modulator that imprints the multiple frequency modes onto an optical carrier. 
To detect the phase shift of each mode, the previously mentioned beatnote is measured by suitable photodiodes and sent to the USRP input ports, yielding both the amplitude and the phase information using IQ detection, before being digitally down converted into the base band. 
A narrow-band polyphase channelizer is used to isolate each frequency component before processing the phase information and generating the error and correction signals. 
The details of our implementation can be found in~\cite{lin2023multi-mode}. 
Compared to using only a single probe mode~\cite{gob17}, this fully-digital multi-heterodyne approach allows decreasing technical noise sources by orders of magnitudes at 1~s time scale or longer and reaching a detection noise (with optical probing power compatible with hours of continuous operation) of $2\times 10^{-16}$ at 1~s, with clear prospects to reach the $10^{-17}$ range or even lower~\cite{lin2023multi-mode}. 

There is however a point of caution due to an unexpected ion-light interaction dynamics in the crystal. We have observed an off-resonant emission when the crystal is probed optically. Whereas the nature of the emission is a subject of ongoing research, the spectral behaviour, which exhibits a bandwidth of approximately 2\,MHz, suggests that a higher heterodyne frequency, typically above 4\,MHz should be used in order to reduce the excess noise related to the emission.

\section{A highly dispersive Eu:YSO slow light crystal cavity}
\label{sec:D2_S4}

\begin{figure}[htbp]
    \centering
    \includegraphics[width=0.9\linewidth]{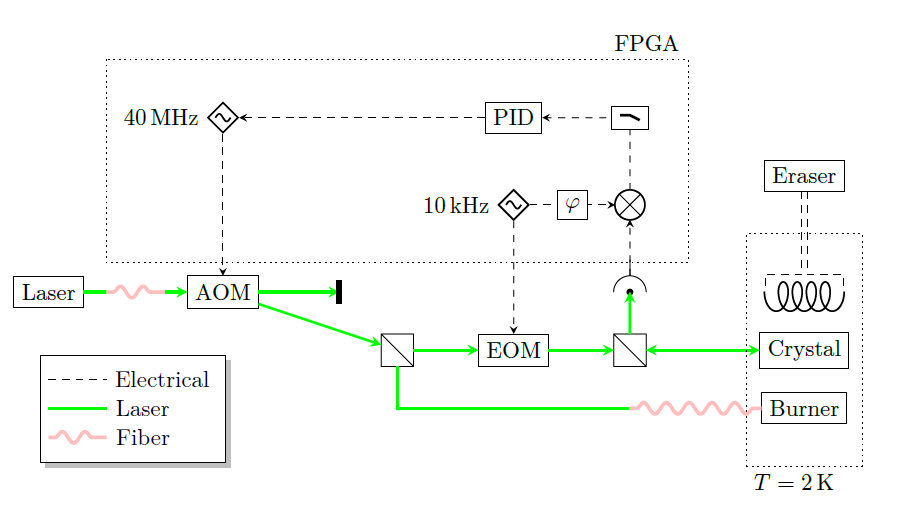}
    \caption{The experimental setup for locking to the cavity. 
		A standard PDH locking loop is used where the error signal is generated digitally by an FPGA. 
		The control signal is fed back to the AOM after passing a numerically controlled oscillator (at 40 MHz) for digital-to-analog conversion. 
		A side-burner which consists of highly diffuse material is located in close proximity to the crystal and is used for hole-burning. 
		To erase holes an eraser circuit is used where resonant magnetic fields can interact with the ions in the crystal in order to reset the hyperfine levels \cite{gus24a}.}
    \label{fig:locking_setup}
\end{figure}

The aim is to significantly enhance the performance of frequency stabilization of lasers using a highly dispersive Eu:YSO slow light crystal cavity. This design leverages unique optical properties of Eu:YSO to achieve a drastic reduction in the group velocity of light, effectively elongating the optical path length within the cavity. This approach inherently has the potential to lower the Brownian noise floor by more than $2.5\times10^5$ times, which is critical for achieving high-frequency stability. For the locking we use a standard Pound-Drever-Hall (PDH) locking loop to stabilize a laser's frequency by locking it to a narrow reflection mode of the dispersive cavity.  
The experimental setup is shown in Fig.\ \ref{fig:locking_setup}. 

Due to the slow light effect, the Eu:YSO cavity will appear very long (equivalent to > 10 km in vacuum). When the laser is locked to the cavity the stability of the cavity is transferred to the laser through the stability criterion ($\delta  \nu /\nu = \delta L/L$), where $\delta \nu$ is the resulting frequency shift due to Brownian noise. This relation shows a proportional decrease in frequency uncertainty if the length $L$ of the cavity is increased. Standard cavities are about $0.1\text{--}0.5\;\mathrm{m}$ long, which highlights the potential for highly dispersive cavities. The method of locking to dispersive cavities using PDH locking loops is described in more detail by Horvath et al. \cite{Horvath2022}

The slow light effect is created by burning away ions in a frequency region, creating a pit in the absorption profile. In these experiments pits could be burnt down to $30\;\mathrm{kHz}$ with high aspect ratio and a resulting slow light factor above $10^5$, making the optical path length long. In order to burn a 40 kHz wide square hole in the absorption profile of the europium ions, the ions were excited using a secant pulse. The pulse parameters were 9.25 kHz/ms frequency scan bookended by two halves of a 10 kHz wide, 600 $\mu$s long hyperbolic secant pulse shape. This pulse was repeated $16\times10^{3}$ times, interspersed with relaxation time of 0.1 ms.

An essential aspect of the cavity design is impedance matching and managing background absorption. The cavity used in this work was designed to ensure impedance matching, so that losses  match the transmission through the front mirror (10 \%). The main source of loss is due to off-resonant residual absorption from the ions on the side of the pit. As the absorption losses in the cavity depend on the hole-width, there is some degree of freedom to adjust the losses in the cavity to make the impedance match. By increasing the hole-width, the residual absorption should decrease as there is less off-resonant absorption (resulting in lower intra-cavity round trip losses) where the opposite is true for making narrower holes.

To obtain the optimal matching conditions the reflection mode was frequency scanned in order to measure the minimum reflectivity $R_{min}$ for different pit widths. Theoretically, the reflectivity for different absorption losses is given by
\begin{equation}\label{eq:Rcav}
    \begin{array}{ccc}
        R_\text{min} &  = &R_1\left|1 - \frac{1-R_1}{\sqrt{\frac{R_1}{R_2}}\exp(\alpha L_\mathrm{RT})-R_1}\right|^2\\
        \alpha_\mathrm{IM} L_\mathrm{RT} &  = &\frac{1}{2}\ln\left(\frac{R_2}{R_1}\right)\
    \end{array}
\end{equation}
Here $R_1 = 0.9$ and $R_2 = 0.99$ is the reflectivity mirrors and $L_\mathrm{RT} = 21$ mm is the round trip loss length of the cavity. An unknown portion of the light was not coupled to the cavity and caused an unknown constant reflectivity offset $R_{nc}$, this parameter is roughly estimated by the lowest measured $R_{min}$. As intracavity losses should decrease with widening pits, wider pits (above 70 kHz) were found to be below the optimum impedance matching $\alpha_\textrm{IM}L$. This means that the pit should have slightly higher losses for perfect matching and that the optimum pit width is below 70 kHz.  The theoretical minimum losses given by off-resonant excitation lets us conclude that impedance matching happens close to $40\text{--}60\;\mathrm{kHz}$. The minimum losses for a $40\;\mathrm{kHz}$ hole was estimated to be about 0.06 in this case (above $\alpha_\textrm{IM}L$). This minimum absorption can be calculated as 
\begin{equation}
    \alpha_\text{min}= \frac{2}{\pi}\alpha_0\arctan \left( \frac{\Gamma_h}{\Gamma_\text{pit}}    \right)
\end{equation}
where $\Gamma_h$ is the homogeneous linewidth and $\Gamma_\text{pit}$ the width of the pit.

\subsection{Evaluation of the lifetime of the spectral transmission} \label{sec:rf-erasure}

Evaluating the lifetime of the spectral transmission window in the Eu:YSO slow light cavity is crucial for understanding its long-term stability. The spectral transmission window, created through hole burning, determines the cavity's ability to maintain a stable frequency over extended periods. The lifetime of this window is closely linked to the onset of excessive drift in the Allan deviation, a measure of frequency stability over time.

Higher locking powers, while initially beneficial for stabilizing the laser, can lead to shorter locking times. This is because increased power levels tend to degrade the hole more quickly, causing hole asymmetries that lead to mode drifts. As the laser frequency is locked to these modes, any drift in the mode translates directly to frequency instability. See section \ref{cavity_parameters} for further discussion on drift rates.

The overlapping Allan deviation $\sigma$ was measured for different averaging periods $\tau$ (between $10^{-2}$ to $10^2$) and locking powers (at 2.5 and 18 nW). The minimum Allan deviation was reached at $\tau \approx 10$ s, reaching $6 \times 10^{-14} \pm 9\times 10^{-17}$. Since there was no fiber noise cancellation technique in use fiber noise is likely to be dominating, thus limiting the absolute value of the frequency stability. 
Nevertheless, a relative comparison can be made to praseodymium where it was found that the Allan deviation for europium shows nearly an order of magnitude improvement in frequency stability, even though the locking power is three orders of magnitude lower than that required for Praseodymium \cite{Horvath2022}. 
As mentioned above, lower locking power allows for significantly longer locking times by reducing the long-term drift rate. 
More information will be available in Gustavsson et. al \cite{Gustavsson}. 

To achieve the lowest drift possible, it is necessary to lock to a mode which is in the center of the hole (due to drift, as described above). The procedure to get the mode centered was to first burn a hole and then measure which position the mode had relative to the center. If the mode was off-center the hole was erased using an eraser circuit using tunable resonant magnetic fields from a coil wrapped around the crystal, targeting the hyperfine transitions of the two isotopes $\text{Eu}^{151}$ and $\text{Eu}^{153}$. A new hole could then be burnt at a shifted frequency to make the mode centered inside the hole. The erasure procedure is required because the hole lifetime (relaxation time of the ground states) has been reported to be up to 48 days~\cite{Oswald2018}. 

The cryogenic RF resonator design featured in-situ tuning for both matching and resonator frequency at 1.5 K. 
This could be achieved by employing capacitor banks using low loss and high-power MEMS switches, capable of handling voltages exceeding 100 volts. 
For more details and motivation behind the capacitor switch bank resonator, refer to Lindén et. al \cite{LindenDesign}.

The pulses were designed to be $\pi/2$-pulses as this was found to be efficient in terms of redistributing the ions in the different hyperfine levels. 
The RF scan range was set to be approximately three times the measured spin broadening of the transition. Full erasure of the hole took less than 10 min, by running RF-pulses on each hyperfine transition repeatedly. 
RF scans were performed in the range of $34\text{--}120$~MHz with a scan range going from approximately 0.8 to 4.5~MHz, depending on the broadening of the transition. 
Majority of the time spent for full erasure (about 60 \%) was on the largest level splitting transition ($\approx 119.2$~MHz) for isotope 153, which also had the largest observed spin broadening. 
For more discussion regarding the pulse parameters and hole erasure performance, refer to Lindén et. al \cite{LindenAtomic}.

In the hole erasure experiments an average energy of 0.47 J was required per pulse sequence sent for hole erasure. The energy required to optically erase holes is given by
\begin{equation}
    E_\text{opt} = h \nu \times N_\text{ions} \times n_\text{scan}
\end{equation}
where $N_\text{ions} \approx 400\times10^{18}$, $\nu \approx 500$ THz is the optical frequency, and $n_\text{scan} \approx 5$ is how many scans are required to shuffle the ions. The resulting energy is about $E_\text{opt} = 663$ J, or close to $1.4\times10^3$ times as much energy. In practice, RF-erasure should allow for both faster and more efficient resetting of the hyperfine states. In fact, utilizing RF fields resonant with these transitions should, in principle, not result in heating losses due to a largely symmetric distribution  between the hyperfine levels. Thus, it is as likely for the ions to be absorbed as it is to undergo stimulated emission. There will still be joule heating due to imperfections in the RF-resonator, however. This is a technical limitation rather than a fundamental one.

\subsection{Identification of the critical slow light cavity parameters} \label{cavity_parameters}

Identifying important locking parameters of the Eu:YSO slow light cavity is essential for understanding and enhancing its performance. We have been focusing on several key metrics: cavity mode linewidth, lifetime, and drift. These parameters are crucial for evaluating the cavity's ability to maintain laser frequency stability over time.

In initial experiments, the Eu:YSO cavity demonstrated a cavity line-width of 1.67 kHz $\pm$ 0.213 kHz in reflection and 3.02 kHz $\pm$ 0.291 kHz in transmission. Such narrow modes indicates a group refractive index $n_g>10^5$ where the group velocity is given by $v_g = c/n_g$. Comparing these parameters with those of an existing Pr:YSO (praseodymium-doped yttrium orthosilicate) slow light cavity, the Eu:YSO system shows a much higher group retardation factor giving it a distinct advantage as the reflection modes are $>10^{2}$ times narrower. The narrowest modes measured in the case of praseodymium was 275 $\pm$ 2 kHz. As the optical path length is $L_{opt} = L n_g$ and the FSR of a cavity is given as
\begin{equation}
    \text{FSR} = \cfrac{c}{2L_{opt}}
\end{equation}
then the mode linewidth will be scaled down with the group refractive index as we have
\begin{equation}
    \nu_{linewidth} =  \cfrac{\text{FSR}}{\mathcal{F}}
\end{equation}
where $\mathcal{F}$ is the finesse. This means that the FSR and mode linewidth will be scaled down by the same slow light factor.  

To measure drift rates two Pound-Drever-Hall locks were established in sequence for 500 s each where each sequence used different locking powers. The low power case (about 2.5 nW) and high power case (18 nW) resulted in cavity mode drifts of 3.66 Hz/s and 16.2 Hz/s respectively. This can be compared to a praseodymium cavity which had a drift rate of 550 $\pm$ 60 Hz/s, showing the stability and locking lifetime of the Eu:YSO system. 

The drift, as discussed in \cite{Horvath2022}, is expected to arise from two sources: off-resonant excitation and hyperfine cross-relaxation. The motivation behind transitioning to europium was largely due to the substantial decrease in the latter, indicating that off-resonant excitation would likely become the primary cause of frequency drift. Hence, it's imperative to maintain precise control over the mode’s frequency to approach the theoretical zero drift point, which occurs at the center of the pit.

To conclude, the lower drift of Eu:YSO cavities results in longer locking times and better frequency stability over different averaging times. A higher fractional frequency stability can be achieved  (almost a magnitude in improvement) and longer locking times has been reached ($\tau > 100$~s).

\section{Summary and conclusion}
\label{sec:D2_conclusion} 

This report summarizes the work carried out in Task 1.2 within WP1 of the EMPIR-project
NEXTLASERS. The aim of the task has been to study various influences on laser frequency
stabilisation based on spectral hole burning in rare-earth doped crystals. In section \ref{sec:D2_S1} the work on thermal noise of rare-earth ion-based frequency
references based on SHB technique is presented. 
For the first time mechanical loss measurements on YSO crystals were performed. The results at 52~K indicate, that thermal noise of the SHB setup at 0.1~K will contribute well below $10^{-17}$ to the instability. 

In section \ref{sec:D2_S2} the environmental influences on the laser frequency stability was investigated. 
It has been found, that electric and magnetic fields and  temperature fluctuations will not be limiting the stability at the $10^{-17}$ level.

In section \ref{sec:D2_S3} detection schemes for rare-earth ion-based frequency references were evaluated, leading to estimated fractional frequency instabilities of $1\times 10^{-17}$ at 100~s averaging time.

In section \ref{sec:D2_S4} work on a highly dispersive Eu:YSO slow light crystal cavity is described. It was found that dispersive cavities, utilizing the narrow linewidth of Europium ions doped in to YSO, had a magnitude in frequency stability improvement over Praseodymium, reaching an Allan deviation of $6\times10^{-14}$. In addition, Europium offers lower frequency drifts of 3.66 Hz/s and 16.2 Hz/s at 2.5 nW and 18 nW respectively as lower locking powers can be used due to a substantial decrease in the hyperfine cross-relaxation process. By the use of optimized hole burning sechyp-pulses, a diffuse side burner, and hole erasure circuit the locking procedure could be improved further.

\clearpage

\graphicspath{{D3_vibrations/figures}}

\chapter{Low frequency vibration isolation}
\label{ch:vib_isolation}

\authorlist{%
Marcin Bober$^1$,
Yann Le Coq$^2$,
Bess Fang$^2$, 
Thomas Fordell$^3$, 
Kalle Hanhij\"arvi$^3$, 
Michael Hartman$^2$,
Sofia Herbers$^4$, 
Jan Kawohl$^4$, 
Clément Lacroûte$^5$, 
Thomas Legero$^{4,\dag}$, 
Xiuji Lin$^2$,
Thomas Lindvall$^3$,
Jacques Millo$^5$, 
Piotr Morzy\'nski$^1$, 
Mateusz Naro\.znik$^1$,  
Stephan Schiller$^6$,  
Uwe Sterr$^4$, 
Rodolphe Le Targat$^2$, 
Anders E. Wallin$^3$, 
Eugen Wiens$^6$, 
Jialiang Yu$^4$, 
Micha\l{} Zawada$^1$
}

\affil{1}{\UMKaff}
\affil{2}{\OPaff}
\affil{3}{\VTTaff}
\affil{4}{\PTBaff}
\affil{5}{\FEMTOaff}
\affil{6}{\HHUDaff}

\corr{thomas.legero@ptb.de}

\chapstart
This chapter describes improved mitigation of vibrations at low frequencies, actively by using state-of-the-art seismometers, tilt meters and interferometric leveling systems, involving multi-degree of freedom servo control and suspension systems and by feedforward techniques. 

\section{Introduction}
\label{sec:D3_intro} 

For short averaging times (< 1s) the frequency stability of state-of-the-art laser systems stabilized to optical cavities \cite{mat17a} is mostly limited by environmental vibrations typically on the order of about 10~µg/$\sqrt{\mathrm{Hz}}$. 
The situation becomes even worse for cryogenic systems, where periodic vibrations arise from the pulse-tube cryo-systems. 
The vibrations significantly degrades the performance of cavity systems cooled to a few Kelvin and compromises the spectral properties of spectral-hole-burning (SHB) crystals \cite{coo15,gob17}.

The impact of these perturbations can be minimized to a certain extent by optimizing the cavities and mounts for low sensitivity to accelerations and rotational movements. Utilizing high-performance seismometers and tiltmeters, we have studied and measured the vibration sensitivities of different room-temperature and cryogenic cavities and achieved residual sensitivities in the range of $10^{-11}$/g.

To realize laser systems with $10^{-17}$ fractional frequency instability additional passive or active vibration isolation (AVI) is necessary (section \ref{sec:Vibration_Control}). 
However, commercially available AVI systems show sufficiently good suppression of vibrations only in the frequency band between 10 Hz and 100 Hz. 
Their performance is quickly degrading outside this range. Furthermore, the available systems are not designed for situations when the source of vibration is part of the system as in the case with closed-cycle cryocoolers. 

Thus multi-input/multi-output (MIMO) servo systems using low noise sensors can be used to implement additional active vibration control and enhance the performance of commercial AVI systems in the frequency range of about 1 mHz – 100 Hz, e.g. using micro-controller based systems.

In addition feedforward techniques can be implemented, where accelerations of the cavities are measured and the vibration-induced frequency fluctuations are corrected to the laser frequency (section \ref{sec:Feed_Forward}). 
This technique considerably improves the performance of active vibration control. 
It can also be used as an alternative for systems that are not suitable for active feedback. 
We have presented examples of feedforward methods on different cavity systems where a reduction of the vibration sensitivity by a factor of ten could be demonstrated. 

In section \ref{sec:Cryostats} we show how to measure the vibrations in cryogenic systems close to the cavity or obtain information of the movement of a SHB crystal by optical measures. 
We present a closed-cycle cryosystem based on a combination of pulse tube cryocooler and continuous flow of helium gas for a silicon cavity at 124 K. 
The combination with active vibration control and feedforward will potentially allow a substantial suppression of vibration induced frequency fluctuations of the cryogenic silicon cavity down to $10^{-17}$ also for Fourier frequencies above 1~Hz. 

\section{Characterization of Sensors and Cavities}\label{sec:Vibration_Control}

To suppress remaining low-frequency vibrations of commercial AVI systems, active vibration control systems can augment the commercial systems at frequencies below a few hertz. 
Suitable, well-characterised sensors for measuring accelerations and inclinations are an important prerequisite for the operation of these systems (section \ref{sec:Sensors}). 
Transfer functions, like the sensitivity of the frequency of cavity systems to translations and rotations (section \ref{subsec:vib_transfer}), as well as from excitation voltages to movements of the AVI systems needs to be carefully measured. 
The setup and performance of the additional active vibration control is discussed in section \ref{sec:active-vibration-control}.

\subsection{Acceleration sensors and tiltmeters} \label{sec:Sensors}

This section focuses on the sensors dedicated to measure motions, which create forces acting on the cavities that change their lengths and their resonance frequencies. 
Similarly, these forces also can change the frequency of spectral holes via deformations of the crystal lattice (see section \ref{sec:D2_S2}). 
The objectives are the integration of such sensors in a closed-loop systems or for feedforward compensation to reduce the motion of the cavities or the impact on the optical frequency. 
The sensors are used at room temperature and are placed outside the cavity vacuum systems. 
In general a small footprint is required due to the high density of optic and photonic components used for the realization of the stabilized lasers.

Few sensors have been identified to measure three translations along orthogonal axis such as seismometers (or velocimeters). 
For the rotations relative to the horizontal, suitable tiltmeters have been identified. 
The sensors has been tested and characterized in the various experimental setups. 

\begin{figure}[hbt]
\centering
    \includegraphics[width=1\textwidth]{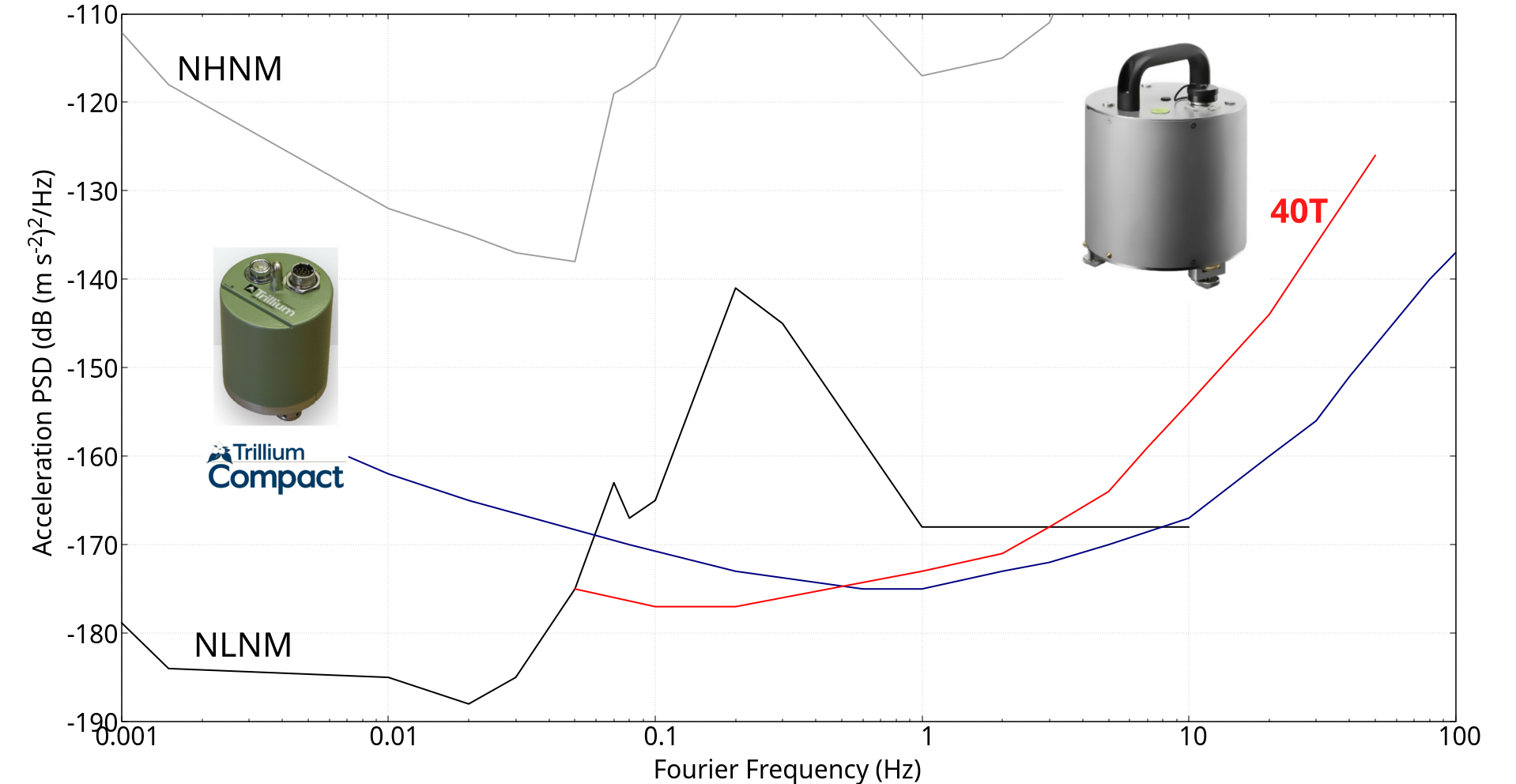}
    \caption{Expected vibration noise of \textit{40T Güralp} (in red) and \textit{Trillium Compact Nanometrics} (in blue) seismometers. The new low noise model (NLNM, in black) and the new high noise model (NHNM in grey) for ambient Earth noise is also shown for comparison \cite{pet93}.}
    \label{fig:CNRS_sismo-noise}
\end{figure}

\subsubsection{Sensors identified}

Low noise acceleration (or velocity) measurement in the Fourier frequency range from about 0.01~Hz to a few tens of hertz is a core activity of seismology that has lead to the development of low noise seismometers \cite{str08}. These instruments are also suited to the problem of accelerations experienced by ultra-stable optical cavities, which facilitates identification. 

The two seismometers tested and identified by CNRS and PTB are the \textit{40T} produced by \textit{Güralp} (UK) and the \textit{Trillium Compact} manufactured by \textit{Nanometrics} (CA). 
Their properties listed in the Table \ref{tab:motion_sensors} are similar. 
One can notice the compactness and the lightness of the Canadian sensor with a slightly lower cutoff frequency. 
The expected noise of each seismometer is an estimation provided by the manufacturer since getting an environment with such a low noise to perform the measurement is almost impossible even with correlation techniques. 
Unfortunately, the acceleration noise in common laboratories is a few orders of magnitude higher. 
The expected noise is shown on Fig.\ \ref{fig:CNRS_sismo-noise} and similar in middle of the band where the new low noise model (NLNM) predicts the highest perturbation. 
Below $\approx 0.5$~Hz, the 40T sensor shows the best performances but data are provided only down to $\approx 50$~mHz. 
For sensing accelerations close to the cavity, the smaller Trillium Compact has been chosen by PTB and VTT.

The limited sensitivity of the built-in accelerometers in active vibration isolation platforms (AVI) leads to small horizontal movements and tilt motion of the platform when the isolation is active. 
This effect is limiting the performance of the AVI in the horizontal directions (see Fig.\ \ref{fig:PTB_Feedback_Si}). 
For sensing the inclination of the AVI, PTB has tested tiltmeters from Lippmann Geophysical Measuring Instruments. 
The chosen "Platform Tiltmeter" is a 2-axes sensor that outputs voltages proportional to the tilt angles. The sensitivity is 209.49 µrad/V and 210.53 µrad/V respectively. The manufacturer cannot supply information on the frequency response of the tiltmeters.
As the sensors use an internal pendulum whose deflection provides the measurement of the tilt relative to the gravitational field, the measured frequency responses show a resonance of the pendulum at around 3 Hz.

\begin{table}[b]
    \centering
    \renewcommand{\arraystretch}{1.2}
	\begin{tabular}{| c | c | c | c | c |}
		\hline
        Type & Company & Footprint, size and weight & Bandwidth & Noise \\  \hline 
        40T & Güralp & \o 17~cm $\times$ 20~cm, 7.1~kg & 17~mHz to 100~Hz & -173 \\
        Trillium Compact & Nanometrics & \o 9~cm $\times$ 13~cm, 1.2~kg & 8~mHz to 100~Hz & -175  \\ \hline
	\end{tabular}
	\caption{Some characteristics of the identified 3-axes seismometers to measure residual motion of the cavities. The level of noise is in unit of the power spectral density of acceleration ($\textrm{dB(m s}^{-2})^2/\textrm{Hz}$) for a Fourier frequency of 1~Hz.}
	\label{tab:motion_sensors}
\end{table}

\subsubsection{Sensor characterization: sensitivity, noise}

The two seismometers were used to characterize the acceleration noise with the CNRS cavity stabilized laser experiment. Briefly, the frequency of laser source used is stabilized on cryogenic silicon cavity cooled to 17~K. The pulse tube based cryostat housing the cavity is set on an optical table isolated from the ground by active vibration isolation supports (\textit{AVI-200-M, Table Stable}). Vibrations rising from the pulse tube are decoupled as much as possible from the table.

We compared the vibration noise measured simultaneously in the same condition with the two seismometers listed in Tab.\ \ref{tab:motion_sensors}. The two seismometers have been placed on the optical table around the cryostat. Agreement between the measurement for the same axis is varying with the Fourier frequency. 
At very low frequency we often observe less than 2~dB for an absolute level of the noise of between -140 to -$120 \textrm{\,dB\,(m\,s}^{-2})^{2}/\textrm{Hz}$ function of the conditions. 
At higher frequency, typically between 10~Hz and 100~Hz the discrepancy can reach 20~dB. 
We do not believe in such a large difference between the two calibrated sensors. 
One explanation could arise from the two different positions due to the high space requirement on the table: resonances, rotations and the complex mechanical heavy structure could affect the noise differently at different positions.

\subsection{Vibration sensors for cryogenic environments}
\label{sec:cryo-vibration-sensors}

The seismometers described so far are not suitable to operate under vacuum and cryogenic temperatures. 
If vibrations and tilt have to be measured close to a cryogenic cavity or to SHB crystals, specific sensors have to be used.

\begin{figure}[htbp]
	\centering
	\includegraphics[width=12cm]{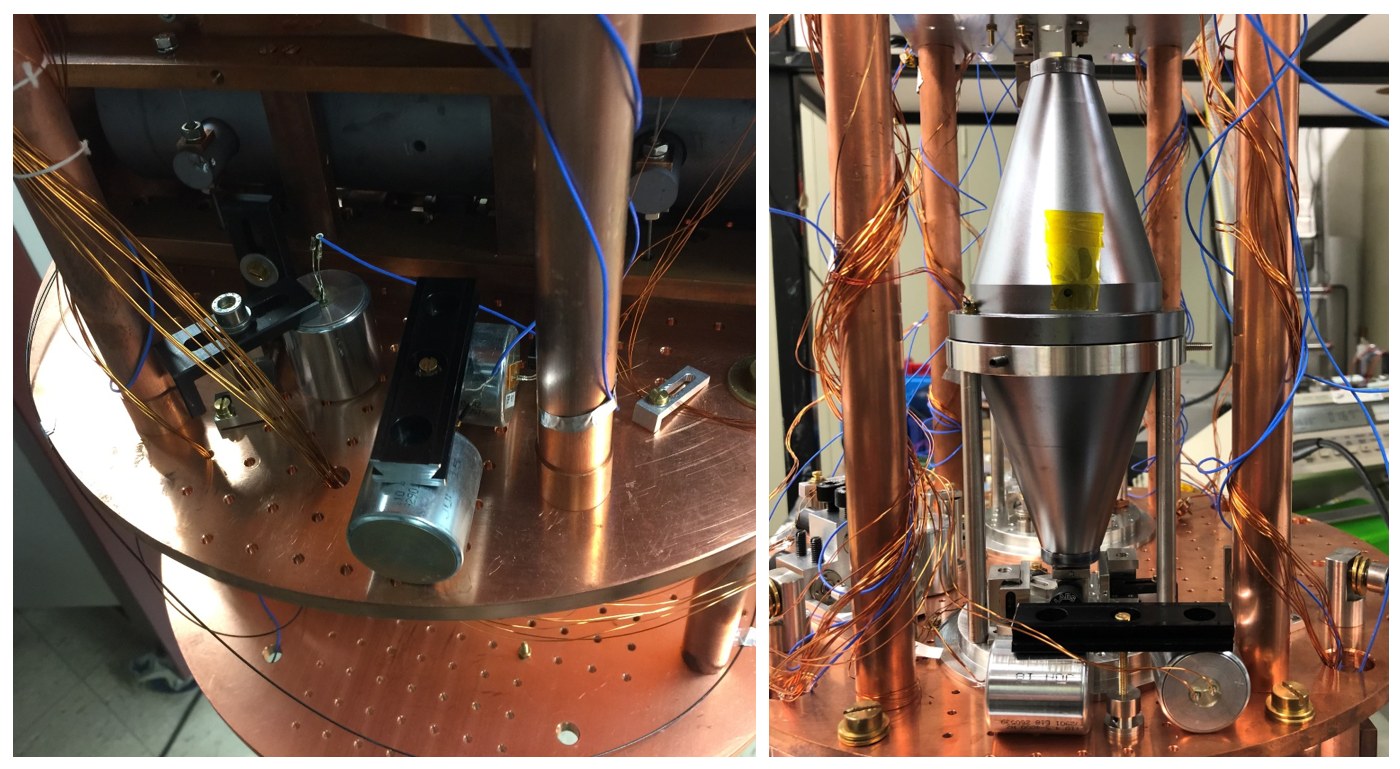}
	\caption{Geophone installations next to optical resonators in a 1.5 K closed-cycle cryostat at HHUD.}
	\label{fig:HHUD_Cryo_Geophones_Setup}
\end{figure}

\begin{figure}[htbp]
	\centering 
	\includegraphics[width=10cm]{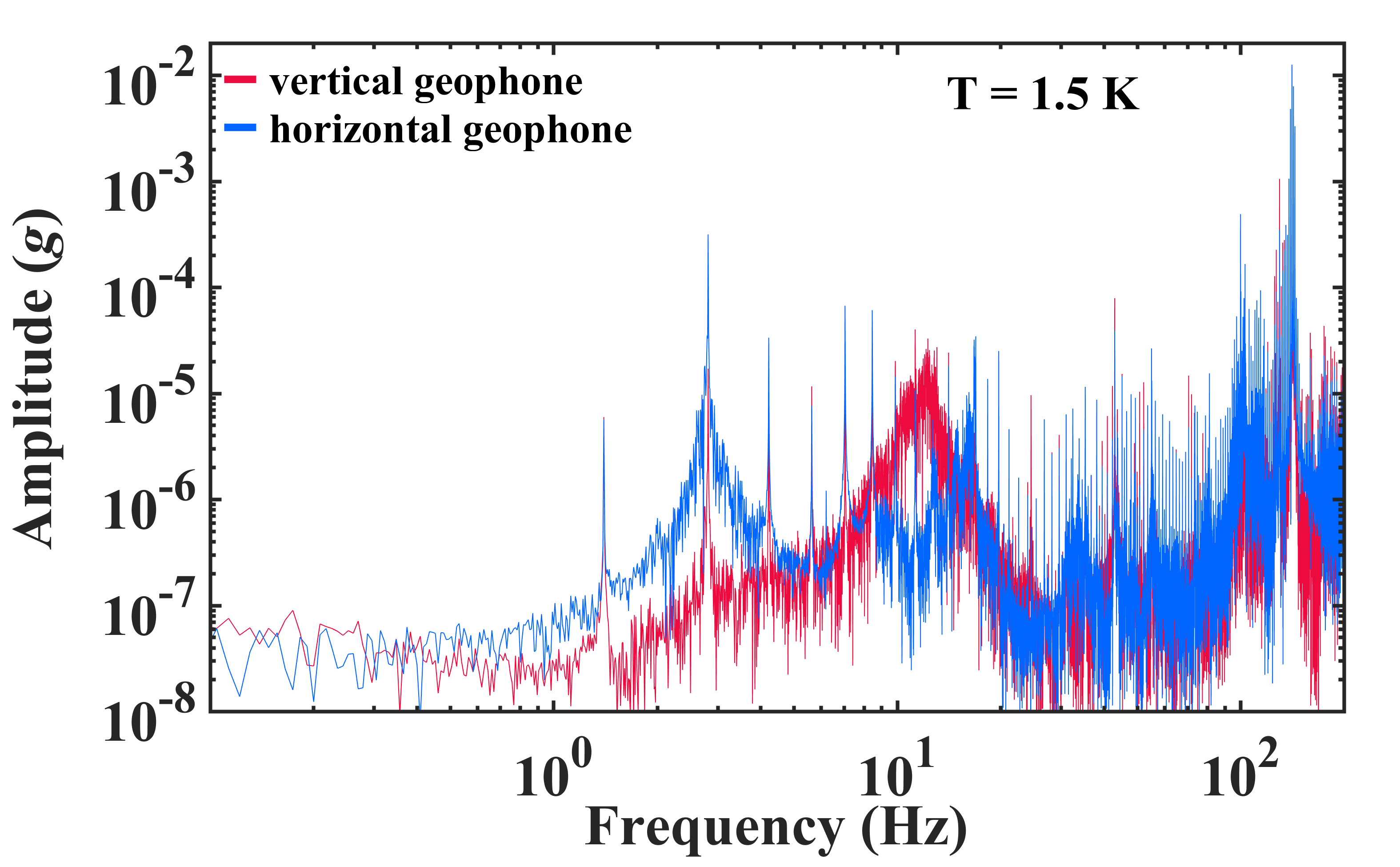}
	\caption{Spectrum of vibrations induced by a closed-cycle cooler at HHUD measured with calibrated geophones at 1.5 K.}
	\label{fig:HHUD_Acceleration_Spectrum}
\end{figure}

As an example, at HHUD a set of commercial geophones (GS-11D from Geospace Technologies) was used to monitor vibrations inside a 1.5 K closed-cycle cryostat equipped with a pulse tube cooler and a Joule-Thomson stage  (see Fig.\ \ref{fig:HHUD_Cryo_Geophones_Setup}). 
These geophones are inexpensive sensors with a natural frequency of 4.5 Hz available in vertical and horizontal configurations. 
To increase the sensitivity of geophones in the frequency range below their natural frequency a suitable amplifier was designed and built. 
To use them as acceleration sensors they were first calibrated at room temperature in the frequency range from 1~Hz to 200~Hz by comparing their output voltage with the output of a calibrated acceleration sensor, while attached next to each other on a shaker platform. 
After cooling down to 1.5 K geophones were recalibrated using the procedure outlined in \cite{kan05a}. 
Using calibrated geophones a spectrum of vibrations induced by a closed-cycle cooler was measured (see Fig.\ \ref{fig:HHUD_Acceleration_Spectrum}). 

\subsection{Optical sensing of cavity position and velocity}
\begin{figure}[t!]
\centering
    \includegraphics[width=0.5\textwidth]{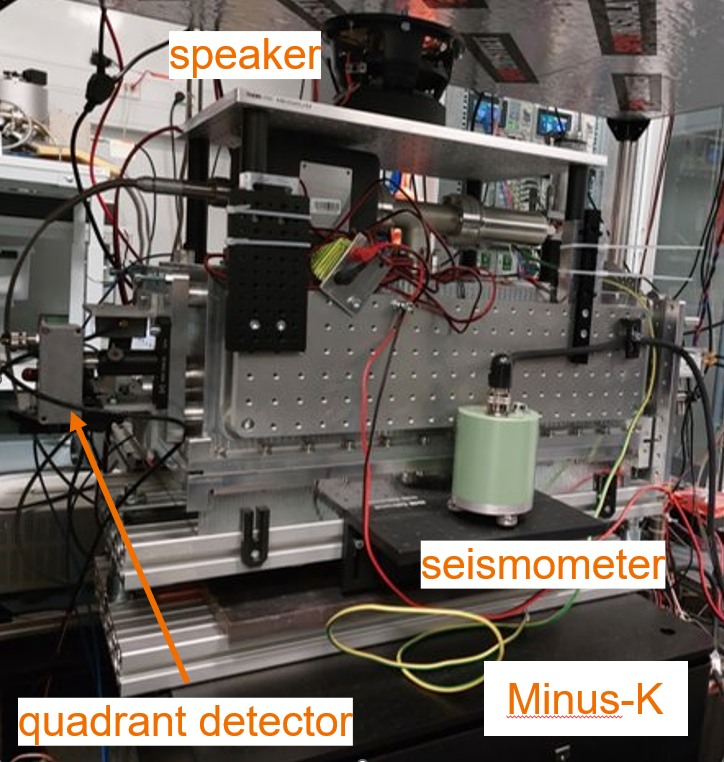}
    \hspace{1cm}
    \includegraphics[width=0.2\textwidth]{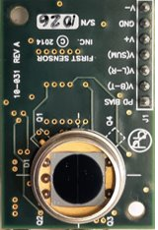}
    \caption{Left: Vacuum system of the 300-mm horizontal cavity at VTT on a passive Minus-K vibration isolation platform. Right: Commercial four-quadrant detector (First Sensor).}
    \label{fig:VTT_ULE30cm_quadDet}
\end{figure}

\mbox{} 
As an alternative to external sensors also the position of the laser beam transmitted by the cavity can be used to obtain information on cavity movements. 
In a setup at VTT a quadrant photodetector in the transmitted path was used to sense position changes of a horizontal 300-mm ULE glass cavity (Fig.~\ref{fig:VTT_ULE30cm_quadDet}). 
The cavity transmission is very weak, only a few hundred nW, and off-the-shelf detectors were unsuitable due to their low transimpedance gain (typically 10~k$\Omega$). 
Instead of making a custom circuit board, we chose to modify a commercial detector (First Sensor QP50-6-18U-SD2 detector) by increasing the transimpedance gain to 10~M$\Omega$. 
With the detector mounted on the platform, it sees cavity movement with respect to the vacuum chamber while the seismometer detects movements of the entire (floating) platform. 

\begin{figure}[t!]
\centering
    \includegraphics[width=0.49\textwidth]{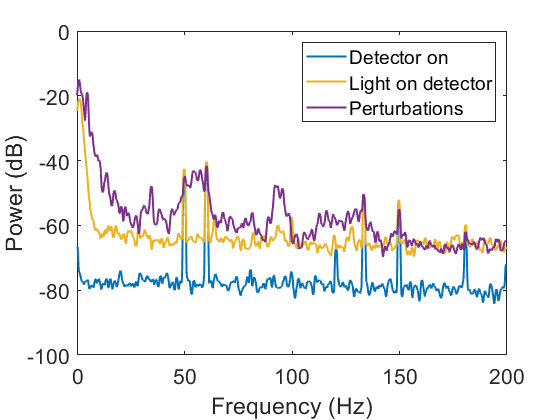}
    \includegraphics[width=0.49\textwidth]{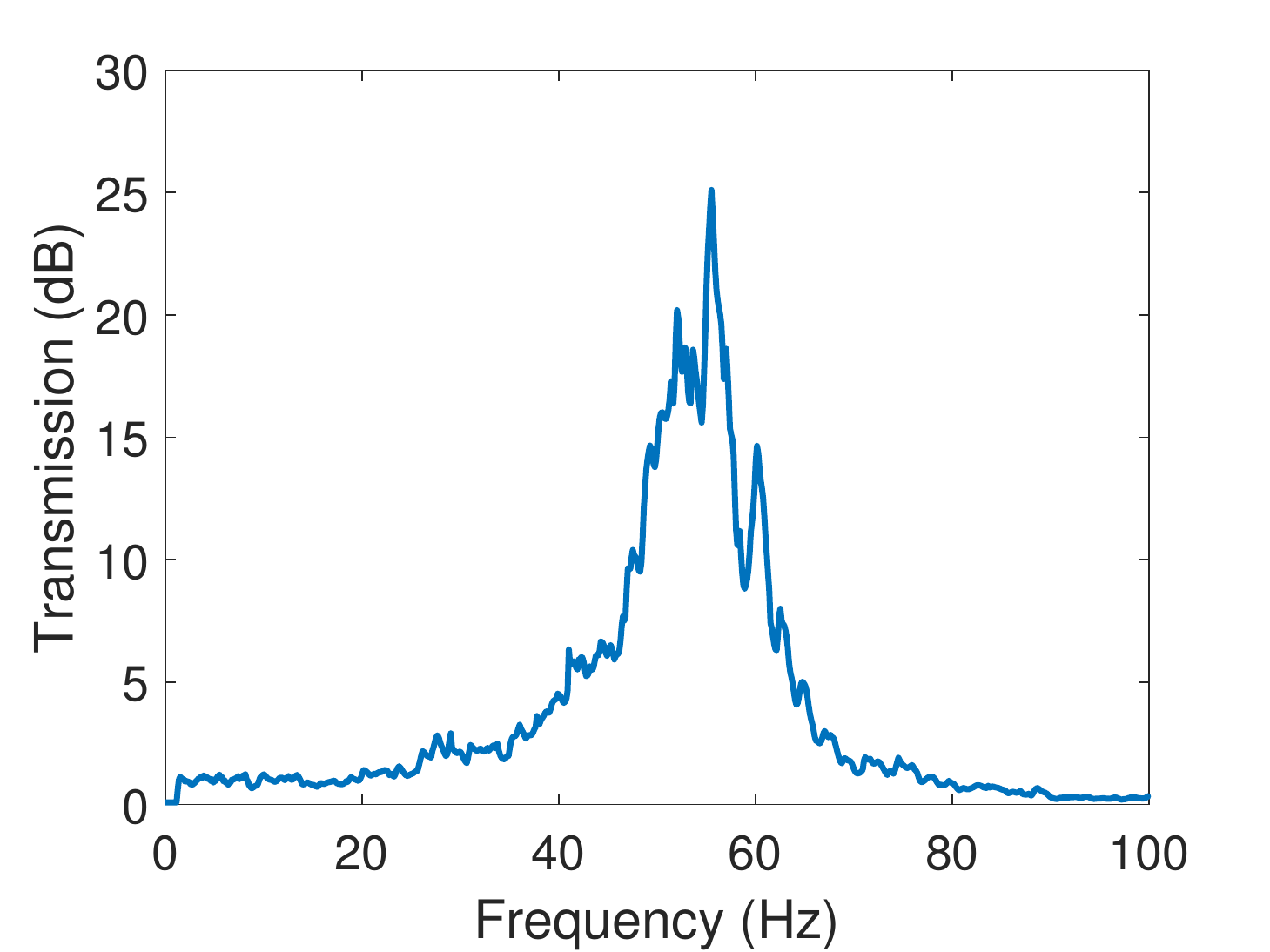}
    \caption{Left: Spectral content of the vertical signal from the quadrant photodetector with light on detector (yellow) and with small induced vibrations on the minus-K platform (purple). Right: A dummy mass equal in weight to the 300-mm cavity resting on four viton spheres forms a damped resonator and shows a clear increase of vibrations at the mechanical resonance frequency. }
    \label{fig:VTT_ULE30cm_quadDet_FFT}
\end{figure}

Fig.~\ref{fig:VTT_ULE30cm_quadDet_FFT} (left) shows the spectral content at low frequencies of the vertical component of the position sensor signal. 
The narrow peaks are electrical interference, but with light on the detector (yellow) also some broader spectral features can be seen at 93~Hz and 127~Hz that could indicate cavity vibrations relative to the platform. 
When the minus-K platform is tapped lightly (purple), the signal increases strongly particularly around 34~Hz, 55~Hz, 93~Hz and 130~Hz. 
The increase at 55~Hz corresponds well with the expected resonance of the damped mechanical resonator formed by the cavity resting on the four viton spheres. 
On the right in Fig.~\ref{fig:VTT_ULE30cm_quadDet_FFT} is the measurement of vibrations of a dummy mass equal in weight to the cavity and resting on similar viton spheres. 
A huge increase in vibrations at the resonance can be seen, which is in good agreement with the quadrant detector spectrum. 
The other resonance frequencies seen by the detector are more difficult to pinpoint, but they are also most likely related to problems with the cavity support, which is explained in more detail in subsection \ref{subsec:FF_resonance_VTT} below.

\subsection{Transfer functions of active vibration isolation systems}

Adding a correction to a commercially available active vibration isolation system requires the ability to modulate the acceleration with an external signal. CNRS has measured the transfer function of the \textit{AVI-200-M} (\textit{Table Stable}) actuators. For this measurement an optical table and an extra load laid on the \textit{AVI-200-M} elements (total load of $\sim200$~kg). Acceleration are induced using dedicated connections on the system while a seismometer (\textit{40T, Güralp}) measure the response. This test is performed without a complex experiment (without cryostat or rigid helium tubes for example) on the table, the center of mass is close to the optical table (about 10~cm) and mass is uniform. A FFT analyzer send a chirped sinewave signal to the actuator and analyze the signal of the relevant direction from the seismometer.

\begin{figure}[t]
\centering
    \includegraphics[width=1\textwidth]{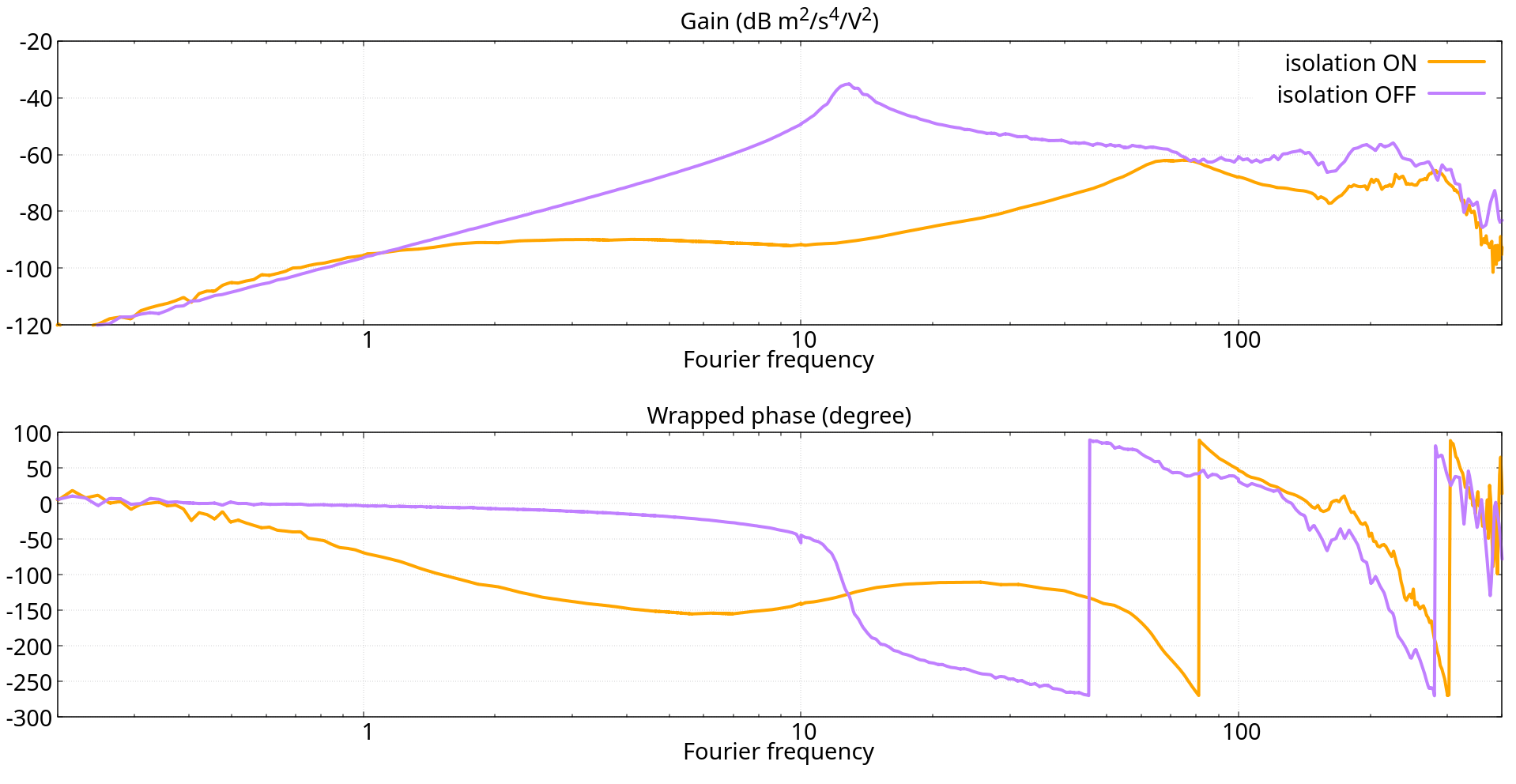}
    \caption{Bode diagram (top gain, bottom phase) of transfer function between the vertical actuators (through input dedicated connectors) and the seismometer \textit{40T Güralp} (vertical axis). The purple curve is the transfer function when the active isolation is disabled, the orange curve when the active isolation is enable.}
    \label{fig:CNRS_TF_AVI200}
\end{figure} 

A Bode diagram of the transfer function for the vertical axis is shown in Fig.\ \ref{fig:CNRS_TF_AVI200} when the response, in velocity, is converted to accelerations. We observe the effect of the active control (isolation on) with a resonance and the corresponding phase shift close to 10~Hz. The gain at 0.1~Hz is $-120 \textrm{\,dB m}^2/\textrm{s}^{4}/\textrm{V}^2$ with a slope of $+30$~dB per decade leading to a dynamic of 90~dB when isolation is enable. When the isolation is disable the maximum gain is at $\simeq 80$~Hz (close to $-60 \textrm{\,dB m}^2/\textrm{s}^{4}/\textrm{V}^2$). The small gain and the limited maximum voltage allowed to modulate the \textit{AVI-200-M} elements, strongly limit the dynamic of the compensation to reject vibrations at low frequency. Since the identified seismometers are sensitive enough and fully operational in this frequency range, the limitation is rising from the actuators of the commercial system. Another way to induce motion of the optical table standing on the \textit{AVI-200-M} elements with a larger gain should benefit to the vibration noise by improving control loop.

\subsection{Transfer functions of cavity systems to accelerations}
\label{subsec:vib_transfer}

Several cavity systems have been set up and improved within the NEXTLASERS project and investigated for their sensitivities to vibrations.
For the feedforward method, it is important to understand the dynamical response of the whole setup to external perturbations. 
By artificially disturbing the experimental setup it is determined how specific frequencies and amplitudes of perturbation changes the laser frequency. 
Then, one may correct these changes with, e.g., a feedforward correction to the laser frequency by an acoustic-optical modulator.

The systems that are presented here as examples include a horizontal 10~cm long cavity made of ULE glass at UMK, 
a new 40~cm long room-temperature ULE glass cavity at OBSPARIS,
a horizontal 30~cm long cavity and a 17.6~cm long vertical cavity designed for a transportable clock laser system both at VTT. 
At PTB the vibration sensitivity of a cryogenic silicon cavity with crystalline AlGaAs mirrors operated at 124K \cite{yu23a} was analyzed.
Most of these cavities were designed to reach ultimate performance with projected frequency instabilities in the mid or low $10^{-17}$ range.

\begin{figure}[t!]
    \centering
    \includegraphics[width=0.65\linewidth]{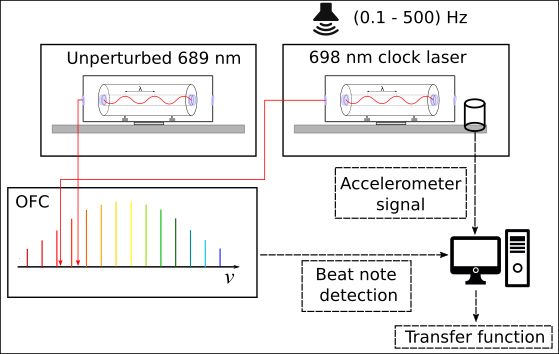}
    \caption{Scheme of the transfer function measurements at UMK. The 698~nm laser cavity breadboard is disturbed by a modified speaker. The measurements are performed by observing the quality of the stability transfer of the 698~nm cavity to an unperturbed 689~nm cavity with an optical frequency comb.}
    \label{fig:UMK_Transfer_function_diagram}
    \vspace{0.5cm}
    \includegraphics[width = 0.7\linewidth]{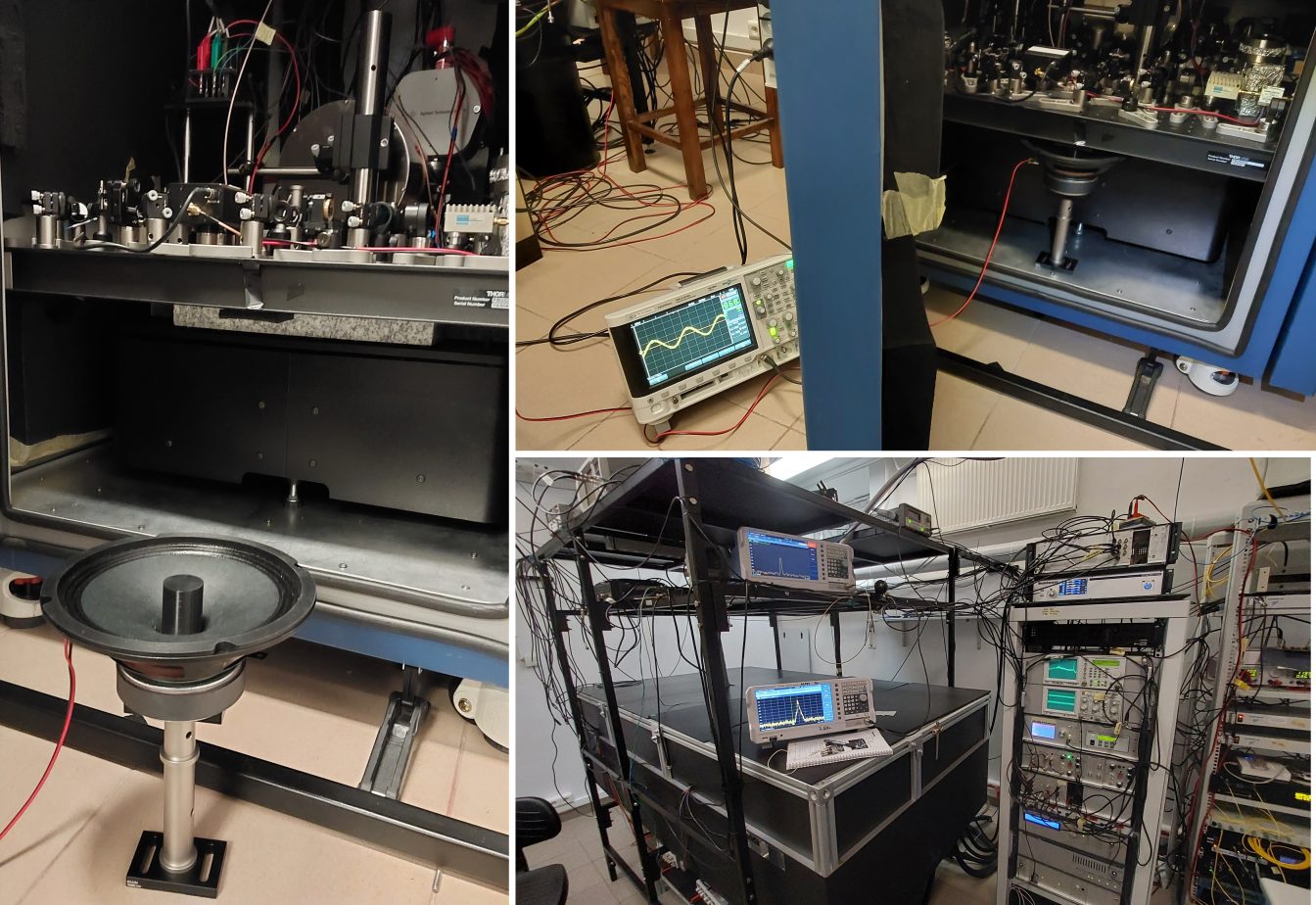}
    \caption{{(Left)} Photo of a 3D-printed post glued to the speaker's membrane, which served as a vibration generator, alongside an opened acoustic box of the 698~nm ultra-stable cavity. {(Top right)} Photo of the ongoing measurements where the speaker is placed below the breadboard of the cavity. The oscilloscope shows a sinusoidal signal recorded by an accelerometer placed on the breadboard. {(Bottom right)} Photo of the spectrum analyzer displaying the detected frequency beatnote in the comb room.}
    \label{fig:UMK_Photo_transfer_fuction_measurements}
\end{figure}

\begin{figure}[b!]
    \centering
    \includegraphics[width = 0.6\linewidth]{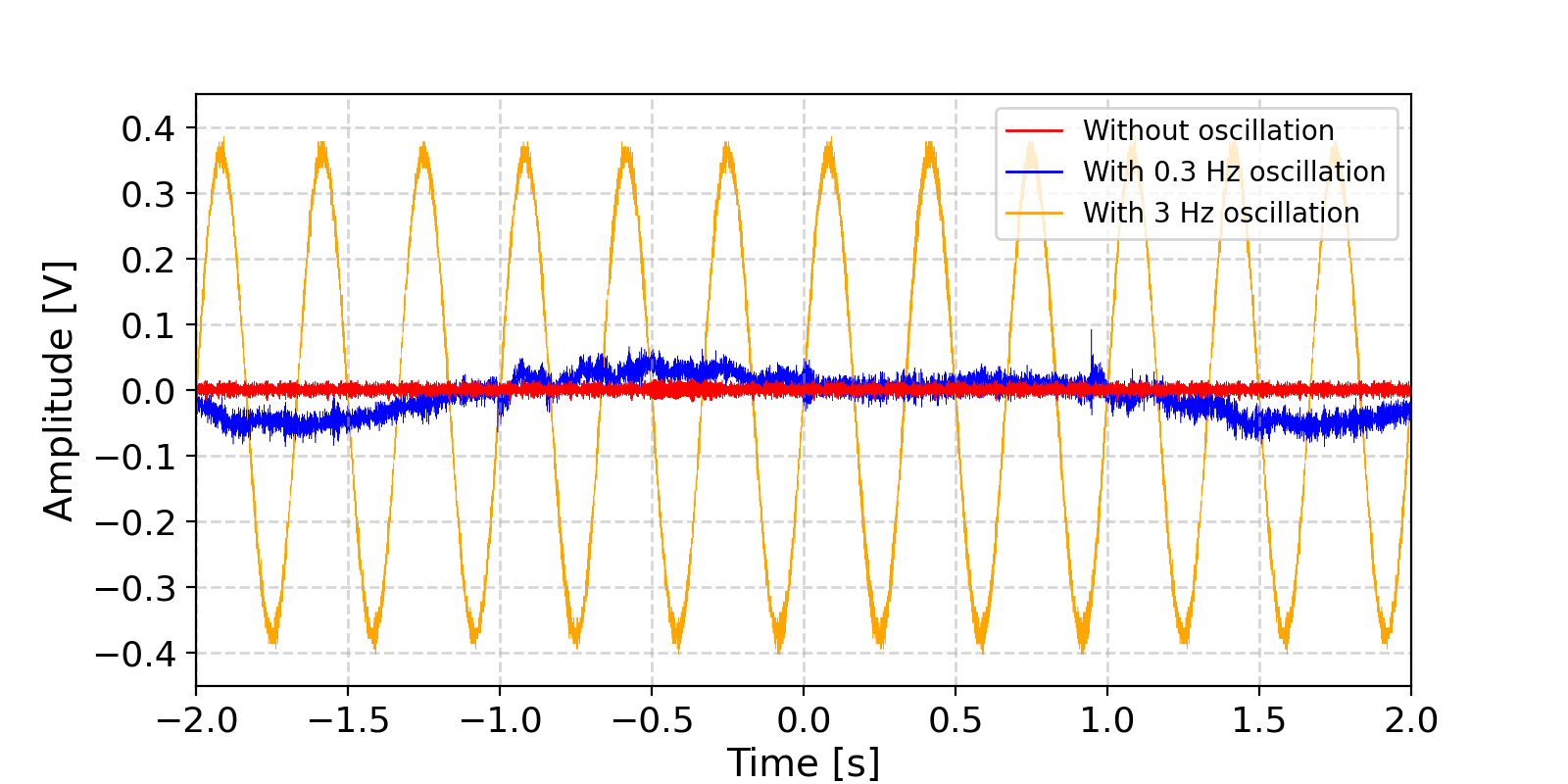} 
    \caption{Two exemplary sinusoidal signals from the accelerometer located on the 698~nm cavity's breadboard, together with the background signal (without oscillation).}
    \label{fig:UMK_Transfer_function_accelerometer}
    \vspace{0.5cm}
    \includegraphics[width = 0.6\linewidth]{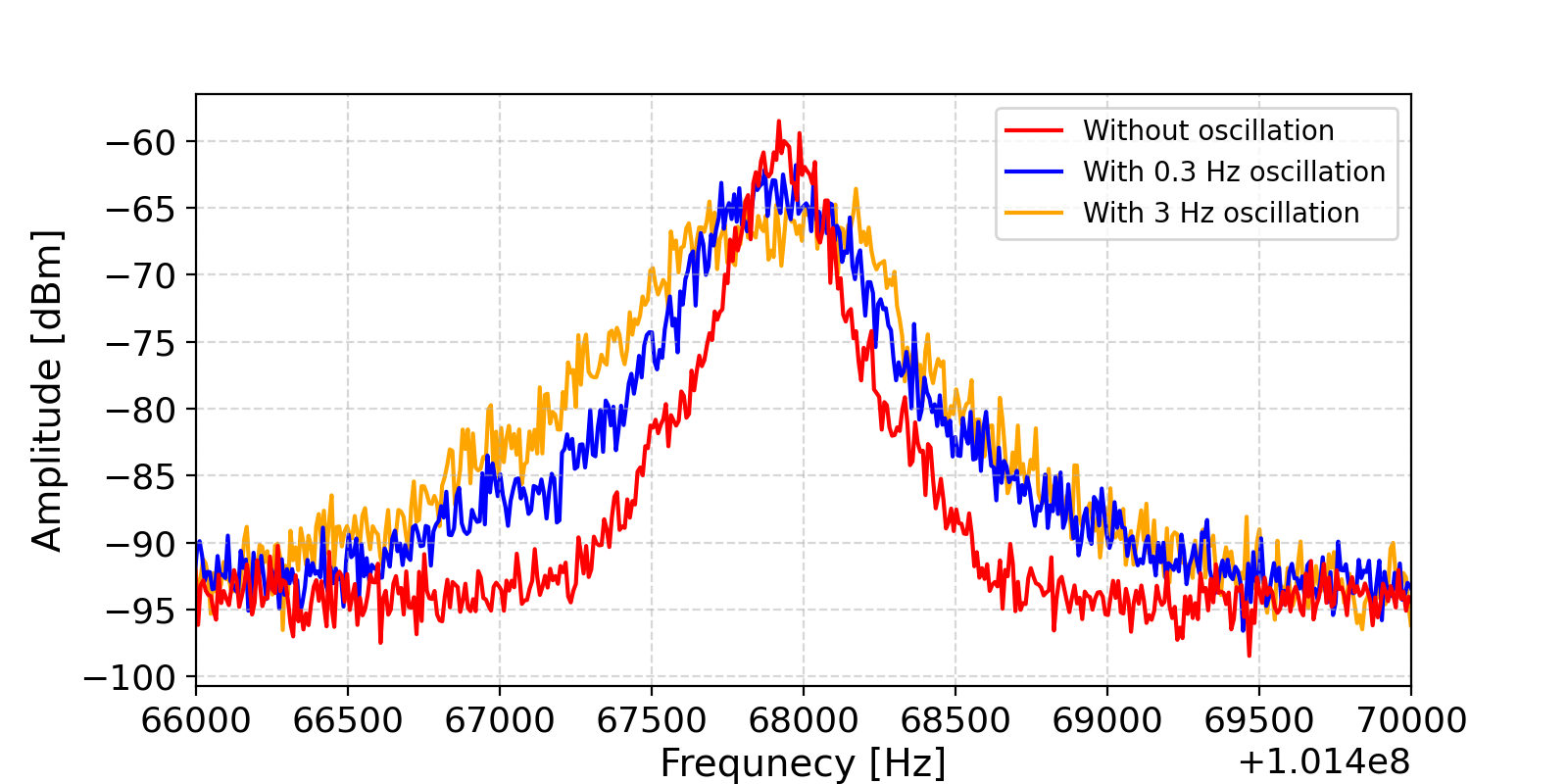} 
    \caption{The influence of 698~nm cavity 0.3~Hz and 3~Hz vibrations on the 689~nm laser beatnote broadening.}
    \label{fig:UMK_Transfer_function_beatnote}
    \vspace{0.5cm}
    \includegraphics[width = 0.7\linewidth]{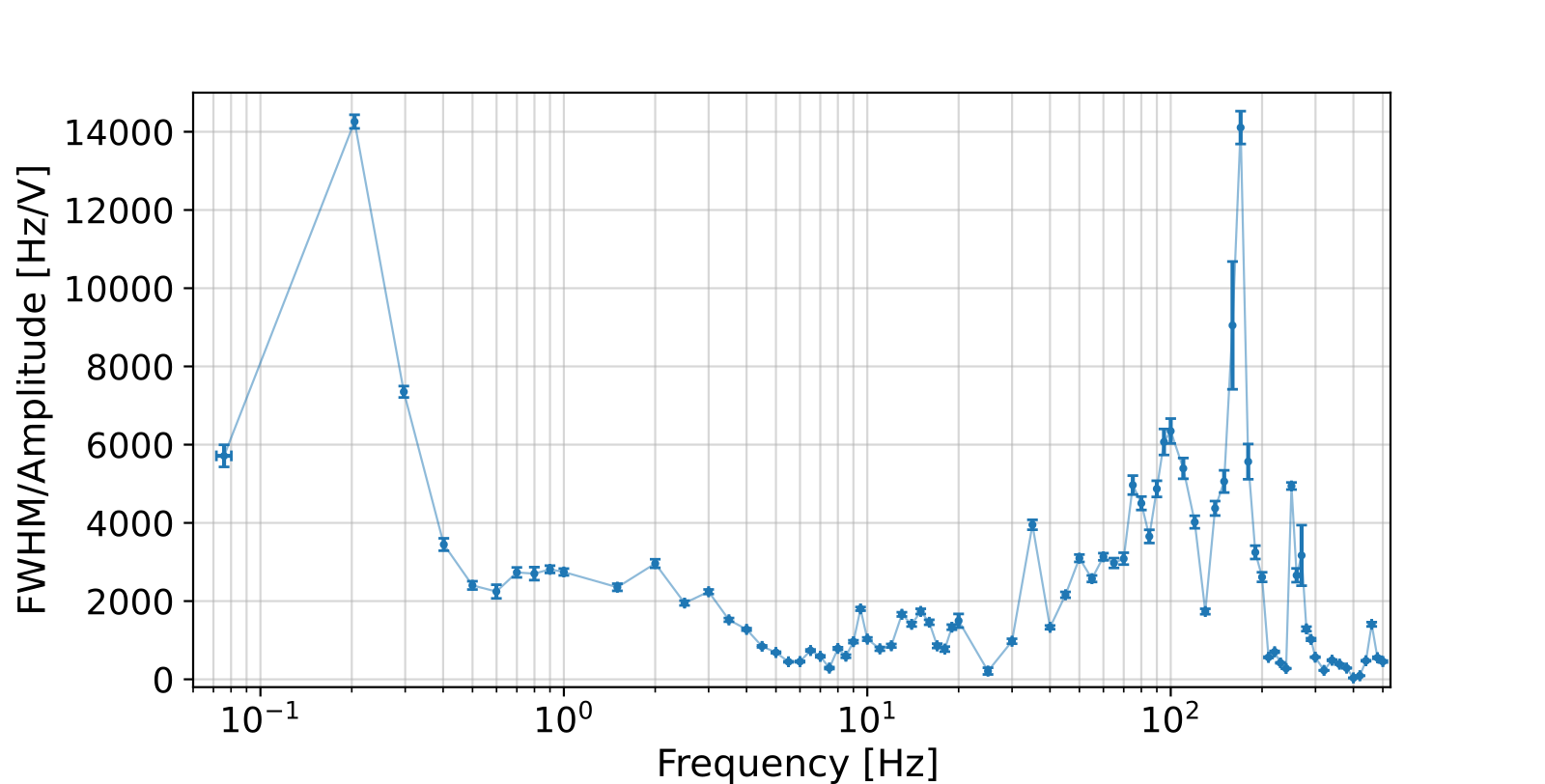} 
    \caption{The results of transfer function measurements as a ratio of a FWHM of the beat note [Hz] signal between unperturbed 689~nm laser and optical frequency comb locked to the perturbed 698~nm cavity, and the accelerometer signal amplitude [V], in the range of 0.1 - 500~Hz.}
    \label{fig:UMK_Transfer_function_Lorentz}
\end{figure}

\subsubsection{Performance of a horizontal 10 cm ULE cavity at UMK}\label{sec:UMK-cavity}
The idea of the transfer function measurement is depicted in Fig.~\ref{fig:UMK_Transfer_function_diagram} for the measurements at UMK. 
Here the response function of a horizontal 10 cm long ULE glass cavity made by Stable Laser
Systems was measured.
A $0.1-500$~Hz periodic and sinusoidal disturbance of the cavity breadboards is induced using a modified Visaton BG 20 speaker. 
A 3D-printed post was glued to the speaker's membrane, and this assembly was then attached to the cavity breadboard. 
The complete measurement setup is presented in  Fig.~\ref{fig:UMK_Photo_transfer_fuction_measurements}.

The induced vibrations are observed with an accelerometer to record the signal amplitude and to verify the quality of the periodic signal (Fig.~\ref{fig:UMK_Transfer_function_accelerometer}).
A 698~nm diode laser is stabilized to the 698~nm cavity. 
Its frequency is compared to a 689~nm laser that is locked to an unperturbed ultra-stable 689~nm cavity via an optical frequency comb for stability transfer (see Chapter \ref{cha:trans}). Mechanical vibration of the 698~nm cavity results in a beat note signal broadening, visible on a spectrum analyser (see Fig.~\ref{fig:UMK_Transfer_function_beatnote}), which is directly proportional to the absolute value of the transfer function from  vibrations to cavity resonance frequency. 

Fig.~\ref{fig:UMK_Transfer_function_Lorentz} presents the  results of transfer function measurements as a ratio of a FWHM of the beat note signal between unperturbed 689~nm laser and optical frequency comb locked to the perturbed 698~nm cavity, and the accelerometer signal amplitude, in the range of 0.1 - 500~Hz.

\subsubsection{Performance of a horizontal 40 cm ULE cavity at OBSPARIS}\label{sec:OBSPARIS-cavity}

The OBSPARIS metrological chain relies on 1542~nm ultrastable lasers, in order to reference the frequency combs (optical measurements, as well as low-noise microwave generation) and the national optical fiber link. 
A new 1542~nm laser is under construction, it is based on a 40~cm-long ULE glass spacer, featuring crystalline coatings deposited on fused silica substrates, leading to a finesse of 300 000. While its theoretical noise floor is at the mid-$10^{-17}$ level, an intermediate evaluation of the stability against an independent cavity lead to $4.0\times10^{-16}$ at ~1s. This limit is set by vacuum fluctuations and by the lack of acoustic isolation.

Considering the ULE spacer has an aspect ratio very different from 1 ($\sim\frac{1}{4}$), the cavity will experience vibrations that are likely to limit its level of performance. In order to fight this effect, the setup was equipped with two seismometers able to sense the instantaneous velocities around the cavity itself. The chosen strategy is to calibrate the sensitivity coefficients, to calculate on-the-fly the frequency shift resulting from vibrations and to add by feedforward a correction on the 'useful' beam sent towards the frequency combs. This approach aims at avoiding active feedback to the heavy ($~\sim500$~kg) platform supporting the cavity. Assuming vibrations kick in at the few $10^{-15}$ level for a well-isolated Fabry-Perot cavity, both sensitivity coefficients and absolute velocities must be known with 1$\%$ accuracy in order to reach the $10^{-17}$ stability range.

\begin{figure}[t!]
\centering
    \includegraphics[width=1\textwidth]{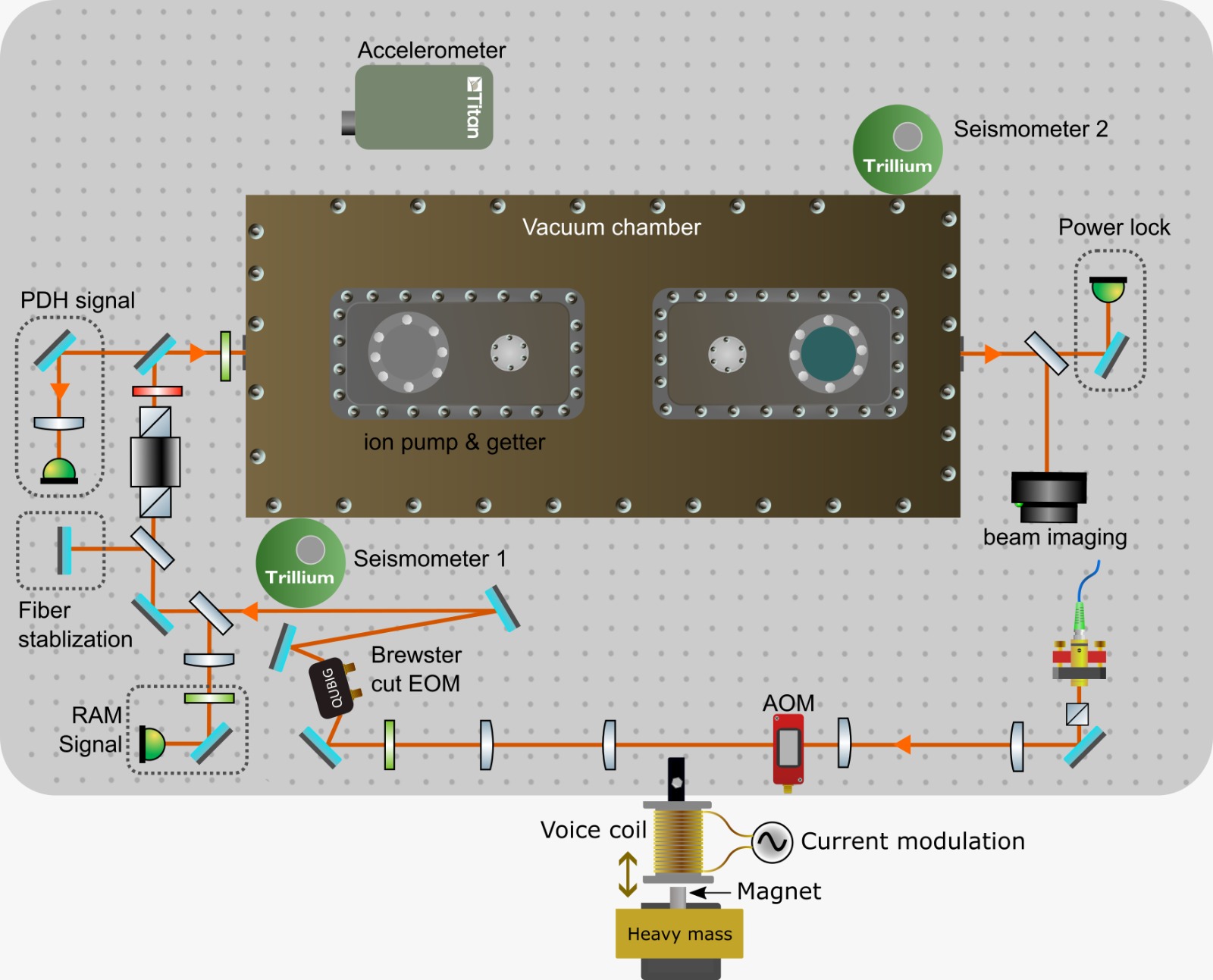}
    \caption{(Left) Diagram of the experimental setup, illustrating the positions of the two seismometers relative to the cavity and the voice coil system used to excite the platform. (Right) Geometry of the system and the configuration of the axes.}
    \label{fig:OBSPARIS_Setup}
\end{figure}

In order to distinguish translations and rotations, both affecting the cavity, we make use of two seismometers oriented towards the same direction, located symmetrically with respect to the mechanical center of the long cavity, but not along its optical axis (Fig.\ \ref{fig:OBSPARIS_Setup}). 
The velocities recorded simultaneously by the two instruments, 
$\vv{V_1}=(v_{1,x}, v_{1,y}, v_{1,z})$ and 
$\vv{V_2}=(v_{2,x}, v_{2,y}, v_{2,z})$, can be reinterpreted as the sum of:

\begin{itemize}
    \item rotations about the axis (x, y, z, crossing at mechanical center of the cavity): 
    $\vv{V_R}=\left(\vv{V_1}-\vv{V_2}\right)/2$
    \item translations along the axis (x, y, z): 
    $\vv{V_T}=\left(\vv{V_1}+\vv{V_2}\right)/2$
\end{itemize}

Translational acceleration is represented by a single vector, while rotational accelerations are assumed to be small enough to be independent one from the another. 
We therefore include in our study Euler accelerations (leading to transverse deformations of the spacer), such as 
$\vv{a_{R,z}}=-\ddot{\theta_z}\cdot\vv{e_z}\times \vv{R}$ 
to describe the effect of rotation about the z axis, where $\theta_z$ is the angle change about the z axis and $\vv{R}$ is the vector between the z axis and seismometer 1.

\begin{figure}[b!]
\centering
    \includegraphics[width=1\textwidth]{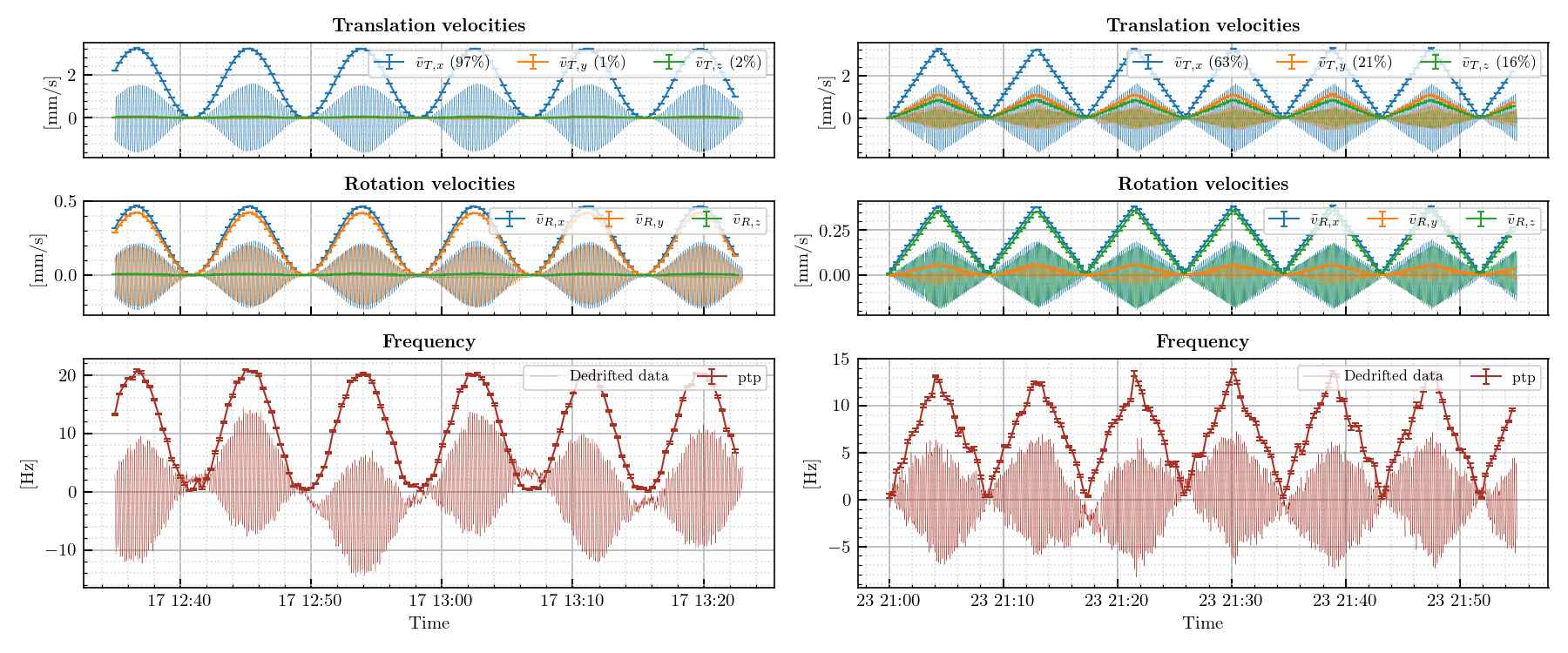}
    \caption{Examples of two configurations of excitations of the platform, with a sinusoidal modulation featuring an amplitude that is itself varying (according to a sinus or a triangle) to test the linearity of the model. (Left) Excitation mostly along the x-axis, (Right) mixed excitation. For each example, both translation and rotation velocities are plotted, along with the peak-to-peak detection done on packets of 20 seconds. The bottom row shows the measured frequency response of the beatnote count. We also show the percentage contribution of each axis with respect to the total excitation.}
    \label{fig:OBSPARIS_preview_modulation_seismos_configurations}
\end{figure}

In order to determine the corresponding sensitivity coefficients, we induce purposely slow sinusoidal accelerations ($f_\mathrm{mod}=$0.1~Hz) to excite the system, with the help of a magnet pulling/pushing a voice coil attached to the optical bench and fed by a modulated current. 
We measure the corresponding translation/rotations together with the resulting variation on the frequency of the cavity. 
Two examples are represented on Fig.~\ref{fig:OBSPARIS_preview_modulation_seismos_configurations}, as to yield distinct configurations enabling the measurements of the 2 vectors 
$\vv{k}=\left( k_x, k_y, k_z\right)$ and 
$\vv{\ell}=\left(\ell_x, \ell_y, \ell_z \right)$ 
to model the impact of motion on the frequency: 

\begin{equation}
   \delta\nu(t)=\vv{k} \cdot \vv{a_T} + \vv{\ell} \cdot \vv{a_R}
   \label{eq:response_freq_acc}
\end{equation}

In order to derive coefficients, we implement peak detectors to derive the resulting modulation amplitudes of both velocities and frequency response. 
The resulting data is then concatenated with results from other excitation configurations and passed through a least-square optimization process to find the coefficients that best solve Eq.\ref{eq:response_freq_acc} and therefore that best cancel the induced frequency modulation. 
Using nine different modulation configurations similar to the cases shown in Fig.~\ref{fig:OBSPARIS_preview_modulation_seismos_configurations}, we derive the optimal $\vv{k}$ and  $\vv{l}$. 
It appears than indeed the use of the two seismometers to take into account both translations and rotations is the most effective, allowing to reject the modulation by a factor $\approx 30$. 
In this case, the first evaluation of the coefficients reads, in $\mathrm{(m/s^2)^{-1}}$:

\begin{table}[h!]
\renewcommand{\arraystretch}{1.2}
\centering
\begin{tabular}{|c|c|c|}
\hline
                         \textbf{$k_x$}           & \textbf{$k_y$}          & \textbf{$k_z$}          \\ \hline
$+4.8 \times 10^{-11}$  & $+1.3 \times 10^{-10}$  & $+2.0 \times 10^{-11}$   \\
\hline
\textbf{$l_x$}           & \textbf{$l_y$}          & \textbf{$l_z$} \\
\hline
$+4.9 \times 10^{-11}$ & $-9.1 \times 10^{-13}$ & $-6.4 \times 10^{-10}$ \\
\hline
\end{tabular}
\end{table}

\begin{figure}[t!]
\centering
    \includegraphics[width=1\textwidth]{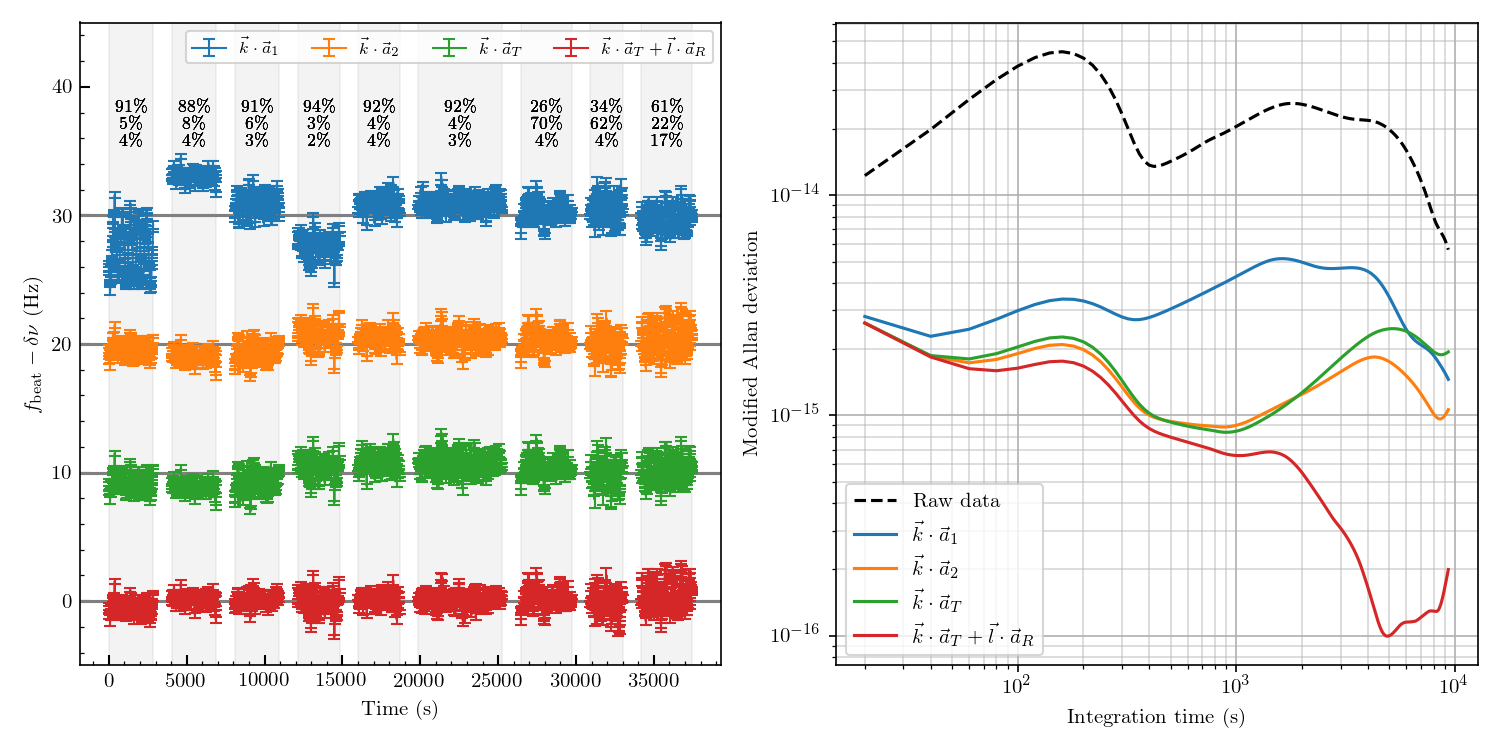}
    \caption{(Left) Distribution of the residual amplitudes of frequency excursions after the optimization of the sensitivity coefficients. 
		Each configuration is highlighted in gray, with the distributions of the acceleration vector components along the seismometer axes shown in percentages. 
		The datasets are labeled according to the specific signals used for optimization and have arbitrary offsets for visualization purposes. 
		Blue: use of seismometer 1 only to fit the data, Yellow: use of seismometer 2 only, Green: with both devices, use of $V_T$ data only, and Red: with both devices, use of both $V_T$ and $V_R$ data. 
		(Right) The corresponding Modified Allan deviation for each dataset. 
		The black curve shows frequency variations of the cavity without correction, while the colored curves correspond to datasets shown on the left. 
		The fit of both translational and rotational data allows us to reject the effect of vibrations by a factor of~30.}
    \label{fig:OBSPARIS_correlation_results_sum}
\end{figure}

While the relative statistical uncertainty of the $\vv{k}$ coefficients dropped to approximately 0.5\% as we added more variations in the excitation configuration, the $\vv{l}$ coefficients remain poorly constrained (around 10\% relative uncertainty). 
Future work will include a better knowledge of the geometrical factors of the setup, and a study of possible drifts of the coefficients over time.

\subsubsection{Performance of a horizontal 300 mm and a vertical 176 mm cavity at VTT}

\begin{figure}[t!]
\centering
    \includegraphics[width=0.25\textwidth]{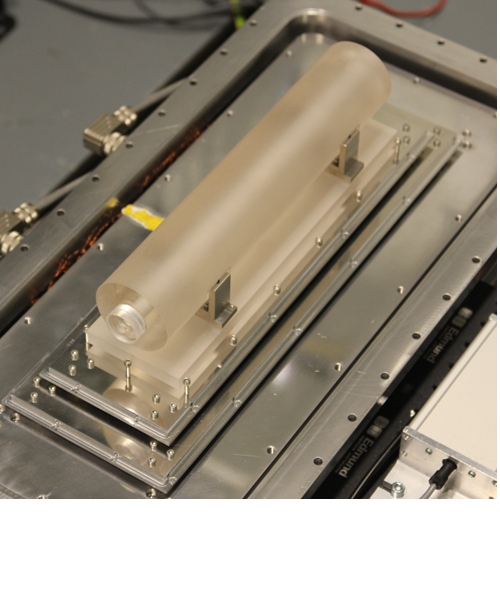}
    \includegraphics[width=0.74\textwidth]{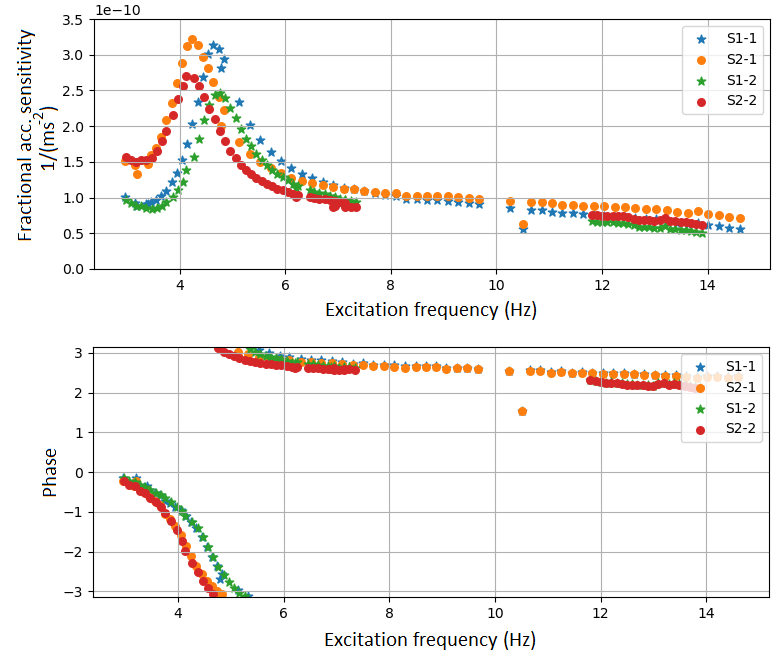}
    \caption{Left: 300-mm ULE glass cavity. Top right: Acceleration sensitivity (amplitude) in the vertical direction measured with two seismometers (S1 and S2) located on either sides of the vacuum chamber. Bottom right: Phase of the frequency shift.}
    \label{fig:VTT_ULE30cm}
\end{figure}

The horizontal, 300-mm ULE cavity at VTT is similar to but smaller than the cavity presented in \cite{hae15a}. 
The cavity is supported by four viton spheres, and one end of the cavity uses a self-balancing mount. 
The vacuum chamber sits on top of a floating minus-K vibration isolation platform, which made accurate determination of the vibration sensitivity tricky. 
Two similar seismometers (Trillium Compact) were placed on different sides, and vibrations were excited by a speaker on top of the vacuum chamber (Fig.~\ref{fig:VTT_ULE30cm_quadDet}). 
For the vertical direction, results are given in Fig.~\ref{fig:VTT_ULE30cm} for two sets of measurements using both seismometers (S1 and S2). 
A strong resonance around 4.5~Hz is evident, and this is very problematic for feedforward corrections as discussed in Section \ref{sec:feedforward_VTT} below. 
There is a significant amount of optics attached to the chamber and many wires for the vacuum system, AOMs, EOM and detectors and, unfortunately, the cause of this resonance could not be determined. A possible reason is a resonance of the minus-K platform, which causes a coupling of the axis, distorting the measurement. After all, the cavity and the seismometers are situated in different places and do not always see similar acceleration, e.g, when the platform is rotating. 

\begin{figure}[t!]
\centering
   \includegraphics[width=0.5\textwidth]{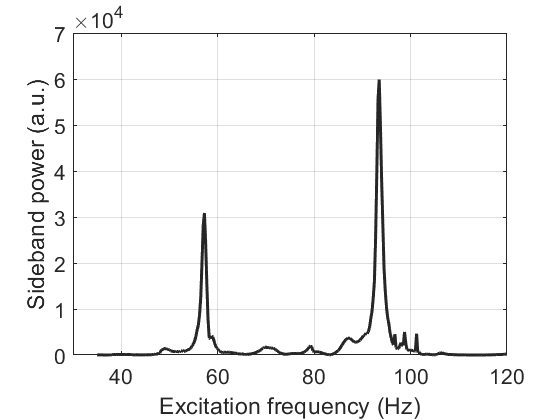}
   \caption{Laser sideband power at the excitation frequency of the loudspeaker. Strong resonances at 57 and 93~Hz indicate problems with the cavity support structures.}
   \label{fig:VTT_ULE30cm_HF}
\end{figure}

\begin{figure}[b!]
\centering
    \includegraphics[width=0.18\textwidth]{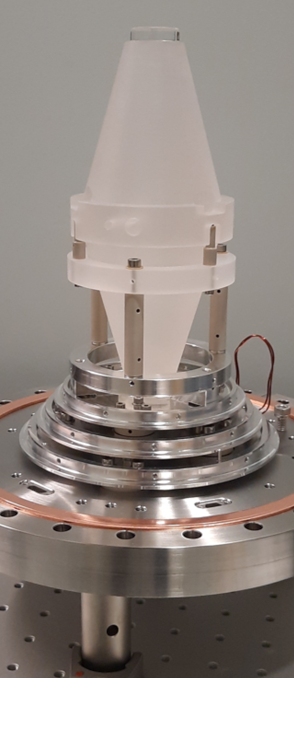}
    \includegraphics[width=0.71\textwidth]{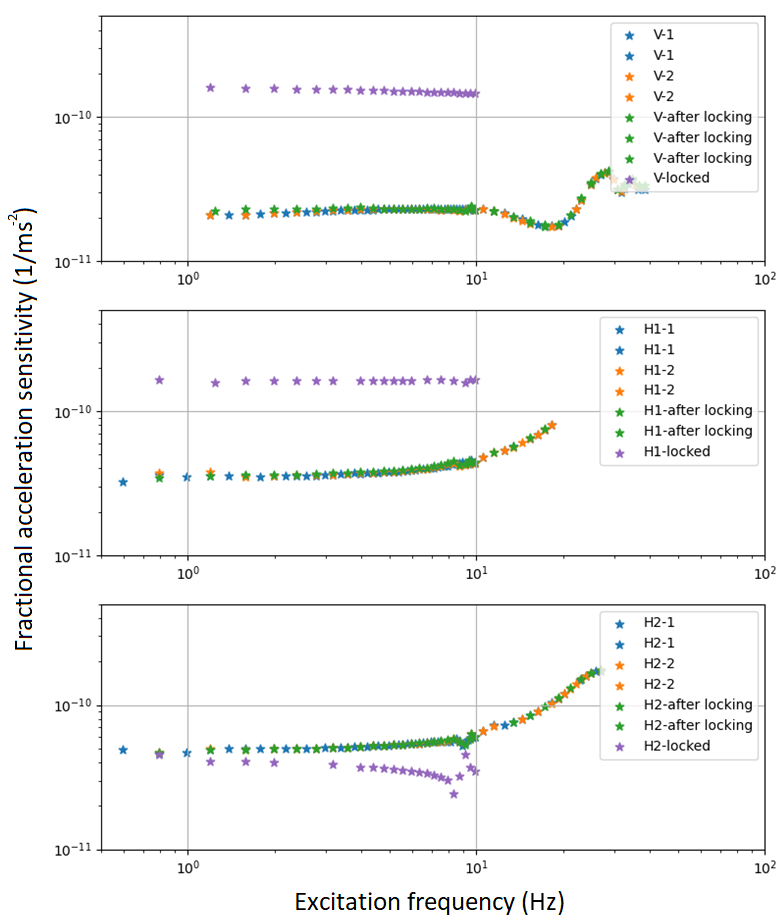}
    \caption{Vertical 176-mm ULE cavity at VTT and three sets of measurements of the fractional acceleration sensitivities in the vertical (V) and horizontal directions (H1 and H2). Also included are measurements when the cavity is locked for transportation (purple).}
    \label{fig:VTT_ULE18cm}
\end{figure}

Beyond 15~Hz, the acceleration sensitivity starts rising again, but the rising values should be treated with caution since the coupling between the axis is also getting stronger. At still higher frequencies, now looking only at the power in the generated sidebands (Fig.~\ref{fig:VTT_ULE30cm_HF} right panel) two very strong resonances appear at 57~Hz and 93~Hz. The first one corresponds well with the expected mechanical resonance of the damped mechanical oscillator formed by the cavity resting on the viton spheres as discussed earlier. 
The 93~Hz resonance is also evident in the quadrant detector signal (Fig.~\ref{fig:VTT_ULE30cm_quadDet_FFT} left) and is most likely another resonance of the cavity mounting, indicating that the mounting should be improved in the future.

The second cavity at VTT is a vertical 176-mm ULE glass cavity that has a manual locking mechanism for transportation (Fig.~\ref{fig:VTT_ULE18cm}). This cavity rests on three viton spheres, and the vacuum chamber is on top of a TS-140 active vibration isolation platform. In this case vibrations could easily be induced in all three directions by applying sinusoidal voltages to the external modulation input ports of the platform. A Trillium Compact seismometer was used to determine the resulting acceleration. Fig.~\ref{fig:VTT_ULE18cm} shows four sets of measurements for the vertical (V) and the horizontal (H1 and H2) directions. As expected, the sensitivity was degraded when the transportation lock was used (purple); however, upon release, the sensitivity returned back to the prior values. The increase toward higher frequencies could be due to an approaching mechanical resonance or due to increased coupling between the axis.

\subsubsection{Performance of a vertical 21 cm cryogenic silicon cavity at PTB}
\label{sec:PTB-silicon-cavity}

\begin{figure}[tbp]
\centering
    \includegraphics[width=0.32\textwidth]{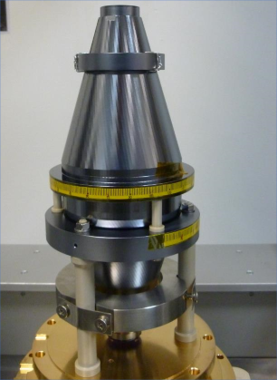}
    \includegraphics[width=0.67\textwidth]{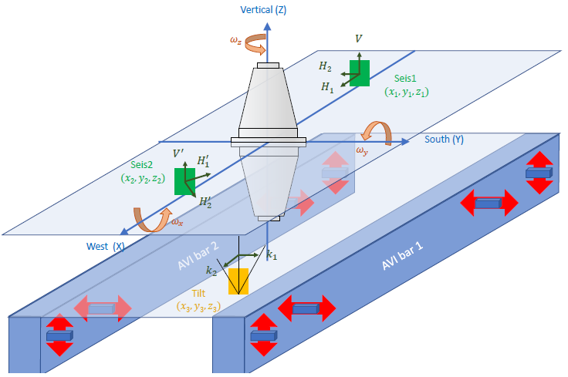}
    \caption{Left: Single crystal silicon cavity of PTB on the tripod support frame. Right: Schematic view of the cryogenic silicon cavity on the active vibrations bars with two seismometers and a tiltmeter: each AVI bar has four actuators with their orientations marked with red arrows. The two Trillium Compact 3-axes seismometers (seis1 and seis2) are mounted at position ($x_1$, $y_1$, $z_1$) and ($x_2$, $y_2$, $z_2$), a tiltmeter with two axes $k_1$, $k_2$ is mounted at position ($x_3$, $y_3$, $z_3$).}
    \label{fig:PTB_Si5}
\end{figure}

PTB is operating two single-crystal silicon resonators at 124 K. The first one (Si2) utilizes conventional dielectric coatings \cite{mat17a}, the second one (Si5) is equipped with crystalline AlGaAs coatings \cite{yu22, yu23}. 
Both systems are working at a wavelength of 1542~nm. 

To characterize the new cavity system Si5 with respect to vibrations and rotational movements, two Trillium compact seismometers and a 2-axes Lippman tiltmeter as described in  section \ref{sec:Sensors} are used. 
The knowledge of the vibration sensitivities is crucial for the study of active vibration control and the realization of feedforward techniques. 
Both techniques are mandatory to mitigate vibrations below $1 \cdot 10^{-17}$ for averaging times <~1~s.  

The shape of the vertical cavities and their support frame (Fig.\ \ref{fig:PTB_Si5}) are optimized to realize minimal vibration 
The tripod support frame takes advantage of the three-fold symmetry of the silicon crystal, as the crystalline [111] direction is parallel to the optical axis of the cavity. 
The vertical vibrational sensitivity was optimized by carefully rotating the resonator around its optical axis \cite{hag13a} . 

\begin{figure}[tbp]
    \centering
    \includegraphics[width=\textwidth]{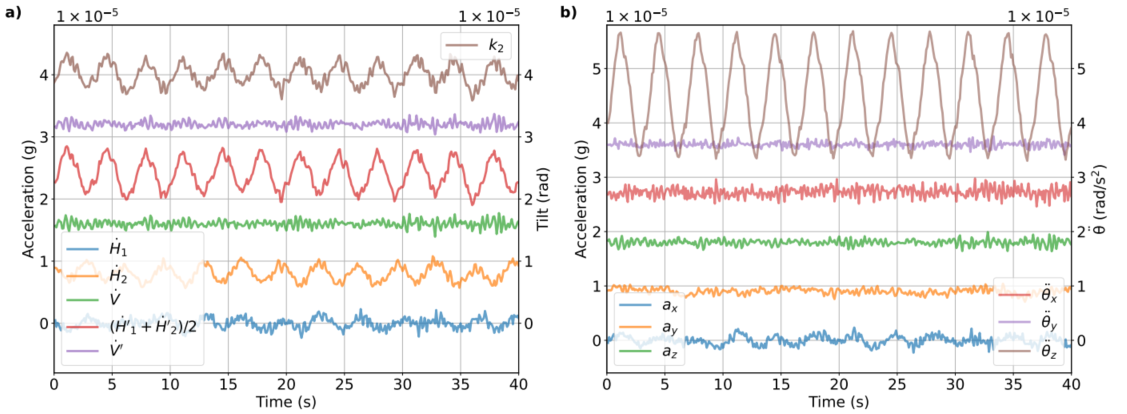}
    \caption{(a) Signals from the two seismometers and the tiltmeter for an excitation frequency of 0.3~Hz. 
    (b) Translational and rotational accelerations of the cavity setup calculated with Eq. \ref{eq:motion_matrix}.}
    \label{fig:PTB_Si5-signals}
\end{figure}

The silicon cavity, including the vacuum chamber and the optics, is mounted on two parallel anti-vibration isolation (AVI) bars (Table Stable AVI-200), such that the mid-plane of the cavity is close to the active plane of the AVI-platform (Fig.\ \ref{fig:PTB_Cryo_Si}). 
With the Table Stable Exciter Box the AVI bars can also be used to accelerate the cavity-setup in different directions. 
As the different axes cannot be excited with very good mutual suppression, the cavity typically exhibit both translational and rotational motions. 
Therefore, two 3-axes seismometers (Trillium Compact) and one tiltmeter (Lippmann) are needed to get complete information about the translational and rotational movements of the cavity. 
Fig.\ \ref{fig:PTB_Si5} gives a schematic overview about the setup. With the coordinates of the seismometers seis1 (-0.15, 0, 0) and seis2 (0.30, 0, 0.07) and the tiltmeter (0.21, 0.19, -0.23) a full-rank matrix containing both vibrational and rotational motions is derived \cite{yu23}:

\begin{equation}
\begin{bmatrix}
a_x-g\theta_y \\
a_y+g\theta_x\\
a_z\\
\ddot{\theta}_x\\
\ddot{\theta}_y\\
\ddot{\theta}_z
\end{bmatrix}
=\frac{1}{\mathcal{R}_\mathrm{seis}}
\begin{bmatrix}
 -1 & 0 & 0 & 0 & 0 & 0  \\
0 & 0.651 & 0 & -0.272 & 0 & 0.077 \\
0 & 0 & -0.667 & 0 & -0.333 & 0  \\
0 & -0.709 & 0 & 2.837 & 0 & 3.546 \\
0 & 0 & -2.222 & 0 & 2.222 & 0  \\
0 & 2.325 & 0 & -1.812 & 0 & 0.512
\end{bmatrix}
\begin{bmatrix}
\dot{H}_1\\
\dot{H}_2\\
\dot{V} \\
(\dot{H}_1'+\dot{H}_2')/2\\
\dot{V}'\\
k_2\cdot\mathcal{R}_\mathrm{seis}/\mathcal{R}_\mathrm{tilt}
\end{bmatrix}
\label{eq:motion_matrix}
\end{equation}

Here, $H_1$, $H_2$, $V$ and $H'_1$, $H_2'$, $V'$ are the output signals of the two accelerometers, $k_2$ the signal for tilt around $x$ of the tiltmeter, and $\mathcal{R}_\mathrm{seis}$ is the response of the seismometer. 
The position of sensors are given by
($x_1$, $y_1$, $z_1$) and ($x_2$, $y_2$, $z_2$). The accelerations and the tilt angles are denoted by $a_i$ and $\theta_i$ with $i \in ( x, y, z)$.
Note, that in general accelerations parallel to the AVI platform (along $x$ and $y$) cannot be distinguished from platform tilt ($\theta_y$ and $\theta_x$) that couple gravitational acceleration $g$ to that directions. 

This matrix can be applied directly to the recorded time traces of the seismometer and tiltmeter when exciting the AVI. 
Fig.\ \ref{fig:PTB_Si5-signals} shows a measurement and the derived cavity motion for an excitation frequency of 0.3 Hz \cite{yu23}.  

\paragraph{Vibrational and rotational sensitivities}\mbox{} \\

\begin{figure}[hbt]
    \centering
    \includegraphics[width=0.9\textwidth]{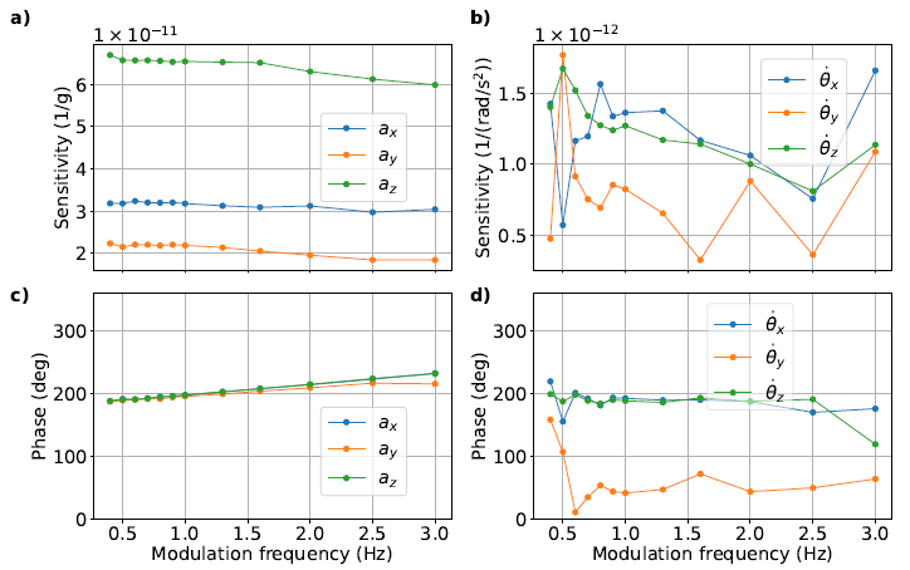}
    \caption{Amplitude and phase of the frequency response of the silicon cavity to translational (a, c) and rotational (b, d) accelerations.}
    \label{fig:PTB_Si5-sensitivity}
\end{figure}

The cavity system was excited with different frequencies in different, predominantly translational and rotational modes. 
From the complex amplitudes of the sensor outputs, the cavity motion was derived according to Eq.\ \ref{eq:motion_matrix}. 
The sensitives of the cavity frequency for the translation and rotational motions are calculated from the frequency data obtained (Fig.\ \ref{fig:PTB_Si5-sensitivity}), 
which gives us the full information of the transfer function. 
For Fourier frequencies below 1~Hz where we expect a frequency independent transfer functions because of the lack of resonances in the cavity mounting. 
The results are summarized in Tab.\ \ref{table:PTB_vib_rot_sen}. 

\begin{table}[!ht]
{
\renewcommand{\arraystretch}{1.2}
\centering
\begin{tabular}{c|c|c}
\hline
Motion & Sensitivity & Phase \\\hline
$a_x$ & $(3.1\pm 1 )\times 10^{-11}$/g & $(194\pm 5 )^\circ$ \\\hline
$a_y$ & $(2.1\pm 1 )\times 10^{-11}$/g & $(192\pm 4 )^\circ$\\\hline
$a_z$ & $(6.5\pm 1 )\times 10^{-11}$/g & $(193\pm 5 )^\circ$\\\hline
$\ddot{\theta_x}$ &  $(1.4\pm 0.6 )\times 10^{-12}\mathrm{/(rad/s^2)}$ & $(198 \pm 54)^\circ$\\\hline
$\ddot{\theta_y}$ &  $(0.9\pm 0.6 )\times 10^{-12}\mathrm{/(rad/s^2)}$ & $(53 \pm 89)^\circ$\\\hline
$\ddot{\theta_z}$ &  $(1.4\pm 0.4 )\times 10^{-12}\mathrm{/(rad/s^2)}$ & $(189 \pm 59)^\circ$\\\hline
\end{tabular}
\caption{Vibrational and rotational sensitivities below 1 Hz excitation frequency of the PTB silicon cavity at 124 K.}
\label{table:PTB_vib_rot_sen}
}
\end{table}

\section{Low-frequency active vibration control} \label{sec:active-vibration-control}

The need for active control of low-frequency vibrations occurs in various fields like gravitational wave detection or atom interferometers. Systems that augment passive platforms with additional active control have been used e.g. for gravimeters \cite{oon22}. 
Here we present additional suppression of low frequency vibrations by augmenting a commercial AVI system used for the silicon cavity Si5 at PTB. 
Compared to using a passive system this approach has the advantage, that it starts already with a stiff system, which is helpful in suppressing perturbances that act directly on the platform, like from connections for cooling the system.

\begin{figure}[bt]
    \centering
    \includegraphics[width=12cm]{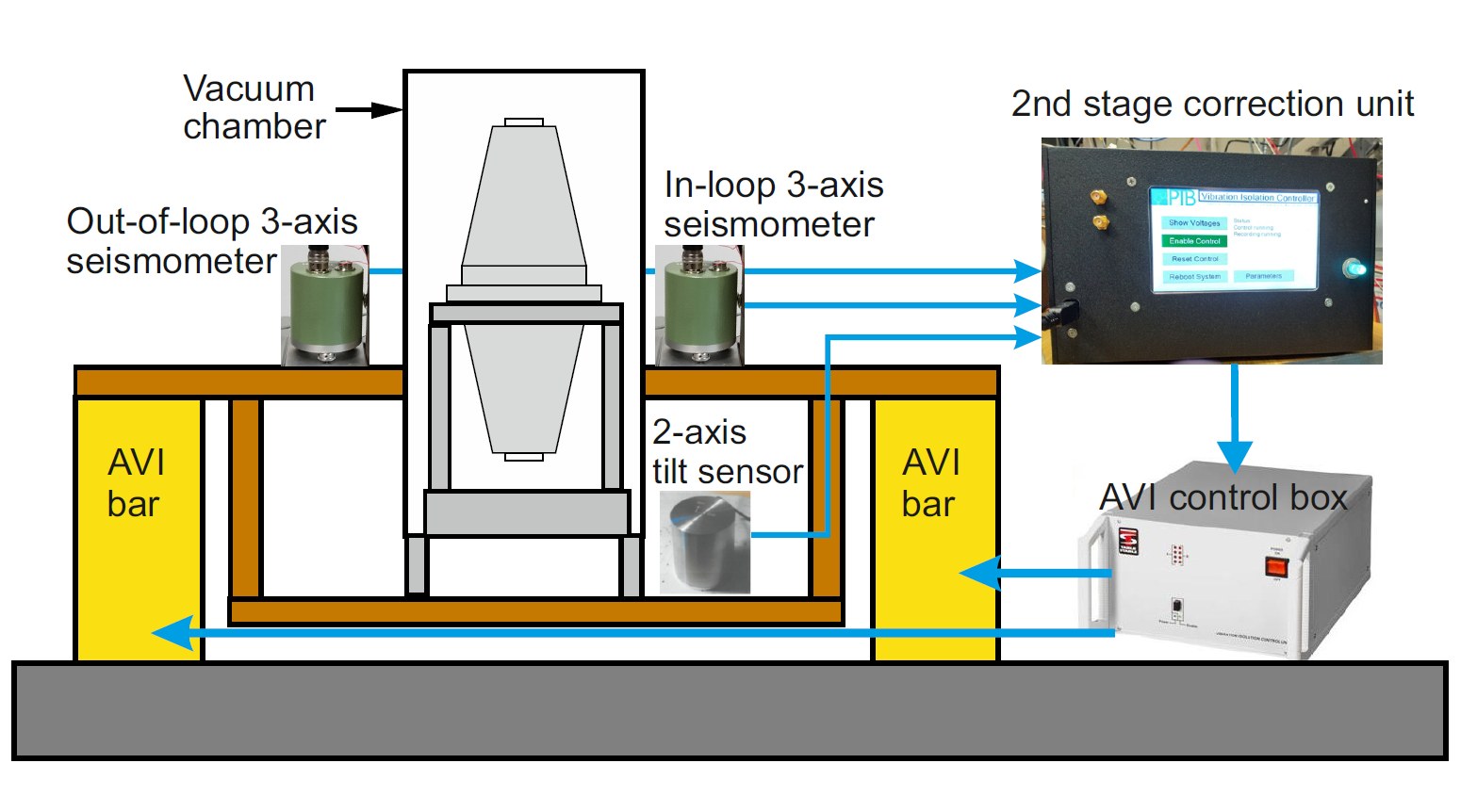}
    \caption{Cryogenic silicon cavity of PTB on the AVI system with the two seismometers and the tiltmeter. 
    Signals from the sensors are processed by a microcontroller-based correction unit and fed back to the AVI bars. 
    }
    \label{fig:PTB_Cryo_Si}
\end{figure}

The setup of the cavity on two AVI bars with two seismometers and one tiltmeter is shown in Fig.\ \ref{fig:PTB_Cryo_Si}. 
A micro-controller based servo electronics (section \ref{sec:hardware_ff}) serves as a second-stage correction unit for reducing the vibration level of the AVI. 
The measured accelerations by one of the seismometers and the tiltmeter are used for feedback via the AVI control box. 
In addition, the signals from the second seismometer are monitored as out-of-loop information.  

\begin{figure}[htbp]
    \centering
    \includegraphics[width=15cm]{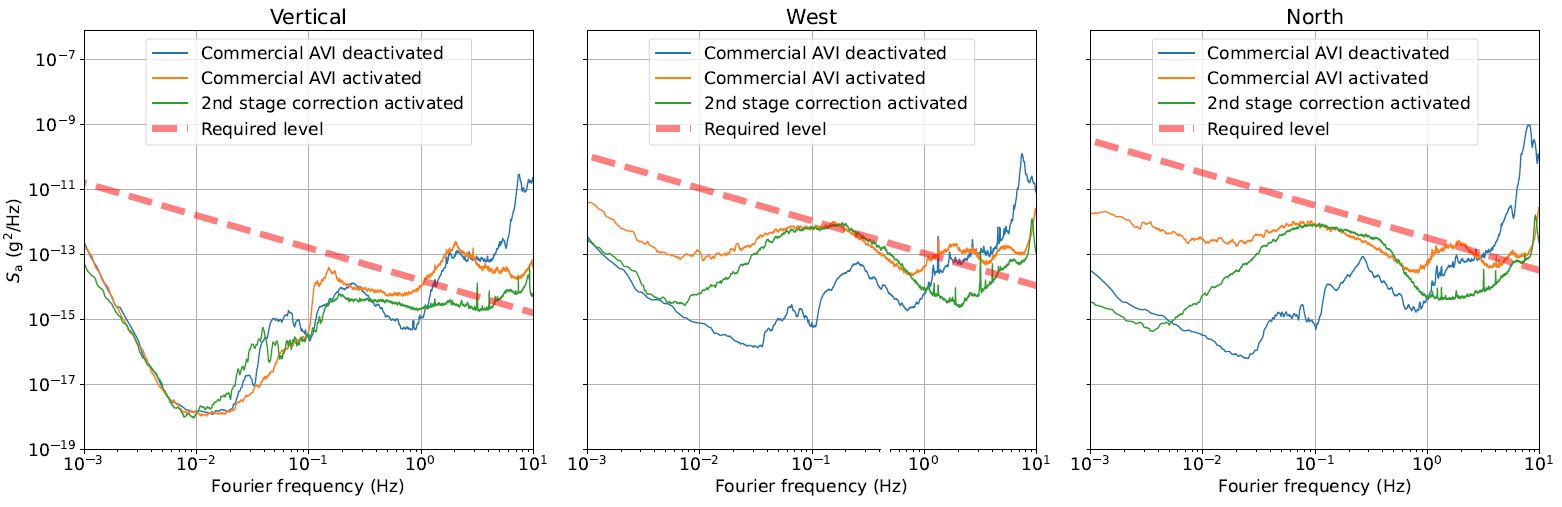}
    \caption{Vibration spectra measured close to the Si cavity. The required level for realizing $1\times 10^{-17}$ fractional frequency stability is given by the dashed red line. The vibration level is measured without any active isolation (AVI deactivated), with the commercial AVI activated and with additional vibration isolation by the second stage correction unit.}
    \label{fig:PTB_Feedback_Si}
\end{figure}

To suppress noise at low frequencies, where the commercial AVI is not effective, the second-stage correction unit \cite{kaw21} provides additional correction signals to the actuators of the commercial, activated AVI-system. 
The unit implements Infinite Impulse Response (IIR) filters for the measured acceleration and tilt signals to provide feedback servo signals to the AVI to correct for vibrations and tilts for Fourier frequencies <5~Hz (Fig.\ \ref{fig:PTB_Feedback_Si}). 
The sensor signals are high-pass filtered and accelerations are fed-back to the translational motion whereas the tilt is fed back to rotational degrees of freedom.

Figure \ref{fig:PTB_Feedback_Si} illustrates the performance of the combined AVI system. 
The vibration level measured close to the cavity without any vibration isolation (AVI deactivated) exceeds the required level for Fourier frequencies > 1Hz. 
Activating the AVI suppresses the vibrations only for frequencies above 5~Hz. 
The limited performance of the in-built seismometers of the AVI system at low frequencies leads to a considerable increase of horizontal accelerations and tilt that leads to excess horizontal noise around 0.1~Hz.
Up to this frequency the required vibration level indicated by the dashed line is reached. 
For higher Fourier frequencies additional suppression of the effects of vibrations on the optical frequency can be achieved by implementing feedforward corrections on the laser frequency as described in section \ref{sec:Feed_Forward}.

\section{Feedforward techniques}
\label{sec:Feed_Forward}

Feedforward strategies aim to predict vibration-induced frequency fluctuations of the cavity from accelerations sensors and then apply the opposite correction to the laser frequency. This technique is necessary to further improve the performance of active vibration control (see section \ref{sec:active-vibration-control}). It can also be a solution for systems with high weight that are not suitable for reasonable active feedback. The efficiency of the feedforward method depends strongly on a reliable description of the transfer function of the cavity frequency on external accelerations.
Sensitivity measurements of a couple of cavity systems within the project has been discussed in section \ref{subsec:vib_transfer}. 

The feedforward technique requires sufficiently fast control electronics to process the measured signals and make corrections to the laser frequency corresponding to the individual transfer functions of the different cavity systems. 
We have developed and tested microcontroller-based control and data processing devices (section \ref{sec:hardware_ff}) and investigated feedforward methods on the room-temperature ULE-glass cavities at UMK (horizontal 10~cm) and at VTT (horizontal 300~mm and vertical 176~mm). 
We also implemented frequency feedforward at the cryogenic silicon cavity (vertical 210~mm long) at PTB. 

With the feedforward method the sensitivities to vibrations could be reduced by up to a factor of ten for Fourier frequencies up to a few ten hertz. 
The combination with active vibration control will allow a substantial suppression of vibration induced frequency fluctuations of the cryogenic silicon cavity down to $10^{-17}$ also for Fourier frequencies above 1~Hz.

\subsection{Hardware for optimised feedforward correction
\label{sec:hardware_ff}}

The implementation of feedforward techniques required the development of suitable electronic hardware to implement the measured transfer functions in real-time and to correct the laser frequency with high precision. 
To achieve sufficient bandwidth, readout- and output rates of up to 10~kSamples/s need to be realized.
Here we give a few examples of suitable micro controller based systems for realizing the required IIR filter functions and direct digital synthesizer (DDS) to generate the correction signal that were investigated during the NEXTLASERS project. 
 
\subsubsection{Use of digital actuators to compensate frequency excursions at UMK}

\begin{figure}[t!]
    \centering
    \includegraphics[width = 0.6\linewidth]{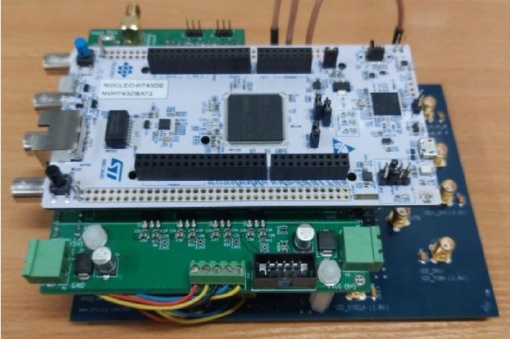} 
    \caption{The RF driver for feedforward correction of  the 698 nm clock laser of the ${}^{88}$Sr optical lattice clocks at UMK}
    \label{fig:UMK_DDS}
\end{figure}

At UMK, the feedforward correction to the 698 nm clock laser of the ${}^{88}$Sr optical lattice clocks is applied by an acousto-optic modulator (AOM). 
The system that drives the AOM consists of a Nucleo-H743ZL2 board with STM32H743ZI micro-controller, AD9912 DDS evaluation board, and a home-made module that provides fast-SPI communication between micro-controller and DDS, an Ethernet port, and programmable analogue and digital ports. 
The module, presented in Fig.~\ref{fig:UMK_DDS}, requires a 5~V power supply and provides all the necessary power for the DDS board. 
If a homemade 1~GHz voltage-controlled crystal oscillator (VCXO) is used as the external reference clock signal, the device may be used as a voltage-controlled generator with digitally tunable centre frequency.

\subsubsection{Testing digital actuators at OBSPARIS}

At OBSPARIS, we have developed a similar system in order to correct the phase of the reference used for stabilizing the light path between the laser and the frequency combs. 
Correcting this phase directly by the inputs yielded by velocities measured by the seismometers allows exploiting the natural integration (accelerations affect the frequency of the laser, velocities affect its phase). 
A DDS9912 with a resolution of 48 bits was prepared as actuator, it is controlled either by an Arduino board whose latency was evaluated to be a few ms at most, or by a Raspberry Pi board featuring a repetition rate $\approx 5 \mu$s and a latency well below 1~ms.

\subsubsection{Using the ARTIQ platform at VTT}

\begin{figure}[t!]
\centering
    \includegraphics[width=0.49\textwidth]{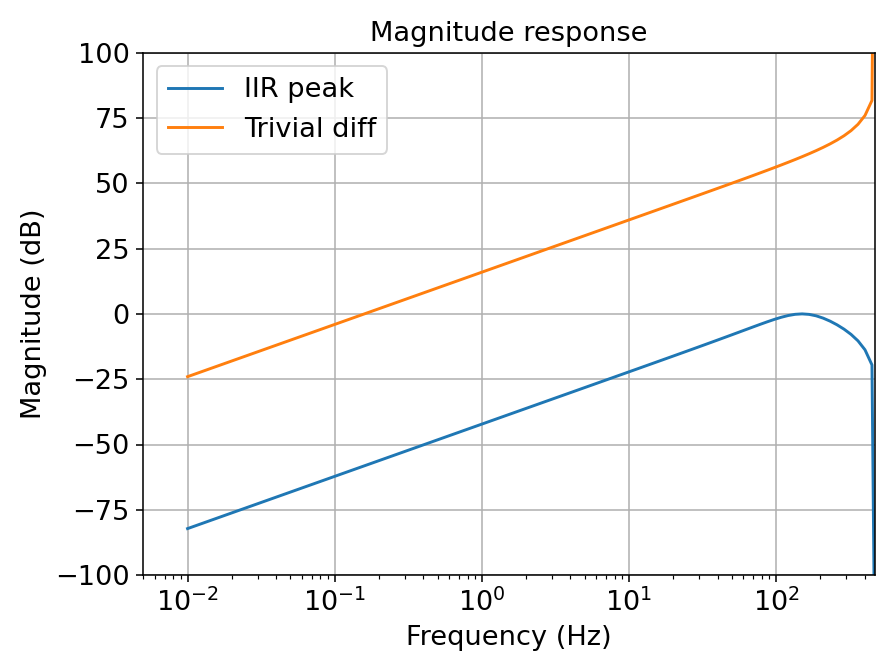}
     \includegraphics[width=0.49\textwidth]{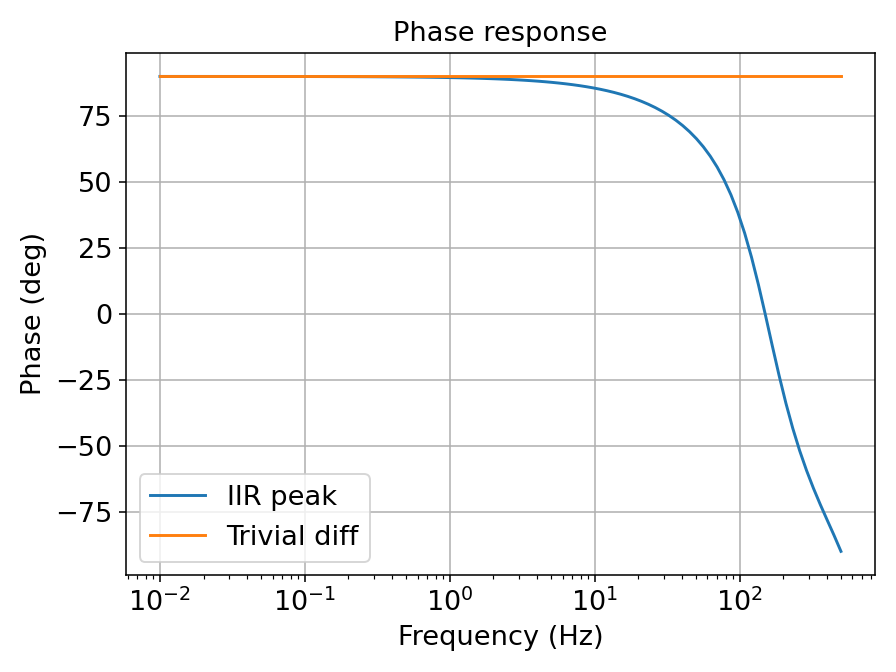}
    \caption{Derivative IIR filter magnitude and phase response when the ARTIQ sample rate is 1~kSample/s.}
    \label{fig:VTT_filter}
\end{figure}

At VTT, the $^{88}$Sr$^+$ single-ion optical clock is controlled using the ARTIQ platform, which is a real-time control system developed for experiments in quantum physics. 
Voltage signals from the seismometer are digitized with the Sinara 5108 sampler and frequency corrections are applied to the drift compensation AOM via the Sinara 4412 direct-digital synthesizer (DDS). 
ARTIQ is programmed by specifying the instant of time when input/output operations are carried out. If the seismometer is read at time $t_0$ and the DDS is updated at $t_1$, then this leaves $t_1-t_0$ time for the calculation. 
The output of the seismometer is velocity while the quantity of interest here is acceleration. Direct derivation is very sensitive to noise and not a working solution for differentiating  seismometer signals. 
For this reason, the derivation was carried out by a second order infinite-impulse response filter. The operations limited the control loop speed to about 1~ms, limiting the possible correction bandwidth to around 10~Hz due to the large phase shifts caused by the filter (Fig.~\ref{fig:VTT_filter}). 
In this work version 1.0 of the ARTIQ FPGA board was used. 
A new version is now available (v.~2.0) that includes an FPGA-based fixed-point unit and a system-on-a-chip that might enable much higher loop rates, increasing the control bandwidth. 

\subsubsection{Digital MIMO controller at PTB}

\begin{figure}[hbt]
\centering
    \includegraphics[width=0.79\textwidth]{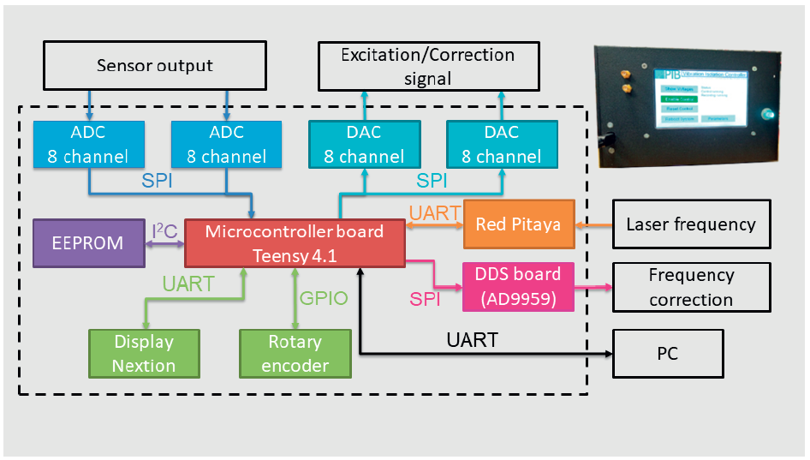}
    \caption{Schematic of a micro-controller based environment for active vibration isolation and frequency feedforward at PTB.}
    \label{fig:PTB_microcontroller}
\end{figure}

PTB has developed and implemented a digital based multiple input / multiple output (MIMO) controller \cite{kaw21}. A schematic of the digital controller is shown in Fig.\ \ref{fig:PTB_microcontroller}. The central element in the system is a Teensy 4.1 micro-controller, based on an ARM Cortex-M7 processor with a clock speed of 600~MHz. 
The performance of the Teensy 4.1  allows the implementation of more complex control algorithms to servo-control the AVI via IIR filter functions in 6 degrees of freedom (section \ref{sec:active-vibration-control}) and in addition perform a simultaneous frequency feedforward with transfer function also implemented by biquad IIR filters (section \ref{sec:Feed_Forward}).

The controller is equipped with 16~ADC input channels with 10~kHz sampling rate and 24~bit resolution for reading the signals from the seismometers and the tiltmeter. 
There are 16 DAC output channels with 10~kHz sampling rate and 16~bit resolution for the feedback to the AVI. 
The controller also contains a StemLabs Red Pitaya\textsuperscript{\textcopyright} System on Chip (SoC) with Field Programmable Gate Array (FPGA) and dual-core processor \cite{RedPitaya} to perform fast high and ultra-resolution frequency measurements, e.g. as needed for determining the transfer function of the cavity (subsection \ref{sec:PTB-silicon-cavity}). 
We have further extended the controller for fast frequency feedforward \cite{and23} by adding a DDS board using the QUAD-DDS AD9959 from Analog Devices to generate the correction frequency for the laser. 

\subsection{Feedforward corrections in combination with passive vibration isolation}
\label{sec:feedforward_VTT}

\begin{figure}[htbp]
\centering
    \includegraphics[width=\textwidth]{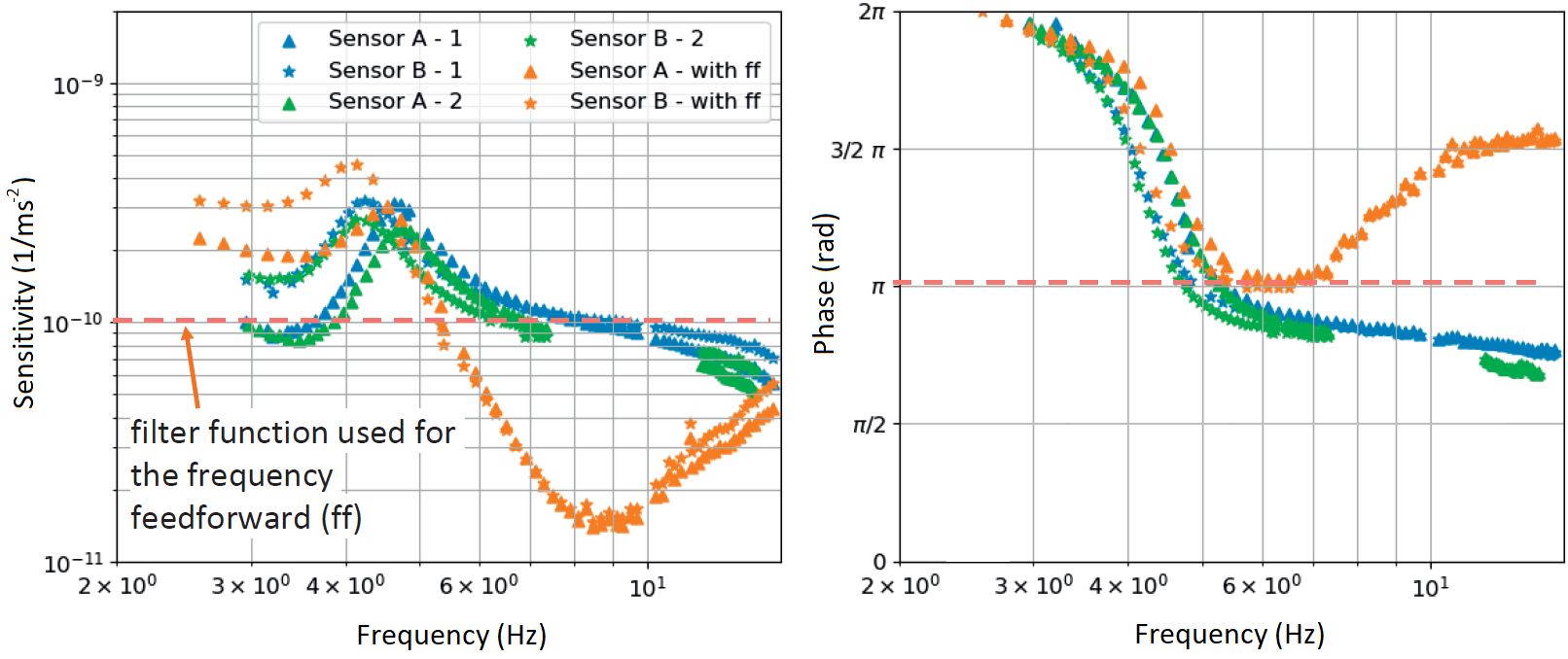}
    \caption{Effect of feedforward (FF) frequency corrections on the fractional acceleration sensitivity of the horizontal 300-mm cavity that is resting on a passive minus-K vibration isolation platform at VTT. Vibrations were induced in the vertical direction via a loudspeaker.}
    \label{fig:VTT_ULE30_FF}
\end{figure}

Contrary to closed-loop control, feedforward correction is sensitive to the accuracy of the correction parameters, since the applied correction should equal exactly the actual shift. 
Mechanical resonances in cavity mounting, therefore, make the required amplitude and phase of the required correction frequency-dependent and an effective feedforward based on platform vibration measurements difficult to realize. 
Consequently, the low-frequency resonance seen in the VTT 30-cm cavity Fig.~\ref{fig:VTT_ULE30cm} is problematic and would require a complicated filter function for correcting it. 
For instance, Fig.~\ref{fig:VTT_ULE30_FF} (brown markers) shows how optimising the feedforward parameters for one frequency (9~Hz in the shown case) increases the vibration sensitivity at lower frequencies close to the resonance. 
It was determined that the possible benefits from seeking an optimized filter function does not outweigh the large amount of work needed; consequently, the development of the feedforward scheme was not pursued further for this cavity.

\subsection{Feedforward corrections in combination with active vibration isolation} 
\label{sec:ff_active}

Based on the developed hardware, the frequency feedforward was combined with commercial AVI systems for Fourier frequencies between 1 mHz and 100 Hz. 

\subsubsection{Feedforward correction tested with a vertical 176~mm cavity at VTT}\label{subsec:FF_resonance_VTT}

The ARTIQ system was also used for testing feedforward frequency corrections to the transportable, vertical 176~mm cavity shown in Fig.~\ref{fig:VTT_ULE18cm}. 
To this end, the active vibration isolation platform was de-activated and external sinusoidal signals were applied to vibrate the cavity. 
The resulting frequency modulation shows up as sidebands in the beat note against another laser. 
The ARTIQ software parameters were then used to minimize the power spectral density in the sidebands.

\begin{figure}[htbp]
\centering
    \includegraphics[width=0.52\textwidth]{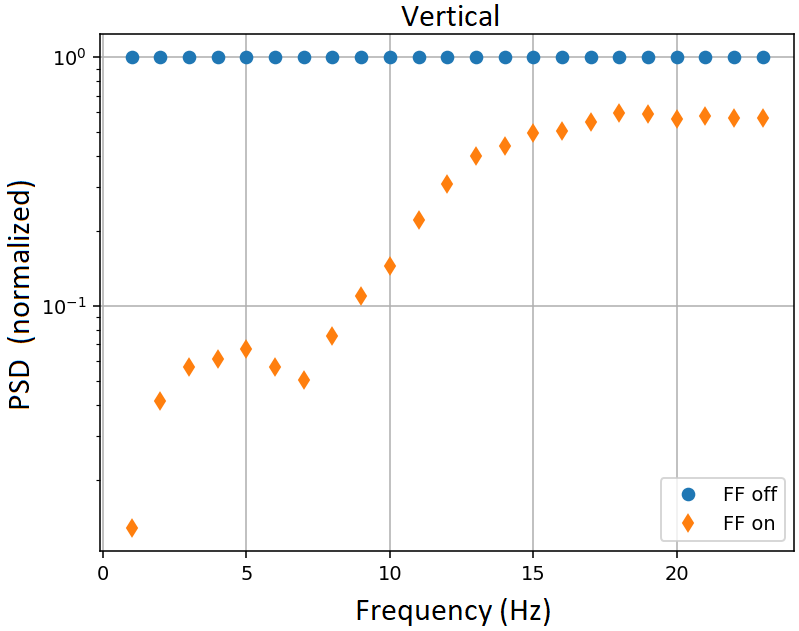}\\
    \vspace{0.5cm}
    \includegraphics[width=0.52\textwidth]{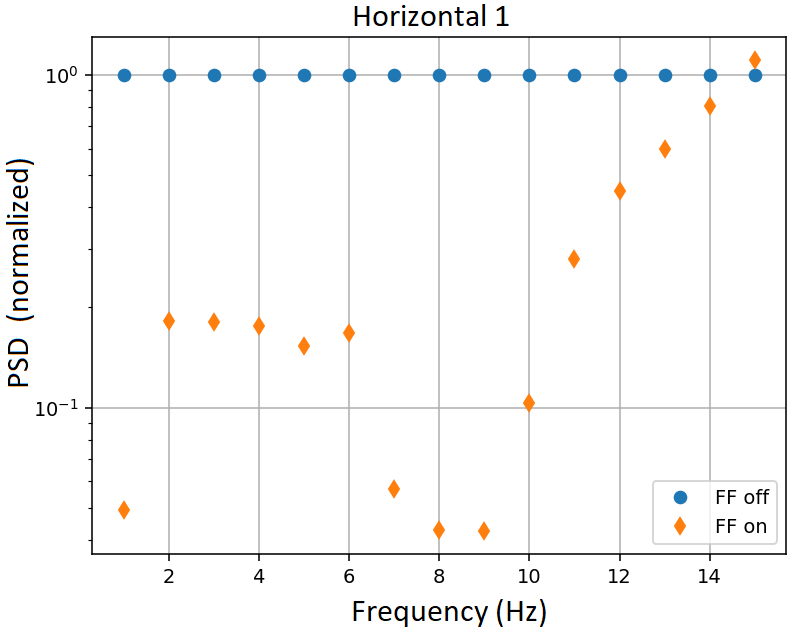}\\
    \vspace{0.5cm}
    \includegraphics[width=0.52\textwidth]{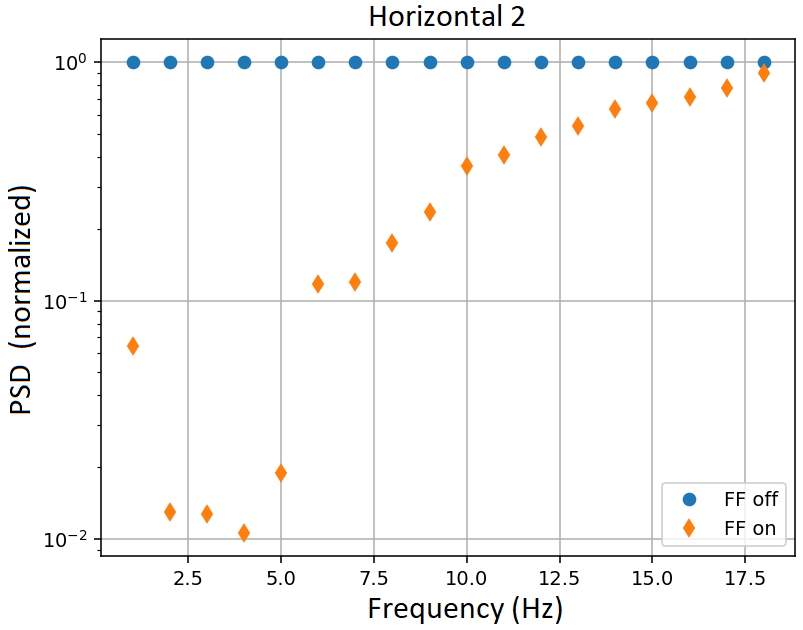}
    \caption{Feedforward frequency corrections to the vertical 176-mm ULE cavity at VTT as a function of platform vibrational excitation frequency. The graphs show the laser sideband power spectral density (PSD) at the excitation frequency with feedforward corrections normalized to the free running case.}
    \label{fig:VTT_ULE18_FF}
\end{figure}

Since the measured acceleration sensitivity does not show any resonances (Fig.~\ref{fig:VTT_ULE18cm}), the feedforward scheme should be easier to implement than for the horizontal cavity (Fig.~\ref{fig:VTT_ULE30_FF}). 
Indeed, results show that in the vertical direction (Fig.~\ref{fig:VTT_ULE18_FF}) and for low-frequencies, a considerable reduction in sideband power can be seen; however, in the "Horizontal 1" direction poorer performance is obtained, possibly due to a minor resonance and/or coupling of the axis due to the fairly light, movable cart that the system is resting on. Since the seismometer is not located at the same position as the cavity, they do not necessarily experience the same acceleration if the excitation is not purely translational. 
In the "Horizontal 2" direction, the results are again quite good except at the lowest frequency (1~Hz), but this value might be compromised by the need to use higher amplitude excitation signals in order to induce a sufficiently strong signal. 

\subsubsection{Feedforward for the 698~nm clock laser cavity at UMK}

\begin{figure}[htbp]
    \centering
    \includegraphics[width = 0.6\linewidth]{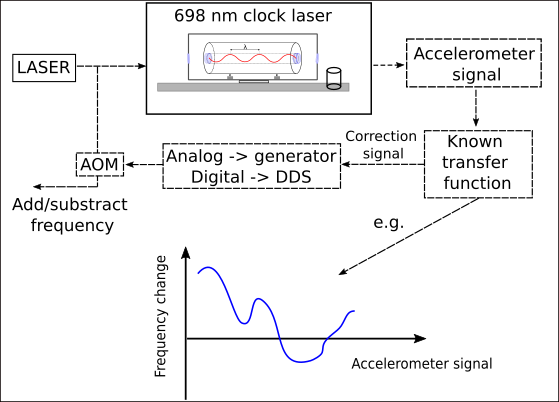} 
    \caption{Schematic diagram illustrating an application of the transfer function to the ultra-stable 698 nm cavity at UMK by sending a feedforward correction to the AOM, which corrects the frequency.}
    \label{fig:UMK_TTF2}
\end{figure}

The transfer function, such as  measured in subsection~\ref{subsec:vib_transfer}, is used to feedforward correction into the clock laser frequency. The Fig.~\ref{fig:UMK_TTF2} present a diagram of feedforward correction system at UMK. The undesirable vibration of the cavity breadboard are measured by an accelerometer. The recorded signal is sent to a microcontroller of the RF driver presented in subsection~\ref{sec:hardware_ff} where the transfer function is applied and a correction is added to the RF signal sent to the AOM.

\subsubsection{Feedforward for the cryogenic silicon cavity at PTB}\label{sec:PTB-feedforward}

\begin{figure}[htbp]
\centering
    \includegraphics[width=0.9\textwidth]{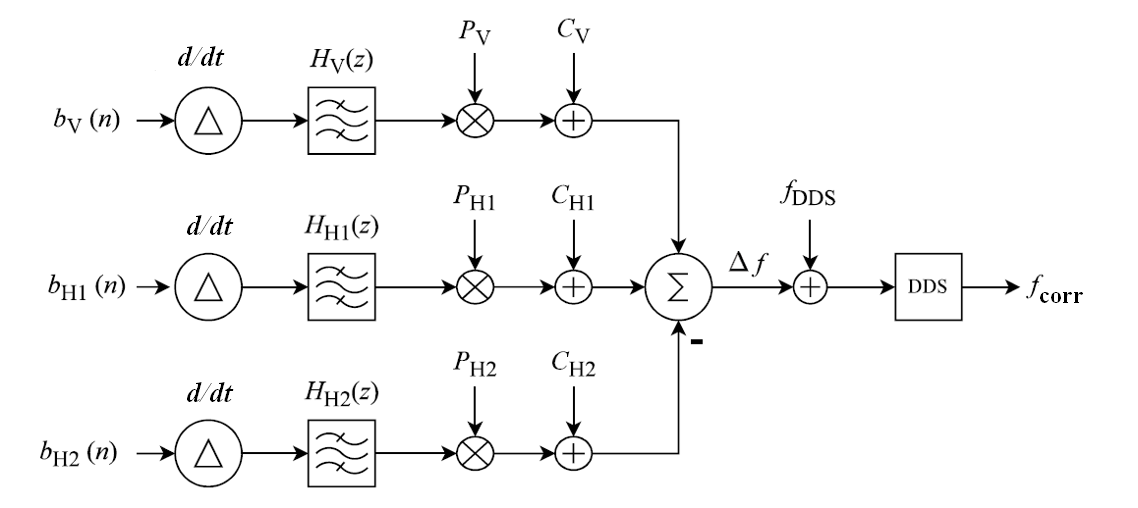}
    \caption{Block diagram of the feedforward correction at PTB. 
    As the output of the employed Trillium compact seismometer is proportional to velocity, the $1^\mathrm{st}$ derivative of the sensor signals ($b_\mathrm{V}, b_\mathrm{H1}, b_\mathrm{H2}$) is calculated before being processed by the individual IIR filters ($H_\mathrm{V}, H_\mathrm{H1}, H_\mathrm{H2}$). 
    After scaling the frequency response by the factors ($P_\mathrm{V}, P_\mathrm{H1}, P_\mathrm{H2}$) the signals are summed to obtain the frequency correction $\Delta f$.
    An offset frequency $f_\mathrm{DDS}$ is added by a DDS that provides the RF signal for driving the AOM to correct the laser frequency.}
    \label{fig:PTB_Feedforward_Filter}
\end{figure}

A first implementation of the frequency feedforward correction was tested, based on the measured transfer function of the silicon cavity (subsection \ref{sec:PTB-silicon-cavity}). 
The block diagram of the feedforward correction is shown in Fig.\ \ref{fig:PTB_Feedforward_Filter}. 
The velocity-proportional signals from the accelerometers are first differentiated to obtain accelerations. 
Then separate digital filters (implementation as biquad IIR filters) for the three directions are applied to model the transfer functions $H_\mathrm{V}$, $H_\mathrm{H1}$ and $H_\mathrm{H2}$ from accelerations to frequency.
The results of the three directions are added together to give the total frequency correction $\Delta f$. 
A center frequency $f_\mathrm{DDS}$ is added, and a DDS finally provides the RF correction frequency $f_\mathrm{corr}$ for driving the AOM. 

\begin{figure}[hbtp]
\centering    
    \includegraphics[width=0.9\textwidth]{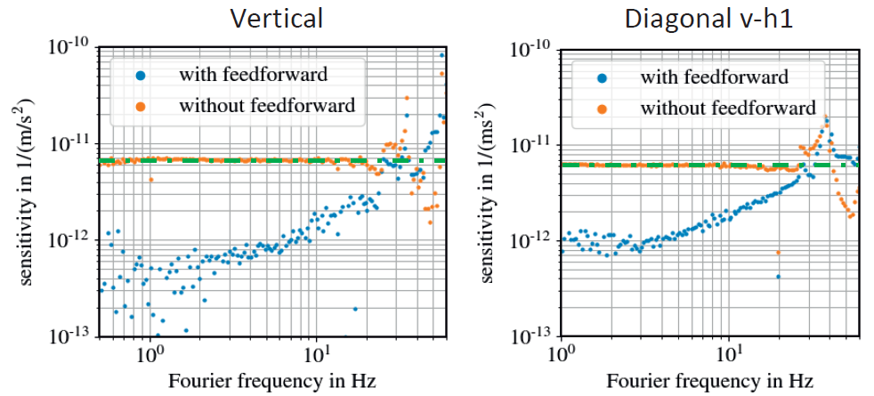}
    \caption{Reduction of the vibration sensitivity of the silicon cavity with feedforward correction in vertical direction (left) and a diagonal between vertical and horizontal direction (right).}
    \label{fig:PTB_Feedforward_Result}
\end{figure}

For the first test, a simple band-pass filter (green dashed line), 10 mHz – 110 Hz, was used. But even with this simple approach, we could demonstrate a reduction of the cavity sensitivity from 1250 Hz/(ms$^{-2}$) to about 150 Hz/(ms$^{-2}$) in vertical direction and from initially 1350 Hz/(ms$^{-2}$) to about 350 Hz/(ms$^{-2}$) in the diagonal v-h1 direction \cite{and23}. This is almost a factor of ten improvement for Fourier frequencies up to about 30 Hz (Fig.\ \ref{fig:PTB_Feedforward_Result}).

One of the main issues in the system is a strong cross-talk between horizontal and vertical directions in the AVI system. 
This is partly caused by the stiff, vacuum insulated hoses for the cold nitrogen gas. 
To improve the feedforward performance further, the cross-talk between the AVI axes must be reduced and more complex filter functions need to be implemented, that also take into account resonances in the transfer function.

\section{Mitigation of vibrations in closed-cycle cryostats}
\label{sec:Cryostats}

Closed-cycle cryostats show strong periodic vibrations affecting the performance of ultra stable cavities and SHB crystals. 
These vibrations can be reduced by active control or by mechanically decoupling the cooler from the frequency reference. 
Here two examples of systems are presented that were investigated in the NEXTLASERS project:

For an Eu:YSO crystal in a dilution cryostat at OBSPARIS, mechanical decoupling turned out to be insufficient for thermal contact and heat evacuation. 
As alternative to measuring vibrations close to the SHB crystal within the cryogenic environment using dedicated sensors (subsection \ref{sec:cryo-vibration-sensors}), optical measurement and cancellation of the residual vibration were tested. 

At PTB a new cryostat for 124~K was tested and the residual vibration noise was measured outside the vacuum chamber close to the cavity. The vibration level of this system is discussed in subsection \ref{sec:PTB-cryostat}.

\subsection{Reducing the sensitivity of SHB crystal to residual vibrations}\label{sec:SHB_vibrations}

In the context of laser frequency stabilization via spectral hole burning (SHB), motion of the rare-earth ion doped crystal presents Doppler-shift induced frequency noise between the reference ions and the probing laser. 
To combat this, a minimal-contact mounting of the crystal was proposed to reduce the transfer of vibrations resulting from the cryostat.  
However, due to limited cooling power of the dilution fridge, the prepared crystal mount with reduced mechanical coupling does not allow sufficient thermal exchange for efficient cool down. 
Instead, we propose to measure and compensate the residual vibration of the crystal optically. 

\begin{figure}[hbt]
	\centering
	\includegraphics[width=\textwidth]{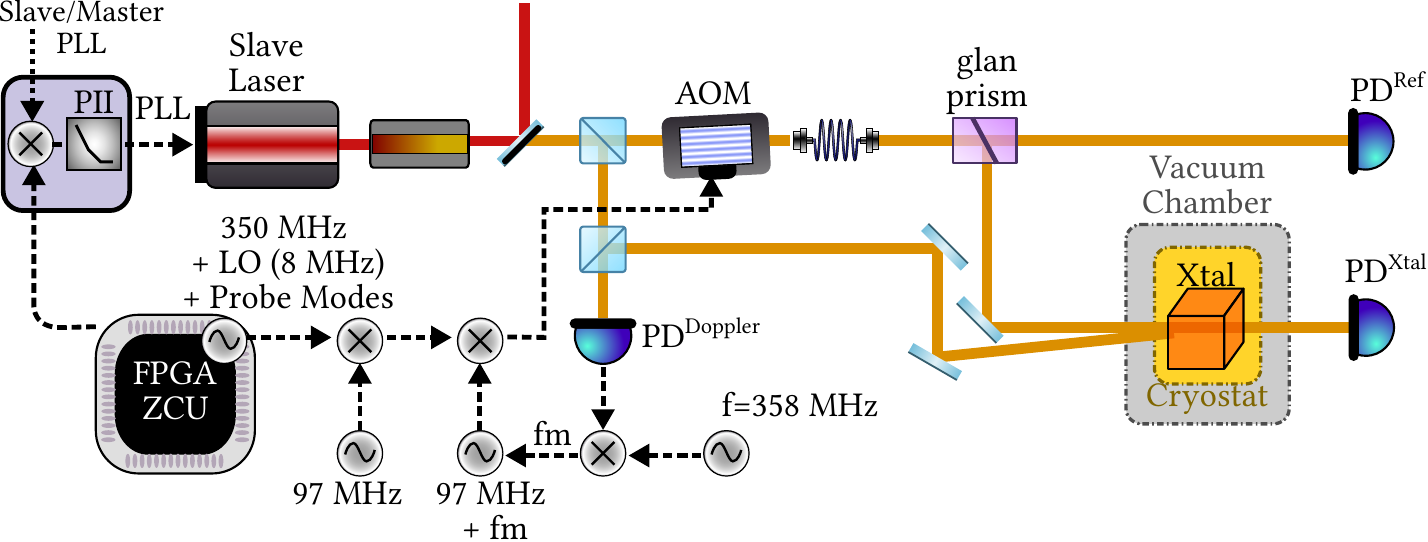}
	\caption{Illustration of first implementation of optical crystal-motion and cancellation scheme for dilution-cooled rare-earth ion spectral-hole burning.}
	\label{fig:SHBxtalDopplerCancellerScheme}
\end{figure}

Figure \ref{fig:SHBxtalDopplerCancellerScheme} shows an illustration of the first iteration of the crystal motion canceling scheme.  In this scheme the light from the pre-stabilized slave laser is frequency shifted by the AOM to produce the frequency spectrum around $\lambda=580\,\mathrm{nm}$ which is desired to probe the spectral holes and subsequently sent to the crystal (Xtal) in the normal fashion for SHB laser stabilization. 

\begin{figure}[htbp]
	\centering
	\includegraphics[width=\textwidth]{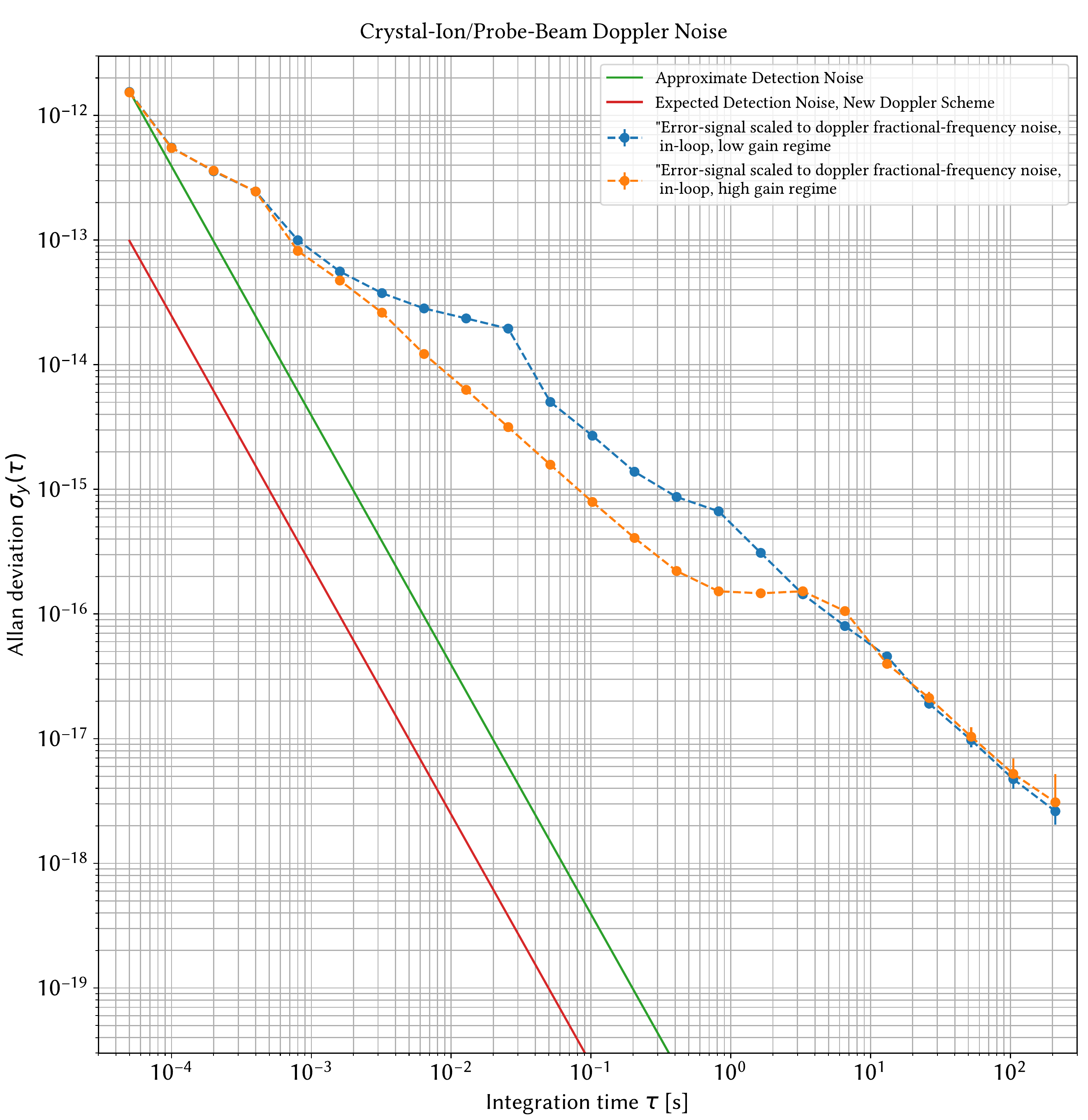}
	\caption{Allan deviation of fractional-frequency noise induced by motion of the spectroscopic crystal in SHB.  The low-gain feedback case (blue) uses the minimum gain to prevent phase wrapping and gives a lower-bound on the uncompensated Doppler noise due to crystal motion. The first implementation of a feedback system reduces the Doppler noise contribution (orange).  The detection noise limit (green) should be reduced (red) using a case optimized photodetector and RF signal chain.}
	\label{fig:SHB_ADEV_Doppler}
\end{figure}

For the crystal motion ($\delta{z(t)}$) detection scheme, we collect the light back-reflected from the crystal which experiences a phase modulation, $\delta{\phi(t)}=\frac{4\pi}{\lambda}\delta{z(t)}$ due to vibration.  The back-reflected light is interfered with a sample of the pre-frequency-shifted laser light to produce an optical beat note around the AOM driving signal frequency, and which contains the crystal motion information.  This beat note is detected at $\mathrm{PD}^\mathrm{Doppler}$ and the crystal motion signal is extracted by demodulation at the shift frequency.  This demodulated signal produces a voltage proportional to the phase error introduced by the motion of the crystal and can be calibrated using the peak to peak voltage of the wrapped signal corresponding to $2\pi$ of wrapped phase.  We record the error signal using a data acquisition card, where we can convert the measured voltage to phase data using this calibration and then differentiate the resulting phase time-series to calculate the frequency error due to crystal motion. A lower bound for crystal-motion-induced frequency-instability is plotted in blue as an Allan deviation of the frequency error in Fig.\ \ref{fig:SHB_ADEV_Doppler}.

The cancellation of the Doppler noise is achieved by the use of the error signal in a feedback system, where the error signal is used to drive a frequency modulation of the AOM frequency shift, correcting the laser frequency incident on the crystal.  
The first iteration of the sensing and control system (Fig.\ \ref{fig:SHBxtalDopplerCancellerScheme}) uses light reflected off-axis from the crystal, naturally producing an non-common-path Mach-Zehnder configuration for the production of the error signal allowing for the detection and reduction (plotted in orange in Fig.\ \ref{fig:SHB_ADEV_Doppler}) of crystal-motion induced Doppler noise, but leaves much optical path uncompensated.  
The success of this first implementation of the SHB crystal sensing and control scheme coupled with the identification of this limitation motivated the start of a currently ongoing optical redesign to allow for the collection of on-axis reflected light.  
Use of the common optical path coupled with a more optimized selection of detection photodiode and electronics provides a near-term prospect of a more complete cancellation of vibration induced frequency noise.

\clearpage

\graphicspath{{D4_decoupling/}}
\chapter[Optical references in closed cycle cryostats]{Optical references integrated into closed cycle cryostats for continuous cryogenic operation of SHB and optical cavities at 124 K, 4 K and below}
\label{ch:cryostats}

\authorlist{%
Joannès Barbarat$^1$, 
Yann Le Coq$^2$,
Bess Fang$^2$, 
Christophe Fluhr$^1$, 
Jonathan Gillot$^1$, 
Vincent Giordano$^1$, 
Michael Hartman$^2$,
Sofia Herbers$^3$, 
Yann Kersalé$^1$, 
Clément Lacroûte$^1$,
Thomas Legero$^3$,  
Xiuji Lin$^2$,
Rémi Meyer$^1$, 
Jacques Millo$^1$, 
Daniele Nicolodi$^3$, 
Lars Rippe$^4$ 
Pierre Roset$^1$, 
Uwe Sterr$^3$, 
Rodolphe Le Targat$^2$ 
}

\affil{1}{\FEMTOaff}
\affil{2}{\OPaff}
\affil{3}{\PTBaff}
\affil{4}{\ULUNDaff}

\corr{clement.lacroute@femto-st.fr}

\chapstart
This chapter describes the achieved performances of four optical resonators at temperatures of 124~K and below 2~K, with a focus on the optical properties and performances of the resonators in closed-cycle cryocoolers.

Possible implementations of temperature control and passive vibration compensation in closed cycle cryocoolers are investigated. Cryogenic cooling fundamentally reduces the thermal noise of Fabry-Perot cavities and the linewidth of spectral holes in doped crystals. However, continuous and low maintenance cooling requires closed cycle coolers that produce vibrations and temperature fluctuations detrimental to the operation of low noise devices. The behaviour of these systems is investigated, and methods are developed and implemented that enable cryogenic cooling without degrading the performance of the reference systems.

In these closed-cycle cryocoolers, Fabry-Perot cavities at 124~K, 4~K and 0.4~K were investigated, with optical Finesses of 352 600 and 410 000, as well as spectral holes in an europium doped crystal at 0.3~K with linewidths of order 1~kHz. 
Measurements of the temperature sensitivities of the resonators were performed, as well as a preliminary estimate of their fractional frequency stabilities.

\section{Introduction}
We have set up four cryostats that allow efficient temperature regulation for four cryogenic references. 
These cooling machines are optimized to passively reduce the vibration level and to enable continuous, low-maintenance operation at cryogenic temperatures and active temperature regulation of the resonators.

For a 124~K silicon cavity, a cold He-gas circulation system was installed. A closed-cycle pulse-tube cryocooler is used to cool down a heat exchanger as shown Fig.\ \ref{fig:He-ptb}. Vibrations from the pulse-tube cooler are decoupled from the cavity by flexible hoses that are damped at intermediate positions. 

With a closed-cycle dilution cryostat for a sub-kelvin silicon cavity a minimum temperature of 15~mK was reached and stabilised operation at 20~mK was achieved. Similarly sub~2-K cryostats for the spectral-hole burning setups have been installed for first tests. Temperatures of 1.5~K and 80~mK have been achieved, and temperature stabilization at 400~mK with 30~µK fluctuations was demonstrated.

\section{Description of the resonators}

\subsection{124~K Fabry-Perot cavity}
PTB operates two single-crystal silicon Fabry-Perot cavity at the zero-CTE temperature of 124 K.
One system (Si2) is cooled by cold nitrogen gas that is evaporated from a LN2 dewar. 
It employs dielectric SiO$_2$/Ta$_2$O$_5$ mirror coatings.
A circulating He cryocooler was designed and assembled by PTB, as described above. 
It now operates the second similar silicon system (Si5) \cite{yu23} where it replaced the LN2 based cooling system.
This cavity employs crystalline GaAs/AlGaAs Bragg reflectors \cite{col16}.

\subsection{Fabry-Perot cavity below 1~K }

The cavity at CNRS is composed of a single-crystal silicon cylindrical spacer and two single-crystal silicon mirror substrates \cite{bar24}. The dimensions of the cylindrical spacer are 180~mm in length and 200~mm in diameter. We chose the highest possible length given the available space in the cryocooler. The aspect ratio allows us to reduce the radial acceleration sensitivity by cancelling tilts from radial accelerations for this simple cylindrical geometry \cite{jak2009}. The weight of this spacer is about 13 kg. This high spacer mass might help to improve the passive filtering of temperature fluctuations, and to reduce the long-term frequency drift.

The optical axis of the spacer is aligned to the [111] crystalline axis. The mirror substrates also have their optical axis aligned to the [111] crystalline axis.  Both substrates and the spacer include a flat for proper alignment during optical contacting. A custom contacting mask was fabricated to ensure proper centering of the mirrors and proper alignment of the crystalline axis.

The mirrors use Al$_{0.92}$Ga$_{0.08}$As/GaAs crystalline coatings specified for use at 17~K.  The crystalline coatings are optically contacted to the substrates. Their slow axis is indicated by a flat, which presents a misalignment of less than 6$^{\circ}$ with regards to the substrates flat, coming from the manufacturing processes. The crystalline orientation of the coatings is most likely similar to those presented in \cite{yu23a}, with the [100] direction of GaAs normal to the mirror surface.

The cavity mount is made in duralumin in order to reduce its weight. It is attached below the mixing plate (MX) by four M6 screws to minimize thermal resistance thanks to the high strength produced. The cavity is held horizontally on 3 contact points in order to avoid hyper static equilibrium. Nothing else has yet been implemented to optimise the accelerometric sensitivity of the spacer.

\begin{figure}[h!]
		\center
		\includegraphics[width=6.5cm]{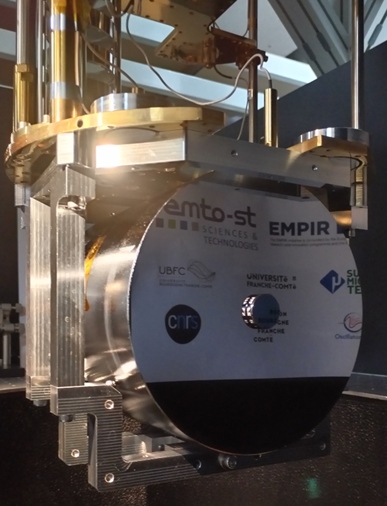}
	\caption{Single-crystal silicon cavity held in its support. The support is directly attached to the MX plate. One of the three supporting spheres is shown on the left bottom corner.}
	\label{Fig:mount}
\end{figure}

Fig.\ \ref{Fig:mount} shows the cavity held in its support, attached to the MX plate of the dilution cryocooler. One of the three supporting spheres can be seen at the bottom-left. A second one is located in the same plane transverse to the optical axis, while the third one is at the center of the second supporting arch.

\subsection{Spectral hole burning in doped crystal at 2~K}

For slow light locking with highly dispersive cavities we used a europium doped YSO crystal at a concentration of 1~at\% with isotopes 151 and 153 in their natural abundance. The crystal was mounted in a cryostat (Optidry, MyCryofirm) inside a cell kept around 2~K using a Joule-Thomson loop. 
Thermal contact to the crystal was achieved by a 0.1 mbar helium environment inside the cell. 
The crystal dimension was $14 \times15\times21\   \mathrm{mm}^3$ cut so that the corresponding crystal orientation was $D_1$, $b$, and $D_2$. 
The locking beam was orthogonal to $D_1$, $b$ whereas the burning was orthogonal to $D_1$, and $D_2$, using a diffuse burner mounted on the side. The side burner was used to avoid standing wave interference patters resulting from the mirrors, which cause inhomogeneous burning. 

The crystal was initially mechanically polished to minimize the wedge angle between the mirror faces (8 µrad) and afterwards made optically flat by superpolishing them with Ion Beam Forming to approximately 1 nm RMS. 
Front and back mirrors were coated using Ion Beam Sputtering to 90 \% and 99 \% reflectivity, respectively.

\subsection{Spectral hole burning in doped crystal below 1~K}

At OBSPARIS, a Eu:YSO crystal, grown using the Czochralski method from the melt, with doping concentration of 0.1~at\% and the Eu$^{3+}$ ions in their natural abundance, is mounted on the custom dilution stage added to a pulse-tube cryocooler (Optidry from Pasqal-MyCryoFirm). The crystal measures 8\,mm$\times$8\,mm$\times$4\,mm, with the shorter edge cut parallel to the crystal $b$ axis, which is also the direction of propagation of the interrogation laser beams. The thermal contact between the crystal and the cold supporting plate is ensured by a thin layer of silver loaded lacquer. 

\section{Passive vibration rejection} \label{sec:D4_S1}

Closed-cycle cryocooling usually relies on the use of pulse-tube or Gifford-McMahon cryocoolers. They employ compressed helium gas as working medium, which avoids the regular refills required by liquid nitrogen or liquid helium bath. 
Both cost and maintenance are therefore reduced, and experiments are not disrupted by frequent temperature changes impaired by the refills. 
One drawback of e.g. pulse-tube coolers is the intrinsic vibration noise close to 1~Hz impacted by the He pulsing in the cryocooler tubes.

As illustrated in this section, the main approach to reduce the impact of these vibrations is to physically decouple the moving parts, located in the cryocoolers ``cold head'', from the main experimental chamber. 
Further mechanical decoupling can also be implemented inside the cryocoolers main chamber where the optical references are located.
Mechanical decoupling of the noisy cold head from the sensitive cavity can be achieved by flexible copper or carbon links \cite{cap06, wan23, val24} or by heat conduction by a buffer gas \cite{wan10b}.

\subsection{Pulse-tube cryocoolers}
\label{sec:PTB-cryostat}

\begin{figure}[hbt!]
    \centering
    \includegraphics[width=15cm]{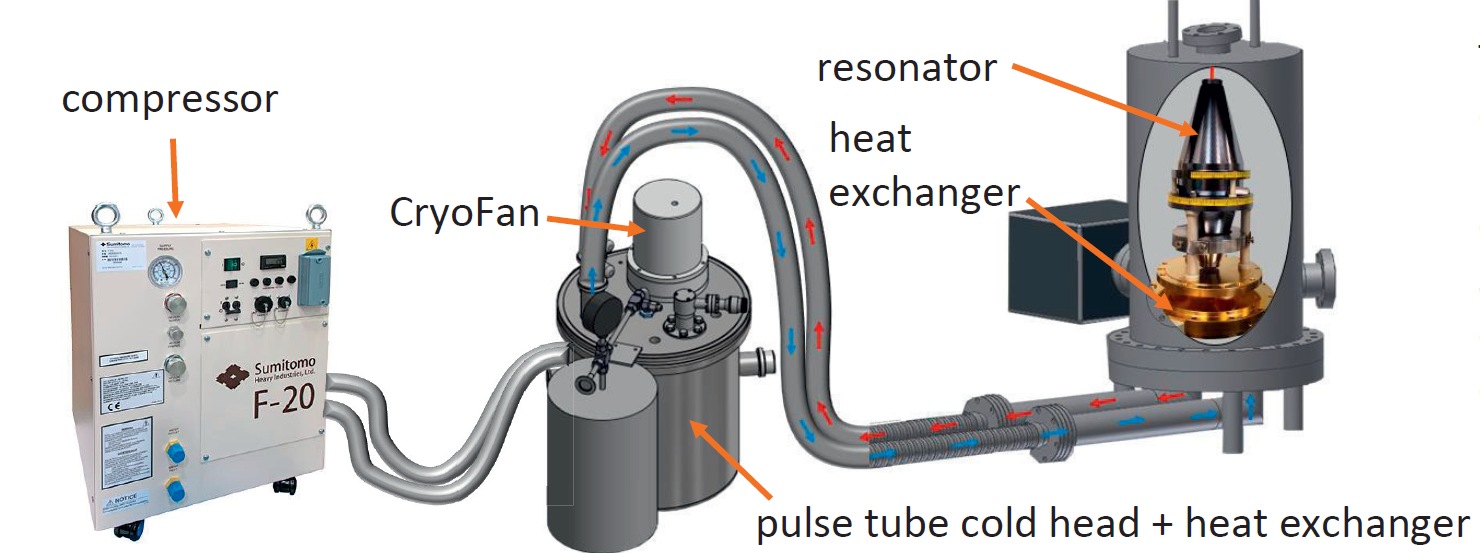}
    \caption{Schematics of the closed-cycle cooling of a Fabry-Perot cavity at PTB at 124~K.}
    \label{fig:He-ptb}
\end{figure}

We first illustrate passive vibration decoupling implemented for single-crystal silicon cavities. 
A new cooling package based on cold He gas circulation was designed for the Si5 cavity described in \cite{yu22}.  
The gas is cooled by a single stage pulse-tube cooler and it is circulated with a cryofan \cite{cryofan} between a heat exchanger at the cold head and a second one at the bottom of the outer heat shield inside the vacuum chamber of the resonator. 
The Si resonator chamber is decoupled from the cooling package by flexible, vacuum isolated hoses. 
The cooling package vibrations are damped by rigidly attaching the flexible hoses to the granite table that supports the Si resonator chamber on an active vibration-isolation system.
The noisy compressor supplying the pulse tube cooler with high-pressure He is located in the basement below the lab room.
The general scheme is shown Fig.\ \ref{fig:He-ptb}.

Fig.\ \ref{fig:PTB_Feedback_Si} compares vibrations of the old cooling device employing cold nitrogen gas from a Dewar with liquid nitrogen gas (left) and of the new cryo system with the pulse-tubed cryo-cooler (right). We measured the vibration level on the AVI platform near the cavity and found no degradation in vibration noise due to the pulse tube cooler. In the horizontal directions, the vibration level is identical to level in the previous arrangement. We even see a small improvement in the vertical direction between 0.1~Hz and 1~Hz.
As shown in subsection \ref{sec:4_silicon}, the temperature stability could be improved compared to the previous system. 
\\[5pt]

\begin{figure}[htbp]
\centering
    \includegraphics[width=17cm]{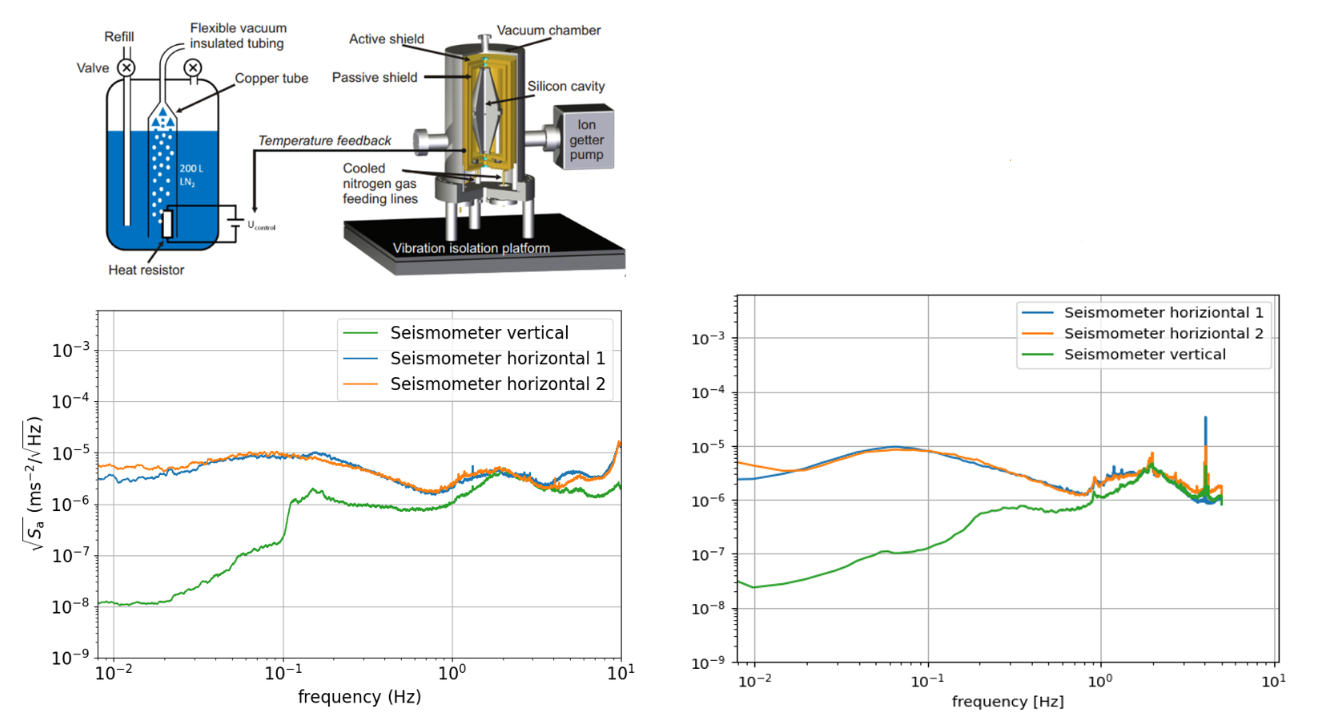}
    \caption{Comparison of the former, continuous nitrogen gas flow cooling system (left) and the pulse-tube based cryo-system (right) for cooling a 21~cm long silicon cavity to 124~K at PTB. 
    The vibration levels in both systems are comparable, with the new system even performing slightly better in the vertical direction.}
    \label{fig:PTB_vibrations_cryostat}
\end{figure}

\begin{figure}[hbt!]
    \centering
    \includegraphics[width=10cm]{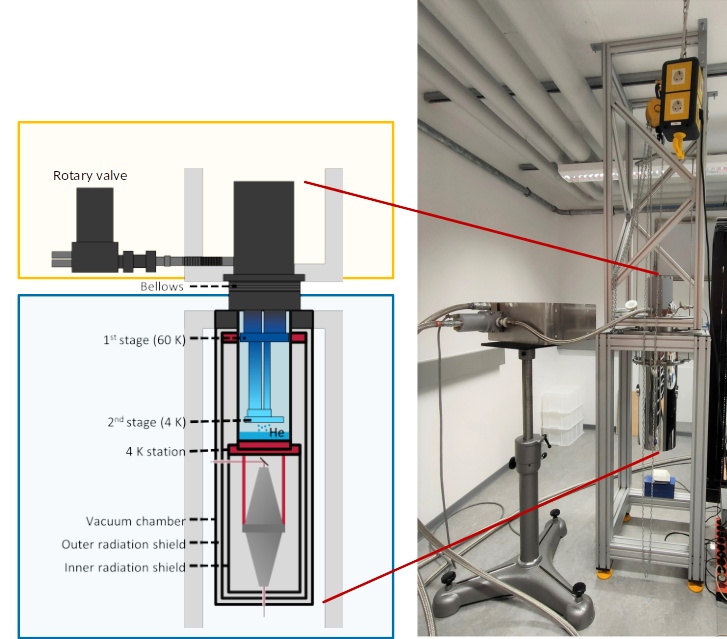}
    \caption{Pulse-tube cryostat for a 4~K Fabry-Perot cavity at PTB.}
    \label{fig:ptb-4K_cryo}
\end{figure}

To operate another cavity at 4~K, a commercial two-stage pulse-tube cryostat (Cryomech PT407-RM) \cite{cryomech} was employed. 
Mechanical decoupling between the cryocooler's cold head and the cavity setup is ensured by using a liquid-He heat exchanger. 
In addition, the remote rotary valve (Fig.\ \ref{fig:ptb-4K_cryo}) decouples its vibrations from the cyostat. 
This type of cooler was successfully employed in operation of a cryogenic sapphire microwave resonator \cite{wan10b,har10a,har12}. 
To improve mechanical decoupling between cold head and cavity, the cryocooler is rigidly mounted to the lab ceiling while the experimental chamber is held by a rigid structure standing on the lab floor. 
Residual mechanical coupling remains through the bellow between the cold head and the main vacuum chamber, that effectively reduces the longitudinal coupling, but is still rather rigid in the radial directions.
\\[5pt]

\begin{figure}[h]
    \centering
    \includegraphics[width=8cm]{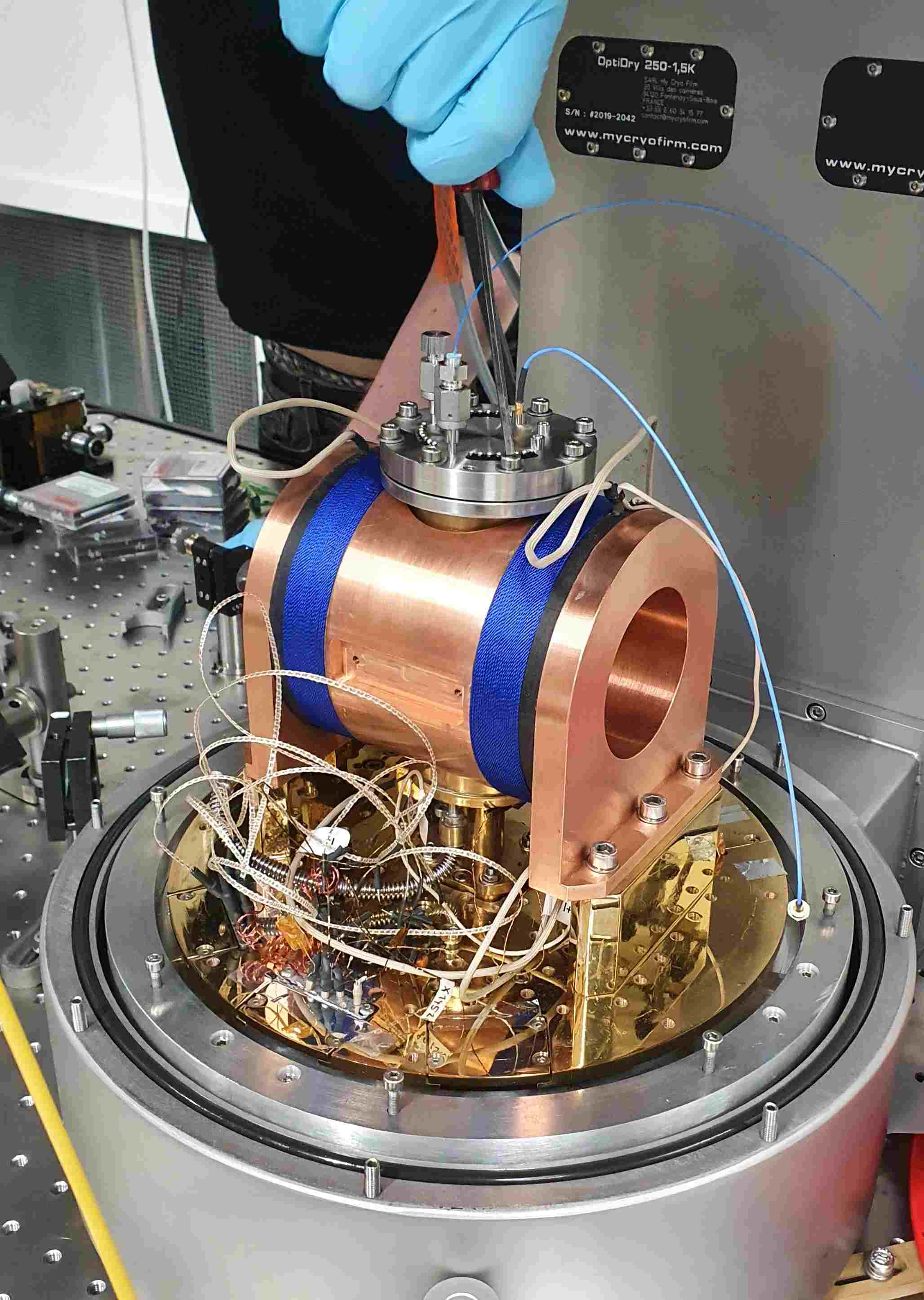}
    \caption{Closed-cycle cryostat at ULUND.}
    \label{fig:ulund-cryo}
\end{figure}

A similar approach was implemented in a 4~K closed-cycle cryostat designed to minimize vibrations at ULUND, shown Fig.\ \ref{fig:ulund-cryo}.  The cryostat has an additional Joule-Thomson loop, which continuously cools a hermetically closed cell to below 2~K, which hosts a Eu:YSO crystal for spectral hole burning laser stabilization.

\begin{figure}[h]
    \centering
    \includegraphics[width=12cm]{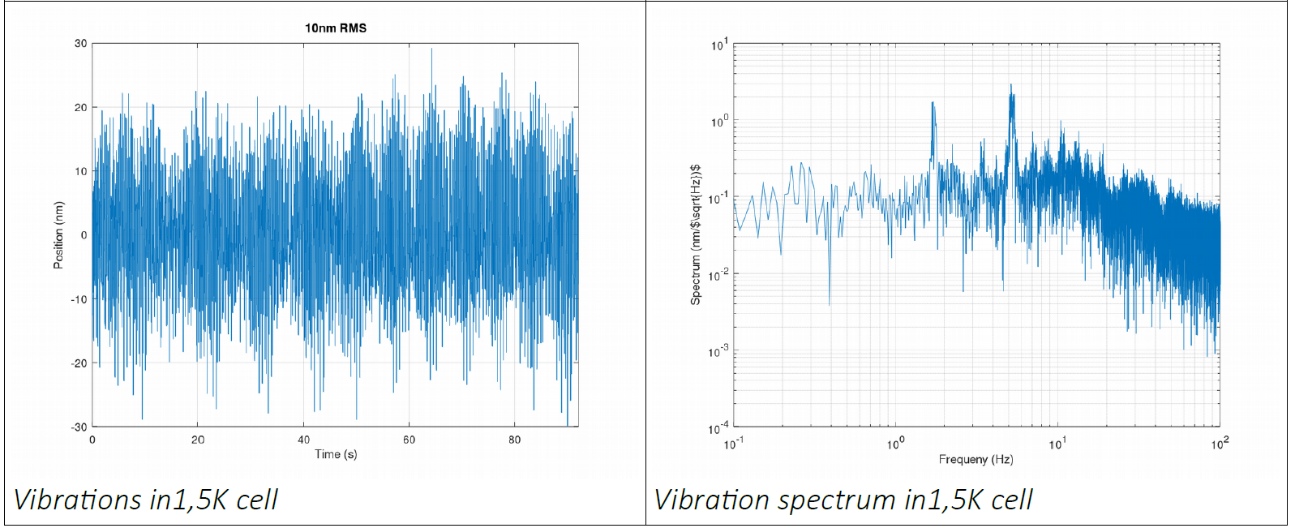}
    \caption{Measured vibrations in the ULUND cryocooler by the manufacturer, measured at room temperature. Left: displacement versus time. Right: displacement spectrum.}
    \label{fig:ulund-vibs}
\end{figure}

The vibrations in the 1.5 K cell were measured by the manufacturer at 10.0 nm RMS (see Fig.\ \ref{fig:ulund-vibs}). However, unexpectedly large acoustic vibration and movement signatures of acoustic noise compared to previously reported levels measured by the manufacturer were observed at cold temperatures. Possible improvement of the crystal mounting structure is being investigated. Decoupling of the vibrations using a kapton foam pillow inside the cryostat was also investigated, which considerably decreased these vibrations.

\subsection{Dilution cryostats}

The question of vibrations management in dilution cryostats, which are necessary to reach temperatures below 1~K, is essentially the same than with higher-temperature pulse-tube cryocoolers. 
Both systems include a pulse-tube cryocooling system which induces vibrations. 
The main difference is the additional cooling stages inside the experimental chamber, which may require cascaded decoupling techniques.

A first example is a Bluefors dilution cryocooler \cite{cryofan,bluefors} setup for the operation of a single-crystal silicon cavity below 1~K. 
The cryocooler is composed of 4 main stages, shown in Fig.\ \ref{Fig:cooldown}: the 50~K stage, the 4~K stage, the still stage and the mixing chamber stage (noted MX). 
The two first stages are based on one closed cycle $^4$He pulse-tube cryocooler. 
Compression-extension cycle at $\sim 1.4$~Hz of the gas inside "tubes" is generated with a compressor unit together with a rotating valve (three positions: open to high pressure, open to low pressure or closed). 
The first two stages reach temperatures of about 50~K and 3~K.
Then, a Joule-Thomson stage can help reaching temperatures around 1.3~K or below with the appropriate conditions. 
To reach lower temperatures, a last stage that uses a dilution of $^3$He in $^4$He is necessary. 
The plate connected to the mixing chamber can reach a temperature as low as a few mK.

The MX stage is mechanically decoupled from these vibrations through several stages: mechanical grounding of the rotation valve and He hoses through high masses; pneumatic isolation of the cryostat from the cryocooler head; flexible copper braids inside the cryostat that decouple the pulse-tubes vibrations from the science stages.

\begin{figure}[h]
    \centering
    \includegraphics[height=5cm]{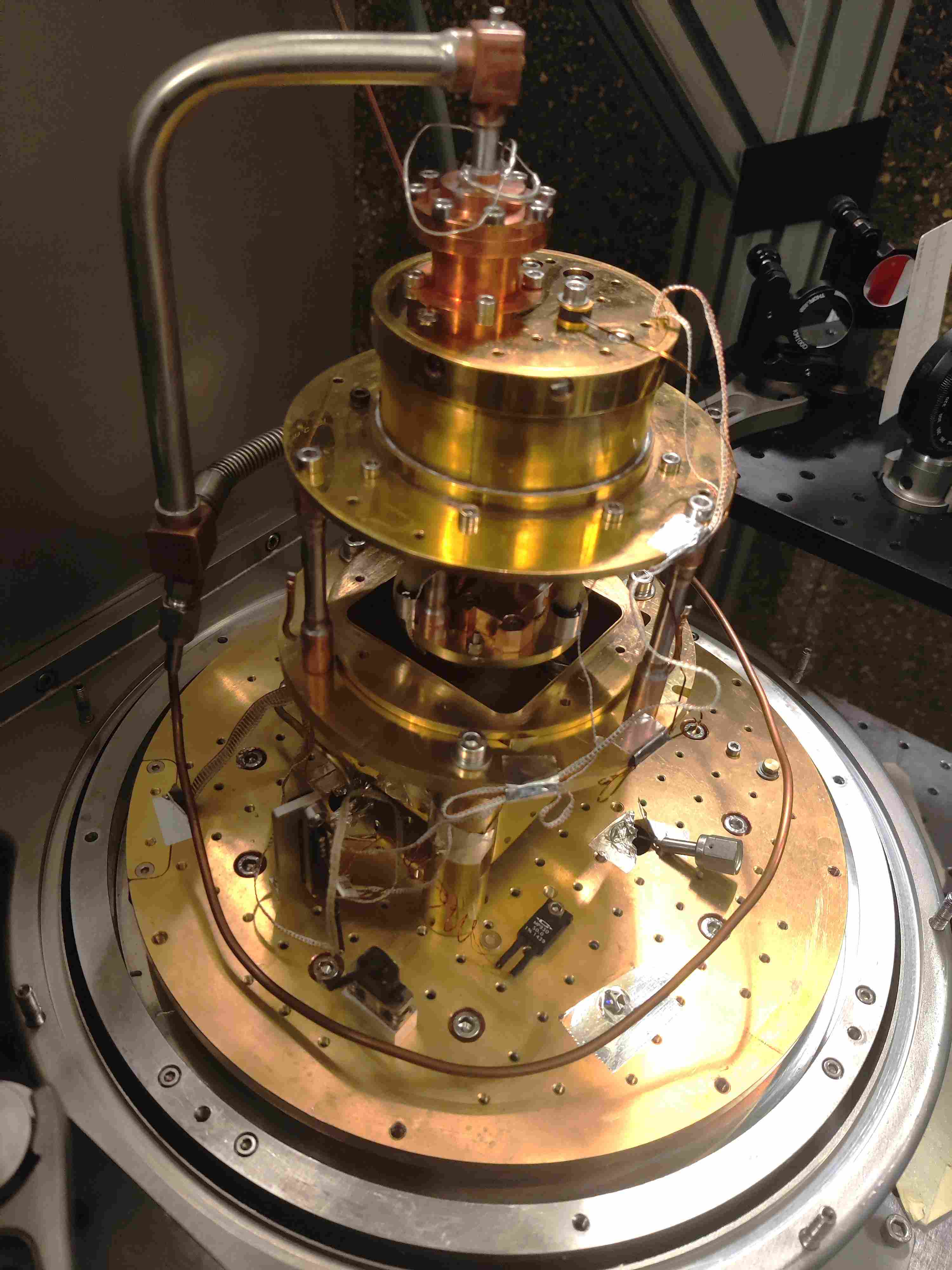}
    \includegraphics[height=5cm]{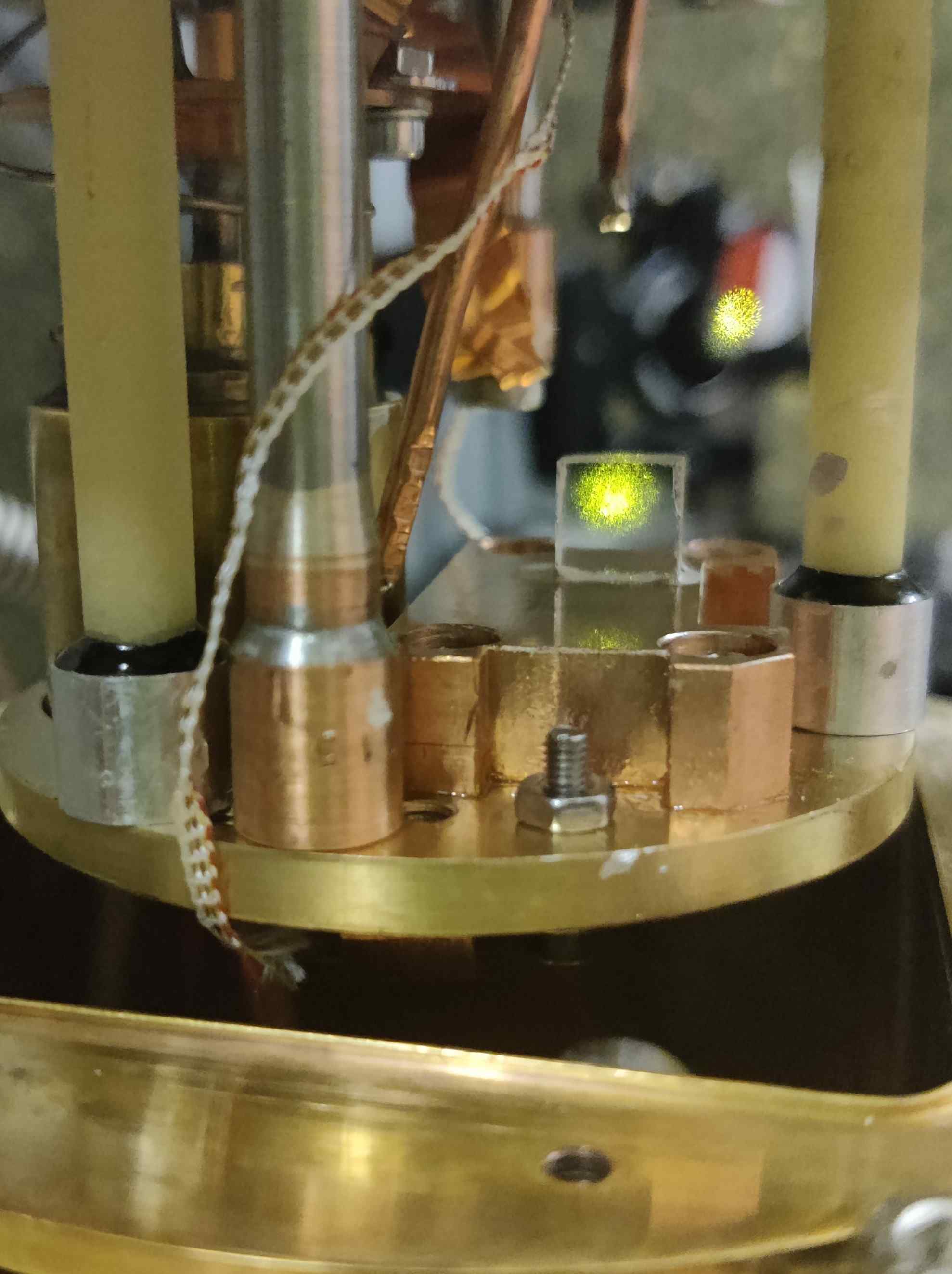}
    \caption{Left: custom fitted dilution stage in a pulse-tube cryocooler. Right: doped Eu:YSO crystal on its holder.}
    \label{fig:obs-cryo}
\end{figure}

A same approach can be deployed for SHB references in dilution cryostats. 
A custom dilution stage was integrated into a pulse-tube cryocooler to cool down a Eu:YSO crystal below 1 K., shown in Fig. \ref{fig:obs-cryo} left.

The vibration decoupling of the injection line of the dilution circuit is ensured by thermalizing the arriving He (to be cooled down) on the cold head before the annealed copper braids to reduce the coupling of the vibration noise onto the part of the cryostat on which the crystal is rigidly fixed. 
For the pumping line, since thermalization is less of a concern, the vibration decoupling is ensured outside the cryostat, on a structure made of aluminium profiles that is decoupled from the cryostat through a soft bellow, which also isolates the vibrations from the rotating valve. 
The residual vibration seen by the crystal is evaluated optically to be on the level of $10^{-16}$ or below at 1~s when converted through the Doppler effect to a fractional frequency instability. 
Active compensation will become relevant once the performance of the frequency stabilization on spectral holes reaches this level and is detailed in chapter \ref{ch:vib_isolation}.

\section{Temperature stability}
\label{sec:D4_TStab}

\subsection{Single-crystal silicon cavities operated at 124~K}
\label{sec:4_silicon}

\begin{figure}[hbt]
    \centering
    \includegraphics[width=7cm]{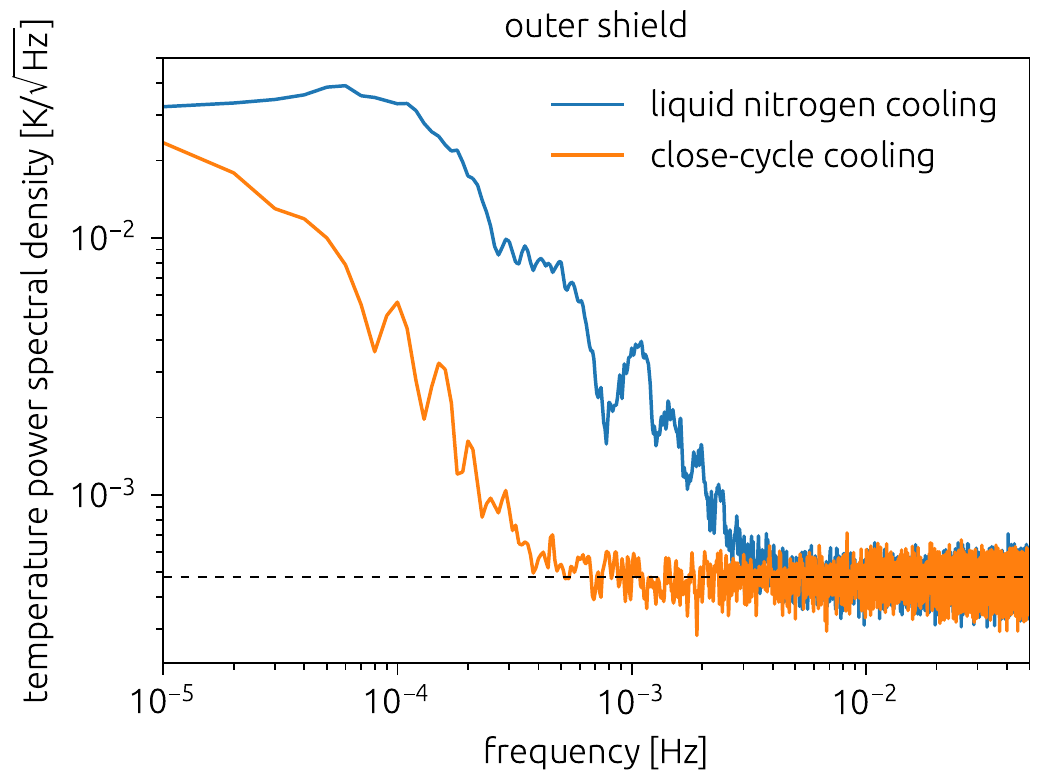}
    \includegraphics[width=7cm]{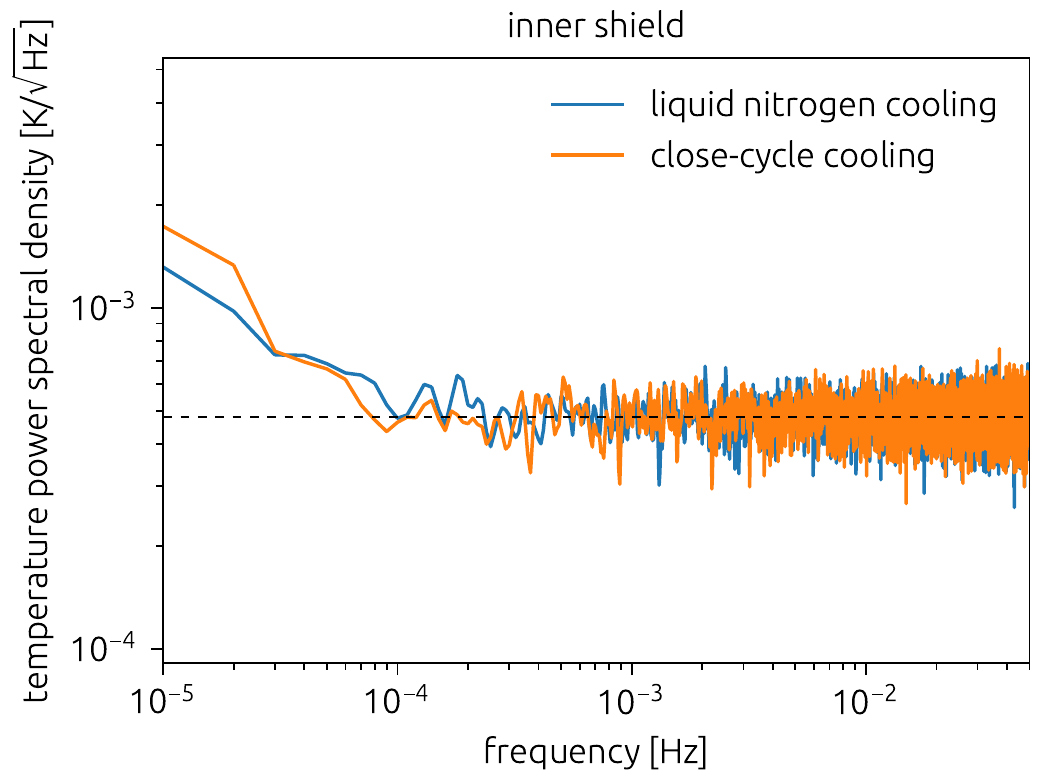}
    \caption{Temperature stability of the outer (left) and inner (right) heat shield of the PTB Si5 cavity with cooling using liquid nitrogen and close-cycle cooling.}
    \label{fig:ptb_temperature}
\end{figure}

Two 124~K silicon Fabry-Perot resonators (Si2 and Si5) were initially operated using cold nitrogen gas that was evaporated from a liquid nitrogen bath \cite{kes12a,hag13a}, before Si5 was integrated in the pulse-tube cryocooler described above. The temperature of the cryostat outer heat shield is controlled by adjusting the flow rate of He with the rotation speed of the cryofan. 
A better temperature stability could be achieved than with the liquid nitrogen cooled system, that was regularly perturbed by liquid nitrogen refills Fig.~\ref{fig:ptb_temperature}.

\subsection{Single-crystal silicon cavities operated below 1~K}

As mentioned above, operation of optical references in dilution cryocoolers requires the use of additional temperature stages. 
In the  Bluefors dilution cryocooler described in the previous section, cylindrical aluminium shields are thermalized on each stage, except for the MX, to significantly suppress radiative heat exchange from the room temperature environment to the cavity.

Horizontally-oriented optical access is granted through 4 pairs of AR-coated viewports. The window of the vacuum chamber is made of BK7 while thermal shields are equipped with Si0$_{2}$ windows. 
The radiated infra-red power from the 300~K, 50~K and 4~K windows is estimated to be  below 1~µW at the MX stage, well below the available cooling power at 100~mK.

\begin{figure}[htbp]
		\center
		\includegraphics[width=10cm]{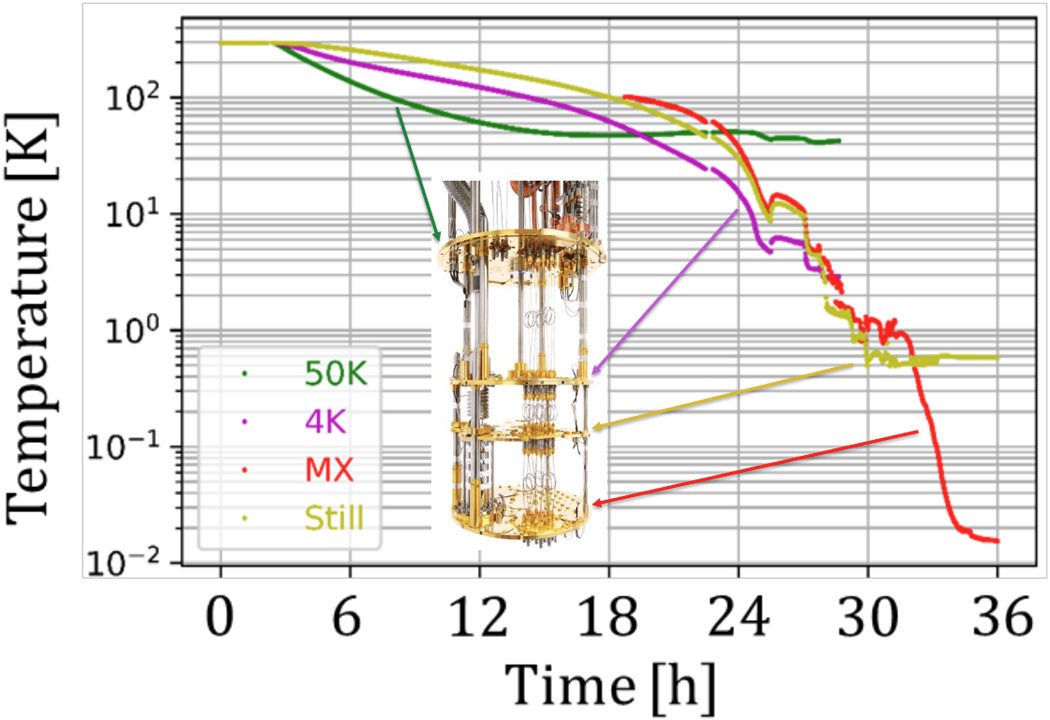}
	\caption{Cooldown of the cryocooler. The insert shows the four cryocooler stages. The unloaded MX stage reaches the sub-100~mK range in 36 hours.}
	\label{Fig:cooldown}
\end{figure}

The cooldown procedure to reach the mK temperature range takes approximately 36 hours without thermal load (cavity) inside and is divided in three main phases (Fig.\ \ref{Fig:cooldown}). 
The first one is the pulse-tube $^4$He cryocooling procedure, of roughly 24 hours; the dilution refrigerator can then be started and the $^3$He/$^4$He mixture is first condensed and cools down to about 0.8~K in around 6 hours at the still stage; the $^3$He/$^4$He mixture then separates in two phases and temperatures below 100 mK are reached within roughly 6 hours.

\subsubsection{Temperature monitoring}

Temperature of each stage is monitored by at least one sensor (see Tab.\ \ref{tab:temp_sensors}). First stage of the pulse tube (50~K) is monitored with a common PT100 sensor and only used in case of cooling issue. 
Cernox\textregistered \ sensors are used for the 4~K stage and the still plate with a calibration range down to 0.1~K.

Two additional sensors help monitoring the MX plate and cavity mount temperatures: a calibrated RuO$_{2}$ RX-102B and an uncalibrated RX-202A-CD, with a working range down to 0.01~K. 
All theses probes are monitored by a multiplexed AC resistance bridge and a temperature controller. 
To prevent the thermal disturbance caused by the reading of the probe, measurements are performed with a ratio of 3 seconds over 10 seconds.

\begin{table}[htbp]
    \centering
    \renewcommand{\arraystretch}{1.1}
    \caption{Temperature sensors installed in the cryostat. LS stands for LakeShore, cal. for calibration, and uncal. for uncalibrated.}
    \begin{tabular}{|c|c|c|c|c|}
        \hline
        {Stage} & {type} & {ref.} & {Cal. from} & {to} \\
        \hline
        50~K & PT100 & -- & 310~K & 20~K \\
        4~K & Cernox\textregistered & CX-1010 (LS) & 310~K & 0.1~K \\
        still & Cernox\textregistered & CX-1010 (LS) & 310~K & 0.1~K \\
        MX & RuO$_{2}$ & RX-102B (LS) & 100~K & 0.010~K \\
        MX & RuO$_{2}$ & RX-102B (LS) & 100~K & 0.010~K \\
        Cavity mount & RuO$_{2}$ & RX-202A-CD (LS) & uncal. & uncal. \\
        \hline
    \end{tabular}
    \label{tab:temp_sensors}
\end{table}

\subsubsection{Temperature measurements}

Preliminary cooling tests for characterizing the cryocooler and temperature probes were conducted. The first one, in an unloaded configuration, showed a lowest achieved temperature of 15~mK, which can be compared to the lowest temperature achieved of 7~mK without windows. 
A second run was performed with the cavity support and a copper part attached to the MX plate where a calibrated RuO$_{2}$ RX-102B and an uncalibrated RX-202A-CD were installed. 
The measured temperature of this copper block is 20 mK. The temperatures measured by the RuO$_{2}$ sensor on the copper part and the RuO$_{2}$ sensor on the mixing plate agree within 1 mK.

\begin{figure}[h!]
		\center
        \includegraphics[width=12cm]{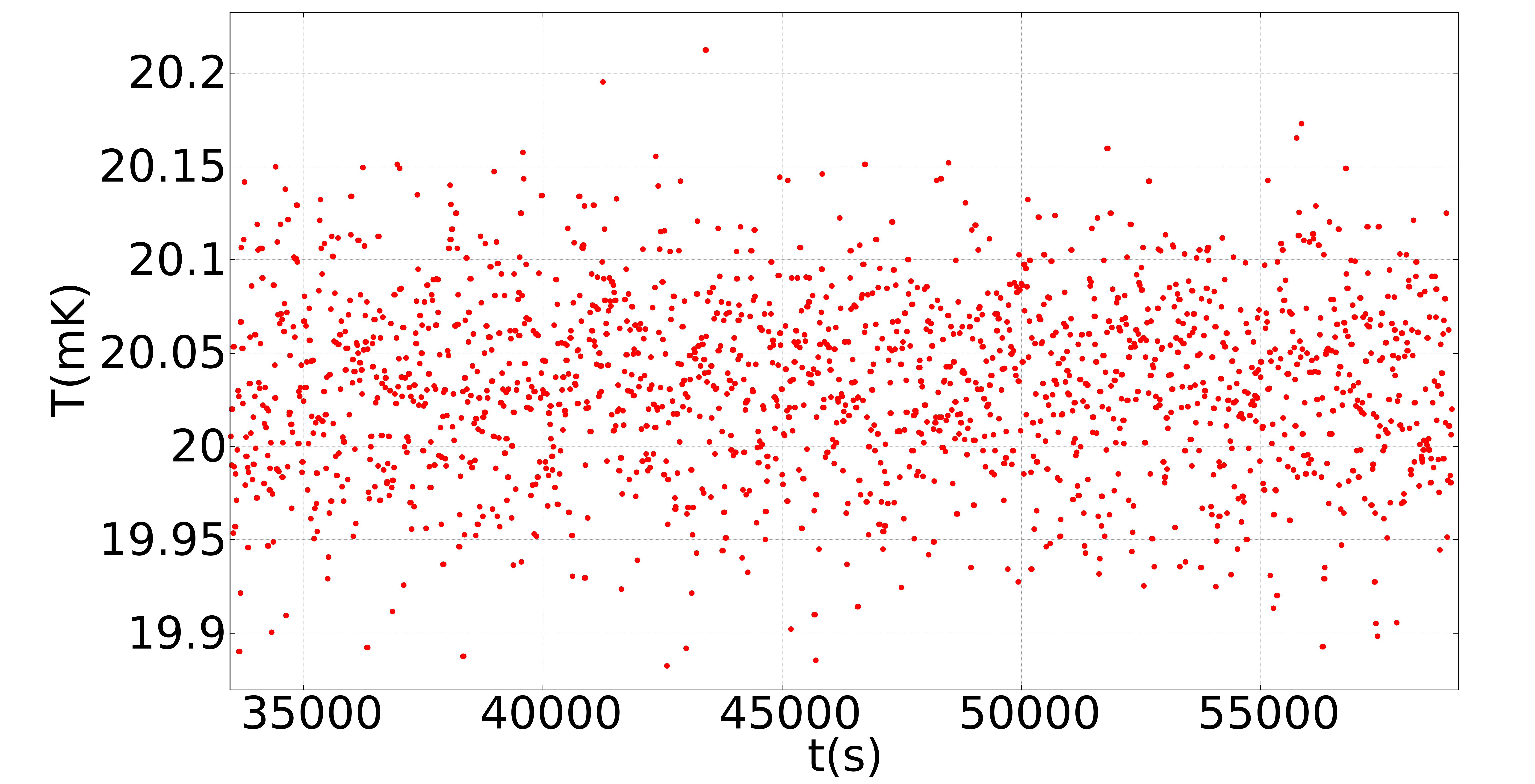}
       \includegraphics[width=12cm]{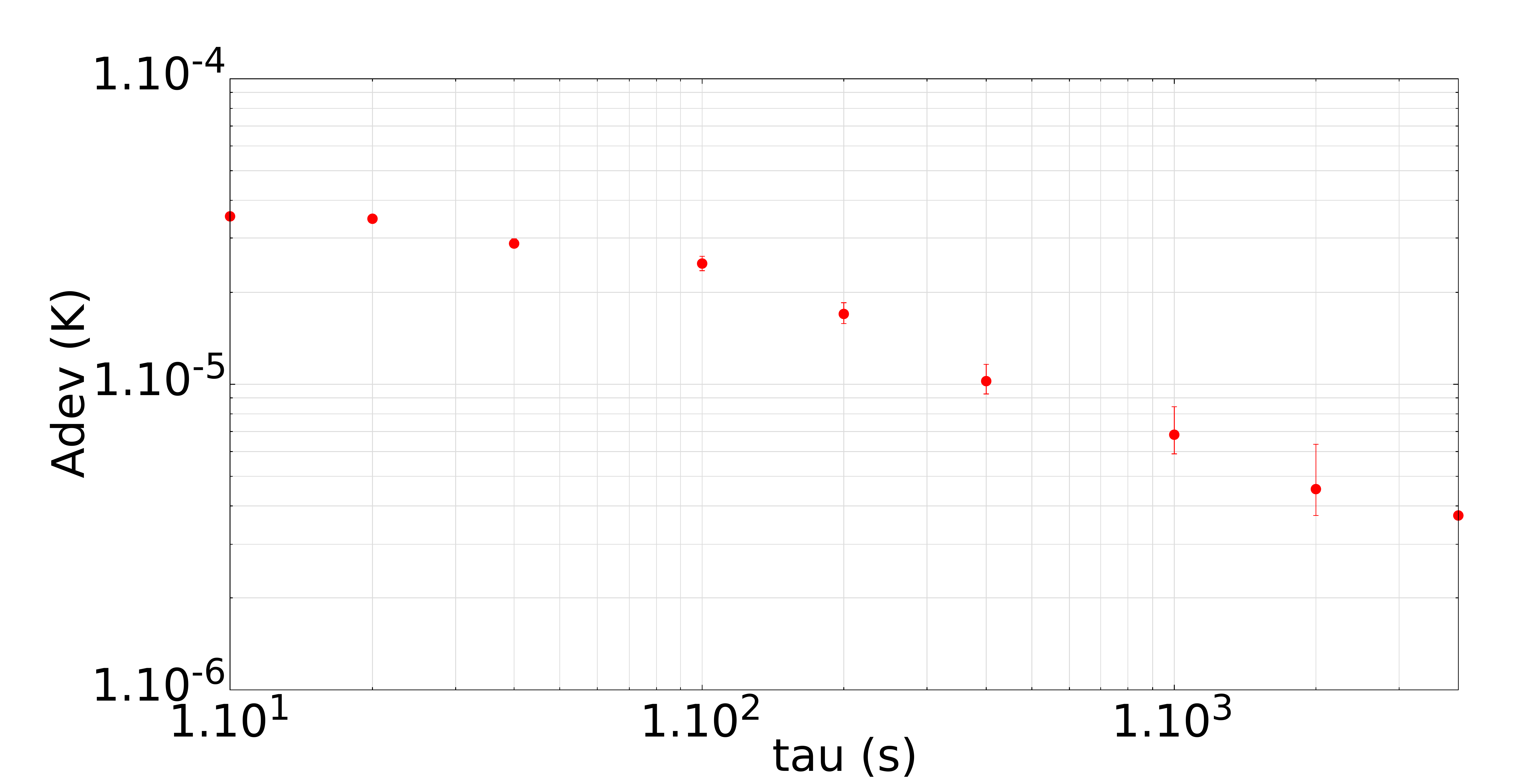}
	\caption{Top: Temperature of the MX stage when temperature stabilization is activated. This is an in-loop measurement. Bottom: Allan deviation of the temperature data.}
	\label{Fig:tempe}
\end{figure}

A test of the temperature control at the MX stage when loaded with the cavity mount. 
The temperature was stabilized to 20~mK with fluctuations below 0.1~mK, as measured by the in-loop sensor (Fig.\ \ref{Fig:tempe}).

\subsection{Spectral-hole-burning crystals operated below 1~K}
\label{sec:D4_SHB_1K}

\begin{figure}[h]
    \centering
    \includegraphics[height=5cm]{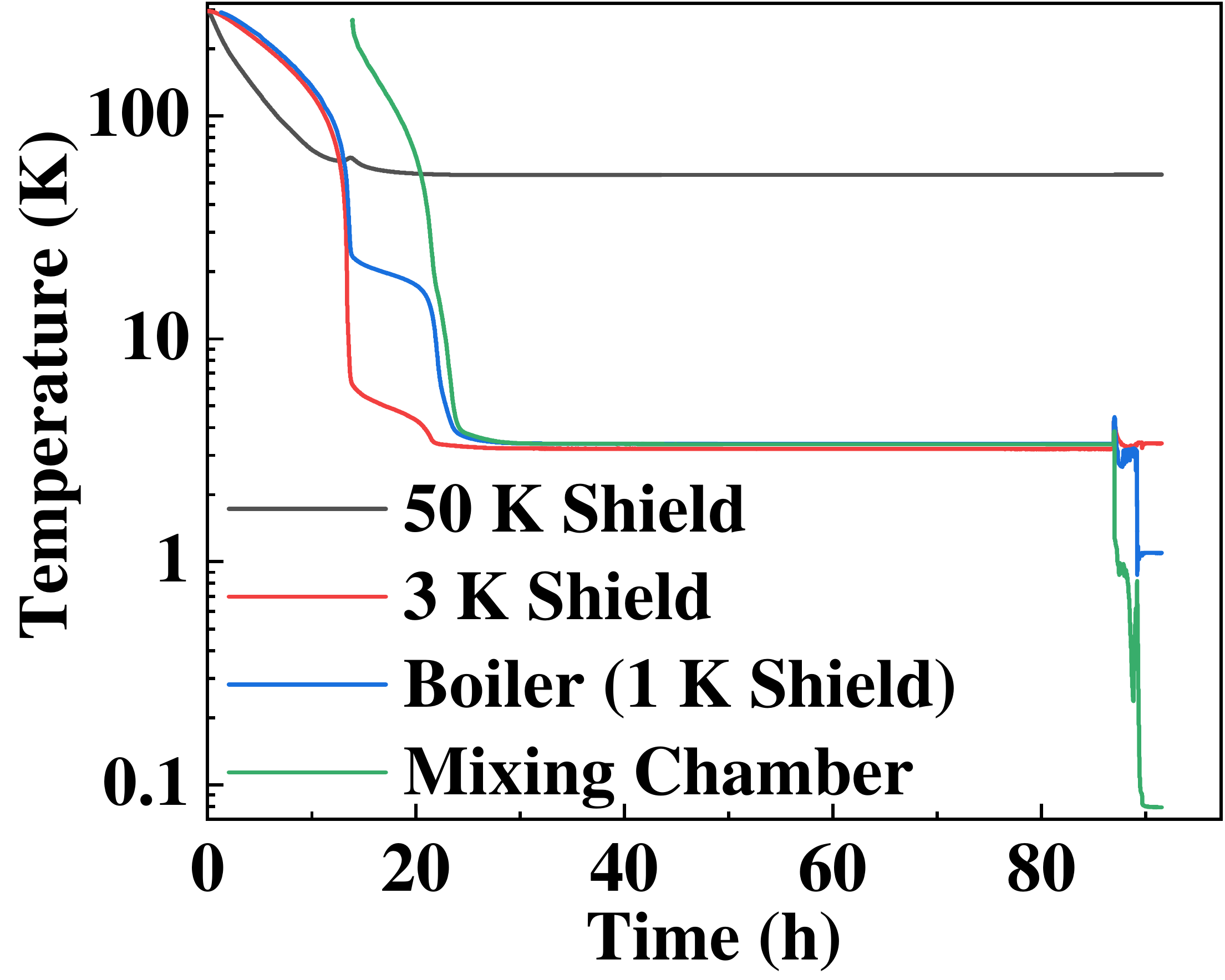}
    \caption{A typical cool down curve of the cryostat and the dilution at OBSPARIS.}
    \label{fig:obs-cryo2}
\end{figure}

The custom dilution stage integrated into a pulse-tube cryocooler for a Eu:YSO crystal below 1 K., shown in Fig.\ \ref{fig:obs-cryo2} left, exhibits a temperature-dependent cooling power. 
The obtained cooling power is below 50 µW for temperatures below 1~K. 
When loaded with a crystal and holder, see Fig.\ \ref{fig:obs-cryo2} middle, the cooldown time is slightly below 30 hours from 300 K to 4~K, and about 3 hours from 4~K to the sub-100~mK range, as seen in Fig.~\ref{fig:obs-cryo2} right. 
Note that the $\sim 50$ hours of waiting time shown in the plot is artificial since a manual activation is required to start the dilution process. 
The lowest achievable temperature is around 80 mK, and is dependent on the relative concentration of $^3$He/$^4$He in the mixture, which has been seen to be modified following incidents of leakage.

\subsubsection{Temperature stability}

\begin{figure}[h!]
    \centering
    \includegraphics[width=12cm]{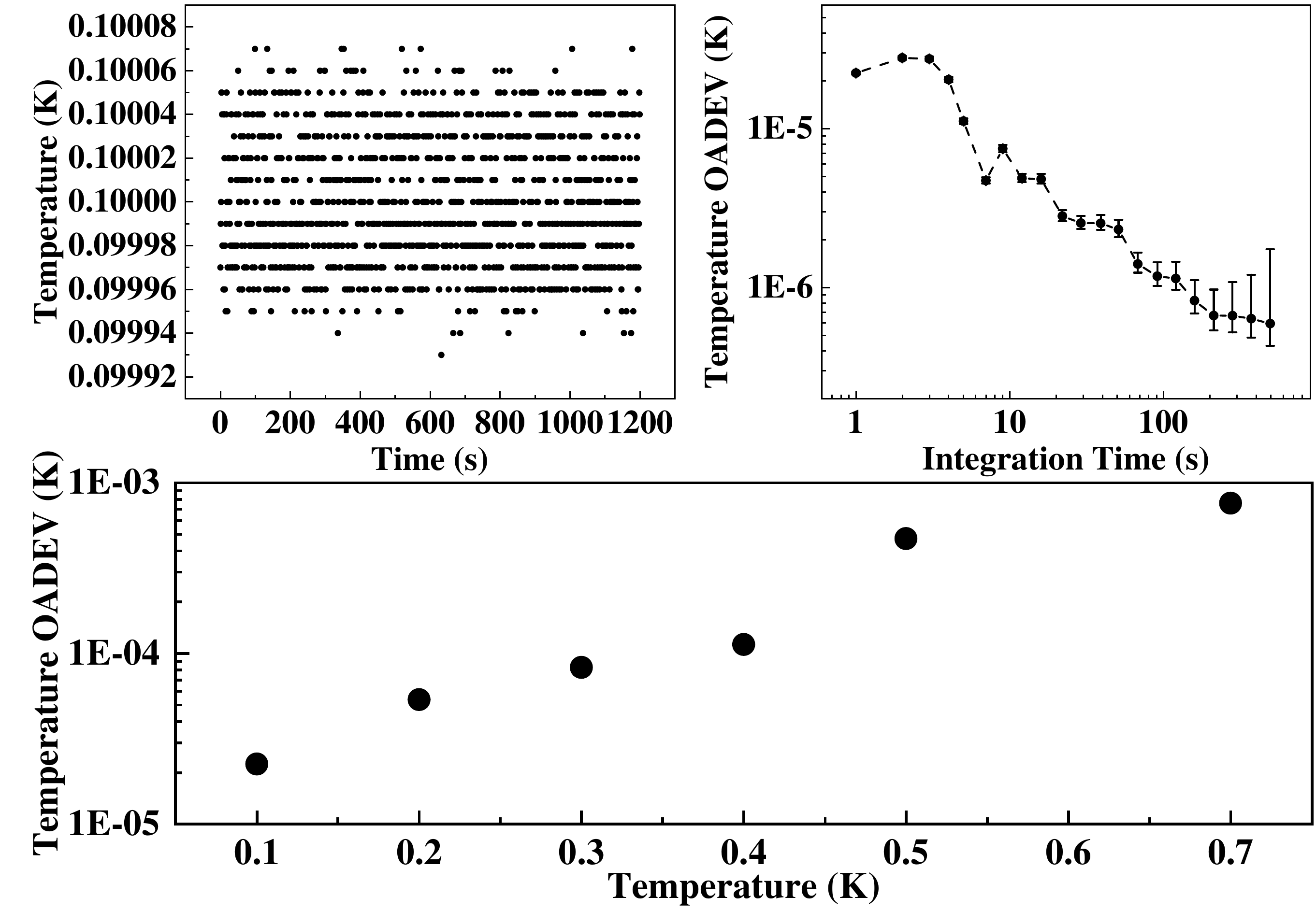}
    \caption{Temperature stability at the dilution stage on OBSPARIS cryocooler. 
		The temperature is stabilized at 100~mK with a locking bandwidth of about 1~Hz. 
		Top left: temporal temperature data. 
		Top right:  overlaping Allan deviation (OADEV) of the temperature. 
		Bottom: OADEV at 1 s averaging time at different temperature values.}
    \label{fig:obs-temp}
\end{figure}

The temperature stability is measured and controlled with a Lakeshore 350 controller at four stages: two Carbon Ceramic Sensors (Temati) at the 50~K and 4~K stages of the pulse tube, a ruthenium oxide sensor on the still (reaching about 1 K), and a Cernox sensor on the cold plate in thermal contact with the mixing chamber. 
We use the cernox sensor for temperature stabilization. A typical time trace and the respective in-loop stability when there is active stabilization at 100~mK are shown in Fig.~\ref{fig:obs-temp} top plots. 
We notice that the temperature read-out is approaching its digitization limit but still acceptable at this temperature, which is the lowest rated working temperature for the Lakeshore 350 controller declared by its manufacturer. 
For the lack of supplementary thermometry connections, no out-of-loop temperature measurements has been carried out to date.  
However, since the locking bandwidth is about 1~Hz, the actual gain of the servo lock at 1~Hz is minimal, so that the ADEV at 1~s is still a realistic estimation of the actual temperature noise present on the cold plate. 

It is worth noting that this temperature noise is temperature dependent, and increases with the absolute temperature. The bottom plot in Fig.~\ref{fig:obs-temp} shows the ADEV at 1~s between 100~mK up to 700~mK, beyond which our dilution stage cannot ensure a stable operation. There appears to be two different trends : a roughly linear increase between 100~mK and 400~mK, and a higher noise level for higher temperatures. 

The dilution system achieves continuous and reliable operation within usual circumstances. 
Fragility was found in the liquid He circuitry, and several incidents of (reparable) leakage were caused by power outage, pulse tube failure, or false manipulation. 
Fail safe features are in the process of being improved.

\clearpage

\graphicspath{{D5_characterization/}}
\section{Linewidths of the optical references}
\label{sec:D5_S1}

\subsection{Finesse of the 124~K Fabry-Perot cavity}
\begin{figure}[hbt!]
    \centering
    \includegraphics[width=7cm]{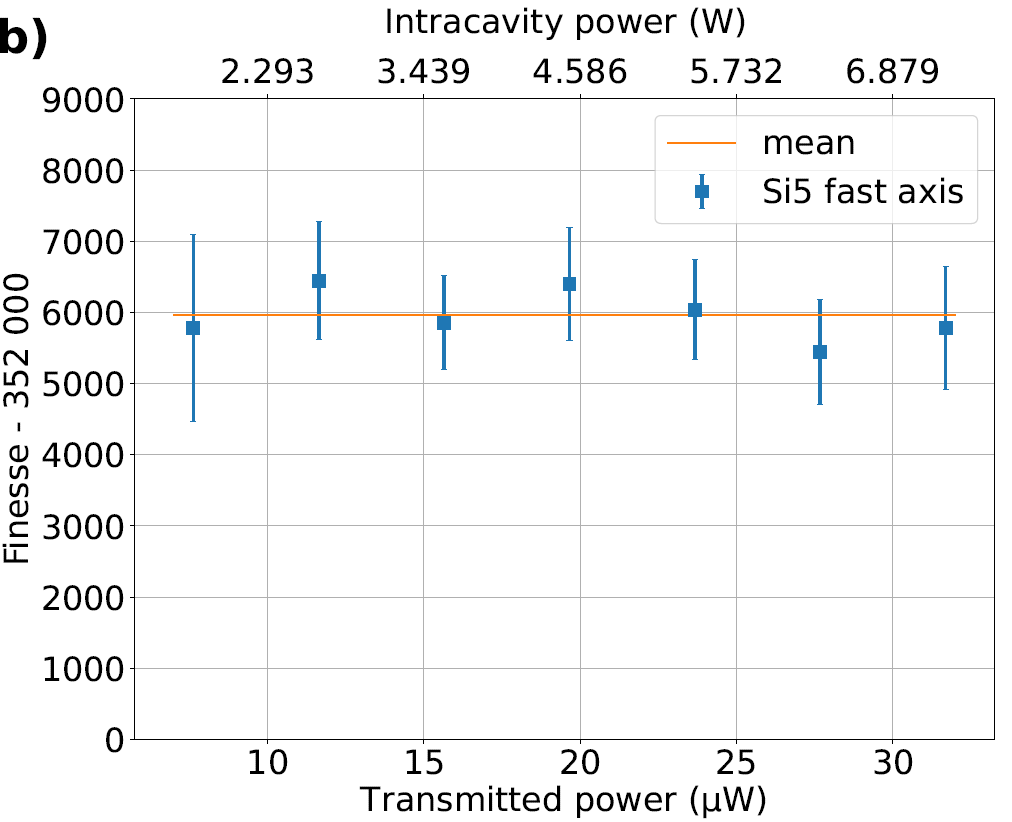}
    \caption{Finesse measurement of the Fabry-Perot cavity with GaAs/AlGaAs mirrors at PTB as function of the transmitted power, measured at a temperature of 124~K. Finesse values are obtained by the ringdown technique.}
    \label{fig:finesse_ptb}
\end{figure}

Before the cooling was switched to the cryocooler, the optical properties of the Si5 cavity were characterized. 
As the semiconductor-based AlGaAs coatings might show multi-photon absorption, the finesse of the resonator was measured versus the intracavity optical power. 
The intracavity power can be obtained from the transmitted power \cite{yu23a}, using a mirror transmission of $P_\mathrm{intra} / P_\mathrm{trans} = 0.229~$W/$\mu$W. 
No dependency was observed within the measurement range, as pictured in Fig.\ \ref{fig:finesse_ptb}. 
Given the spacer length of 21.2~cm, the average measured finesse of 358\,000 corresponds to a resonance linewidth of about 2~kHz.

\subsection{Finesse of the sub-kelvin Fabry-Perot cavity}
An initial cooldown of the cavity to 3~K was performed. The closest temperature sensor to the cavity is located on the aluminum support frame, below the resonator. The frame reached a temperature of 3.6~K in about 8.5 days. The cooldown was therefore much longer than previous unloaded cooldowns, which lasted less than 2 days. The final temperatures at the cavity frame and the mixing plate are very similar, with less than 1 \% difference. As discussed below, we think that the frame temperature is a good indicator of the cavity temperature in this range.

The dilution cycle was engaged in April 2024. A minimum temperature of 340~mK was reached at the cavity frame. Cooling down from 3.6~K to 340~mK takes roughly 4 hours, as all heat capacities and resistances drop significantly at these temperatures. In this temperature range, there is a high gradient between the mixing plate temperature, which is above the cavity support frame and cools down to 34~mK, the frame top which reaches 200~mK, and the frame bottom that stabilizes to 340~mK. We consider that the frame bottom temperature indicates a lower bound of the cavity actual temperature, which is unknown. The main explanation for this unexpected high gradient is a missing window in the cryocooler, that leads to a direct exposure of the cavity input mirror to the 4~K shield radiation. This window had to be temporarily removed for alignment purposes, and will be placed back in future experiments. 

\subsubsection{Finesse measurements}

\begin{figure}[h!]
	\centering
	\includegraphics[width=10cm]{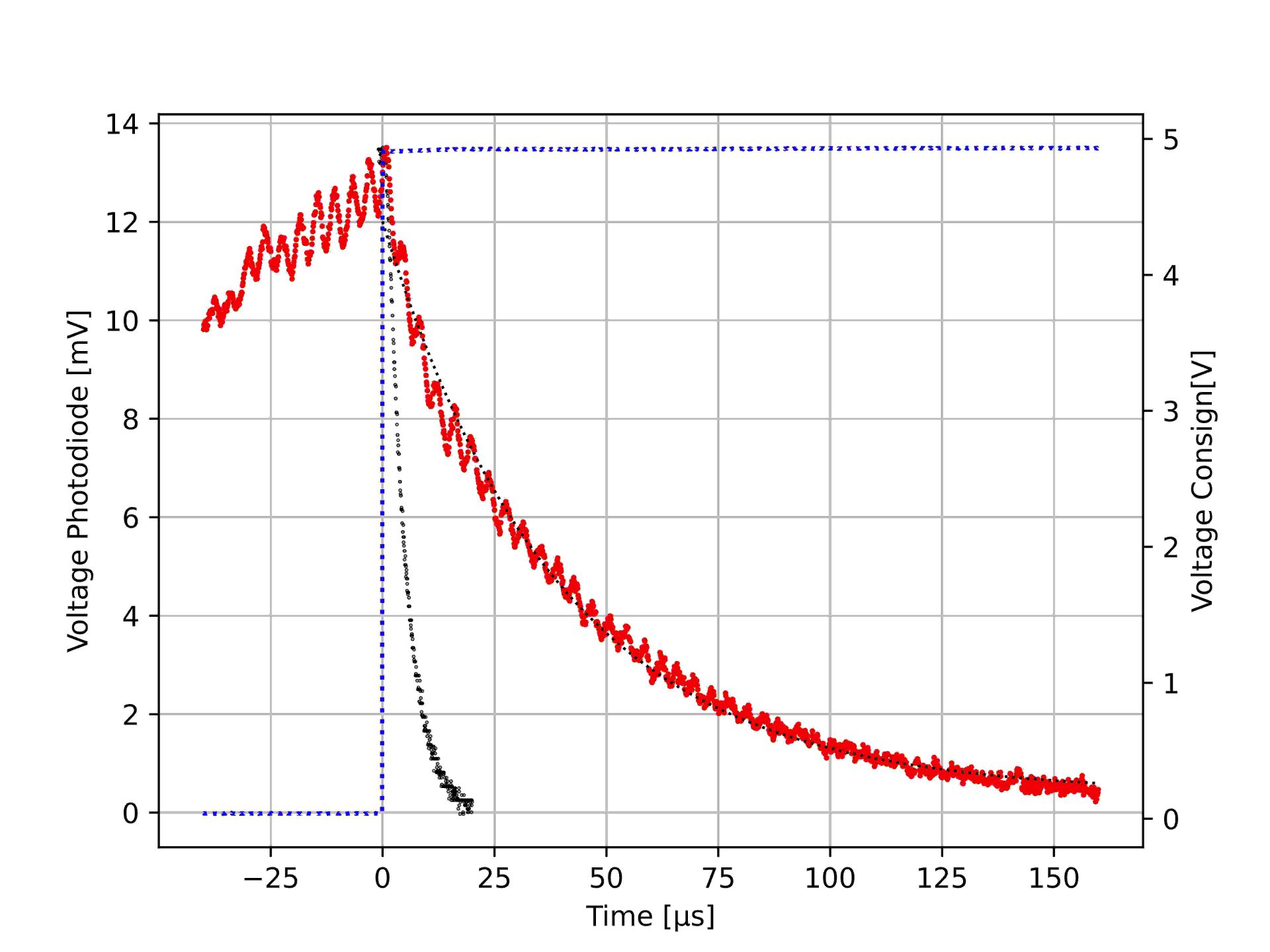}
	\caption{Finesse measurement of the single-crystal silicon cavity at room temperature. 
	Red: cavity transmission. A beatnote is clearly visible that is a signature of birefringence splitting. 
	Black: photodiode response. 
	Blue dotted line: voltage control of the RF switch that turns the light off.}
	\label{fig:ringdown_cnrs}
\end{figure}

The assembled cavity (Fig.\ \ref{Fig:mount}) shows a finesse of about 220\,000 at room temperature (Fig.\ \ref{fig:ringdown_cnrs}) determined by ring-down measurement. Finesse in vacuum and at low temperatures is expected to be higher, as we face residual absorption in air at wavelengths close to 1542~nm and coating layer thickness is optimized for 17~K.

To simultaneously measure the resonance splitting caused by the coatings birefringence, we injected the cavity with a linearly-polarized light with a projection on both crystalline axis. 
This results in a beatnote signal superimposed to the typical exponential decay of the ring-down measurement. 
The TEM00 mode splitting due to the birefringence of the coatings is of about 250 kHz (Fig.\ \ref{fig:ringdown_cnrs}), in good agreement to the birefringence measured in other cavities using AlGaAs mirrors \cite{yu23a,ked23}.

\begin{figure}[hbt!]
    \centering
    \includegraphics[width=7cm]{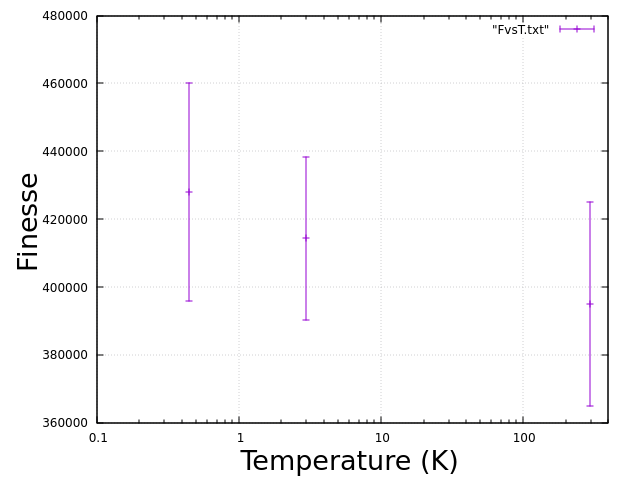}
    \includegraphics[width=7cm]{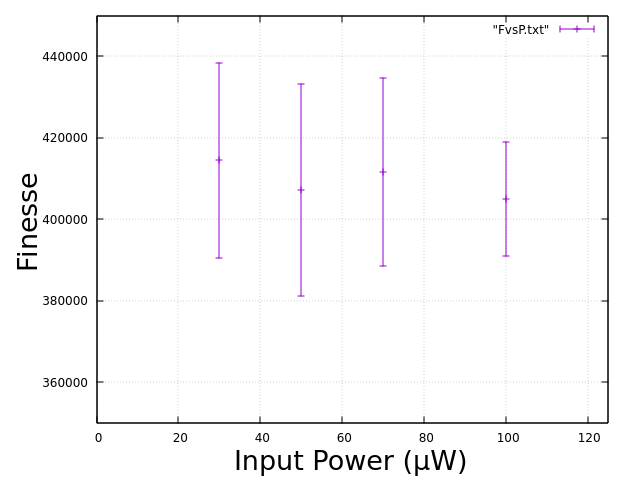}
    \caption{Finesse measurement of the Fabry-Perot cavity at CNRS. Finesse values are obtained by the ringdown technique. Each point is the mean of ten measurements, error bars are one standard deviation. Left: Finesse versus temperature. Right: Finesse versus input power to the cavity.}
    \label{fig:finesse_dicav_1}
\end{figure}

The cavity finesse increased to 400\,000 upon pumpdown to ultra-high vacuum. We attribute this increase to losses in ambient pressure due to absorption lines of water close to our operating wavelength of 1542~nm. We measured the cavity finesse at different temperatures, as shown Fig.~\ref{fig:finesse_dicav_1}. As expected, there is a noticeable increase at lower temperature.

The finesse was also measured for different input optical powers to the cavity, at a temperature of 3~K. Similarly to the measurements by PTB, there is no measurable variation of the finesse over the measured range.

\subsection{SHB reference below 1~K}
\subsubsection{Width of the holes and frequency discriminator}
A linearly polarized Gaussian beam is used to imprint spectral hole patterns in this setup, as a polarization aligned to the D1 dielectric axis of the Eu:YSO crystal maximizes the signal. 
Using a constant optical power and staying below the threshold of power broadening, which was experimentally seen beyond a peak intensity of 180\,nW/cm$^{2}$, a systematic characterization of the effect of the burn duration, which is proportional to the total deposited energy, on the spectral hole line width and the frequency discriminator has been performed on the crystal substitution site 1. 
Below 1~K, the FWHM is around 1 to 2 kHz. The frequency discriminator can reach a maximum value of around 0.6~mrad/Hz at 100-300~mK.

Both the linewidth increases with longer burn durations. 
The minimum line width at temperature $T < 1$\,K was measured in the absence of a polarizing magnetic field, and by optimizing burn parameters. 
FWHM values as low as 600\,Hz were obtained. 
Even though this is remarkably low, the decreased value of the frequency discriminator rules out the use of such spectral holes for high-performance metrology applications.

For laser frequency stabilization, suitable burning parameters were found such that 0.6 mrad/Hz slope and FWHM of 1.2~kHz is achieved \cite{lin24a}.

\begin{figure}[hbt!]
    \centering
    \includegraphics[width=7cm]{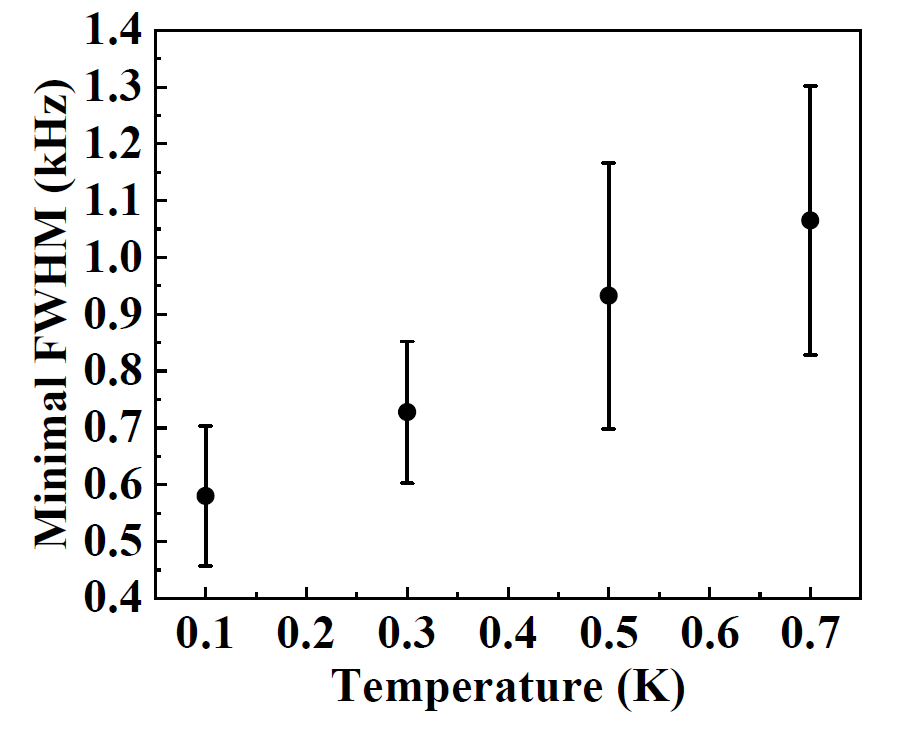}
    \caption{Lowest achieved FWHM of the spectral hole versus temperature.}
    \label{fig:min_fwhm_shb}
\end{figure}

The crystal substitution site 2 is known to exhibit narrower spectral holes with weaker signals at pulse-tube temperatures, giving comparable frequency discriminator values. We have not been able to characterize its behaviour at sub 1 K, due to technical issues encountered in our dilution cryostat. However, we do not expect significant improvement in frequency discriminator, which is the key parameter for laser stabilization.

\section{Temperature sensitivity of the resonators}
\subsection{124~K Fabry-Perot cavity}
\begin{figure}[hbt!]
    \centering
    \includegraphics[width=8cm]{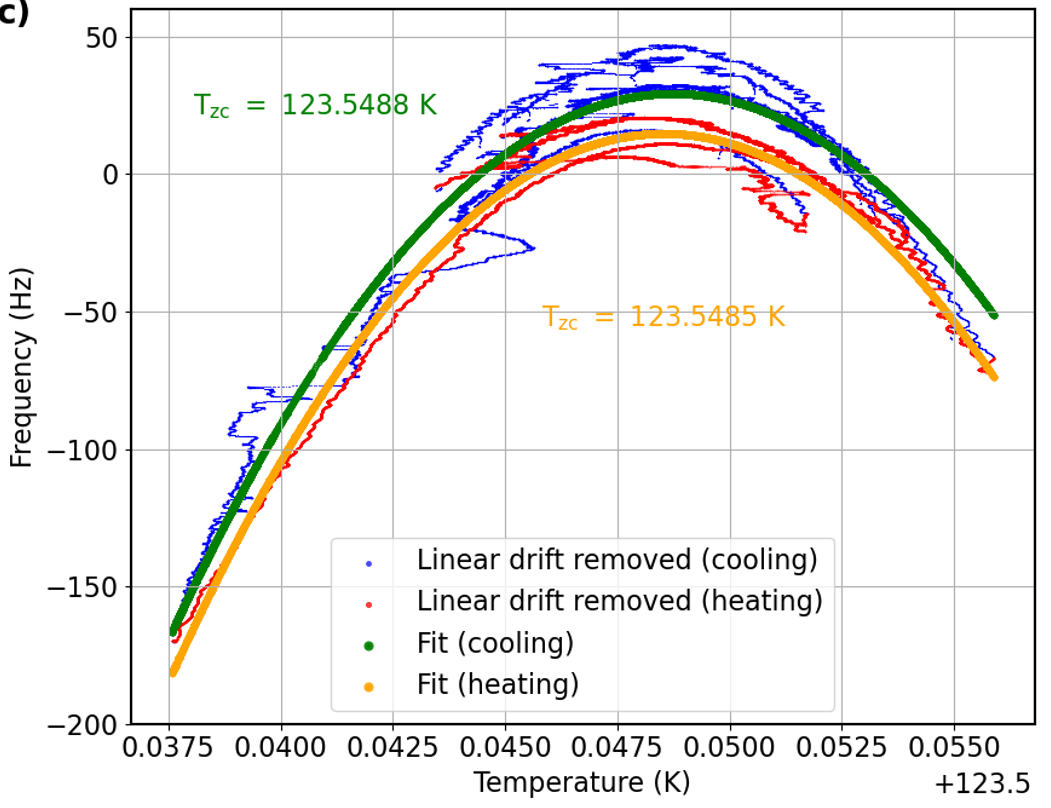}
    \caption{Frequency of Si5 Fabry-Perot cavity versus cavity temperature. The red and blue curves show the frequency change during heating and cooling, while the green and orange curves are quadratic fits to the data for cooling and heating respectively.}
    \label{fig:cte_ptb}
\end{figure}

The coefficient of thermal expansion (CTE) of the cavity Si5 has also been measured around its zero-crossing point, as shown in Fig.\ \ref{fig:cte_ptb}. 
The temperature was changed by about $\pm 20$~mK up and down during one day. 
The plot shows the resonator resonance frequency change versus the cavity temperature. 
After removal of a frequency drift over time, the experimental data was fitted with a quadratic function.
This resulted in a linear CTE $\alpha$ around the crossover temperature $T_0$, with $\alpha(T)=-a(T-T_0)$, $T_0 = 123.549 (5)$~K and $a\approx1.655{\times}10^{-8}$/K$^2$. 
The values slightly differ between cooling and heating, possibly because of additional contributions from the mounting structure or temperature inhomogeneities.

\subsection{Fabry-Perot cavity below 1~K coefficient of thermal expansion}

A laser is locked to the single-crystal silicon resonator using the Pound-Drever-Hall technique. The frequency of the laser is monitored through a beatnote with a laser stabilized to another optical reference cavity. The fractional frequency stability of this reference at 1~s is $\sim 6{\times}10^{-15}$. At longer integration times, its frequency is corrected to cancel the drift by a comparison with the phase of a hydrogen maser signal through an optical frequency comb. This enables measurements over long times without contribution from the optical reference to the measured drift.

The PDH lock could not be maintained continuously during the initial cooldown from room temperature, as the frequency change is significant and would require a greater dynamics. However the beatnote with the frequency reference was intermittently measured. We could register a sign change in the frequency variation with temperature at 16.98~K, very close to the expected temperature of 17~K. As discussed above, the frame temperature seems to provide a good estimate of the cavity temperature in this temperature range.

\begin{figure}[hbt!]
    \centering
    \includegraphics[width=10cm]{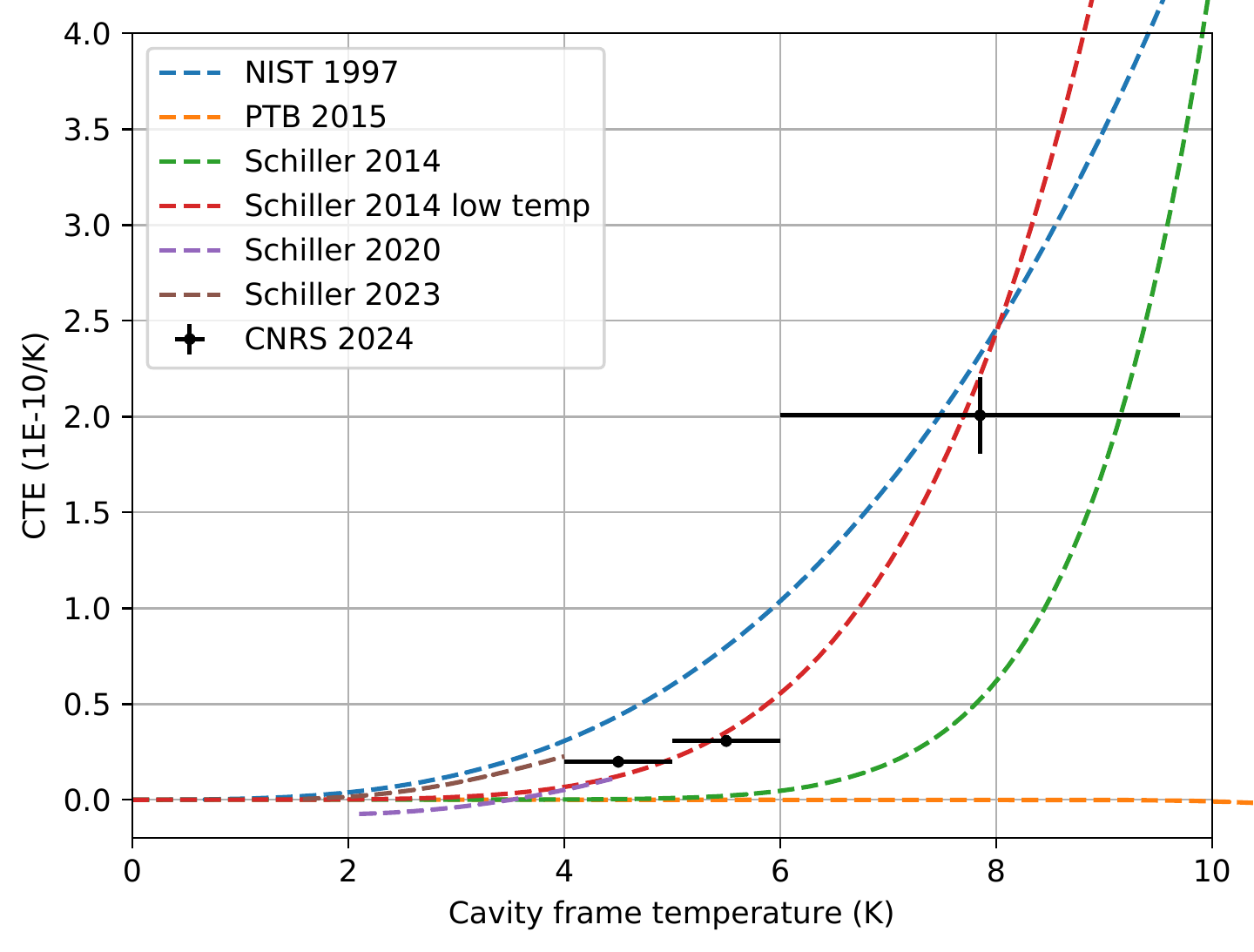}
    \caption{Preliminary measurement of the CNRS cavity above 4~K. NIST 1997 is from \cite{white1997}, PTB 2016 is from \cite{mid15}, Schiller 2014 and Schiller 2014 low temp are from \cite{wiens2014}, Schiller 2020 is from \cite{wiens2020}, Schiller 2023 is \cite{wiens2023}.}
    \label{fig:cte_dicav}
\end{figure}

Three temperature steps were performed, from 4 to 5~K, 5 to 6~K, and 6 to 9.7~K, while monitoring the cavity frequency change. The resulting CTE estimation is shown Fig.\ \ref{fig:cte_dicav}, along with several models from the literature \cite{white1997, mid15, wiens2014, wiens2020, wiens2023}. Our results are in good agreement with measurements from the Schiller group, and will be refined in future experiments.

\subsection{Temperature sensitivity of spectral holes}

The expected scaling of the optical frequency shift $f_{\rm shift} = \nu-\nu_0$ of a given spectral hole is $f_{\rm shift} = \alpha T^4$, where $\alpha$ is a constant of proportionality, with a sensitivity $\frac{df_{\rm shift}}{dT} = 4\alpha T^3$, stipulated by the two-phonon Raman process~\cite{konz2003temperature}.

An initial measurement at pulse-tube temperatures was performed, yielding $\alpha = $ 71(3) Hz\,K$^{-4}$ and 253(4) Hz\,K$^{-4}$ for site 1 and site 2 respectively. 
These numbers imply an estimated sensitivity of $\approx$ 0.3~Hz/K around $T\approx100$~mK according to the above mentioned scaling law in the best scenario (site 1, which is less sensitive to temperature changes).

\begin{figure}[hbt]
    \centering
    \includegraphics[width=10cm]{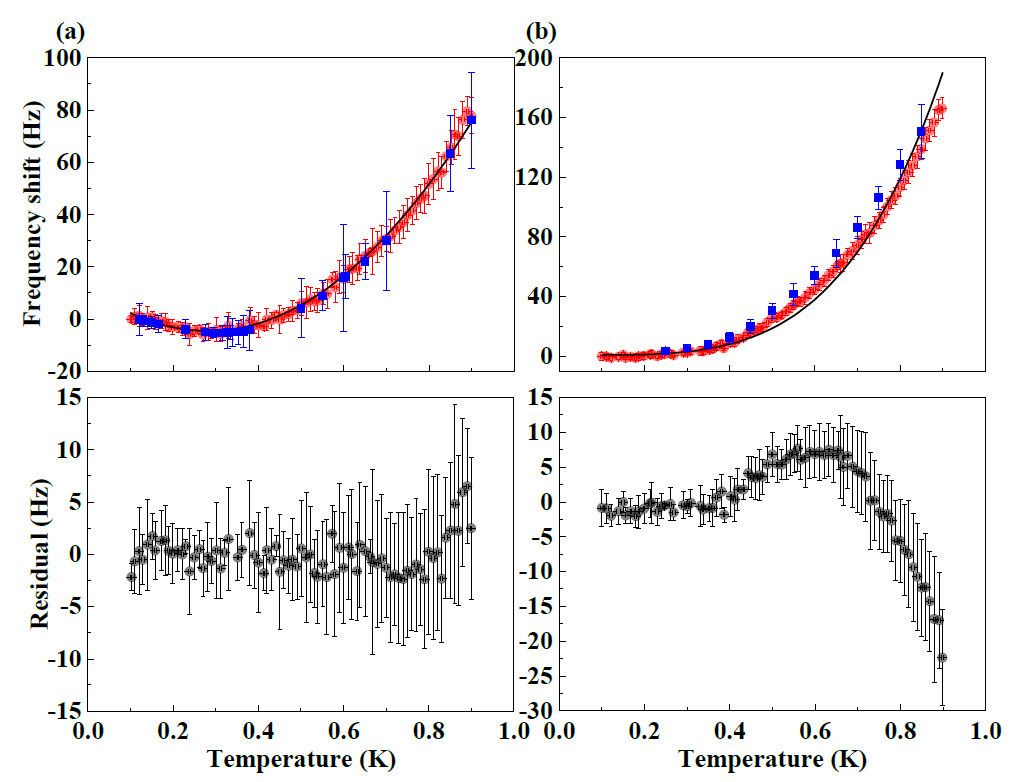}
    \caption{Top: The measured frequency shifts and curve fits for crystal site 1 (a), using a parabolic fit exhibiting a local minimum and site 2 (b) , using a $\alpha T^4+{\rm constant}$ fit. The center of a spectral hole pattern is recorded when modulating the temperature set point ($\blacktriangle$/blue), and when ramping the temperature through all achievable values in the sub 1 K range ($\bullet$/red). 
    Bottom: residuals from curve fit. The error bars correspond to the statistical standard deviations.}
    \label{fig:tempe_shb}
\end{figure}

Low-temperature measurements were performed at both crystal substitution sites \cite{lin24}, as shown in Fig.\ \ref{fig:tempe_shb}. 
Two different behaviours were observed.

Site 2, which corresponds to Fig.~\ref{fig:tempe_shb} (b): a $f_{\rm shift} = \alpha T^4$ behaviour is observed, with $\alpha =$ 274(11)~Hz\,K$^4$. This is consistent with the previously cited $\alpha =$ 253(4)~Hz\,K$^4$ within $2\sigma$ and with the data reported by NIST~\cite{tho13}.

Site 1, which corresponds to Fig.~\ref{fig:tempe_shb} (a): the data obtained with this site matches a quadratic model, with an equivalent zero-CTE temperature of around 290 mK. The physics that leads to this temperature-sensitivity cancellation is not well understood, but can certainly lead to even lower linearized sensitivity to temperature changes.

\section{Expected performances and preliminary stability measurements of the cryogenic optical frequency references }

\subsection{Fabry-Perot cavity at 124~K}

\subsubsection{Vibration influence}
The Si5 cavity is mounted on an anti-vibration isolation (AVI) system.
The sensitivity of the resonator to accelerations in the vertical and in the horizontal East-West and North-South direction was determined by exciting the AVI system along different directions using its control input.

The sensitivities below 1 Hz Fourier frequency are summarized in Tab.\ \ref{table:vib_rot_sen}:
\begin{table}[!ht]
{
\centering
\renewcommand{\arraystretch}{1.2}
\begin{tabular}{c|c}
\hline
Motion & Sensitivity \\\hline
East-West & $(3.1\pm 1 )\times 10^{-11}$/g  \\\hline
North-South & $(2.1\pm 1 )\times 10^{-11}$/g  \\\hline
vertical & $(6.5\pm 1 )\times 10^{-11}$/g \\\hline
\end{tabular}
\caption{Vibrational sensitivities at Fourier frequencies below 1 Hz of Si5 measured at 124 K.}
\label{table:vib_rot_sen}
}
\end{table}

\begin{figure}[hbt!]
    \centering
    \includegraphics[width=16cm]{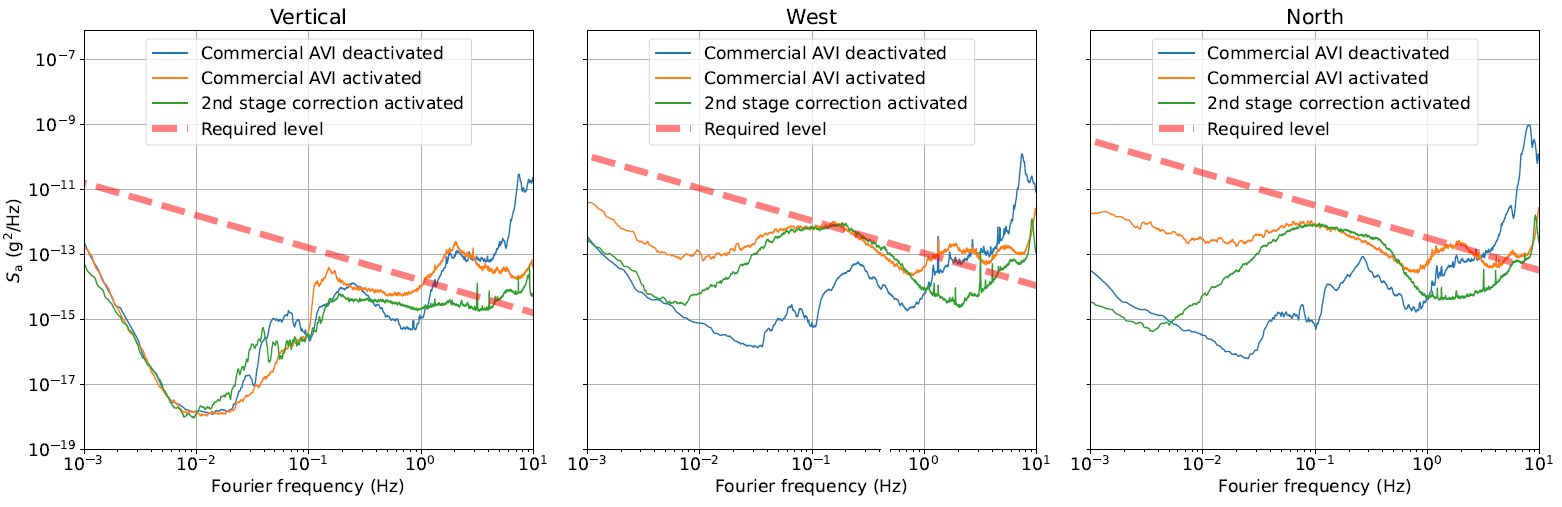}
    \caption{Vibrations at cavity with the LN2 cooling system. The dashed lines indicate the level for $1\times10^{-17}$ flicker instability. Levels without the commercial AVI, with commercial AVI and with the additional low-frequency stage activated.}
    \label{fig:LN2}
\end{figure}

\begin{figure}[hbt!]
    \centering
    \includegraphics[width=16cm]{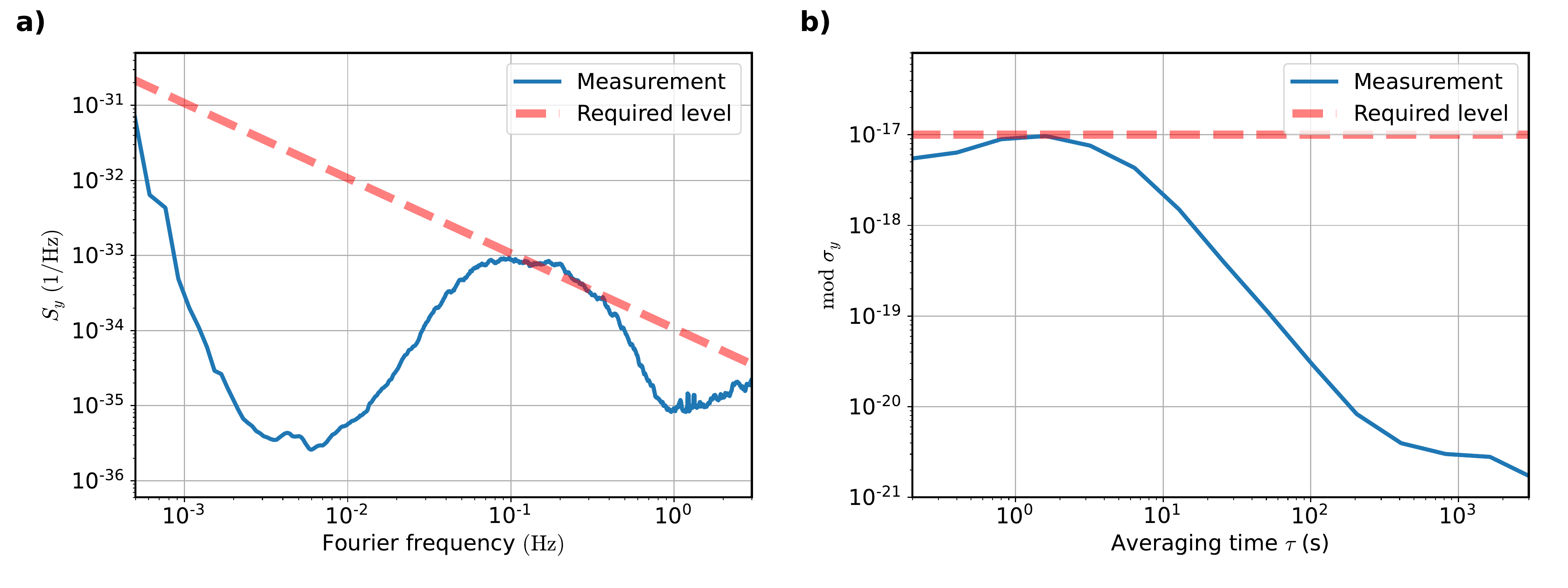}
    \caption{Calculated frequency instability induced by vibrational and rotational accelerations in PSD (a) and MADEV (b).
    The required level for $1 \times 10^{-17}$ flicker noise is included for reference (from \cite{yu23}).}
    \label{fig:vib_stab_ptb}
\end{figure}

With these sensitivities and the measured noise spectra (Fig.\ \ref{fig:LN2}) the frequency noise and the corresponding Allan deviation of Si5 with the LN2 based cooling are shown in (Fig.\ \ref{fig:vib_stab_ptb}).

\begin{figure}[hbt!]
    \centering
    \includegraphics[width=9cm]{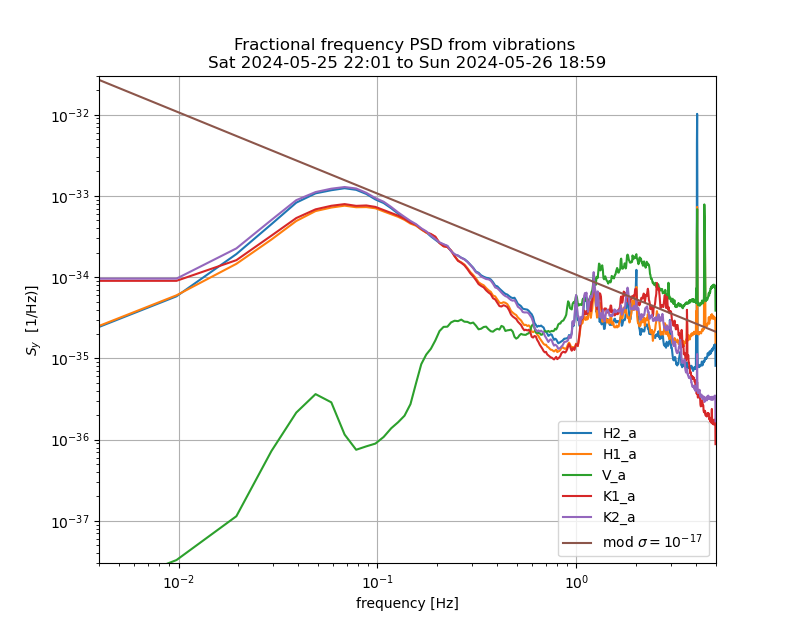}
    \caption{Vibration-induced frequency fluctuations of the Si cavity with the closed-cycle cooling system. The brown line indicate the level for $1\times10^{-17}$ flicker instability. The other lines correspond to vertical and two horizontal directions and to the signals from the tilt sensor. Only the commercial AVI system was activated.}
    \label{fig:PTB_CC}
\end{figure}

Later, the cooling was replaced with closed-cycle cryo-cooler, leading to essentially the same vibration levels (Fig.\ \ref{fig:PTB_CC}).

\subsubsection{Projected performances}
\begin{figure}[hbt!]
    \centering
    \includegraphics[width=12cm]{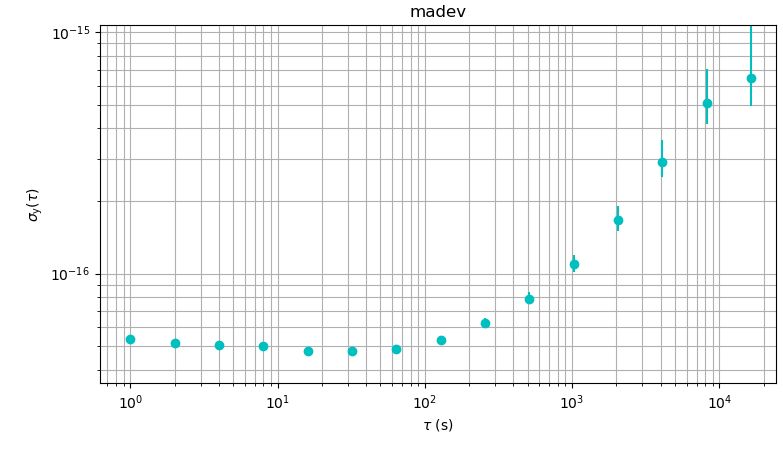}
    \caption{Fractional frequency stability between Si2 and Si5 expressed in modified Allan deviation. Si5 is operated with the closed-cycle cooler.}
    \label{fig:PTB_MADEV_Si2_Si5}
\end{figure}

The temperature noise of the closed-cycle cryo-cooler has been measured, as documented above. 
Based on the measured temperature noise and assuming, that the temperature is within $\pm 10$~mK of the inversion point, we can infer that the temperature fluctuations will impact the cavity with a level of less than $10^{-17}$ for averaging times up to a few hundred seconds.
Thus, the closed-cycle operation will maintain the performances of the cavity obtained with the liquid nitrogen bath, at $4\times 10^{-17}$ as measured in comparison with Si2 (Fig.\ \ref{fig:PTB_MADEV_Si2_Si5}).
For two systems with similar stability, the measured instability of the difference of $5\times 10^{-17}$ between 1~s and 100~s would correspond to individual instabilities of $3.5\times 10^{-17}$. 

\begin{figure}[!ht]
\centering\includegraphics[width=\textwidth]{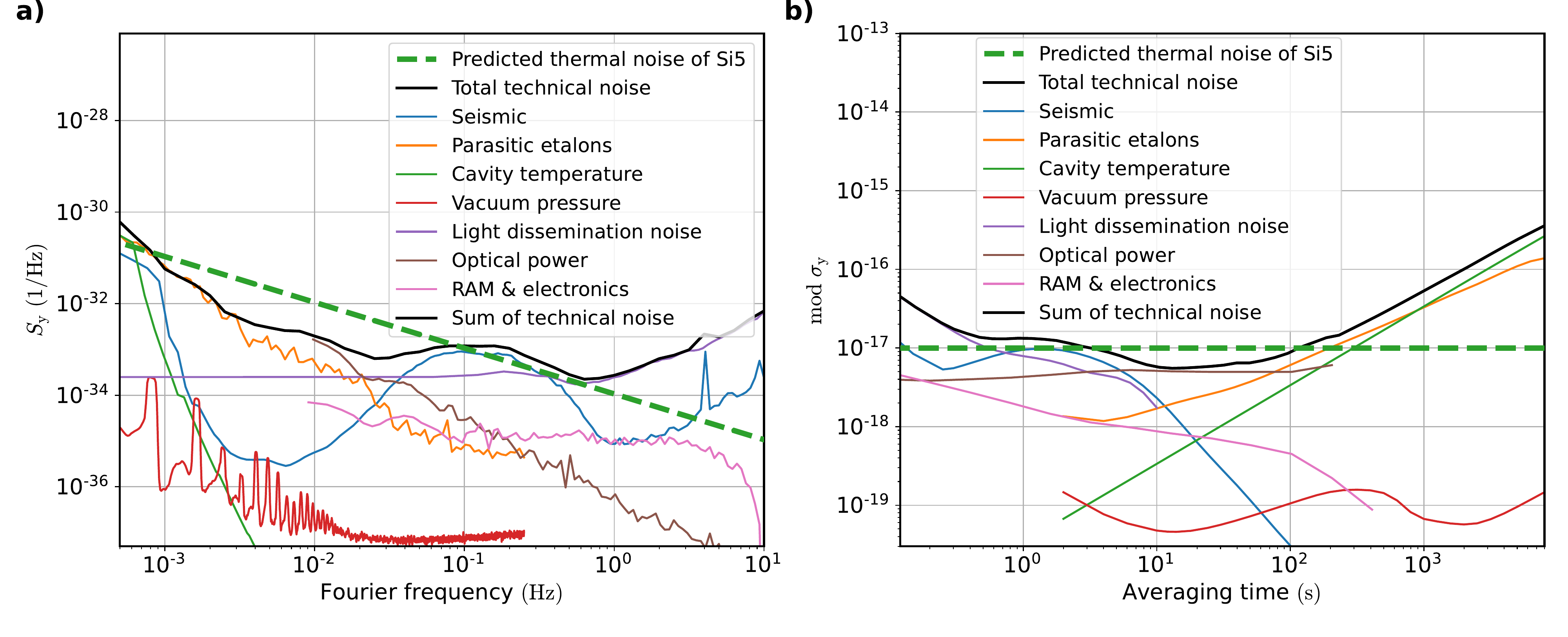}
\caption{Technical noise contributions and their sum in PSD (a) and MADEV (b) (from \cite{yu23}). 
The required level (green dashed) for the $1\times 10^{-17}$ Brownian thermal noise level of AlGaAs coatings is indicated by the green dashed line. }
\label{fig:sum_tech}
\end{figure}

If the coating noise could be suppressed further, the technical noise contributions could support instabilities of $1\times 10^{-17}$ (Fig.\ \ref{fig:sum_tech}).

\subsection{Fabry-Perot cavity below 1~K}
\subsubsection{Projected contribution of temperature fluctuations}
\begin{figure}[hbt!]
    \centering
    \includegraphics[width=10cm]{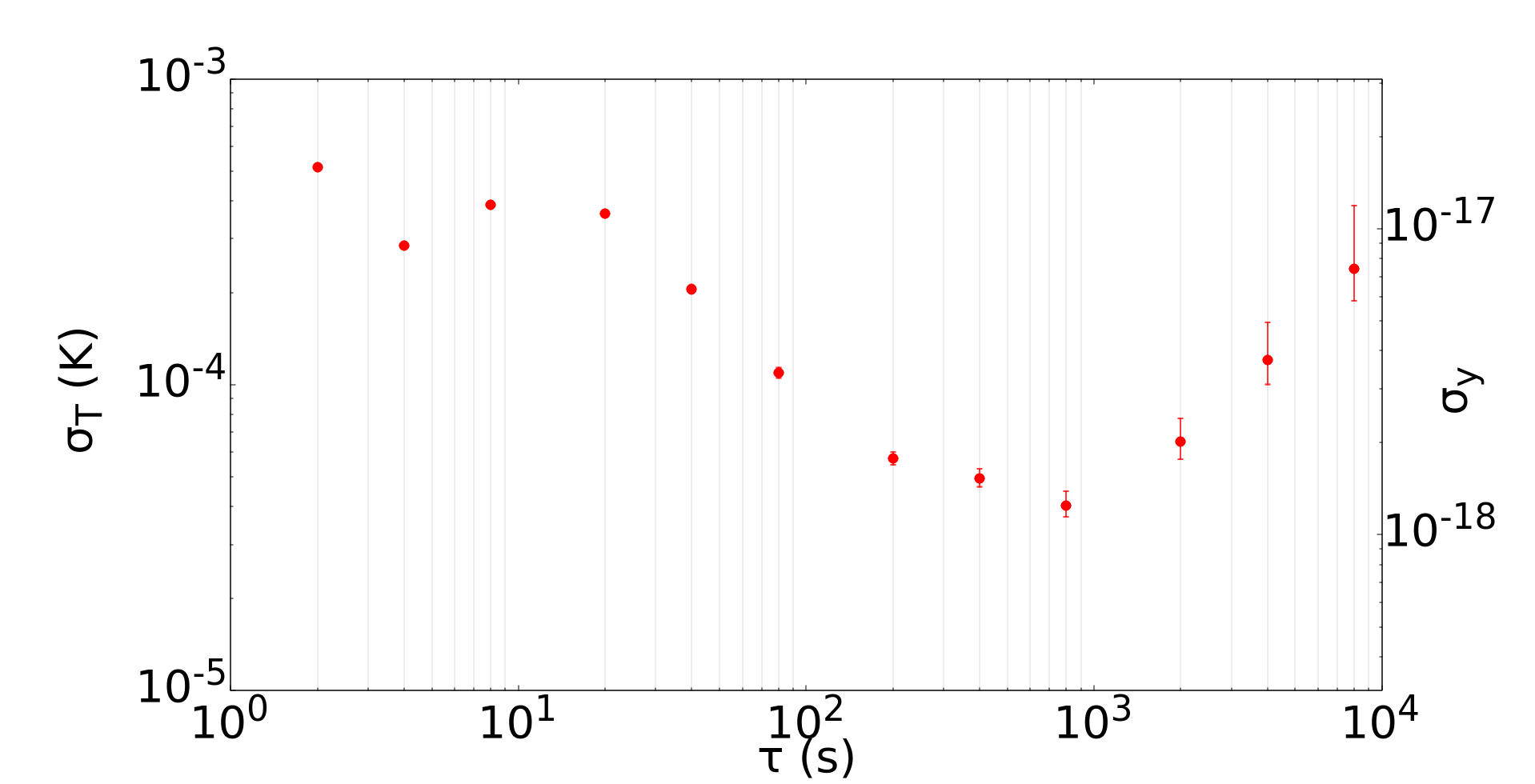}
    \caption{Temperature stability of the cavity mount (left axis) and expected contribution to the laser fractional frequency stability (right axis) of the CNRS cryogenic cavity.}
    \label{fig:stab_tempe_dicav}
\end{figure}

We have monitored the temperature of the frame bottom to estimate its temperature stability. When operating at an average temperature of 0.4~K, we observe temperature fluctuations of order 0.5~mK. Using the estimated CTE from \cite{lyon2008} at 0.4~K, we plot the temperature stability and its expected contribution to the frequency stability Fig.~\ref{fig:stab_tempe_dicav}.

For this measurement, there is no temperature lock engaged and the equilibrium temperature is set by the combination of the cryocooler cooling power, the optical power coupled to the cavity, and radiation from the 4~K shield. The temperature stability can therefore be greatly improved in future experiments. Under those operating conditions, we obtain a projected contribution from temperature fluctuations of less than $1.5{\times}10^{-17}$ at all times.

\subsubsection{Preliminary measurement of the fractional frequency stability of the cavity}

\begin{figure}[hbt!]
    \centering
    \includegraphics[width=10cm]{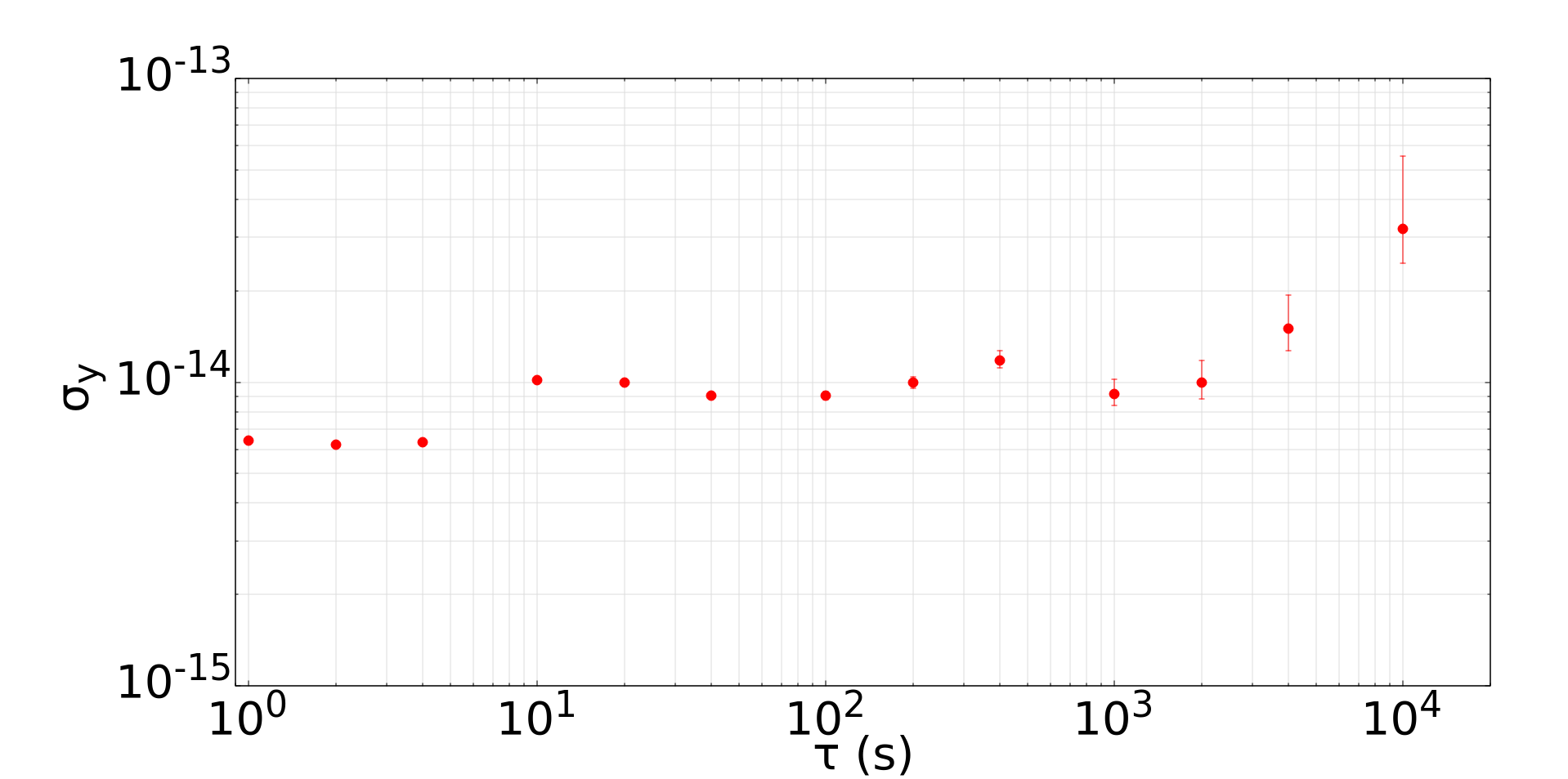}
    \caption{Fractional frequency stability of the CNRS Fabry-Perot cavity, measured at 3.6~K. The measurement is performed by beatnote against another, room-temperature Fabry-Perot cavity.}
    \label{fig:stab_dicav}
\end{figure}

An initial measurement of the laser fractional frequency stability has been performed at 3.6~K. It is important to point out that only the PDH was setup at the time of the experiment. No power stabilization or residual amplitude modulation cancellation were yet setup. The result is shown Fig.\ \ref{fig:stab_dicav}. 

Below 10~s, the measurement is limited by the room-temperature cavity that was used to perform the beatnote. Between 10 and 1000~s, we observe a $10^{-14}$ floor, followed by an increase compatible with a random walk. This behaviour needs to be analyzed.

\subsection{SHB below 1~K}
\subsubsection{Projected performances of spectral holes at cryogenic temperatures}

\begin{figure}[hbt!]
    \centering
    \includegraphics[width=10cm]{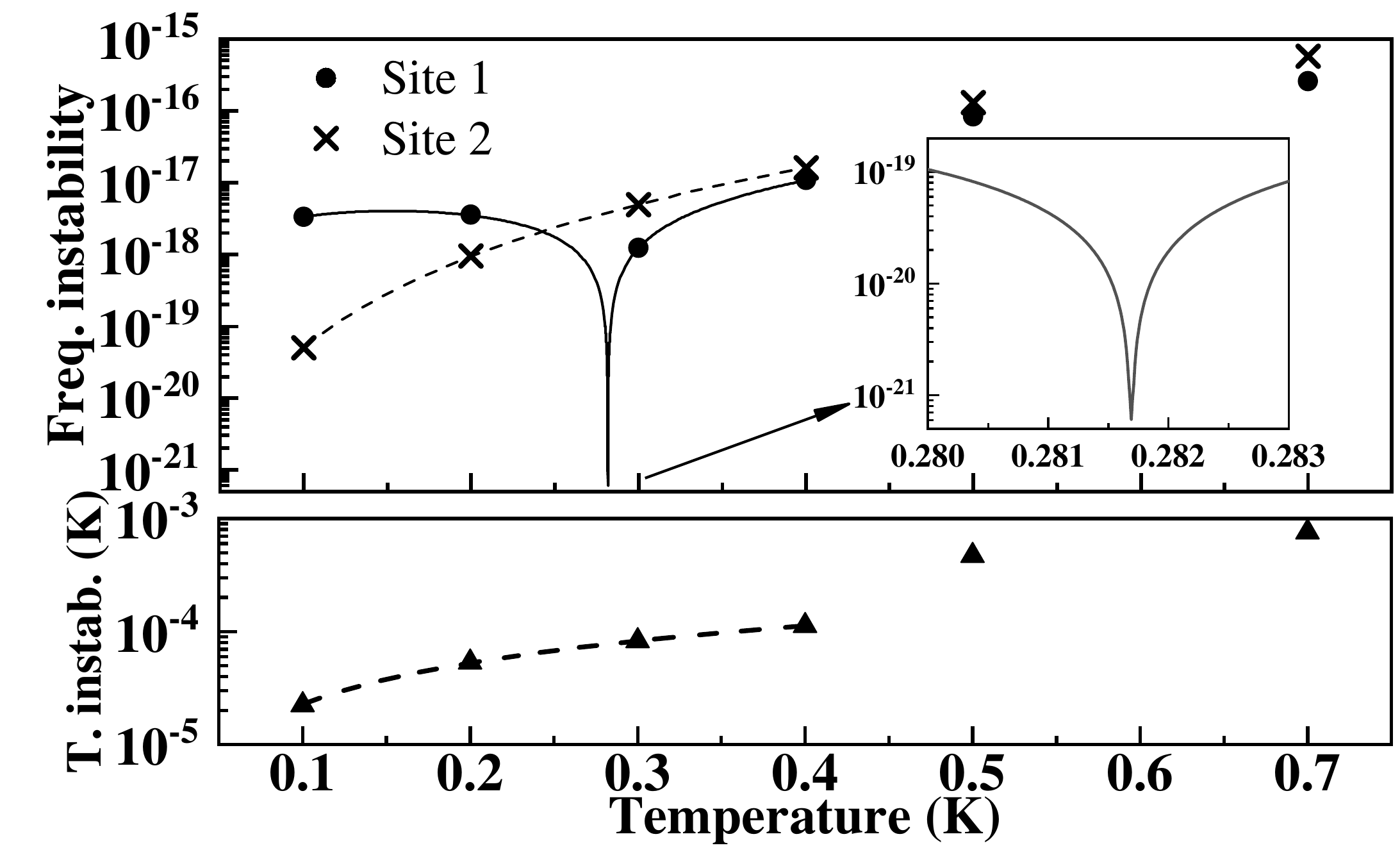}
    \caption{\label{fig:shb_T2f} Bottom: Measured temperature instabilities ($\blacktriangle$) fitted linearly (dashed line). 
		Top: Corresponding projected temperature induced fractional frequency instabilities at 1\,s. 
		The inset shows the details around the zero-frequency-drift temperature point. 
		The (statistical) error bars are smaller than the size of the markers.}
\end{figure}

The achieved temperature instabilities scale linearly with temperature in the range from 0.1 K to 0.4 K.  
This provides a means to estimate the temperature instability $\sigma_T$ at 1\,s averaging time for any temperature $T$ in this interval. 
The projected frequency instability due to temperature fluctuations are presented in Fig.~\ref{fig:shb_T2f}. 
The best scenario, corresponding to spectral holes at site 1 at the equivalent zero-CTE temperature, indicates a projected frequency instability as low as $6\times 10^{-22}$ at 1 s, which is compatible if not lower than the best scenario at 4\,K in a helium gas cell~\cite{zhang2023first}, and certainly well below any other noise sources present in the current system. 

The residual vibrations and their management is presented in chapter \ref{ch:vib_isolation}. 
The equivalent Doppler noise is estimated experimentally at $5\times 10^{-16}$ at 1\,s in fractional frequency, using a Mach-Zehnder interferometer. 
Whereas the Doppler effect can be cancelled optically down to about $1.5\times 10^{-16}$ by modifying the phase of the interrogation laser, the underlying movement of the crystal also implies a strain induced noise at a level below $10^{-17}$. 
These numbers are likely to become a technical limitation after the detection noise, and will require further efforts for reducing the motion of the crystal.  

\subsubsection{Preliminary measurement of the optical reference frequency stability}

Parameters have been optimized to obtain the best possible fractional frequency stability while rejecting temperature fluctuations. 
Site 1 is used, close to the temperature inflexion point of 290~mK, to reduce the impact of the temperature noise.
The burning parameters were optimized such that a 0.6 mrad/Hz slope is achieved.

\begin{figure}[hbt!]
    \centering
    \includegraphics[width=7cm]{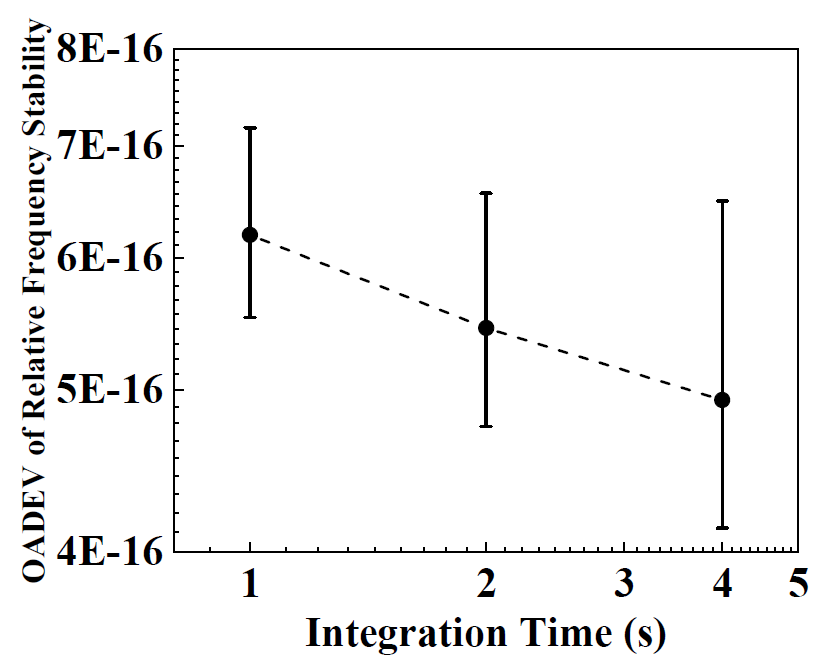}
    \caption{Preliminary fractional frequency stability measurement. The overlapping Alan deviation of an optical beatnote at 1.5 µm between the SHB-stabilized laser and an optical reference is shown versus integration time.}
    \label{fig:stab_shb}
\end{figure}

A preliminary fractional frequency instability measurement, evaluated through a beatnote between the SHB-stabilized laser and a Fabry-Perot cavity stabilized laser (optical reference at 1.5~µm), is shown in Fig.\ \ref{fig:stab_shb}. The resulting instability is $6{\times}10^{-16}$ at 1~s, which corresponds to $4{\times}10^{-16}$ for the SHB laser after subtracting about $4{\times}10^{-16}$ for the reference cavity based on independent evaluations.

\clearpage

\graphicspath{{D6_transfer/figures}}
\chapter[Stability transfer]{Stability transfer from ultrastable lasers to wavelengths applicable to optical clocks}
\label{cha:trans}

\authorlist{%
Erik Benkler$^1$, 
Cecilia Clivati$^{2,\dag}$, 
Thomas Fordell$^3$, 
Kalle Hanhijärvi$^3$, 
Angelina Jaros$^1$, 
Alessia Penza$^2$, 
Benjamin Pointard$^4$, 
Matias Risaro$^2$, 
Paolo Savio$^5$, 
Uwe Sterr$^1$ 
Rodolphe Le Targat$^4$, 
Anders E. Wallin$^3$ 
}

\affil{1}{\PTBaff}
\affil{2}{\INRIMaff}
\affil{3}{\VTTaff}
\affil{4}{\OPaff}
\affil{5}{\LINKSaff}

\corr{c.clivati@inrim.it}

\chapstart
\section{Introduction}
As various novel techniques aim at pushing  laser instability to \SI{e-17}{} and beyond, methods  must be developed to enable such a superior stability being  exploitable in targeted applications, especially in the context of atomic physics, fundamental metrology and spectroscopy.  The best ultrastable lasers demonstrated so far operate at wavelengths optimized for stability of the references (spectral holes in crystals or high-finesse resonators with low-thermal noise mirrors) that generally differ from those actually needed, e.g.,  to interrogate optical clock transitions in atoms or ions. Moreover, some applications (e.g. optical clock ensemble operation) require several ultrastable lasers at different wavelengths simultaneously.
Thanks to its broadband spectrum, the optical comb can act as a transfer oscillator,  
enabling to convert the spectral properties of a reference laser to other spectral domains with no degradation.  The transfer oscillator approach can be used to stabilize one laser (the slave laser, e.g. at a wavelength of interest for optical clocks) to another (the master, e.g. a \SI{e-17}{}-instability  laser) in a feedback-loop. 
This operation can be performed for multiple slave lasers simultaneously, allowing a single, 
high performance reference laser to be maintained in a laboratory and leveraging its  use for many  experiments, in a modular approach.

\begin{figure}
\centering
\includegraphics[width=0.8\textwidth]{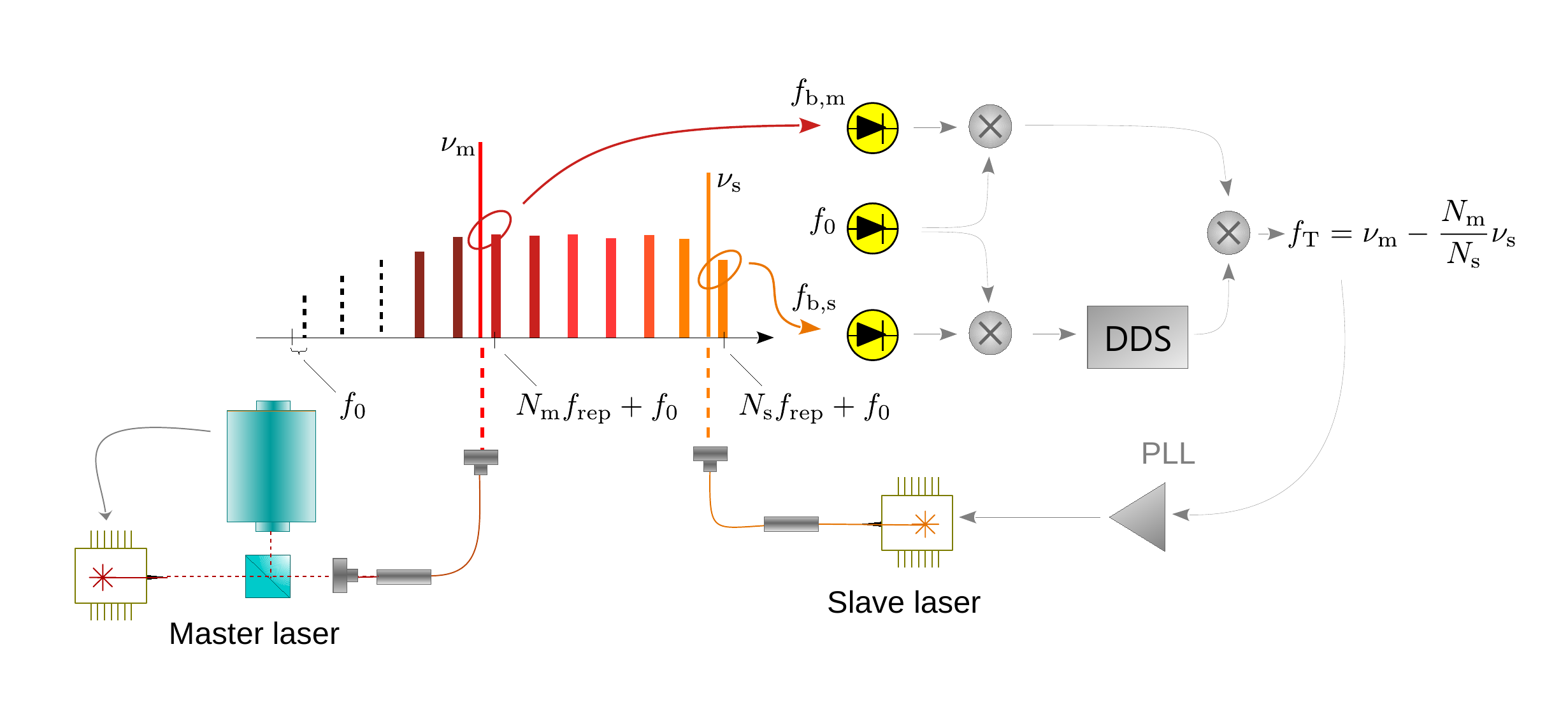}
    \caption{Scheme of the transfer oscillator principle. \label{fig:transfer_oscillator} }
\end{figure}

The  transfer oscillator principle has been extensively described in the literature \cite{telle} and consists in deriving the  relative frequency fluctuations of two lasers in different spectral domains by processing the beatnote of each individual laser with the closest comb tooth in such a way that the comb noise is rejected from the measurement.   
Briefly, we assume the master and slave lasers frequencies $\nu_\text{m}$ and $\nu_\text{s}$ to be described by the usual comb equation \cite{diddams}:
\begin{equation}
    \begin{split}
\nu_\text{m} &=N_\text{m} f_\text{rep} + f_0 + f_\text{b,m}    \\
\nu_\text{s} &=N_\text{s} f_\text{rep} + f_0 + f_\text{b,s}    \\
\end{split}
\end{equation}
where $f_\text{rep}$ is the comb repetition rate, $f_0$ the carrier-to-envelope offset, $f_\text{b,m}  $ and $f_\text{b,s}$ the beatnotes of the master and slave lasers with  the closest comb tooth of order $N_\text{m}$ and $N_\text{s}$. An RF signal with frequency $f_\text{T}$ (usually known as transfer beatnote) can be obtained by combining the three RF signals  $f_0$, $f_\text{b,m}  $ and $f_\text{b,s}$ with proper scaling such that:
\begin{equation}
\label{eq:transfer_beat}
f_\text{T} =(f_0 + f_\text{b,m})- \frac{N_\text{m}}{N_\text{s}} (f_0 + f_\text{b,s}) = \nu_\text{m}-\frac{N_\text{m}}{N_\text{s}}\nu_\text{s}
\end{equation}
It is straightforward to see that fluctuations in $\nu_\text{m}$ and $\nu_\text{s}$ are mapped into fluctuations of $f_\text{T}$, the latter with a scaling term $N_\text{m}/N_\text{s}$ that accounts for the spectral separation from the master, while fluctuations in comb parameters $f_\text{rep}$ and $f_0$ are rejected from the measurement.  
The transfer beatnote can be then used to feed a control loop that stabilizes the slave onto the master. 
This scheme can be implemented with real RF signals, e.g. using analog mixers to perform sums and differences and prescalers or Direct Digital Synthetizers for non-integer scaling \cite{johnson} (see Fig.\ \ref{fig:transfer_oscillator}), or by their real-time sampling and processing. 

For supporting newly-developed ultrastable lasers, the spectral transfer process must not contribute an instability  higher than $\SI{1e-17}{1/\sqrt{\tau/s}}$, $\tau$ being the measurement time in seconds. 
This requires addressing  critical aspects of the implementation, such as the delivery of radiation to the optical comb using fibers, non-common optical paths on the different comb branches, detection and in general electronics noise. In addition, for the scheme to be applicable to multiple laser simultaneously, a suitable design of the electronics must be considered to contain the growth of hardware complexity.

This chapter gives an overview of these aspects, describing their impact in terms of contributed noise and instability, as well as implementation strategies and methods to  mitigate or suppress it. 
Section \ref{sec:delivery} addresses the limits of  light-delivery systems based on optical fibers; section \ref{sec:noncommon_optical_paths} discusses the use of single-branch optical combs and associated issues; sections \ref{sec:improving_SNR}, \ref{sec:digital} and \ref{sec:distributing} address electronics noise in the detection and synthesis chains, introducing dedicated advanced signal processing and showing typical results of the optimized stability transfer from ultrastable lasers to wavelengths applicable to optical clocks. 

\section{Delivery of laser beams to the comb}
\label{sec:delivery}
Reference lasers, optical clocks and frequency combs are usually installed in different laboratories, with radiation distributed among them via  fibers. 
Their optical length needs to be actively stabilized and this is commonly done  via Doppler noise cancellation schemes. 
Most designs consist in a largely unbalanced Michelson interferometer, in which part of the light delivered to the comb is reflected back and compared to the input signal, to measure the noise accumulated in a round-trip. This is then cancelled either by direct actuation on the fiber physical length or on the carrier phase using an acousto-optic modulator (AOM) \cite{williams2008,rauf2018}. 
For achieving the lowest uncertainty, not only the fiber length variations but the entire optical path until the point where radiation is combined to the comb,  usually with a free-space interferometer, must be stabilized. 
Free-space paths are less subject to temperature changes, acoustic noise, pressure and humidity than optical fibers. Nevertheless, deformation of the optical bench geometry or the air refractive index \cite{ben19} can contribute additional instability. Undesired  fluctuations of the short, reference arms at the launching end of the Michelson interferometer contribute as well, as they are indistinguishable from the actual path to be stabilised. 
Finally, while it is often convenient to use polarization-maintaining fibers to avoid beatnotes fading, one must ensure that the polarization of injected light on  the forward and backward path is precisely aligned to the principal fiber axes, to avoid introducing non-reciprocal and fast-varying birefringent noise on the fibers.  

\begin{figure}   \centering \includegraphics[width=0.8\textwidth]{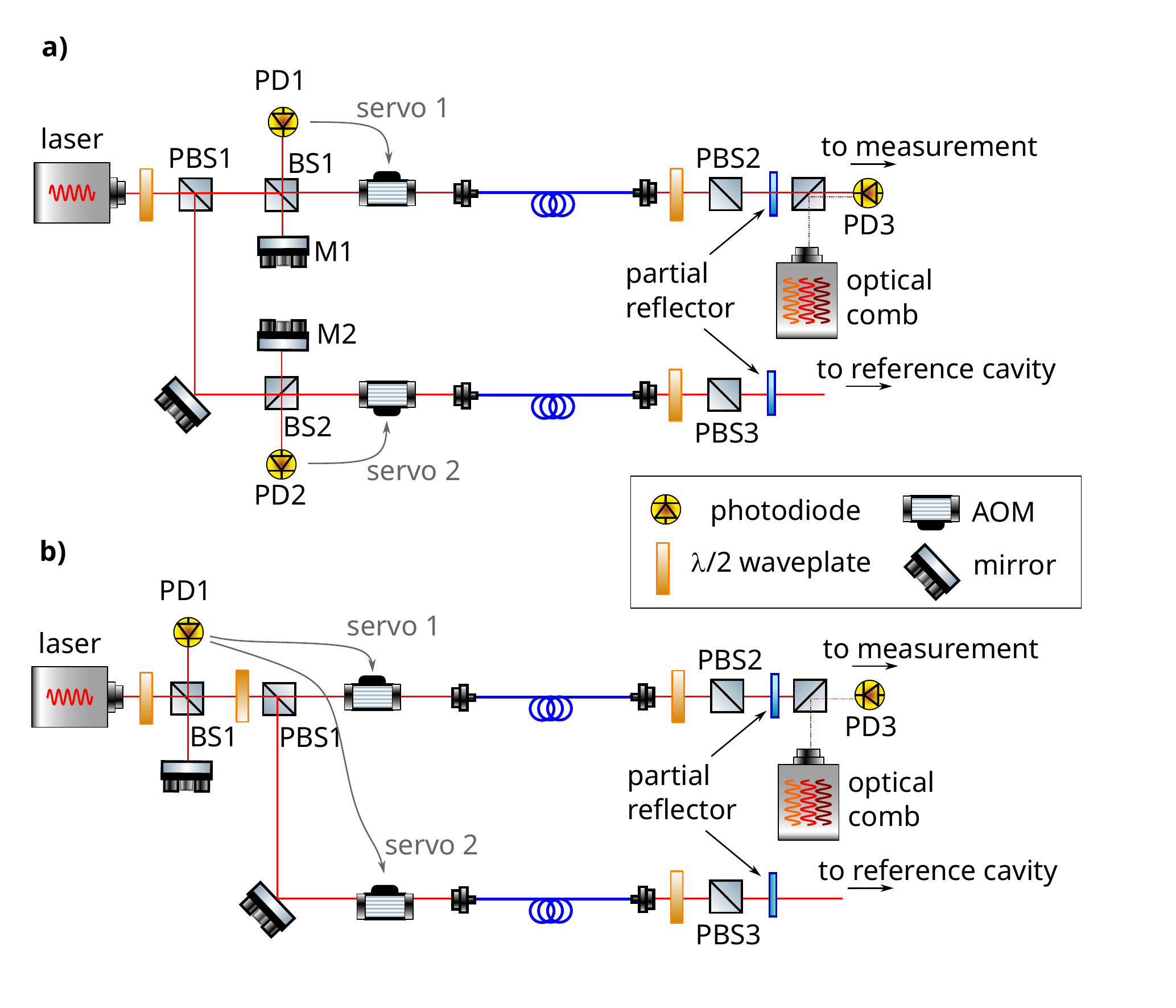}
\caption{Setup for  fiber-delivery systems using separate (a) or common (b) reference mirrors at both sites. M: mirror, PBS: polarizing beam splitter, BS: beam splitter, PD: photodiode. It is important to note that if polarization maintaining fibers are used, it is advised to clean polarization before the backward launch, using a $\lambda/2$ waveplate and a PBS (PBS2 and PBS3).\label{fig:links}}
\end{figure}

\begin{figure}
\centering
\includegraphics[width=0.67\textwidth]{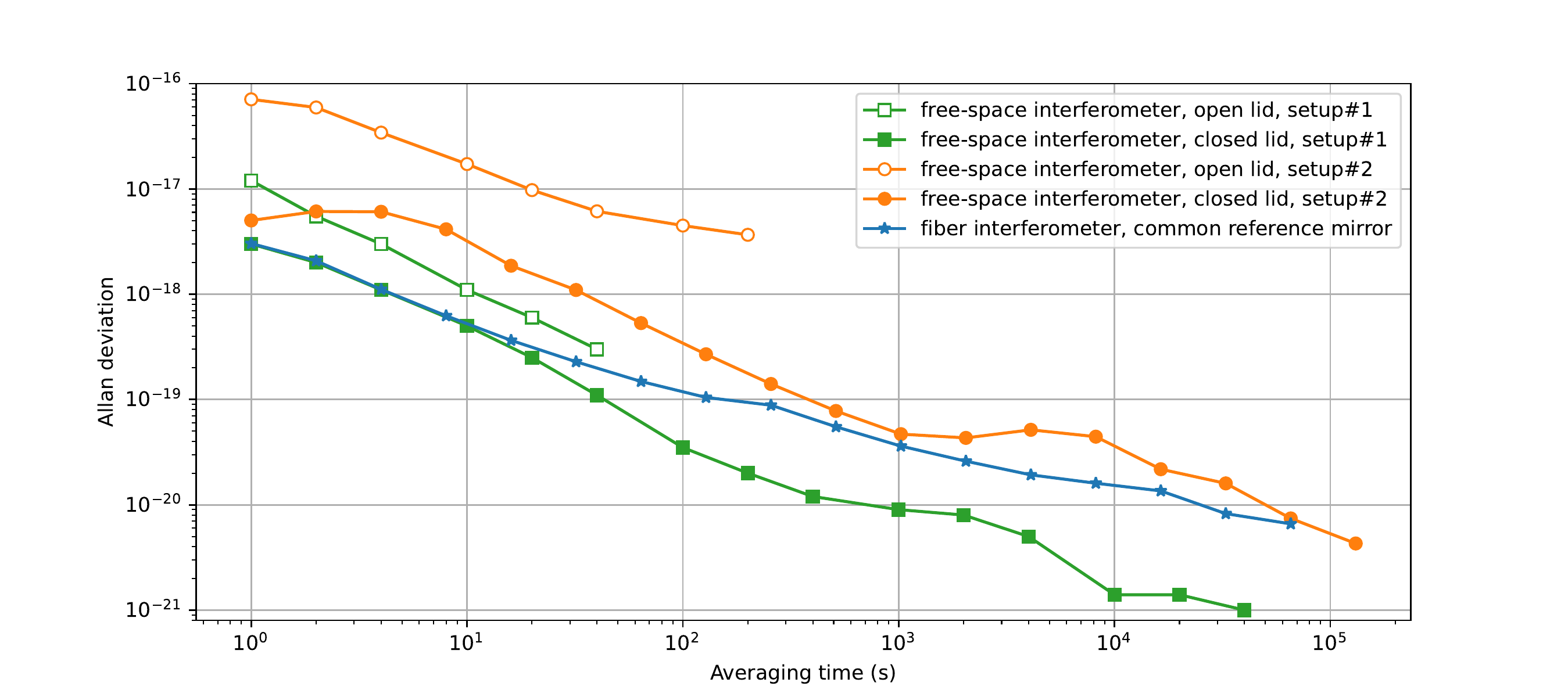}
\caption{Left: the frequency instability expressed as Allan deviation for the fiber delivery systems in various  configurations. Orange open squares and green open circles: separate reference mirrors and open lids. Orange filled squares and green filled circles: the same, with closed lids; blue diamonds: common-reference mirror setup. 
\label{fig:links_adev_psd}}
\end{figure}

Fig.\ \ref{fig:links}a depicts a common scheme for fiber-based delivery of radiation. 
Here, a laser is split into two beams and enters two independent  Michelson interferometers, one directed to the comb for the stability transfer and another to the reference cavity (in the case of the master laser) or to the target application (in the case of the slave laser). 
The split ratio is governed by adjusting the $\lambda/2$ waveplate in front of polarization beam-splitter PBS1. Each interferometer consists in a non-polarizing beam splitter (BS1 or BS2), that further splits the radiation into two beams and sends part of it to a mirror (M1 and M2) and then to a  photodiode (PD1 or PD2), while injecting the rest into the fiber. 
At the fiber output, a partial reflector  sends part of the signals back to the transmitting end for the  optical path stabilization and should thus be placed as close as possible to the end of the delivery path.  The two AOMs at the transmitting end are the actuators for the fibers' stabilization loops (servo 1 and servo 2). If polarization-maintaining fibers are used, it is advised to clean polarization at the output via a $\lambda/2$-waveplate followed by a PBS (PBS2 and PBS3), to avoid the fiber itself to introduce additional birefringence.
Fig.\ \ref{fig:links_adev_psd} shows typical results of the stabilization loops, and have been obtained by combining two beams, delivered across  \SI{100}{m} stabilized fibers, onto a common photodiode and using a common partial reflector for the two return signals. 
When the setup is simply mounted on an optical bench, with no special care to protect it from air currents and acoustic noise, we observe an  instability ranging between a few parts in \SI{e-17}{} and \SI{1e-16}{} at \SI{1}{s} (open squares and open circles). 
The major contribution to it are fluctuations on the reference arms PBS1-M1 and PBS2-M2, which are not common between the two Michelson interferometers. 
The actual value  strongly depends on the laboratory environment and is in no-case compatible with the delivery of state-of-the-art ultrastable lasers. When the stabilization optics is protected by a box, the instability improves to a few parts in \SI{e-18}{}, 
with residual differences depending on the quality of the acoustic shielding (filled squares and circles in Fig.\ \ref{fig:links_adev_psd}). 

To further reduce the amount of non-common optical paths and ensure a better control of fluctuations, it is worth considering a modified design of the stabilization interferometer (Fig.\ \ref{fig:links}b), in which a single reference mirror at the launching end is employed, common to the two paths and whose instability is thus common-mode rejected. 
The blue curve in Fig.\ \ref{fig:links_adev_psd} shows the corresponding achieved instability. On the short term it is similar to that achievable with the best acoustic enclosure, while a slight degradation on the long term might be due to uncontrolled polarization rotations that are not stabilized by the Doppler noise cancellation scheme \cite{clivati2020}. Still, an ultimate instability of about \SI{1e-20}{} is attained on hour-long integration times with no resolved offsets. Noteworthy, the setup of Fig.\ \ref{fig:links}(b)  has been obtained by mounting the interferometer at the launching end (beam splitter and reference mirror) with fiber-coupled components, further highlighting the robustness to environmental-induced perturbations. 

Overall, fiber-based light delivery systems appear to be a critical component of the  spectral transfer chain. Although fiber stabilization is today a well-established technique, applied in many laboratories for laser distribution from campus-scales to continental distances, ensuring contributed instability lower than \SI{1e-17}{} is not obvious for operational systems, with practical limitations set by unavoidable non-common paths subject to pressure and temperature fluctuations of the air refractive index, as well as seismic and acoustic noise deforming the optical bench. 

Their instability contribution can be kept  below \SI{1e-17}{} at \SI{1}{s} only with careful design and passive shielding of the interferometer and minimization of optical paths. As the actual performance is strongly related to the laboratory environment and subject to change upon time, the design of fluctuation-insensitive interferometers with common reference arms  is  strongly recommended.

\section{Single-branch optical combs}
\label{sec:noncommon_optical_paths}
The most common comb technology is based on Er:doped fiber, that covers  a native spectrum in the 1530-\SI{1560}{nm} range. 
This spectrum is usually amplified with an Er:doped fiber amplifier (EDFA), broadened to a span between 1 and \SI{2}{\micro m} with the help of highly nonlinear fibers (HNLF) to enable self-referencing,  and can be further extended to reach other regions of interest. For instance,  wavelength-conversion modules based on second-harmonic-generation (SHG) crystals can convert it to visible wavelengths, usually on dedicated arms covering a few nm.
In all cases, it is important to minimize the differential optical path travelled by radiation at the reference and target wavelengths; to the first order, this requires comb light in the two spectral regions to be extracted from the same comb output,  as non-correlated phase fluctuations of different comb branches has been shown to contribute  instability higher than \SI{5e-17}{} \cite{ben19,barbieri2019}.
Alternative approaches have been investigated, consisting in mutual stabilization of different comb branches \cite{rolland2018} or real-time phase tracking of their differential variations \cite{giunta2019}.

\begin{figure}
\centering
\includegraphics[width=0.7\textwidth]{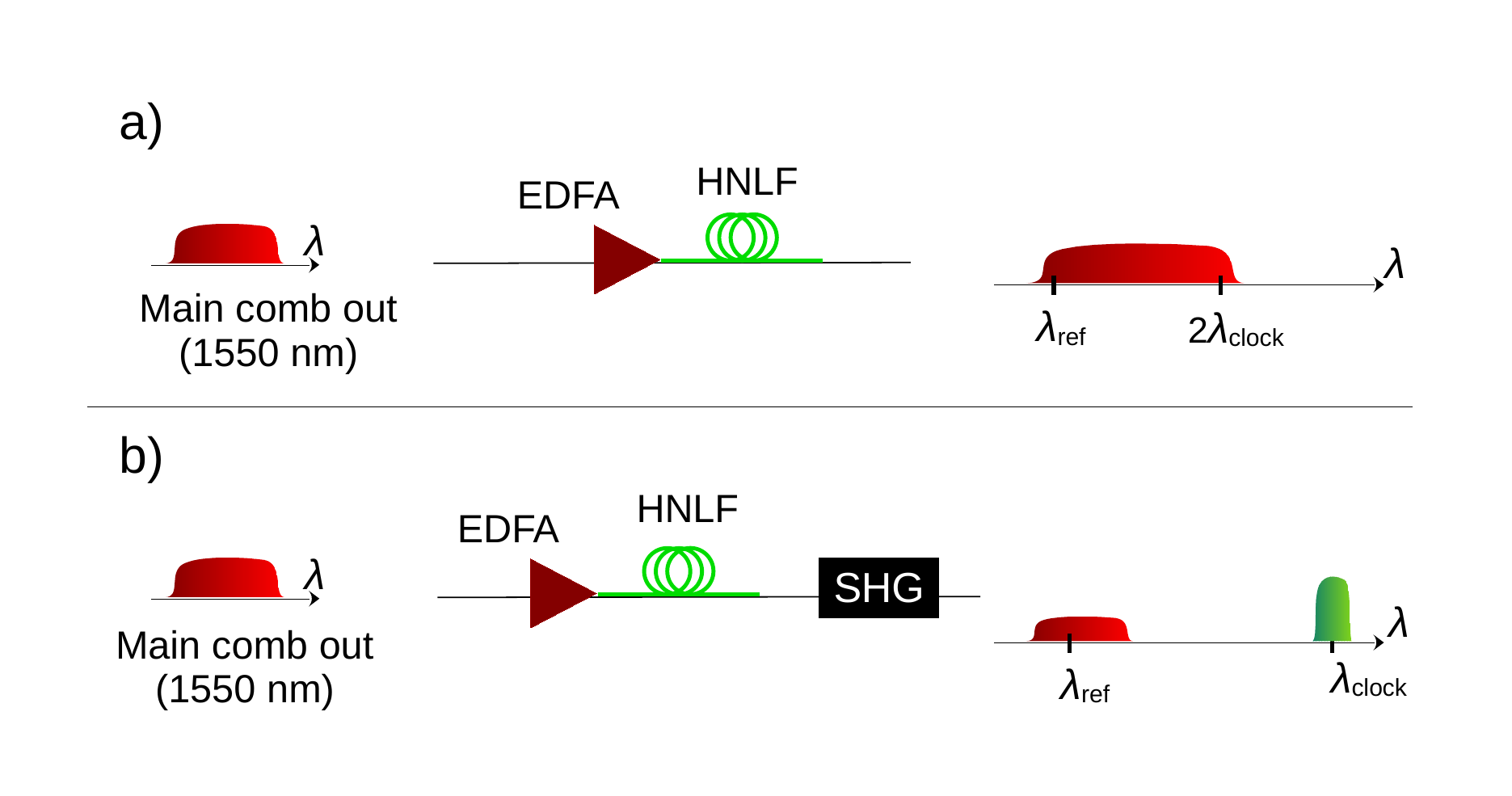}
\caption{Spectral broadening strategies with SHG or nonlinear fiber \label{fig:spectral_broadening_strategies}}
\end{figure}

Focusing on the implementation of single-branch measurement setups, two approaches can be exploited (see Fig.\ \ref{fig:spectral_broadening_strategies}):
\begin{enumerate}
\item If clock radiation in the visible domain is produced by SHG of an infrared seed laser (e.g. \SI{1396}{nm}$\rightarrow$\SI{698}{nm} or \SI{1156}{nm}$\rightarrow$\SI{578}{nm} for Sr and Yb lattice clocks respectively), it is possible to implement the transfer oscillator scheme in the infrared domain, e.g. beating the seed laser with an octave-spanning comb output, with no need for generating visible comb light. The subsequent duplication process of the cw-laser beam has been demonstrated to contribute negligible  instability \cite{yeaton2012}. 

\item A part of the comb light is passed through a SHG crystal designed for operation at the clock wavelength in the visible domain and beaten to the clock laser at the same wavelength. The beatnote between the infrared reference laser and the comb instead exploits the leakage of light seeding the SHG crystal.
\end{enumerate}

\begin{figure}
\centering    \includegraphics[width=0.8\textwidth]{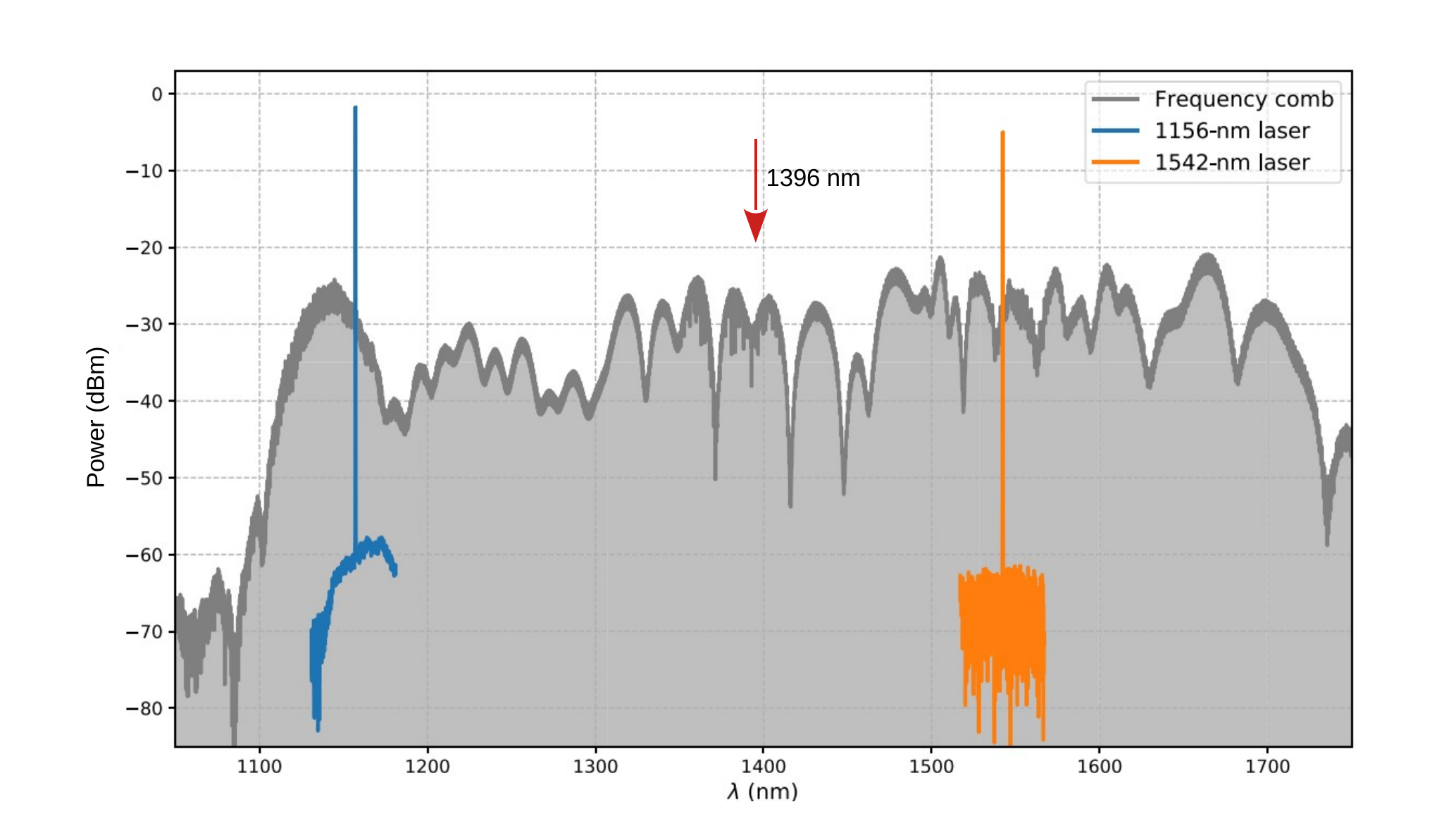}
\caption{The broadband-spanning spectrum produced by sending a seed  Er:comb output at \SI{1560}{nm} to an Erbium fiber amplifier and then through a highly nonlinear fiber. 
\label{fig:comb_spectrum_INRIM}}
\end{figure}

Approach (1) guarantees higher flexibility, as the octave-spanning comb output between \SI{1}{\micro m} and \SI{2}{\micro m} naturally covers the regions of interest for sub-harmonics of many optical clock wavelengths \cite{leopardi2017}.
As an example, the typical output spectrum obtained by amplifying a portion of the comb at \SI{1560}{nm} and broadening it with a highly-nonlinear Raman-shifting fiber is shown in Fig.\ \ref{fig:comb_spectrum_INRIM}. 
However, an aspect of concern is the available power at the wavelengths of interest. 
As a rule of thumb, a minimum SNR of  \SI{80}{dB} in \SI{1}{Hz} bandwidth is needed for reliable and cycle-slip-free processing of the comb/laser beatnotes. 
Fine tuning of the  pump current of the erbium-doped fiber amplifier and frequent adjustment may be thus required to simultaneously ensure sufficient comb power at the two or more comb wavelengths involved in the stability transfer.

With this configuration, we obtained instabilities at the level of \SI{7e-18}{} at \SI{1}{s} for the transfer beatnote between \SI{1156}{nm} ($\nu_\text{m}$ = \SI{259}{THz}) and \SI{1542}{nm} ($\nu_\text{s}$ = \SI{194}{THz}). 
These wavelength regions represent some of common use in metrological and fundamental physics laboratories: the former is the sub-harmonic of the $^1P_0-^3P_1$ transition of $^{171}$Yb; the latter is used for long-distance fiber distribution of accurate optical frequencies among remote laboratories. 
Other wavelengths of interest, e.g. the sub-harmonic of other Yb and Sr clock transitions, fall inside this range as well.

Approach (2) requires to  exploit the infrared light leaking from the SHG crystal, and is thus particularly suited for reference lasers operating around the \SI{1560}{nm} region which is that of the primary comb and is always present in the comb spectrum even when it is optimized for another region. 
This may pose some  design constraints, e.g. to ensure that sufficient power at the reference wavelength  is available out of the crystal. 
In addition, thermal effects occurring when the crystal is hit by comb pulses with higher peak intensity have been suspected to contribute additional instability and degrade the overall spectral transfer performances.
For instance, in \cite{herbers2019}, additional phase noise has been observed  when comparing optical frequency doubling of a single-frequency laser or a comb in a in a periodically-poled lithium niobate (PPLN) crystal. 
Although the source of this additional noise in the comb doubling remained unclear, one difference between frequency doubling a single frequency laser and a multitude of comb lines is that the latter is actually rather sum-frequency generation (SFG) between the comb lines within the PPLN phase-matching bandwidth instead of a pure SHG process. Furthermore, the comb light within the phase-matching bandwidth may have a higher relative intensity noise (RIN) than the single-frequency laser, and conversion of this RIN to phase noise (AM-PM conversion) could explain the observed higher phase noise. 
More generally, AM-PM conversion could as well lead to additional noise during other nonlinear processes like four wave mixing etc., which are exploited for extension of a frequency comb spectral envelope to other wavelengths. 
This additional noise would directly limit the quality of a spectral purity transfer.  
E.g., for a stability transfer from a fundamental comb generated in the 1.55~$\mu$m wavelength region to a target wavelength around the clock transition of neutral Sr atoms near 698~nm, the fundamental comb is first broadened to 1400~nm by nonlinear processes in a highly nonlinear fiber, and part of the broadened comb is then frequency-doubled in an SHG crystal to get sufficiently strong comb lines in the target spectral range near 698~nm. 
Such effects may explain why similar levels of noise added in spectral purity transfer were observed by different groups \cite{nic14,ben19}. An ultimate characterization of this effect requires the comparison of stability transfer performed on distinct yet identical comb branches, one of which deterministically amplitude-modulated, and has not been achieved yet.
Still, we quantified an upper limit for this effect to be $2\times10^{-18}$ at \SI{1}{s}, by doubling the low-wavelength comb tail from \SI{2}{\micro m} to \SI{1}{\micro m}.

Overall, the difference in performances between approach (1) and (2) is mostly due to technical effects like residual uncompensated paths in the beam-delivery systems. A choice among these two  should thus be based on the actual spectral regions of interest for a specific laboratory.
 
\section{Improving the SNR of comb beatnotes}
\label{sec:improving_SNR}
As discussed in section \ref{sec:noncommon_optical_paths}, spectral transfer with instability below \SI{1e-17}{} can only be obtained using optical combs in single-branch configuration, i.e. where radiation at the master and slave wavelength is produced and travels through the same optical elements. 
In many cases, this limits the available power of relevant comb teeth and as a result beatnotes between lasers and comb can sometimes feature a SNR close to the critical threshold of 80 dBc/Hz required for cycle-slip free operation and robust phase-lock on MHz-wide bandwidths \cite{risaro2022,sinclair2015}. 
Tracking oscillators, often employed as clean-up filters, suffer from the same limitation, introduce  complexity to the setup and contribute additional noise to the system on the medium term \cite{ben19}. 

A useful solution to improve SNR is gated beatnote detection. 
This approach was first proposed in 2013 by Desch\^enes et al. \cite{deschenes2013} and relies on the observation that beatnotes between a cw-laser and the comb have a pulsed nature in the temporal domain, while the most impacting noise processes, such as photodiode thermal noise or the cw laser shot noise, are stationary. 
Their impact can therefore be reduced if detection is activated only in the temporal windows in which the signal is present,  avoiding noise power generated outside those windows to affect the measurement. 
Gated detection can be implemented in different ways, which are summarized below. 
\begin{enumerate}
    \item 
The first demonstration \cite{deschenes2013}  made use of hardware components such as mixers, delay lines and an additional fast photodiode; in this approach a gating signal, obtained by sending the comb beam to a fast photodiode, was multiplied to the comb/laser beatnote on a microwave-mixer. 
The pulses corresponding to the beatnote and gating signals were synchronized to a few tens of ps using a precisely-matched delay line.
\item
A second option consists in modulating either the cw-light beam or both the cw-laser and the comb with a fast amplitude modulator (e.g. exploiting lithium niobate technology). 
The modulator driving signal is the electrical pulse train generated by photodetecting the comb pulses with a high-speed, high-power photodetector. 
\item
A third, novel approach is based on off-the-shelf Track and Hold Amplifiers (THA) that implement electrical gating \cite{risaro2022}. Track-and-hold amplifiers can operate in the so-called ``transparent mode'', letting the input signal to pass, or in the ``hold mode'', storing the sampled voltage in a capacitor until the next sampling window is triggered. 
The clock signal controlling  the gate window  can be provided by a DDS whose internal clock is locked to the same \SI{10}{MHz} signal referencing the comb.
For optimal performances, the maximum hold time should match the pulse separation of the optical comb. Commercial comb models usually have $f_\text{rep}$ between \SI{100}{MHz} and \SI{250}{MHz}. The latter corresponds to a pulse separation of \SI{4}{ns} that can be conveniently covered by dual-stage THAs (e.g.
model HMC1061LC5 by Hittite), composed by two cascaded THAs featuring a hold time of \SI{2}{ns} each.

\end{enumerate}

From the implementation perspective, electrical gating (i.e. approach 1 or 3) is preferable to optical gating (approach 2). 
In fact, while optical gating can enable very short aperture windows, it only rejects the laser's shot noise, leaving the thermal noise level of the subsequent detection chain unaffected and thus not exploiting the full advantage of gated detection. In addition, lithium niobate modulators tend to have large insertion losses of 4 dB or more, and phase shifts induced by the modulator might compromise accuracy. 
Focusing on the electrical gating approaches,  (3) represents an effective implementation possibility of (1), with small form factor and low power consumption, suitable for replication and further digital processing.

\begin{figure}
\includegraphics[width=\textwidth]{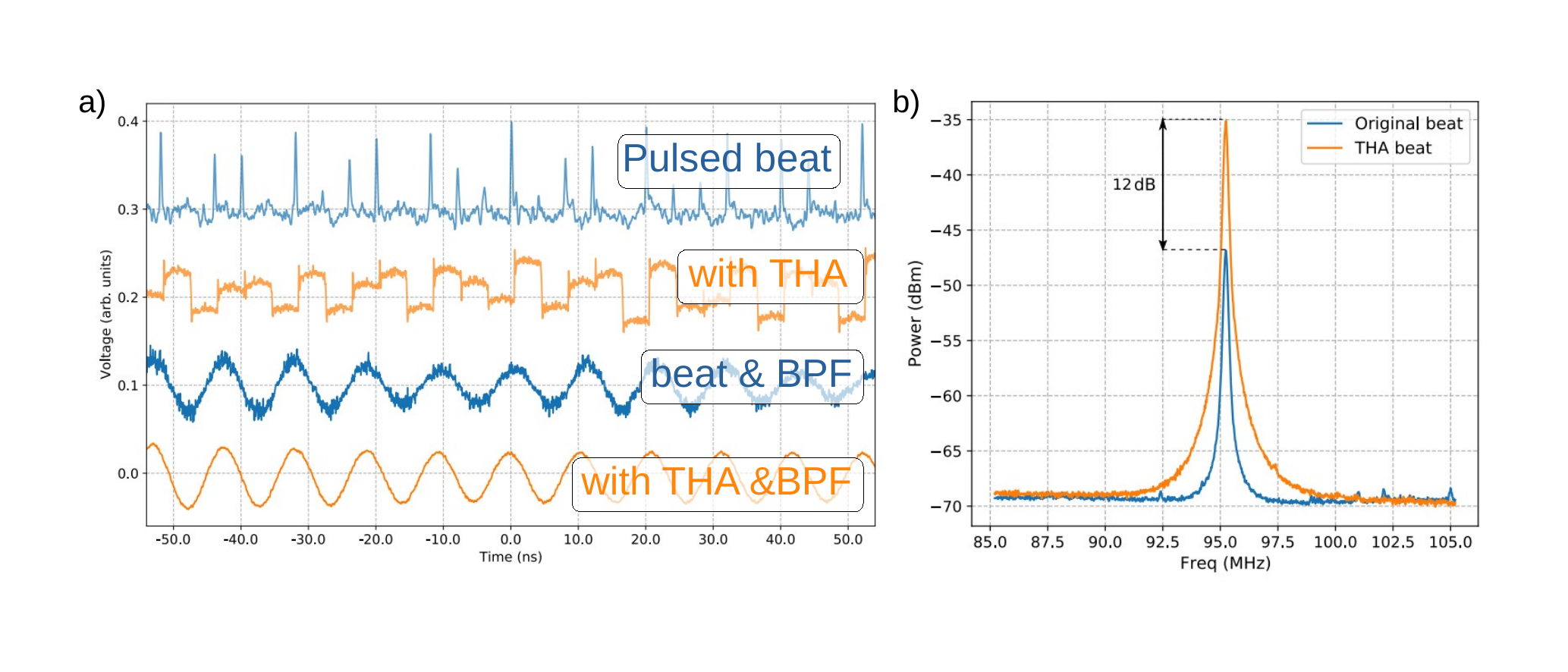}
\caption{(a) Time-domain view of the gated detection
technique. The following traces were recorded with a fast oscilloscope at 25 GS/s (top to bottom): the  photodiode signal corresponding to a beatnote at \SI{95}{MHz} between a cw laser and a comb; the signal resulting from passing the above in a THA; the trace obtained by band-pass filtering the raw signal; the trace obtained by band-pass filtering the THA output with the same band-pass filter (BPF).
(b) RF spectra of the results obtained with
the conventional technique and THA-based gating
(resolution bandwidth 100 kHz). The THA beat shows an improvement of 12 dB with respect to the original beat. \label{fig:THA} }
\end{figure}

Fig.\ \ref{fig:THA}a depicts the gated oscillation concept when implemented using approach 3. 
The first row shows the typical beatnote between a cw-laser and a comb: it appears as a series of pulses with varying amplitude, spacing of \SI{4}{ns} set by the repetition rate of the comb (\SI{250}{MHz}) and duration that depends on the photodiode bandwidth (about \SI{300}{ps} for \SI{2}{GHz}-photodiodes). 

The second row shows the  THA output when the gating windows are adjusted to be synchronous to the signal pulses, with the characteristic step-like behaviour resulting from holding the photodiode voltage for \SI{4}{ns}. 
Traces in the third and fourth rows show the same signals as the first and second row respectively, after band-pass filtering is applied around the beatnote frequency. 
A clear improvement in SNR is observed in the latter case. 
The corresponding RF spectra  are shown in Fig.\ \ref{fig:THA}b. 
While the bare photodiode output shows a \SI{22}{dB} SNR  in a \SI{100}{kHz} bandwidth limited by photodetection noise, a significant improvement is  observed on the THA output, with a  \SI{12}{dB} increase in power with no change to the noise level. 
Similar results can be obtained using other implementation approaches. 
The obtained SNR improvement is in agreement with the expected value of $2b/f_\text{rep}$ where $b$ is the photodiode bandwidth.
 While in principle gated detection can perform even better with artificially-reduced cw-laser power, the shown results are typical for real operating systems where the cw-laser power sent to the comb is in the range  \SI{100}{\micro W}-\SI{1}{mW} and the comb power per tooth is about \SI{10}{nW}. It is important to stress that even an apparently moderate \SI{12}{dB} SNR improvement has a relevant impact on measurements, as cycles slips have an exponential dependence on the SNR \cite{risaro2022,sinclair2015}. In critical conditions, it can drive the system from being at or below the cycle-slip-free threshold to a fully cycle-slip-free region.

 Ultimate limits to the maximum achievable improvement may be represented by the shot noise of comb light which, being pulsed, cannot be reduced by gated detection. Jitter or a mismatch in the aperture time and duration of the gating signal could in principle lead to non optimal performance. However, synchronizing comb pulses within a few tens of picoseconds is enough to implement gated detection of 300-ps-duration pulses, and requires preliminary matching of  the electrical paths to within a few mm. While long term phenomena such as thermal drifts or dephasing effects could also reduce the signal's frequency stability or introduce de-phasing of the beatnotes and gating signals, we observed no degradation to the instability of THA-processed beatnotes down to the limit of \SI{1e-21}{} over several  hours. 

Overall, gated detection can be considered as a useful tool for implementing stability transfer at the level of the best ultrastable cavities in those conditions where the power of the optical comb is critical and may prevent from robust measurements and processing with traditional detection schemes. 

\section{Digital phase-locks for highly 
\label{sec:digital}
performing, flexible and multi-wavelength spectral transfer}
The comb  as a transfer oscillator allows for a single master laser to act as a reference for several slave lasers simultaneously, thus potentially serving multiple  experiments.    
This implies synthesizing and processing  the transfer beatnotes between the master and each slave laser in independent phase-locked-loops. 
When implemented by hardware, each synthesis chain requires several  RF-mixers and direct digital synthesizers, a variable number of band-pass filters, tracking oscillators and possibly additional stages for frequency up- and down-conversion. The overall apparatus results to be hardly reconfigurable and bulky, not practical in  cases where replication for different target wavelengths  is desired, subject to non-linearities or non-idealities of hardware components and always requiring some amount of manpower to be established and maintained. 

Digital electronic platforms are promising in reducing the amount of external components, simplifying the process, improving robustness and flexibility and enabling a number of slave lasers to be  locked to a common master laser in a modular approach. While  high reproducibility, quick design reconfiguration and tight control over operational parameters  are well-known advantages of digital electronics,  important concerns 
are related to bandwidth and resolution considerations set by the sampling rate and effective number of bits in the employed logic, and by the latency. In addition to these, the  design of digital systems usually requires programming abilities that are still not covered by the majority of users and may represent therefore a potential barrier to its introduction. In this respect, both off-the-shelf and customized solutions are now commercially available, requiring different levels of programming abilities. We here overview a possible strategy for transfer beatnote synthesis using a digital electronics platform, describing two implementation solutions. 

In brief, digital implementation requires the beatnote  $f_\text{b,m}$  ($f_\text{b,s}$) between the comb and  the master (slave) laser, and the offset frequency $f_0$ to be synchronously sampled with a fast analog/digital converter (ADC). The phase fluctuations of these signals with respect to nominal values are extracted by real-time I/Q demodulation and then combined with proper scaling according  to Eq. \ref{eq:transfer_beat}. This enables to produce a phase error which is insensitive to the comb noise and proportional to the phase fluctuations of the slave laser relative to the master. This information is converted into a voltage by a digital/analog converter (DAC), integrated with a fast proportional-integrative-derivative (PID) loop filter  and used to phase-lock the slave laser to the master. 

I/Q demodulation is
a common approach to phase detection in digital systems also in the frequency metrology context \cite{tourigny2018,sherman2014,donadello2023}. Let us assume that the beatnote signal entering the ADC is described as a voltage:  
\begin{equation}
\label{eq:raw_v}
    V(t)=A(t)\sin (2\pi f t + \varphi(t))
\end{equation}
where $A(t)$ indicates time-varying amplitude, $f$ is the signal's nominal frequency and $\varphi(t)$ the instantaneous phase. This signal is synchronously demodulated by mixing with two  references:
\begin{equation}
\label{eq:quads}
\begin{split}
    r_\text{I}(t)&= \cos(2\pi ft)\\
    r_\text{Q}(t)&= \sin(2\pi ft)\\
\end{split}
\end{equation}
having the same nominal frequency as the signal, unitary amplitude and $\pi/2$ relative phase shift. 
The mixed signals are low-pass filtered with transfer function
$H_\text{LPF}\{\cdot\}$ at a cutoff frequency $f_\text{BW}$  to extract the \textit{in-phase} and \textit{quadrature}  components:
\begin{equation}
    \begin{split}
        I (t) = H_\text{LPF}\{r_\text{I} (t)V (t)\} &=\frac{1}{2} A(t) \cos(\varphi(t)) \\
        Q (t) = H_\text{LPF}\{r_\text{Q} (t)V (t)\} &=\frac{1}{2} A(t) \sin(\varphi(t)) \\
    \end{split}
\end{equation}
The two are finally combined to retrieve the instantaneous amplitude and phase of the signal:
\begin{equation}
\label{eq:amp_phase}
\begin{split}
        A(t) &= 2 \sqrt{I^2(t) + Q^2(t)}\\
\varphi '(t) &= \arctan(Q(t), I (t))\\
\end{split}
\end{equation}
where $\varphi '(t)$ is a wrapped version of $\varphi(t)$, as the domain of the 
inverse tangent function is defined between $-\pi/2$ and $+\pi/2$ only. $\varphi(t)$ is  obtained from   $\varphi'(t)$ removing discontinuities larger than $\pi$ by applying a suitable unwrap algorithm. 
When $V (t)$ is sampled by the
ADC at frequency $f_\text{s}$, the previous relations remain valid at
discrete instants $n/f_\text{s}$ with $n$ integer, and the low-pass filter is implemented in the form of a finite-impulse-response (FIR) filter, whose number of taps plays a role in defining the overall  latency and computational resources occupied by the process. 
 $f_\text{s}$ can be conveniently set to be equal to the comb $f_\text{rep}$, if the ADC supports it. This  naturally fulfills the Nyquist criterion as  $f_\text{b,m}, f_\text{b,s}< f_\text{rep}/2$ in the first spectral window. In addition,  it allows for direct sample decimation that corresponds  to gating the sampled signal in the digital domain, improving the signals SNR.

\begin{figure}
\centering
    \includegraphics[width=0.8\textwidth]{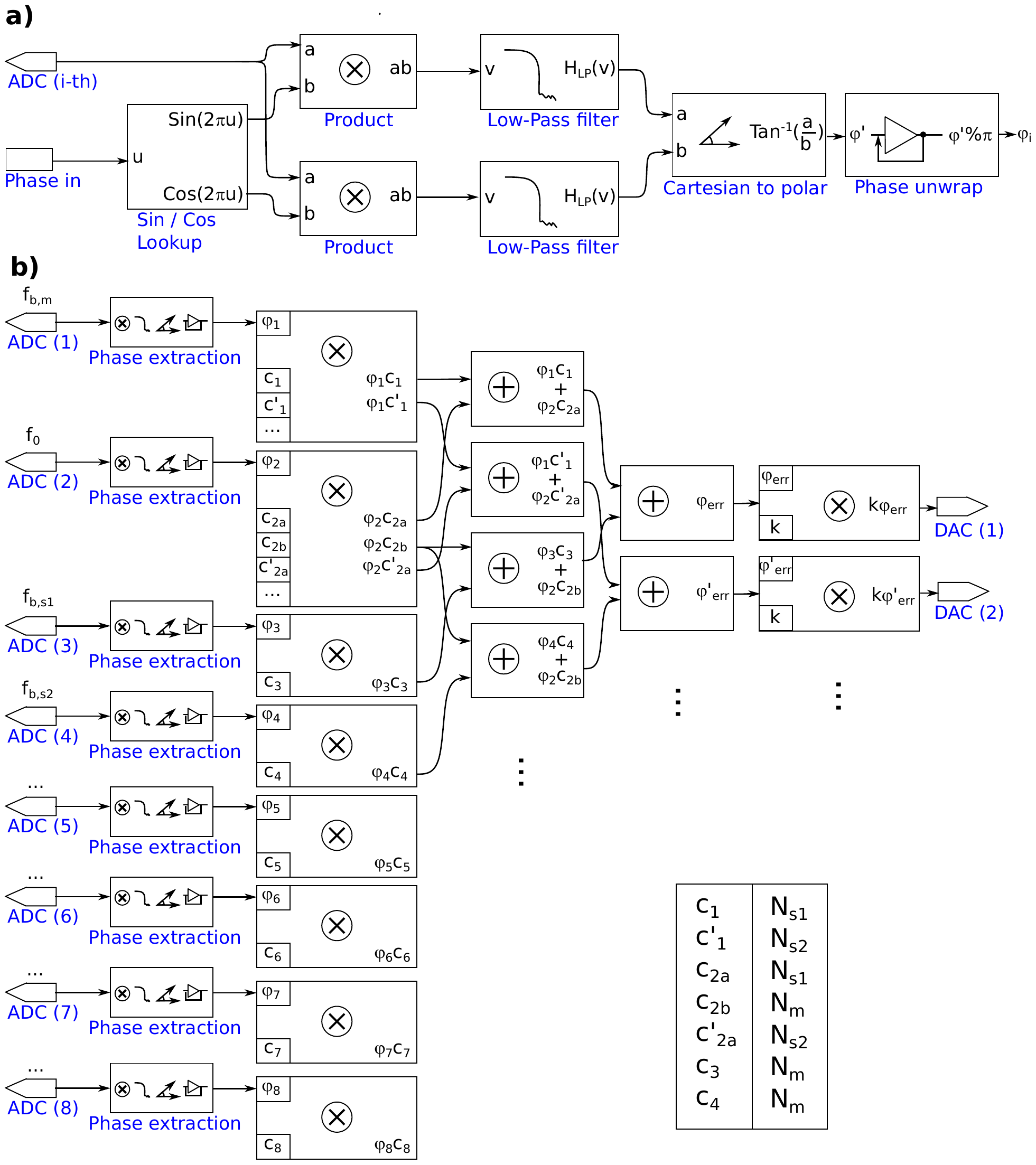}
    \caption{a)The digital signal processing of the $i$-th input channel performed within the FPGA. $u$, $a$, $b$, $v$ represent arbitrary input signals. b) Block diagram showing the combination of sampled phases to obtain the error signals $\varphi_\text{err}$ and  $\varphi'_\text{err}$  used for locking S1 and S2 to the
master. Channels 5 to 8 are not used in this example but can be configured to lock 
 up to 4 additional slave lasers. \label{fig:fpga_design_all_channels_LINKS}}
\end{figure}

The block diagram of the digital processing for each of the input channels is summarized in Fig.\ \ref{fig:fpga_design_all_channels_LINKS}a.
For each channel, a nominal beatnote frequency is defined, arbitrarily chosen within about \SI{100}{kHz} from the actual one; a numerically-controlled local oscillator is  then  synthesized in its two quadratures $r_\text{I}$ and $r_\text{Q}$ at the nominal frequency, and multiplied to the sampled signal. Each output is   passed to a FIR filter and fed to a cordic module that computes the inverse tangent. Finally, the signal is passed to an unwrap module. 
The demodulated phases of each beatnote are linearly combined to produce the phase of the transfer beatnote for locking the slave laser to the master (Fig.\ \ref{fig:fpga_design_all_channels_LINKS}b). 
The combination coefficients depends on the scaling required for each beatnote (hence on the laser frequencies) and  their sign. 
As an example, to derive the error signal $ \varphi_\text{err}$ for locking the slave laser S$_1$ to M, 
the phase of: (i) the beatnote between the master laser and the comb, (ii) the beatnote between S$_1$ and the comb, and (iii) the comb's carrier-envelope offset are obtained from the photodiode signals sampled on ADC channels 1, 2 and 3 (Eq. \ref{eq:raw_v}), processed according to Eqs. \ref{eq:quads}-\ref{eq:amp_phase} and combined according to the formula:
\begin{equation}
\label{eq:transfer_beat2}
    \varphi_\text{err}=k (c_1 \varphi_\text{b,m} + c_{2\text{a}} \varphi_0 + c_{2\text{b}} \varphi_0 + c_3 \varphi_\text{b,s1} )
\end{equation}

where $c_1$, $c_{2\text{a}}$, $c_{2\text{b}}$ and $c_3$ are chosen to fulfil Eq. \ref{eq:transfer_beat} and an overall scaling by $k$ is applied before passing the result to the output DAC.
While a  minimum of 3 ADC channels must be configured for locking a slave laser  to the master, it is enough to configure further ADC channels for each additional slave laser. Channels related to $f_0$ and $f_\text{b,m}$ are common to all and only their scaling coefficients needs to be updated. This considerably reduces the implementation complexity compared to a hardware-based system. 
Coefficient values fulfilling  Eq. \ref{eq:transfer_beat} for slave lasers S1 and S2 are shown in Fig.\ \ref{fig:fpga_design_all_channels_LINKS}b.

Focusing on the digital design details of the implementation, it shall be noted that in rewriting Eq. \ref{eq:transfer_beat2} from \ref{eq:transfer_beat}, all coefficients $c_1 \dots c_4$ must be represented as integer values
as divisions introduce high latencies and non-integer representation of  $N_\text{m}/N_\text{s}$ as a fixed point constant would lead to unacceptable rounding errors and measurable output frequency shifts. 
Typically, this implementation introduces delays in a few tens of clock cycles, i.e. lower than \SI{100}{ns}. 

The phase error signal can be sent to the DAC as it is and used to feed a PID control loop to stabilize the slave laser to the master. This solution is suitable in  cases where the slave laser and the electronics are placed within a few m from the comb, so that the correction can be sent to the slave laser via coax cable. In most common situations, where the master and slave lasers are in different laboratories, 
this approach is not possible, as dc signals are  subject to radio-frequency interference, attenuation and ground loops. In this case, the output stage can conveniently include a  DDS producing a sine-wave whose frequency is continuously adjusted to be $f_\text{out}(t)= f_\text{nom} + c \varphi_\text{err}(t)$, i. e.
deviate from a preset nominal value $f_\text{nom} $ by a small quantity proportional to the instantaneous value of $\varphi_\text{err}$ via a coefficient $c$. At any clock cycle, when a new estimation of $\varphi_\text{err}$ becomes available, the difference with the previous value is computed. Scaling by $c$  is implemented as a shift to keep complexity low, and the DDS frequency is adjusted by updating the corresponding tuning word.  While introducing additional delay in the range of \SI{100}{ns}, this approach enables effective transfer of the error information across a campus scale.

A possible platform for implementation consists in the Ettus Research's RF Network on a Chip (RFNoC), that utilizes signal processing blocks interconnected with optionally runtime configurable bus connections. Blocks can be written in HDL (Verilog, VHDL), or by using Xilinx's Vitis High Level Synthesis (HLS) compiler, which produces FPGA implementation (IP blocks) from C++ source code.
For ease of development, HLS can be used for programming the FPGA.

Ettus USRP3-series X300 Software Defined Radio (SDR) running Xilinx Kintex 7-325T FPGA and  equipped with LFRX and LFTX RF-daughterboards gives 4 input and 2 output channels that are suited for the transfer beat implementation. ADCs and DAC run at $f_{ADC}$ = $f_{DAC}$ = \SI{200}{MHz}, however the RF-daughterboards limit the analog bandwidth to approximately \SI{30}{MHz}. Inputs are IQ-sampled with 16 bits of accuracy. Digitized data is represented as 32-bit integers on the FPGA (16 bits for both real and imaginary parts).

The development environment and ultra-high definition (UHD) device drivers are mostly open source, and are available from Ettus' Github repository. UHD-driver version was 4.4 (Release).

Blocks are interconnected using the AXI bus. While the RFNoC framework supports runtime-configurable bus routing between signal processing blocks, static routing can be chosen to keep the signal processing delay as low as possible. Digital up/down-converters (DUC/DDC) are useful in reducing the processing rate ($f_{proc}$) whenever custom made DSP blocks are unable to handle the full \SI{200}{MHz} due to its relatively long initialization interval. DUC/DUC-blocks also have DDS-functionality for numerically shifting the input frequency. 

This board can be configured with basic expertise in logic programming and supports a short term instability of \SI{1e-17}{} at \SI{1}{s}, thus proving to be a viable solution for stability transfer at an affordable cost. However, the limited number of available input channels does not support multiple slave laser locks, and the limited bandwidth could be problematic in some measurement conditions.  

More complex, yet powerful and fully customizable solutions are also available on the market. 
Among them, is the Zynq UltraScale+ RFSoC ZCU208 FPGA board by AMD-Xilinx,  featuring a \SI{5}{GS/s} ADC and a  \SI{10}{GS/s} DACs, each with 8 parallel channels  and 14 bits of vertical resolution, \SI{50}{Mb} on-chip memory and \SI{8}{GB} external memory (DDR4)  connected to the embedded ARM processor  and to the programmable section of the FPGA. 
The board kit comes with an external clock generation daughterboard (CLK104), used to reference the ADC and DAC sampling clocks to   an external \SI{10}{MHz} signal, and a BALUN board XM655 that passively interfaces on-chip ADC/DAC pins to \SI{50}{\ohm} single ended inputs/outputs.
On the software side, PYNQ package v 3.0.1 is running on the embedded processor, allowing real-time algorithm debug, raw data download and part of the board reconfiguration with a dedicated and user-friendly Python code and Jupyter notebook.
Owing to the high number of ADC and DAC channels, this board
can implement in real time the transfer oscillator principle and enable to phase-lock up to 6 slave lasers to a common reference laser simultaneously. 
When implementing the above-described processing routine at $f_\text{s}$ = \SI{250}{MHz}, the latency can be kept at the level of \SI{1.35}{\micro s}, higher than expected for the board itself, and possibly limited by hardware components like band-pass filters. 
For the board configuration, libraries are already available for most routines (local oscillator synthesis, low-pass filtering and arctan computation), while the unwrap function needs to be specifically designed for this application. Sample code is available at \url{https://git.mycloud-links.com/psavio/nextlasers_fpga} and more details are given in \cite{sav25}.

\begin{figure}
    \centering
    \includegraphics[width=0.58\textwidth]{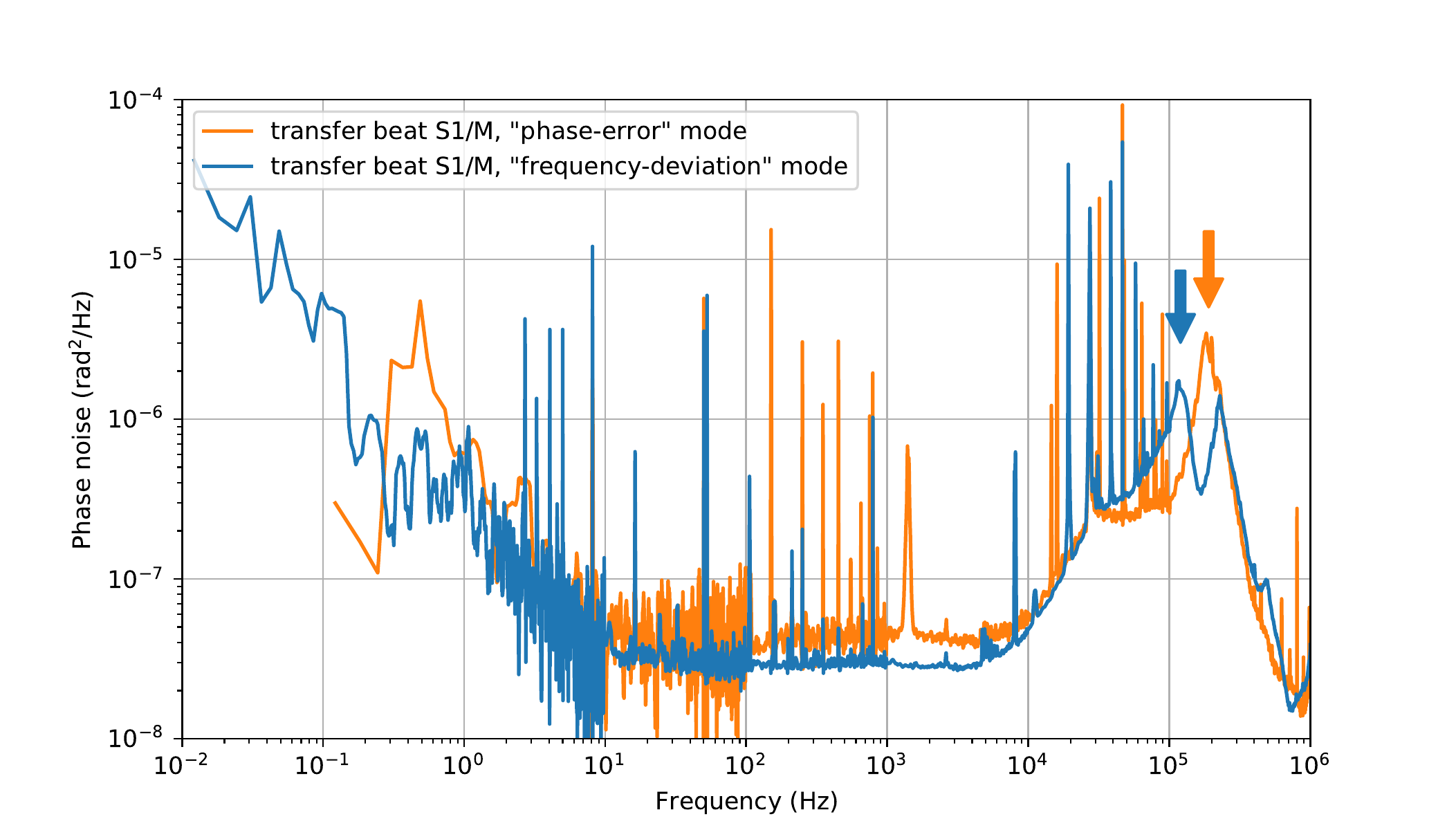}
    \includegraphics[width=0.4\textwidth]{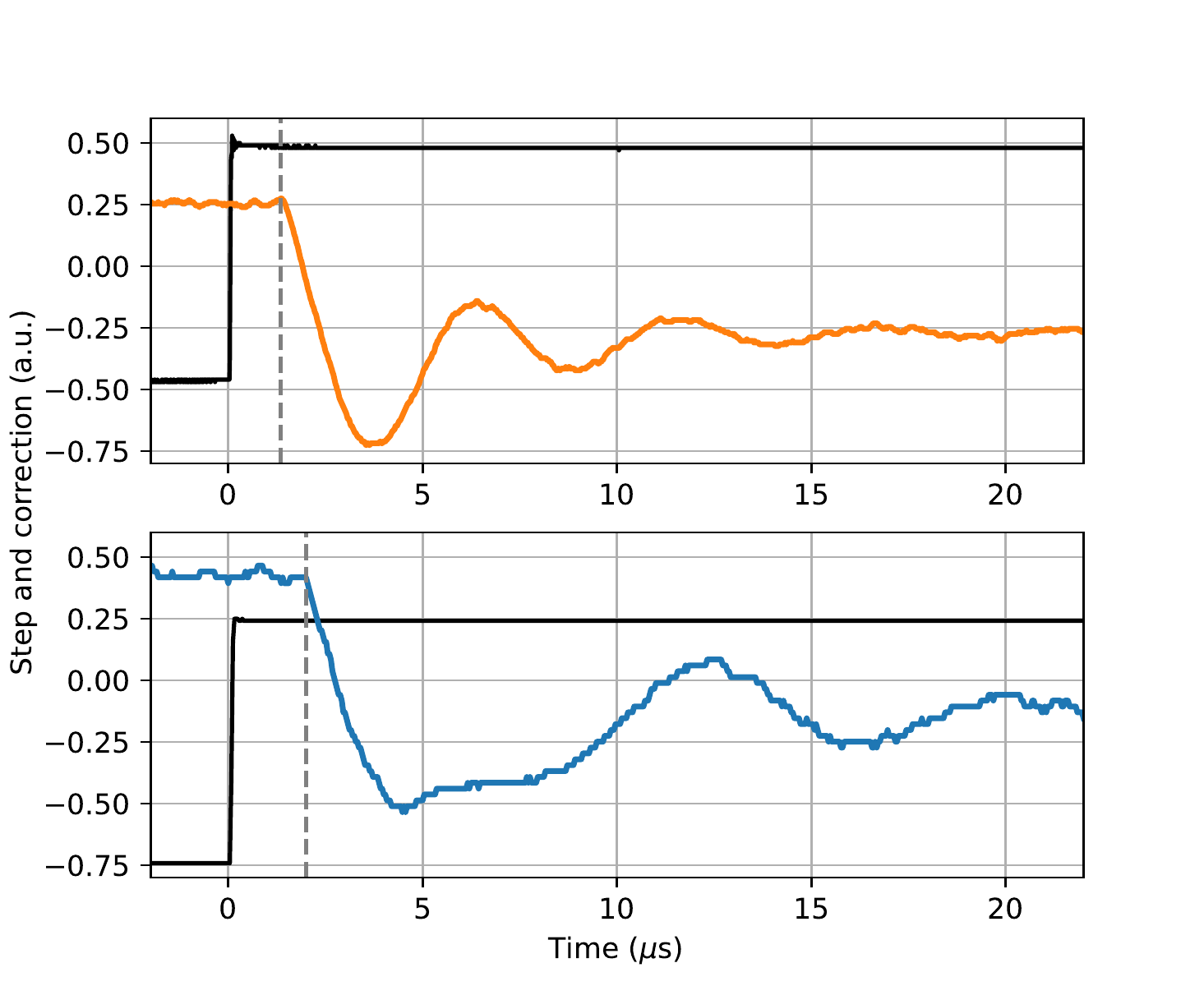}
    \caption{Left: Residual phase noise of the FPGA in the ``phase-error'' (orange) or ``frequency-deviation'' (blue)  mode of the output. 
		Arrows indicate the corresponding loop bandwidth of \SI{190}{kHz} and \SI{120}{kHz}. 
		Right: Measured latency of the FPGA synthesis chain. 
		Top: Response of the loop filter (orange) to a square-wave modulation disturbance applied to the slave-laser frequency when the board is configured with a ``phase-error'' output. 
		Bottom:  Response of the loop filter (blue) when the board is configured with a ``frequency-deviation'' mode. 
		The vertical lines in the two plots indicate the system response latencies of \SI{1.35}{\micro s }  and \SI{2.00}{\micro s } for the two cases. 
		\label{fig:Phase_FM_out_INRIM} }
\end{figure}

The above routine was used to phase-lock different kinds of lasers to an ultrastable cavity-stabilized master using the comb as a transfer oscillator. 
Specifically, we tested it for slaving  a \SI{1542.14}{nm} (\SI{194.400 }{THz}) external cavity diode laser in butterfly package (PLANEX by Redfern Integrated Optics) with an instantaneous linewidth $<$\SI{10}{kHz}, and a continuously-tunable diode laser (CTL-1500 by Toptica) operated at  \SI{1543.51}{nm} (\SI{194.228 }{THz}) with a free-space cavity and linewdith $<$\SI{100}{kHz}. As a reference laser, we used a  \SI{1156}{nm} external-cavity diode laser stabilized to a high-finesse Fabry-Perot cavity. 
Both lasers could be  simultaneously phase-locked to the master laser by configuring  the FPGA board with two sets of coefficients and driving two independent DAC output stages. Coefficient sets were chosen according to the required scaling for each slave and with sign depending on the beatnotes and comb offset-frequency sign (Fig.\ \ref{fig:fpga_design_all_channels_LINKS}b).
Fig.\ \ref{fig:Phase_FM_out_INRIM} shows the in-loop noise floor of the FPGA board in the ``phase-error'' or ``frequency-deviation'' output modes (left)  and corresponding latencies (right). 
In this measurement,  the comb was operated in the narrow linewidth regime, but similar results are obtained with the comb locked in the RF domain. 
In both operation modes, the white phase noise floor is \SI{3e-8}{rad\squared/Hz}, which may be consistent with the raw beatnotes SNR. 
The impact of latencies limiting the locking bandwidth becomes appreciable  at  Fourier frequencies $>$\SI{10}{kHz}. We achieved a locking bandwidth of \SI{190}{kHz} in the ``phase-error'' output mode, and  \SI{120}{kHz} in the ``frequency-output'' mode.  

The overall measured delays are higher than expected for the FPGA alone and could be due to analog components rather then the signal processing. Specifically, the additional delay for the ``frequency-deviation'' mode can be attributed to a low-pass filter in the demodulation of the \SI{10}{MHz} signal carrying the phase-error information, featuring \SI{400}{ns} group delay according to the specification sheet. 
Although replacement of filtering components could  reduce latencies in future implementations, the present results are already adequate for low-noise and robust phase-locks of commercial diode lasers even without any sort of prestabilization.
The corresponding short-term instability contributed by the digital locking chain is at the level of \SI{1e-18}{} at \SI{1}{s} and lower for longer integration times. 
\begin{figure}
\centering
\includegraphics[width=0.48\textwidth]{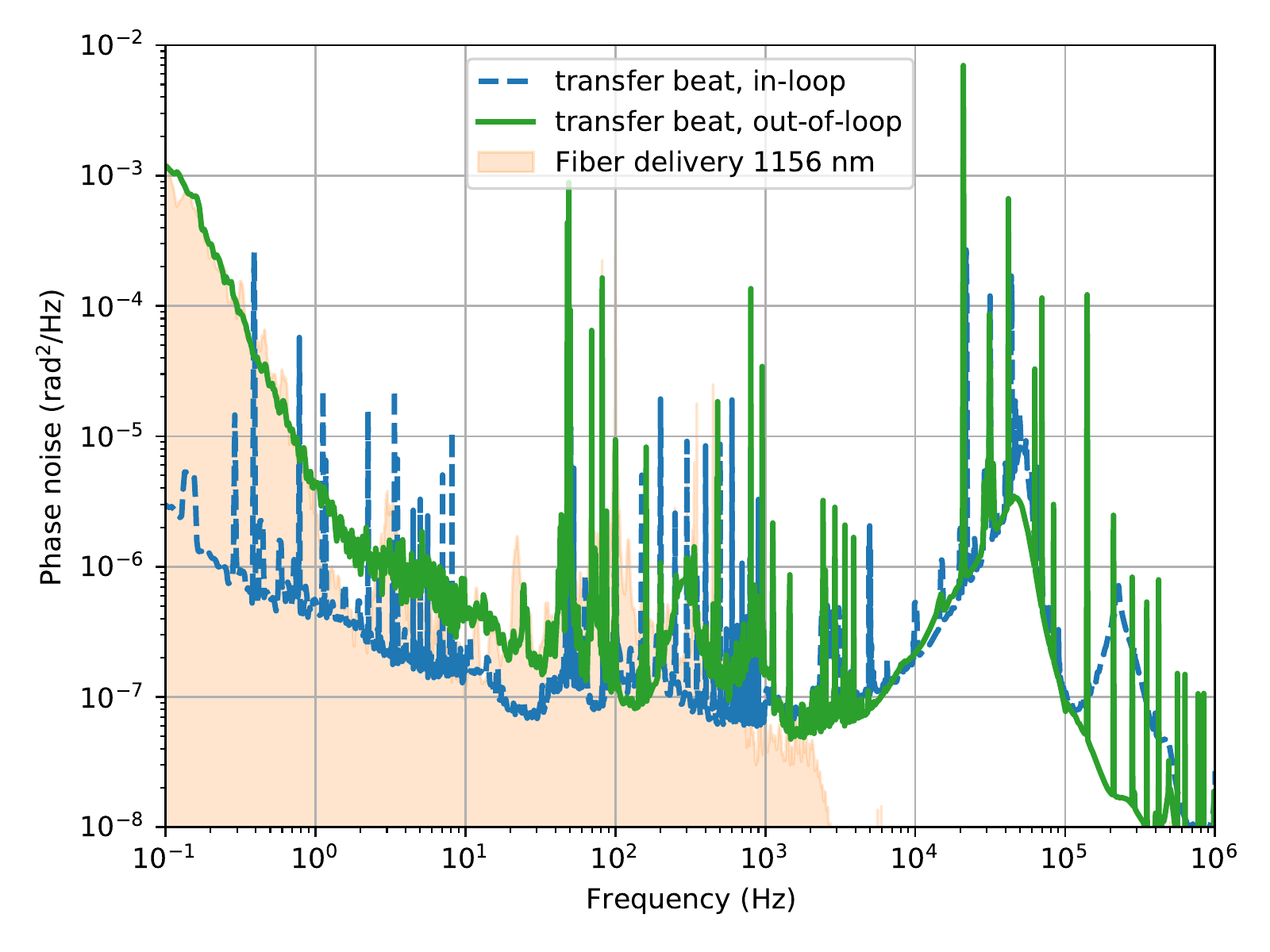}
\includegraphics[width=0.48\textwidth]{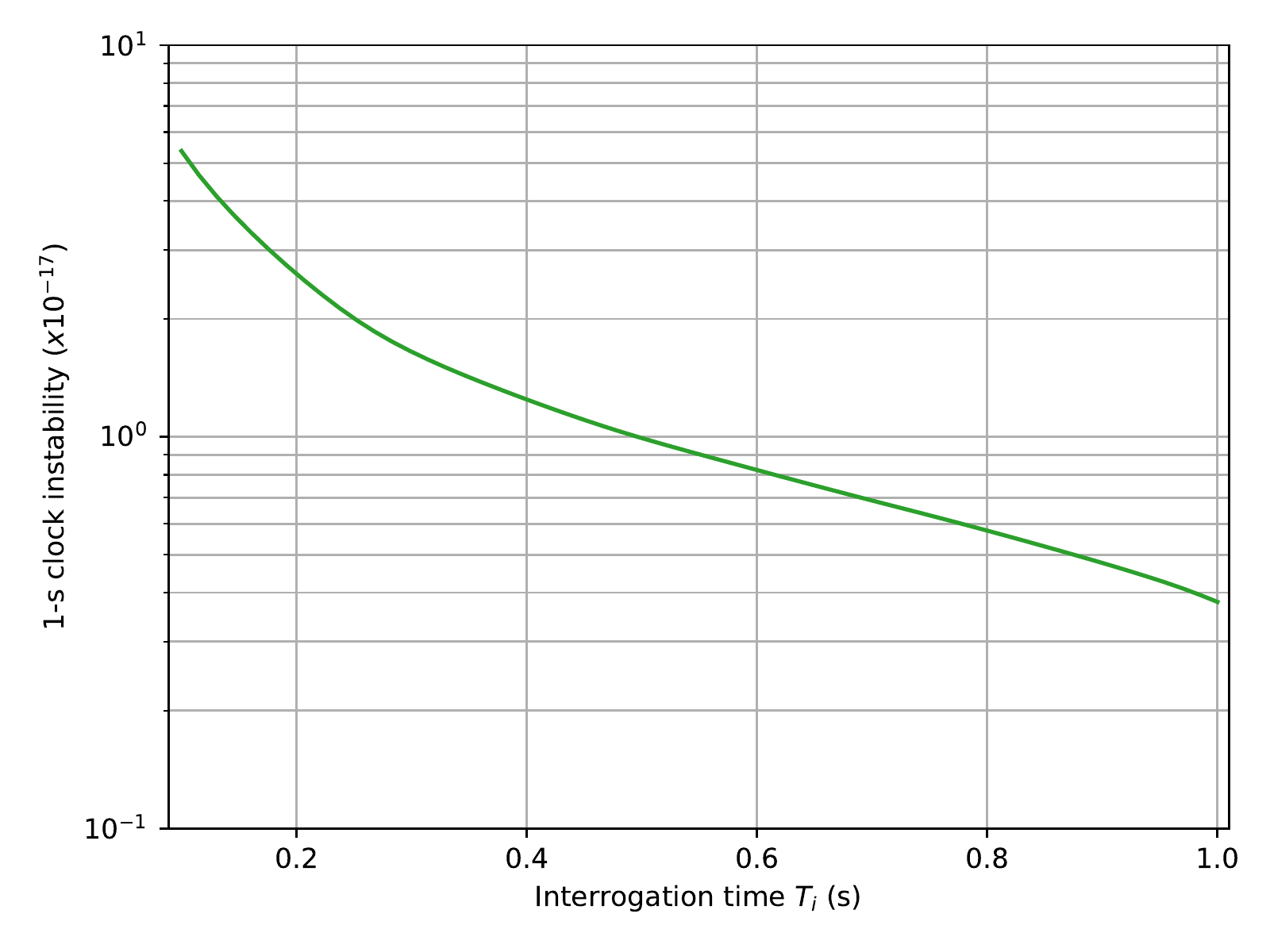}
\caption{Left: phase noise of the spectral transfer between a \SI{1156}{nm} and a \SI{1542}{nm} laser, measured by producing an out-of-loop measurement of the transfer beatnote on a separate comb (green). 
The blue curve and orange shadowed area show respectively the residual noise of the FPGA synthesis chain  and of fiber delivery systems at the two wavelengths. 
Right: Dick-effect instability of an optical clock interrogated by Rabi spectroscopy, as a function of the interrogation time $T_\text{i}$, corresponding to the green trace. \label{fig:outloop_fpga_beat_INRIM}}
\end{figure}
Fig.\ \ref{fig:outloop_fpga_beat_INRIM} (left)  (green curve) shows the noise power spectral density of the \SI{1542}{nm} laser locked to the \SI{1156}{nm} laser using the comb, with the three housed in different laboratories. 
The blue curve represents the noise floor of the digital electronics chain, while the orange shadowed area is the noise contributed by one of the fiber delivery systems. 
We observe that the latter is indeed the major noise source  at Fourier frequencies $<$\SI{1}{kHz} while  the FPGA synthesis chain contributes a barely negligible  noise except for Fourier frequencies higher than \SI{10}{kHz}, where the impact of delay caused by space separation emerges. 

From this spectrum, it is possible to compute the Dick-effect contributed by the synthesis chain in an optical clock interrogation cycle,  under the hypothesis of a noiseless reference laser.  Only the noise at Fourier frequencies corresponding to the clock cycle rate and its  harmonics are relevant to the calculation, with weight that depends on the interrogation time, clock sequence and lock points \cite{santarelli,dick}. 
As an example, we show in Fig.\ \ref{fig:outloop_fpga_beat_INRIM} (right) the estimated Dick effect for Rabi interrogation at a cycle time of \SI{1}{s} and various interrogation times $T_\text{i}$. 
For this estimation, we considered the phase noise of the realized synthesis chain as measured from the out-of-loop comparison, reported in Fig.\ \ref{fig:outloop_fpga_beat_INRIM} (left), in the frequency range \SI{1}{Hz} -- \SI{50}{Hz}. Higher frequencies were found not to contribute significantly. 
The contribution of the transfer beatnote synthesis chains to an optical clock instability is estimated to be \SI{5e-17}{}  for Rabi interrogation at $T_\text{i} = $\SI{100}{ms}. As expected, a longer interrogation time reduces the Dick-effect uncertainty below \SI{1e-17}{}. Such a low Dick-effect-contributed instability is lower than introduced by state-of-the-art interrogation lasers and supports clock ultimate uncertainties at the level of \SI{1e-18}{} to be achieved by integration times as short as \SI{100}{s}.

\section{Distributing the feedback information}
\label{sec:distributing}

As anticipated in previous sections, spatial separation between the optical comb and the lasers to be phase-locked  requires addressing practical concerns such as the implementation of phase-stabilized fiber delivery systems and methods to send the correction information from the point where the transfer beatnote is generated (usually, close to the comb) and the laboratory where the slave laser is housed. 

Besides obvious limitations in the ultimate correction bandwidth introduced by the cables delay (for laboratories separated by 100 m of cable the maximum correction bandwidth is \SI{500}{kHz}) it is relevant to address other aspects that depend on the transfer beatnote generation strategy. In setups where a real RF transfer beatnote is produced exploiting hardware components, this can be sent around directly via coaxial cables. In cases where the slave laser correction signal is produced by an FPGA board, we showed how to address the error signal delivery in a similar way, encoding the information onto a RF sine-wave, as a modulation of the carrier frequency. Still, radio-frequency distribution via coaxial cables may be subject to interferences, noise pick-up and ground effects that can in principle degrade the phase noise of transferred beatnotes. 
As a rule of thumb, the frequency of the transferred beatnote should be adjusted to match important criteria:
\begin{itemize}
    \item the frequency should not be too low, to avoid differential ground effects between the two laboratories that are connected, and to reject the low frequency noise of electronics components such as power supplies,
    \item the transfer beat must be at a position where it can be filtered in order  not to be contaminated by the other peaks, such as comb signals, or modulator parasitic signals, which may be stronger than the relevant beatnote,
    \item the frequency should not be too high in order to avoid temperature-induced delay variations that become relevant for microwaves.
    \item ground decoupling may be necessary to avoid ground-loop issues and parasitic noise.
\end{itemize}

\begin{figure}
\centering
    \includegraphics[width=\textwidth]{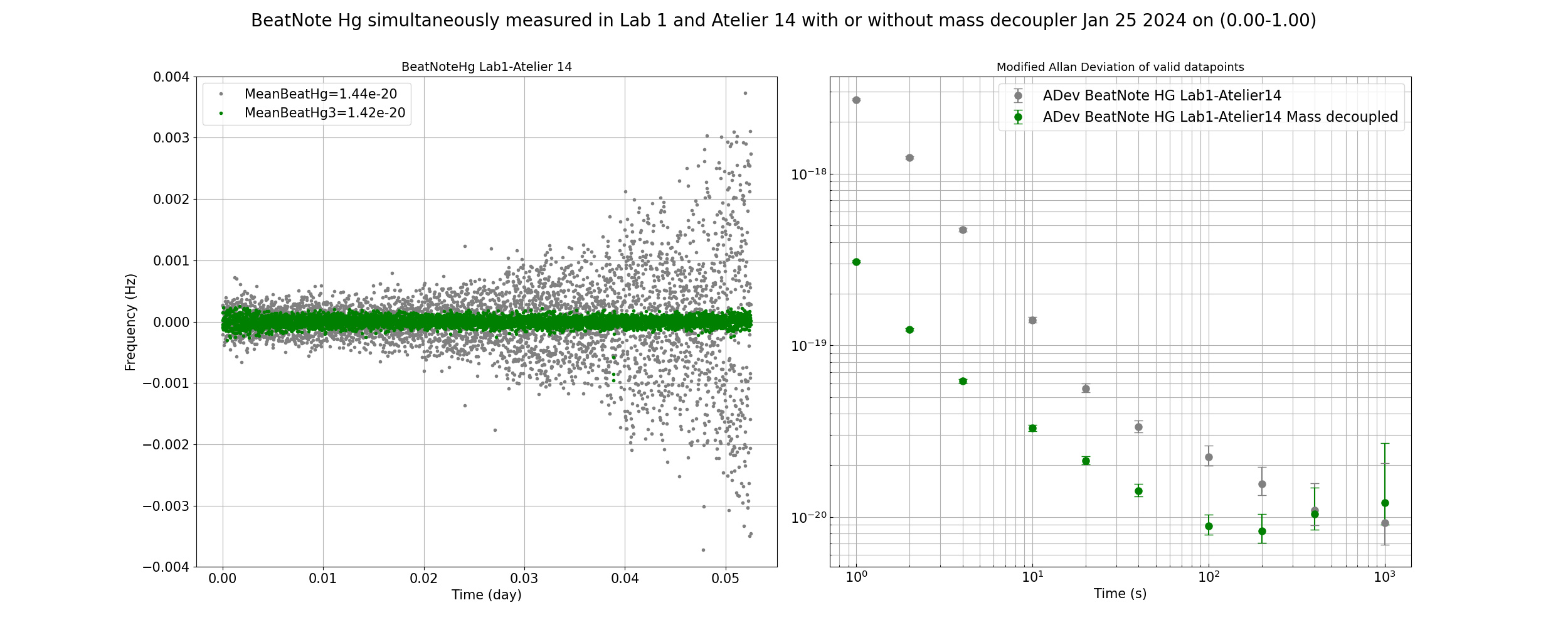}
    \caption{Differential stability of an optical beatnote counted simultaneously by a counter close to photodetection and by a counter installed remotely at 200~m. The use of a ground decoupler (green) improves the stability and the structure of the signal with respect to the case where the signal is transmitted directly (grey).}
    \label{fig:BeatNote_Hg_Lab1_At14}
\end{figure}

Typically, transferring a signal in the range \SI{10}{MHz} -- \SI{100}{MHz} is a convenient compromise, and to this end, offsets in the connections between the optical oscillators and the comb can usually be adjusted accordingly.
As an example, Fig \ref{fig:BeatNote_Hg_Lab1_At14} shows the time record of a RF signal at about \SI{100}{MHz} delivered between distant laboratories connected by 200~m long RF cables featuring SMA connectors. Even though the SNR is changed by less than 4~dB in a band of 1~kHz, differential stability is degraded from a few $10^{-19}$ to a few $10^{-18}$. Additionally, a structure appears in the scattering of the data points, most likely due to a change of SNR or slowly varying stress (temperature, human or city activity) along the cable. Adding a ground decoupler (INMET 8039) allows recovering the initial stability. In both cases, the stability averages down and no systematic offset is resolved down to $1\times10^{-20}$.

\clearpage

\bibliographystyle{apsrev4-2-titles}

\clearpage
\phantomsection
\addcontentsline{toc}{chapter}{Bibliography}
\bibliography{bibliography}

\begin{thebibliography}{125}%
\makeatletter
\providecommand \@ifxundefined [1]{%
 \@ifx{#1\undefined}
}%
\providecommand \@ifnum [1]{%
 \ifnum #1\expandafter \@firstoftwo
 \else \expandafter \@secondoftwo
 \fi
}%
\providecommand \@ifx [1]{%
 \ifx #1\expandafter \@firstoftwo
 \else \expandafter \@secondoftwo
 \fi
}%
\providecommand \natexlab [1]{#1}%
\providecommand \enquote  [1]{``#1''}%
\providecommand \bibnamefont  [1]{#1}%
\providecommand \bibfnamefont [1]{#1}%
\providecommand \citenamefont [1]{#1}%
\providecommand \href@noop [0]{\@secondoftwo}%
\providecommand \href [0]{\begingroup \@sanitize@url \@href}%
\providecommand \@href[1]{\@@startlink{#1}\@@href}%
\providecommand \@@href[1]{\endgroup#1\@@endlink}%
\providecommand \@sanitize@url [0]{\catcode `\\12\catcode `\$12\catcode
  `\&12\catcode `\#12\catcode `\^12\catcode `\_12\catcode `\%12\relax}%
\providecommand \@@startlink[1]{}%
\providecommand \@@endlink[0]{}%
\providecommand \url  [0]{\begingroup\@sanitize@url \@url }%
\providecommand \@url [1]{\endgroup\@href {#1}{\urlprefix }}%
\providecommand \urlprefix  [0]{URL }%
\providecommand \Eprint [0]{\href }%
\providecommand \doibase [0]{https://doi.org/}%
\providecommand \selectlanguage [0]{\@gobble}%
\providecommand \bibinfo  [0]{\@secondoftwo}%
\providecommand \bibfield  [0]{\@secondoftwo}%
\providecommand \translation [1]{[#1]}%
\providecommand \BibitemOpen [0]{}%
\providecommand \bibitemStop [0]{}%
\providecommand \bibitemNoStop [0]{.\EOS\space}%
\providecommand \EOS [0]{\spacefactor3000\relax}%
\providecommand \BibitemShut  [1]{\csname bibitem#1\endcsname}%
\let\auto@bib@innerbib\@empty
\bibitem [{\citenamefont {Schioppo}\ \emph {et~al.}(2022)\citenamefont
  {Schioppo}, \citenamefont {Kronj\"ager}, \citenamefont {Silva}, \citenamefont
  {Ilieva}, \citenamefont {Paterson}, \citenamefont {Baynham}, \citenamefont
  {Bowden}, \citenamefont {Hill}, \citenamefont {Hobson}, \citenamefont
  {Vianello}, \citenamefont {Dovale-\'Alvarez}, \citenamefont {Williams},
  \citenamefont {Marra}, \citenamefont {Margolis}, \citenamefont {Amy-Klein},
  \citenamefont {Lopez}, \citenamefont {Cantin}, \citenamefont
  {\'Alvarez-Mart\'inez}, \citenamefont {Le~Targat}, \citenamefont {Pottie},
  \citenamefont {Quintin}, \citenamefont {Legero}, \citenamefont {H\"afner},
  \citenamefont {Sterr}, \citenamefont {Schwarz}, \citenamefont {D\"orscher},
  \citenamefont {Lisdat}, \citenamefont {Koke}, \citenamefont {Kuhl},
  \citenamefont {Waterholter}, \citenamefont {Benkler},\ and\ \citenamefont
  {Grosche}}]{sch22a}%
  \BibitemOpen
  \bibfield  {author} {\bibinfo {author} {\bibfnamefont {M.}~\bibnamefont
  {Schioppo}}, \bibinfo {author} {\bibfnamefont {J.}~\bibnamefont
  {Kronj\"ager}}, \bibinfo {author} {\bibfnamefont {A.}~\bibnamefont {Silva}},
  \bibinfo {author} {\bibfnamefont {R.}~\bibnamefont {Ilieva}}, \bibinfo
  {author} {\bibfnamefont {J.~W.}\ \bibnamefont {Paterson}}, \bibinfo {author}
  {\bibfnamefont {C.~F.~A.}\ \bibnamefont {Baynham}}, \bibinfo {author}
  {\bibfnamefont {W.}~\bibnamefont {Bowden}}, \bibinfo {author} {\bibfnamefont
  {I.~R.}\ \bibnamefont {Hill}}, \bibinfo {author} {\bibfnamefont
  {R.}~\bibnamefont {Hobson}}, \bibinfo {author} {\bibfnamefont
  {A.}~\bibnamefont {Vianello}}, \bibinfo {author} {\bibfnamefont
  {M.}~\bibnamefont {Dovale-\'Alvarez}}, \bibinfo {author} {\bibfnamefont
  {R.~A.}\ \bibnamefont {Williams}}, \bibinfo {author} {\bibfnamefont
  {G.}~\bibnamefont {Marra}}, \bibinfo {author} {\bibfnamefont {H.~S.}\
  \bibnamefont {Margolis}}, \bibinfo {author} {\bibfnamefont {A.}~\bibnamefont
  {Amy-Klein}}, \bibinfo {author} {\bibfnamefont {O.}~\bibnamefont {Lopez}},
  \bibinfo {author} {\bibfnamefont {E.}~\bibnamefont {Cantin}}, \bibinfo
  {author} {\bibfnamefont {H.}~\bibnamefont {\'Alvarez-Mart\'inez}}, \bibinfo
  {author} {\bibfnamefont {R.}~\bibnamefont {Le~Targat}}, \bibinfo {author}
  {\bibfnamefont {P.~E.}\ \bibnamefont {Pottie}}, \bibinfo {author}
  {\bibfnamefont {N.}~\bibnamefont {Quintin}}, \bibinfo {author} {\bibfnamefont
  {T.}~\bibnamefont {Legero}}, \bibinfo {author} {\bibfnamefont
  {S.}~\bibnamefont {H\"afner}}, \bibinfo {author} {\bibfnamefont
  {U.}~\bibnamefont {Sterr}}, \bibinfo {author} {\bibfnamefont
  {R.}~\bibnamefont {Schwarz}}, \bibinfo {author} {\bibfnamefont
  {S.}~\bibnamefont {D\"orscher}}, \bibinfo {author} {\bibfnamefont
  {C.}~\bibnamefont {Lisdat}}, \bibinfo {author} {\bibfnamefont
  {S.}~\bibnamefont {Koke}}, \bibinfo {author} {\bibfnamefont {A.}~\bibnamefont
  {Kuhl}}, \bibinfo {author} {\bibfnamefont {T.}~\bibnamefont {Waterholter}},
  \bibinfo {author} {\bibfnamefont {E.}~\bibnamefont {Benkler}},\ and\ \bibinfo
  {author} {\bibfnamefont {G.}~\bibnamefont {Grosche}},\ }\bibfield  {title}
  {\emph {\bibinfo {title} {Comparing ultrastable lasers at $7\times 10^{-17}$
  fractional frequency instability through a 2220~km optical fibre network}},\
  }\href {https://doi.org/10.1038/s41467-021-27884-3} {\bibfield  {journal}
  {\bibinfo  {journal} {Nature Commun.}\ }\textbf {\bibinfo {volume} {13}},\
  \bibinfo {pages} {212} (\bibinfo {year} {2022})}\BibitemShut {NoStop}%
\bibitem [{\citenamefont {Nicolodi}\ \emph {et~al.}(2014)\citenamefont
  {Nicolodi}, \citenamefont {Argence}, \citenamefont {Zhang}, \citenamefont
  {Le~Targat}, \citenamefont {Santarelli},\ and\ \citenamefont
  {Le~Coq}}]{nic14}%
  \BibitemOpen
  \bibfield  {author} {\bibinfo {author} {\bibfnamefont {D.}~\bibnamefont
  {Nicolodi}}, \bibinfo {author} {\bibfnamefont {B.}~\bibnamefont {Argence}},
  \bibinfo {author} {\bibfnamefont {W.}~\bibnamefont {Zhang}}, \bibinfo
  {author} {\bibfnamefont {R.}~\bibnamefont {Le~Targat}}, \bibinfo {author}
  {\bibfnamefont {G.}~\bibnamefont {Santarelli}},\ and\ \bibinfo {author}
  {\bibfnamefont {Y.}~\bibnamefont {Le~Coq}},\ }\bibfield  {title} {\emph
  {\bibinfo {title} {Spectral purity transfer between optical wavelengths at
  the $10^{-18}$ level}},\ }\href {https://doi.org/10.1038/nphoton.2013.361}
  {\bibfield  {journal} {\bibinfo  {journal} {Nature Photonics}\ }\textbf
  {\bibinfo {volume} {8}},\ \bibinfo {pages} {219--223} (\bibinfo {year}
  {2014})}\BibitemShut {NoStop}%
\bibitem [{\citenamefont {Benkler}\ \emph {et~al.}(2019)\citenamefont
  {Benkler}, \citenamefont {Lipphardt}, \citenamefont {Puppe}, \citenamefont
  {Wilk}, \citenamefont {Rohde},\ and\ \citenamefont {Sterr}}]{ben19}%
  \BibitemOpen
  \bibfield  {author} {\bibinfo {author} {\bibfnamefont {E.}~\bibnamefont
  {Benkler}}, \bibinfo {author} {\bibfnamefont {B.}~\bibnamefont {Lipphardt}},
  \bibinfo {author} {\bibfnamefont {T.}~\bibnamefont {Puppe}}, \bibinfo
  {author} {\bibfnamefont {R.}~\bibnamefont {Wilk}}, \bibinfo {author}
  {\bibfnamefont {F.}~\bibnamefont {Rohde}},\ and\ \bibinfo {author}
  {\bibfnamefont {U.}~\bibnamefont {Sterr}},\ }\bibfield  {title} {\emph
  {\bibinfo {title} {End-to-end topology for fiber comb based optical frequency
  transfer at the $10^{-21}$ level}},\ }\href
  {https://doi.org/10.1364/OE.27.036886} {\bibfield  {journal} {\bibinfo
  {journal} {Opt. Express}\ }\textbf {\bibinfo {volume} {27}},\ \bibinfo
  {pages} {36886--36902} (\bibinfo {year} {2019})},\ \bibinfo {note} {also see
  erratum \cite{ben20}}\BibitemShut {NoStop}%
\bibitem [{\citenamefont {Li}\ \emph {et~al.}(2022)\citenamefont {Li},
  \citenamefont {Rolland}, \citenamefont {Jiang},\ and\ \citenamefont
  {Fermann}}]{li22a}%
  \BibitemOpen
  \bibfield  {author} {\bibinfo {author} {\bibfnamefont {P.}~\bibnamefont
  {Li}}, \bibinfo {author} {\bibfnamefont {A.}~\bibnamefont {Rolland}},
  \bibinfo {author} {\bibfnamefont {J.}~\bibnamefont {Jiang}},\ and\ \bibinfo
  {author} {\bibfnamefont {M.~E.}\ \bibnamefont {Fermann}},\ }\bibfield
  {title} {\emph {\bibinfo {title} {Coherent frequency transfer with $< 5
  \times 10^{-21}$ stability via a multi-branch comb with noise
  cancellation}},\ }\href {https://doi.org/10.1364/OE.461440} {\bibfield
  {journal} {\bibinfo  {journal} {Opt. Express}\ }\textbf {\bibinfo {volume}
  {30}},\ \bibinfo {pages} {22957--22962} (\bibinfo {year} {2022})}\BibitemShut
  {NoStop}%
\bibitem [{\citenamefont {Xie}\ \emph {et~al.}(2017)\citenamefont {Xie},
  \citenamefont {Bouchand}, \citenamefont {Nicolodi}, \citenamefont {Giunta},
  \citenamefont {H\"ansel}, \citenamefont {Lezius}, \citenamefont {Joshi},
  \citenamefont {Datta}, \citenamefont {Alexandre}, \citenamefont {Lours},
  \citenamefont {Tremblin}, \citenamefont {Santarelli}, \citenamefont
  {Holzwarth},\ and\ \citenamefont {Le~Coq}}]{xie17}%
  \BibitemOpen
  \bibfield  {author} {\bibinfo {author} {\bibfnamefont {X.}~\bibnamefont
  {Xie}}, \bibinfo {author} {\bibfnamefont {R.}~\bibnamefont {Bouchand}},
  \bibinfo {author} {\bibfnamefont {D.}~\bibnamefont {Nicolodi}}, \bibinfo
  {author} {\bibfnamefont {M.}~\bibnamefont {Giunta}}, \bibinfo {author}
  {\bibfnamefont {W.}~\bibnamefont {H\"ansel}}, \bibinfo {author}
  {\bibfnamefont {M.}~\bibnamefont {Lezius}}, \bibinfo {author} {\bibfnamefont
  {A.}~\bibnamefont {Joshi}}, \bibinfo {author} {\bibfnamefont
  {S.}~\bibnamefont {Datta}}, \bibinfo {author} {\bibfnamefont
  {C.}~\bibnamefont {Alexandre}}, \bibinfo {author} {\bibfnamefont
  {M.}~\bibnamefont {Lours}}, \bibinfo {author} {\bibfnamefont {P.-A.}\
  \bibnamefont {Tremblin}}, \bibinfo {author} {\bibfnamefont {G.}~\bibnamefont
  {Santarelli}}, \bibinfo {author} {\bibfnamefont {R.}~\bibnamefont
  {Holzwarth}},\ and\ \bibinfo {author} {\bibfnamefont {Y.}~\bibnamefont
  {Le~Coq}},\ }\bibfield  {title} {\emph {\bibinfo {title} {Photonic microwave
  signals with zeptosecond level absolute timing noise}},\ }\href
  {https://doi.org/10.1038/nphoton.2016.215} {\bibfield  {journal} {\bibinfo
  {journal} {Nature Photonics}\ }\textbf {\bibinfo {volume} {11}},\ \bibinfo
  {pages} {44--47} (\bibinfo {year} {2017})}\BibitemShut {NoStop}%
\bibitem [{\citenamefont {Origlia}\ \emph {et~al.}(2018)\citenamefont
  {Origlia}, \citenamefont {Pramod}, \citenamefont {Schiller}, \citenamefont
  {Singh}, \citenamefont {Bongs}, \citenamefont {Schwarz}, \citenamefont
  {Al-Masoudi}, \citenamefont {D\"orscher}, \citenamefont {Herbers},
  \citenamefont {H\"afner}, \citenamefont {Sterr},\ and\ \citenamefont
  {Lisdat}}]{ori18}%
  \BibitemOpen
  \bibfield  {author} {\bibinfo {author} {\bibfnamefont {S.}~\bibnamefont
  {Origlia}}, \bibinfo {author} {\bibfnamefont {M.~S.}\ \bibnamefont {Pramod}},
  \bibinfo {author} {\bibfnamefont {S.}~\bibnamefont {Schiller}}, \bibinfo
  {author} {\bibfnamefont {Y.}~\bibnamefont {Singh}}, \bibinfo {author}
  {\bibfnamefont {K.}~\bibnamefont {Bongs}}, \bibinfo {author} {\bibfnamefont
  {R.}~\bibnamefont {Schwarz}}, \bibinfo {author} {\bibfnamefont
  {A.}~\bibnamefont {Al-Masoudi}}, \bibinfo {author} {\bibfnamefont
  {S.}~\bibnamefont {D\"orscher}}, \bibinfo {author} {\bibfnamefont
  {S.}~\bibnamefont {Herbers}}, \bibinfo {author} {\bibfnamefont
  {S.}~\bibnamefont {H\"afner}}, \bibinfo {author} {\bibfnamefont
  {U.}~\bibnamefont {Sterr}},\ and\ \bibinfo {author} {\bibfnamefont
  {C.}~\bibnamefont {Lisdat}},\ }\bibfield  {title} {\emph {\bibinfo {title}
  {Towards an optical clock for space: {Compact}, high-performance optical
  lattice clock based on bosonic atoms}},\ }\href
  {https://doi.org/10.1103/PhysRevA.98.053443} {\bibfield  {journal} {\bibinfo
  {journal} {Phys. Rev. A}\ }\textbf {\bibinfo {volume} {98}},\ \bibinfo
  {pages} {053443} (\bibinfo {year} {2018})}\BibitemShut {NoStop}%
\bibitem [{\citenamefont {Milner}\ \emph {et~al.}(2019)\citenamefont {Milner},
  \citenamefont {Robinson}, \citenamefont {Kennedy}, \citenamefont {Bothwell},
  \citenamefont {Kedar}, \citenamefont {Matei}, \citenamefont {Legero},
  \citenamefont {Sterr}, \citenamefont {Riehle}, \citenamefont {Leopardi},
  \citenamefont {Fortier}, \citenamefont {Sherman}, \citenamefont {Levine},
  \citenamefont {Yao}, \citenamefont {Ye},\ and\ \citenamefont
  {Oelker}}]{mil19}%
  \BibitemOpen
  \bibfield  {author} {\bibinfo {author} {\bibfnamefont {W.~R.}\ \bibnamefont
  {Milner}}, \bibinfo {author} {\bibfnamefont {J.~M.}\ \bibnamefont
  {Robinson}}, \bibinfo {author} {\bibfnamefont {C.~J.}\ \bibnamefont
  {Kennedy}}, \bibinfo {author} {\bibfnamefont {T.}~\bibnamefont {Bothwell}},
  \bibinfo {author} {\bibfnamefont {D.}~\bibnamefont {Kedar}}, \bibinfo
  {author} {\bibfnamefont {D.~G.}\ \bibnamefont {Matei}}, \bibinfo {author}
  {\bibfnamefont {T.}~\bibnamefont {Legero}}, \bibinfo {author} {\bibfnamefont
  {U.}~\bibnamefont {Sterr}}, \bibinfo {author} {\bibfnamefont
  {F.}~\bibnamefont {Riehle}}, \bibinfo {author} {\bibfnamefont
  {H.}~\bibnamefont {Leopardi}}, \bibinfo {author} {\bibfnamefont {T.~M.}\
  \bibnamefont {Fortier}}, \bibinfo {author} {\bibfnamefont {J.~A.}\
  \bibnamefont {Sherman}}, \bibinfo {author} {\bibfnamefont {J.}~\bibnamefont
  {Levine}}, \bibinfo {author} {\bibfnamefont {J.}~\bibnamefont {Yao}},
  \bibinfo {author} {\bibfnamefont {J.}~\bibnamefont {Ye}},\ and\ \bibinfo
  {author} {\bibfnamefont {E.}~\bibnamefont {Oelker}},\ }\bibfield  {title}
  {\emph {\bibinfo {title} {Demonstration of a timescale based on a stable
  optical carrier}},\ }\href {https://doi.org/10.1103/PhysRevLett.123.173201}
  {\bibfield  {journal} {\bibinfo  {journal} {Phys. Rev. Lett.}\ }\textbf
  {\bibinfo {volume} {123}},\ \bibinfo {pages} {173201} (\bibinfo {year}
  {2019})}\BibitemShut {NoStop}%
\bibitem [{\citenamefont {Herbers}\ \emph {et~al.}(2022)\citenamefont
  {Herbers}, \citenamefont {H\"{a}fner}, \citenamefont {D\"{o}rscher},
  \citenamefont {L\"{u}cke}, \citenamefont {Sterr},\ and\ \citenamefont
  {Lisdat}}]{her22}%
  \BibitemOpen
  \bibfield  {author} {\bibinfo {author} {\bibfnamefont {S.}~\bibnamefont
  {Herbers}}, \bibinfo {author} {\bibfnamefont {S.}~\bibnamefont {H\"{a}fner}},
  \bibinfo {author} {\bibfnamefont {S.}~\bibnamefont {D\"{o}rscher}}, \bibinfo
  {author} {\bibfnamefont {T.}~\bibnamefont {L\"{u}cke}}, \bibinfo {author}
  {\bibfnamefont {U.}~\bibnamefont {Sterr}},\ and\ \bibinfo {author}
  {\bibfnamefont {C.}~\bibnamefont {Lisdat}},\ }\bibfield  {title} {\emph
  {\bibinfo {title} {Transportable clock laser system with an instability of
  $1.6 \times 10^{-16}$}},\ }\href {https://doi.org/10.1364/OL.470984}
  {\bibfield  {journal} {\bibinfo  {journal} {Opt. Lett.}\ }\textbf {\bibinfo
  {volume} {47}},\ \bibinfo {pages} {5441--5444} (\bibinfo {year}
  {2022})}\BibitemShut {NoStop}%
\bibitem [{\citenamefont {Matei}\ \emph {et~al.}(2017)\citenamefont {Matei},
  \citenamefont {Legero}, \citenamefont {H\"afner}, \citenamefont {Grebing},
  \citenamefont {Weyrich}, \citenamefont {Zhang}, \citenamefont {Sonderhouse},
  \citenamefont {Robinson}, \citenamefont {Ye}, \citenamefont {Riehle},\ and\
  \citenamefont {Sterr}}]{mat17a}%
  \BibitemOpen
  \bibfield  {author} {\bibinfo {author} {\bibfnamefont {D.~G.}\ \bibnamefont
  {Matei}}, \bibinfo {author} {\bibfnamefont {T.}~\bibnamefont {Legero}},
  \bibinfo {author} {\bibfnamefont {S.}~\bibnamefont {H\"afner}}, \bibinfo
  {author} {\bibfnamefont {C.}~\bibnamefont {Grebing}}, \bibinfo {author}
  {\bibfnamefont {R.}~\bibnamefont {Weyrich}}, \bibinfo {author} {\bibfnamefont
  {W.}~\bibnamefont {Zhang}}, \bibinfo {author} {\bibfnamefont
  {L.}~\bibnamefont {Sonderhouse}}, \bibinfo {author} {\bibfnamefont {J.~M.}\
  \bibnamefont {Robinson}}, \bibinfo {author} {\bibfnamefont {J.}~\bibnamefont
  {Ye}}, \bibinfo {author} {\bibfnamefont {F.}~\bibnamefont {Riehle}},\ and\
  \bibinfo {author} {\bibfnamefont {U.}~\bibnamefont {Sterr}},\ }\bibfield
  {title} {\emph {\bibinfo {title} {$1.5~\mu$m lasers with sub-{10 mHz}
  linewidth}},\ }\href {https://doi.org/10.1103/PhysRevLett.118.263202}
  {\bibfield  {journal} {\bibinfo  {journal} {Phys. Rev. Lett.}\ }\textbf
  {\bibinfo {volume} {118}},\ \bibinfo {pages} {263202} (\bibinfo {year}
  {2017})}\BibitemShut {NoStop}%
\bibitem [{\citenamefont {Harry}\ \emph {et~al.}(2006)\citenamefont {Harry},
  \citenamefont {Armandula}, \citenamefont {Black}, \citenamefont {Crooks},
  \citenamefont {Cagnoli}, \citenamefont {Hough}, \citenamefont {Murray},
  \citenamefont {Reid}, \citenamefont {Rowan}, \citenamefont {Sneddon},
  \citenamefont {Fejer}, \citenamefont {Route},\ and\ \citenamefont
  {Penn}}]{har06b}%
  \BibitemOpen
  \bibfield  {author} {\bibinfo {author} {\bibfnamefont {G.~M.}\ \bibnamefont
  {Harry}}, \bibinfo {author} {\bibfnamefont {H.}~\bibnamefont {Armandula}},
  \bibinfo {author} {\bibfnamefont {E.}~\bibnamefont {Black}}, \bibinfo
  {author} {\bibfnamefont {D.~R.~M.}\ \bibnamefont {Crooks}}, \bibinfo {author}
  {\bibfnamefont {G.}~\bibnamefont {Cagnoli}}, \bibinfo {author} {\bibfnamefont
  {J.}~\bibnamefont {Hough}}, \bibinfo {author} {\bibfnamefont
  {P.}~\bibnamefont {Murray}}, \bibinfo {author} {\bibfnamefont
  {S.}~\bibnamefont {Reid}}, \bibinfo {author} {\bibfnamefont {S.}~\bibnamefont
  {Rowan}}, \bibinfo {author} {\bibfnamefont {P.}~\bibnamefont {Sneddon}},
  \bibinfo {author} {\bibfnamefont {M.~M.}\ \bibnamefont {Fejer}}, \bibinfo
  {author} {\bibfnamefont {R.}~\bibnamefont {Route}},\ and\ \bibinfo {author}
  {\bibfnamefont {S.~D.}\ \bibnamefont {Penn}},\ }\bibfield  {title} {\emph
  {\bibinfo {title} {Thermal noise from optical coatings in gravitational wave
  detectors}},\ }\href {https://doi.org/10.1364/AO.45.001569} {\bibfield
  {journal} {\bibinfo  {journal} {Appl. Opt.}\ }\textbf {\bibinfo {volume}
  {45}},\ \bibinfo {pages} {1569--1574} (\bibinfo {year} {2006})}\BibitemShut
  {NoStop}%
\bibitem [{\citenamefont {Granata}\ \emph {et~al.}(2020)\citenamefont
  {Granata}, \citenamefont {Amato}, \citenamefont {Cagnoli}, \citenamefont
  {Coulon}, \citenamefont {Degallaix}, \citenamefont {Forest}, \citenamefont
  {Mereni}, \citenamefont {Michel}, \citenamefont {Pinard}, \citenamefont
  {Sassolas},\ and\ \citenamefont {Teillon}}]{gra20a}%
  \BibitemOpen
  \bibfield  {author} {\bibinfo {author} {\bibfnamefont {M.}~\bibnamefont
  {Granata}}, \bibinfo {author} {\bibfnamefont {A.}~\bibnamefont {Amato}},
  \bibinfo {author} {\bibfnamefont {G.}~\bibnamefont {Cagnoli}}, \bibinfo
  {author} {\bibfnamefont {M.}~\bibnamefont {Coulon}}, \bibinfo {author}
  {\bibfnamefont {J.}~\bibnamefont {Degallaix}}, \bibinfo {author}
  {\bibfnamefont {D.}~\bibnamefont {Forest}}, \bibinfo {author} {\bibfnamefont
  {L.}~\bibnamefont {Mereni}}, \bibinfo {author} {\bibfnamefont
  {C.}~\bibnamefont {Michel}}, \bibinfo {author} {\bibfnamefont
  {L.}~\bibnamefont {Pinard}}, \bibinfo {author} {\bibfnamefont
  {B.}~\bibnamefont {Sassolas}},\ and\ \bibinfo {author} {\bibfnamefont
  {J.}~\bibnamefont {Teillon}},\ }\bibfield  {title} {\emph {\bibinfo {title}
  {Progress in the measurement and reduction of thermal noise in optical
  coatings for gravitational-wave detectors}},\ }\href
  {https://doi.org/10.1364/AO.377293} {\bibfield  {journal} {\bibinfo
  {journal} {Appl. Opt.}\ }\textbf {\bibinfo {volume} {59}},\ \bibinfo {pages}
  {A229--A235} (\bibinfo {year} {2020})}\BibitemShut {NoStop}%
\bibitem [{\citenamefont {Kedar}\ \emph {et~al.}(2023)\citenamefont {Kedar},
  \citenamefont {Yu}, \citenamefont {Oelker}, \citenamefont {Staron},
  \citenamefont {Milner}, \citenamefont {Robinson}, \citenamefont {Legero},
  \citenamefont {Riehle}, \citenamefont {Sterr},\ and\ \citenamefont
  {Ye}}]{ked23}%
  \BibitemOpen
  \bibfield  {author} {\bibinfo {author} {\bibfnamefont {D.}~\bibnamefont
  {Kedar}}, \bibinfo {author} {\bibfnamefont {J.}~\bibnamefont {Yu}}, \bibinfo
  {author} {\bibfnamefont {E.}~\bibnamefont {Oelker}}, \bibinfo {author}
  {\bibfnamefont {A.}~\bibnamefont {Staron}}, \bibinfo {author} {\bibfnamefont
  {W.~R.}\ \bibnamefont {Milner}}, \bibinfo {author} {\bibfnamefont {J.~M.}\
  \bibnamefont {Robinson}}, \bibinfo {author} {\bibfnamefont {T.}~\bibnamefont
  {Legero}}, \bibinfo {author} {\bibfnamefont {F.}~\bibnamefont {Riehle}},
  \bibinfo {author} {\bibfnamefont {U.}~\bibnamefont {Sterr}},\ and\ \bibinfo
  {author} {\bibfnamefont {J.}~\bibnamefont {Ye}},\ }\bibfield  {title} {\emph
  {\bibinfo {title} {Frequency stability of cryogenic silicon cavities with
  semiconductor crystalline coatings}},\ }\href
  {https://doi.org/10.1364/OPTICA.479462} {\bibfield  {journal} {\bibinfo
  {journal} {Optica}\ }\textbf {\bibinfo {volume} {10}},\ \bibinfo {pages}
  {464--470} (\bibinfo {year} {2023})}\BibitemShut {NoStop}%
\bibitem [{\citenamefont {Yu}\ \emph {et~al.}(2023)\citenamefont {Yu},
  \citenamefont {H\"afner}, \citenamefont {Legero}, \citenamefont {Herbers},
  \citenamefont {Nicolodi}, \citenamefont {Ma}, \citenamefont {Riehle},
  \citenamefont {Sterr}, \citenamefont {Kedar}, \citenamefont {Robinson},
  \citenamefont {Oelker},\ and\ \citenamefont {Ye}}]{yu23a}%
  \BibitemOpen
  \bibfield  {author} {\bibinfo {author} {\bibfnamefont {J.}~\bibnamefont
  {Yu}}, \bibinfo {author} {\bibfnamefont {S.}~\bibnamefont {H\"afner}},
  \bibinfo {author} {\bibfnamefont {T.}~\bibnamefont {Legero}}, \bibinfo
  {author} {\bibfnamefont {S.}~\bibnamefont {Herbers}}, \bibinfo {author}
  {\bibfnamefont {D.}~\bibnamefont {Nicolodi}}, \bibinfo {author}
  {\bibfnamefont {C.~Y.}\ \bibnamefont {Ma}}, \bibinfo {author} {\bibfnamefont
  {F.}~\bibnamefont {Riehle}}, \bibinfo {author} {\bibfnamefont
  {U.}~\bibnamefont {Sterr}}, \bibinfo {author} {\bibfnamefont
  {D.}~\bibnamefont {Kedar}}, \bibinfo {author} {\bibfnamefont {J.~M.}\
  \bibnamefont {Robinson}}, \bibinfo {author} {\bibfnamefont {E.}~\bibnamefont
  {Oelker}},\ and\ \bibinfo {author} {\bibfnamefont {J.}~\bibnamefont {Ye}},\
  }\bibfield  {title} {\emph {\bibinfo {title} {Excess noise and photoinduced
  effects in highly reflective crystalline mirror coatings}},\ }\href
  {https://doi.org/10.1103/PhysRevX.13.041002} {\bibfield  {journal} {\bibinfo
  {journal} {Phys. Rev. X}\ }\textbf {\bibinfo {volume} {13}},\ \bibinfo
  {pages} {041002} (\bibinfo {year} {2023})}\BibitemShut {NoStop}%
\bibitem [{\citenamefont {Levin}(1998)}]{lev98}%
  \BibitemOpen
  \bibfield  {author} {\bibinfo {author} {\bibfnamefont {Y.}~\bibnamefont
  {Levin}},\ }\bibfield  {title} {\emph {\bibinfo {title} {Internal thermal
  noise in the {LIGO} test masses: A direct approach}},\ }\href
  {https://doi.org/10.1103/PhysRevD.57.659} {\bibfield  {journal} {\bibinfo
  {journal} {Phys. Rev. D}\ }\textbf {\bibinfo {volume} {57}},\ \bibinfo
  {pages} {659--663} (\bibinfo {year} {1998})}\BibitemShut {NoStop}%
\bibitem [{\citenamefont {Callen}\ and\ \citenamefont {Welton}(1951)}]{cal51}%
  \BibitemOpen
  \bibfield  {author} {\bibinfo {author} {\bibfnamefont {H.~B.}\ \bibnamefont
  {Callen}}\ and\ \bibinfo {author} {\bibfnamefont {T.~A.}\ \bibnamefont
  {Welton}},\ }\bibfield  {title} {\emph {\bibinfo {title} {Irreversibility and
  generalized noise}},\ }\href {https://doi.org/10.1103/PhysRev.83.34}
  {\bibfield  {journal} {\bibinfo  {journal} {Phys. Rev.}\ }\textbf {\bibinfo
  {volume} {83}},\ \bibinfo {pages} {34--40} (\bibinfo {year}
  {1951})}\BibitemShut {NoStop}%
\bibitem [{\citenamefont {Kubo}(1966)}]{kub66}%
  \BibitemOpen
  \bibfield  {author} {\bibinfo {author} {\bibfnamefont {R.}~\bibnamefont
  {Kubo}},\ }\bibfield  {title} {\emph {\bibinfo {title} {The
  fluctuation-dissipation theorem}},\ }\href
  {https://doi.org/10.1088/0034-4885/29/1/306} {\bibfield  {journal} {\bibinfo
  {journal} {Rep. Prog. Phys.}\ }\textbf {\bibinfo {volume} {29}},\ \bibinfo
  {pages} {255--284} (\bibinfo {year} {1966})}\BibitemShut {NoStop}%
\bibitem [{\citenamefont {Abdel-Hafiz}\ \emph {et~al.}(2019)\citenamefont
  {Abdel-Hafiz}, \citenamefont {Ablewski}, \citenamefont {Al-Masoudi},
  \citenamefont {\'Alvarez~Mart\'inez}, \citenamefont {Balling}, \citenamefont
  {Barwood}, \citenamefont {Benkler}, \citenamefont {Bober}, \citenamefont
  {Borkowski}, \citenamefont {Bowden}, \citenamefont {Ciury\l{}o},
  \citenamefont {Cybulski}, \citenamefont {Didier}, \citenamefont
  {Dole\v{z}al}, \citenamefont {D\"orscher}, \citenamefont {Falke},
  \citenamefont {Godun}, \citenamefont {Hamid}, \citenamefont {Hill},
  \citenamefont {Hobson}, \citenamefont {Huntemann}, \citenamefont {Le~Coq},
  \citenamefont {Le~Targat}, \citenamefont {Legero}, \citenamefont {Lindvall},
  \citenamefont {Lisdat}, \citenamefont {Lodewyck}, \citenamefont {Margolis},
  \citenamefont {Mehlst\"aubler}, \citenamefont {Peik}, \citenamefont {Pelzer},
  \citenamefont {Pizzocaro}, \citenamefont {Rauf}, \citenamefont {Rolland},
  \citenamefont {Scharnhorst}, \citenamefont {Schioppo}, \citenamefont
  {Schmidt}, \citenamefont {Schwarz}, \citenamefont {\c{S}enel}, \citenamefont
  {Spethmann}, \citenamefont {Sterr}, \citenamefont {Tamm}, \citenamefont
  {Thomsen}, \citenamefont {Vianello},\ and\ \citenamefont {Zawada}}]{abd19}%
  \BibitemOpen
  \bibfield  {author} {\bibinfo {author} {\bibfnamefont {M.}~\bibnamefont
  {Abdel-Hafiz}}, \bibinfo {author} {\bibfnamefont {P.}~\bibnamefont
  {Ablewski}}, \bibinfo {author} {\bibfnamefont {A.}~\bibnamefont
  {Al-Masoudi}}, \bibinfo {author} {\bibfnamefont {H.}~\bibnamefont
  {\'Alvarez~Mart\'inez}}, \bibinfo {author} {\bibfnamefont {P.}~\bibnamefont
  {Balling}}, \bibinfo {author} {\bibfnamefont {G.}~\bibnamefont {Barwood}},
  \bibinfo {author} {\bibfnamefont {E.}~\bibnamefont {Benkler}}, \bibinfo
  {author} {\bibfnamefont {M.}~\bibnamefont {Bober}}, \bibinfo {author}
  {\bibfnamefont {M.}~\bibnamefont {Borkowski}}, \bibinfo {author}
  {\bibfnamefont {W.}~\bibnamefont {Bowden}}, \bibinfo {author} {\bibfnamefont
  {R.}~\bibnamefont {Ciury\l{}o}}, \bibinfo {author} {\bibfnamefont
  {H.}~\bibnamefont {Cybulski}}, \bibinfo {author} {\bibfnamefont
  {A.}~\bibnamefont {Didier}}, \bibinfo {author} {\bibfnamefont
  {M.}~\bibnamefont {Dole\v{z}al}}, \bibinfo {author} {\bibfnamefont
  {S.}~\bibnamefont {D\"orscher}}, \bibinfo {author} {\bibfnamefont
  {S.}~\bibnamefont {Falke}}, \bibinfo {author} {\bibfnamefont {R.~M.}\
  \bibnamefont {Godun}}, \bibinfo {author} {\bibfnamefont {R.}~\bibnamefont
  {Hamid}}, \bibinfo {author} {\bibfnamefont {I.~R.}\ \bibnamefont {Hill}},
  \bibinfo {author} {\bibfnamefont {R.}~\bibnamefont {Hobson}}, \bibinfo
  {author} {\bibfnamefont {N.}~\bibnamefont {Huntemann}}, \bibinfo {author}
  {\bibfnamefont {Y.}~\bibnamefont {Le~Coq}}, \bibinfo {author} {\bibfnamefont
  {R.}~\bibnamefont {Le~Targat}}, \bibinfo {author} {\bibfnamefont
  {T.}~\bibnamefont {Legero}}, \bibinfo {author} {\bibfnamefont
  {T.}~\bibnamefont {Lindvall}}, \bibinfo {author} {\bibfnamefont
  {C.}~\bibnamefont {Lisdat}}, \bibinfo {author} {\bibfnamefont
  {J.}~\bibnamefont {Lodewyck}}, \bibinfo {author} {\bibfnamefont {H.~S.}\
  \bibnamefont {Margolis}}, \bibinfo {author} {\bibfnamefont {T.~E.}\
  \bibnamefont {Mehlst\"aubler}}, \bibinfo {author} {\bibfnamefont
  {E.}~\bibnamefont {Peik}}, \bibinfo {author} {\bibfnamefont {L.}~\bibnamefont
  {Pelzer}}, \bibinfo {author} {\bibfnamefont {M.}~\bibnamefont {Pizzocaro}},
  \bibinfo {author} {\bibfnamefont {B.}~\bibnamefont {Rauf}}, \bibinfo {author}
  {\bibfnamefont {A.}~\bibnamefont {Rolland}}, \bibinfo {author} {\bibfnamefont
  {N.}~\bibnamefont {Scharnhorst}}, \bibinfo {author} {\bibfnamefont
  {M.}~\bibnamefont {Schioppo}}, \bibinfo {author} {\bibfnamefont {P.~O.}\
  \bibnamefont {Schmidt}}, \bibinfo {author} {\bibfnamefont {R.}~\bibnamefont
  {Schwarz}}, \bibinfo {author} {\bibfnamefont {{\c{C}}.}~\bibnamefont
  {\c{S}enel}}, \bibinfo {author} {\bibfnamefont {N.}~\bibnamefont
  {Spethmann}}, \bibinfo {author} {\bibfnamefont {U.}~\bibnamefont {Sterr}},
  \bibinfo {author} {\bibfnamefont {C.}~\bibnamefont {Tamm}}, \bibinfo {author}
  {\bibfnamefont {J.~W.}\ \bibnamefont {Thomsen}}, \bibinfo {author}
  {\bibfnamefont {A.}~\bibnamefont {Vianello}},\ and\ \bibinfo {author}
  {\bibfnamefont {M.}~\bibnamefont {Zawada}},\ }\href
  {https://doi.org/10.48550/arXiv.1906.11495} {\bibinfo {title} {Guidelines for
  developing optical clocks with $10^{-18}$ fractional frequency
  uncertainty}},\ \bibinfo {howpublished} {arXiv:1906.11495 [physics.atom-ph]}
  (\bibinfo {year} {2019})\BibitemShut {NoStop}%
\bibitem [{\citenamefont {Boyd}\ and\ \citenamefont {Lahaye}(2024)}]{boy24}%
  \BibitemOpen
  \bibfield  {author} {\bibinfo {author} {\bibfnamefont {J.~A.}\ \bibnamefont
  {Boyd}}\ and\ \bibinfo {author} {\bibfnamefont {T.}~\bibnamefont {Lahaye}},\
  }\bibfield  {title} {\emph {\bibinfo {title} {A basic introduction to
  ultrastable optical cavities for laser stabilization}},\ }\href
  {https://doi.org/10.1119/5.0161369} {\bibfield  {journal} {\bibinfo
  {journal} {Am. J. Phys.}\ }\textbf {\bibinfo {volume} {92}},\ \bibinfo
  {pages} {50--58} (\bibinfo {year} {2024})}\BibitemShut {NoStop}%
\bibitem [{\citenamefont {Ito}\ \emph {et~al.}(2017)\citenamefont {Ito},
  \citenamefont {Silva}, \citenamefont {Nakamura},\ and\ \citenamefont
  {Kobayashi}}]{ito17}%
  \BibitemOpen
  \bibfield  {author} {\bibinfo {author} {\bibfnamefont {I.}~\bibnamefont
  {Ito}}, \bibinfo {author} {\bibfnamefont {A.}~\bibnamefont {Silva}}, \bibinfo
  {author} {\bibfnamefont {T.}~\bibnamefont {Nakamura}},\ and\ \bibinfo
  {author} {\bibfnamefont {Y.}~\bibnamefont {Kobayashi}},\ }\bibfield  {title}
  {\emph {\bibinfo {title} {Stable {CW} laser based on low thermal expansion
  ceramic cavity with 4.9 {mHz/s} frequency drift}},\ }\href
  {https://doi.org/10.1364/OE.25.026020} {\bibfield  {journal} {\bibinfo
  {journal} {Opt. Express}\ }\textbf {\bibinfo {volume} {25}},\ \bibinfo
  {pages} {26020--26028} (\bibinfo {year} {2017})}\BibitemShut {NoStop}%
\bibitem [{\citenamefont {Cole}\ \emph {et~al.}(2013)\citenamefont {Cole},
  \citenamefont {Zhang}, \citenamefont {Martin}, \citenamefont {Ye},\ and\
  \citenamefont {Aspelmeyer}}]{col13}%
  \BibitemOpen
  \bibfield  {author} {\bibinfo {author} {\bibfnamefont {G.~D.}\ \bibnamefont
  {Cole}}, \bibinfo {author} {\bibfnamefont {W.}~\bibnamefont {Zhang}},
  \bibinfo {author} {\bibfnamefont {M.~J.}\ \bibnamefont {Martin}}, \bibinfo
  {author} {\bibfnamefont {J.}~\bibnamefont {Ye}},\ and\ \bibinfo {author}
  {\bibfnamefont {M.}~\bibnamefont {Aspelmeyer}},\ }\bibfield  {title} {\emph
  {\bibinfo {title} {Tenfold reduction of {B}rownian noise in optical
  interferometry}},\ }\href {https://doi.org/10.1038/NPHOTON.2013.174}
  {\bibfield  {journal} {\bibinfo  {journal} {Nature Photonics}\ }\textbf
  {\bibinfo {volume} {7}},\ \bibinfo {pages} {644--650} (\bibinfo {year}
  {2013})}\BibitemShut {NoStop}%
\bibitem [{\citenamefont {Cole}\ \emph {et~al.}(2023)\citenamefont {Cole},
  \citenamefont {Ballmer}, \citenamefont {Billingsley}, \citenamefont {Cata\~no
  Lopez}, \citenamefont {Fejer}, \citenamefont {Fritschel}, \citenamefont
  {Gretarsson}, \citenamefont {Harry}, \citenamefont {Kedar}, \citenamefont
  {Legero}, \citenamefont {Makarem}, \citenamefont {Penn}, \citenamefont
  {Reitze}, \citenamefont {Steinlechner}, \citenamefont {Sterr}, \citenamefont
  {Tanioka}, \citenamefont {Truong}, \citenamefont {Ye},\ and\ \citenamefont
  {Yu}}]{col23}%
  \BibitemOpen
  \bibfield  {author} {\bibinfo {author} {\bibfnamefont {G.~D.}\ \bibnamefont
  {Cole}}, \bibinfo {author} {\bibfnamefont {S.}~\bibnamefont {Ballmer}},
  \bibinfo {author} {\bibfnamefont {G.}~\bibnamefont {Billingsley}}, \bibinfo
  {author} {\bibfnamefont {S.~B.}\ \bibnamefont {Cata\~no Lopez}}, \bibinfo
  {author} {\bibfnamefont {M.}~\bibnamefont {Fejer}}, \bibinfo {author}
  {\bibfnamefont {P.}~\bibnamefont {Fritschel}}, \bibinfo {author}
  {\bibfnamefont {A.~M.}\ \bibnamefont {Gretarsson}}, \bibinfo {author}
  {\bibfnamefont {G.~M.}\ \bibnamefont {Harry}}, \bibinfo {author}
  {\bibfnamefont {D.}~\bibnamefont {Kedar}}, \bibinfo {author} {\bibfnamefont
  {T.}~\bibnamefont {Legero}}, \bibinfo {author} {\bibfnamefont
  {C.}~\bibnamefont {Makarem}}, \bibinfo {author} {\bibfnamefont {S.~D.}\
  \bibnamefont {Penn}}, \bibinfo {author} {\bibfnamefont {D.}~\bibnamefont
  {Reitze}}, \bibinfo {author} {\bibfnamefont {J.}~\bibnamefont
  {Steinlechner}}, \bibinfo {author} {\bibfnamefont {U.}~\bibnamefont {Sterr}},
  \bibinfo {author} {\bibfnamefont {S.}~\bibnamefont {Tanioka}}, \bibinfo
  {author} {\bibfnamefont {G.~W.}\ \bibnamefont {Truong}}, \bibinfo {author}
  {\bibfnamefont {J.}~\bibnamefont {Ye}},\ and\ \bibinfo {author}
  {\bibfnamefont {J.}~\bibnamefont {Yu}},\ }\bibfield  {title} {\emph {\bibinfo
  {title} {Substrate-transferred {GaAs}/{AlGaAs} crystalline coatings for
  gravitational-wave detectors}},\ }\href {https://doi.org/10.1063/5.0140663}
  {\bibfield  {journal} {\bibinfo  {journal} {Appl. Phys. Lett.}\ }\textbf
  {\bibinfo {volume} {122}},\ \bibinfo {pages} {110502} (\bibinfo {year}
  {2023})}\BibitemShut {NoStop}%
\bibitem [{\citenamefont {Cole}(2012)}]{col12}%
  \BibitemOpen
  \bibfield  {author} {\bibinfo {author} {\bibfnamefont {G.~D.}\ \bibnamefont
  {Cole}},\ }\bibfield  {title} {\emph {\bibinfo {title} {Cavity optomechanics
  with low-noise crystalline mirrors}},\ }in\ \href
  {https://doi.org/10.1117/12.931226} {\emph {\bibinfo {booktitle} {Optical
  Trapping and Optical Micromanipulation {IX}}}},\ \bibinfo {series} {Proc.
  SPIE}, Vol.\ \bibinfo {volume} {8458}\ (\bibinfo  {publisher} {SPIE},\
  \bibinfo {year} {2012})\ p.\ \bibinfo {pages} {845807}\BibitemShut {NoStop}%
\bibitem [{\citenamefont {Penn}\ \emph {et~al.}(2019)\citenamefont {Penn},
  \citenamefont {Kinley-Hanlon}, \citenamefont {MacMillan}, \citenamefont
  {Heu}, \citenamefont {Follman}, \citenamefont {Deutsch}, \citenamefont
  {Cole},\ and\ \citenamefont {Harry}}]{pen19}%
  \BibitemOpen
  \bibfield  {author} {\bibinfo {author} {\bibfnamefont {S.~D.}\ \bibnamefont
  {Penn}}, \bibinfo {author} {\bibfnamefont {M.~M.}\ \bibnamefont
  {Kinley-Hanlon}}, \bibinfo {author} {\bibfnamefont {I.~A.~O.}\ \bibnamefont
  {MacMillan}}, \bibinfo {author} {\bibfnamefont {P.}~\bibnamefont {Heu}},
  \bibinfo {author} {\bibfnamefont {D.}~\bibnamefont {Follman}}, \bibinfo
  {author} {\bibfnamefont {C.}~\bibnamefont {Deutsch}}, \bibinfo {author}
  {\bibfnamefont {G.~D.}\ \bibnamefont {Cole}},\ and\ \bibinfo {author}
  {\bibfnamefont {G.~M.}\ \bibnamefont {Harry}},\ }\bibfield  {title} {\emph
  {\bibinfo {title} {Mechanical ringdown studies of large-area
  substrate-transferred {GaAs/AlGaAs} crystalline coatings}},\ }\href
  {https://doi.org/10.1364/JOSAB.36.000C15} {\bibfield  {journal} {\bibinfo
  {journal} {J. Opt. Soc. Am. B}\ }\textbf {\bibinfo {volume} {36}},\ \bibinfo
  {pages} {C15--C21} (\bibinfo {year} {2019})}\BibitemShut {NoStop}%
\bibitem [{\citenamefont {Yamamoto}\ \emph {et~al.}(2006)\citenamefont
  {Yamamoto}, \citenamefont {Miyoki}, \citenamefont {Uchiyama}, \citenamefont
  {Ishitsuka}, \citenamefont {Ohashi}, \citenamefont {Kuroda}, \citenamefont
  {Tomaru}, \citenamefont {Sato}, \citenamefont {Suzuki}, \citenamefont
  {Haruyama}, \citenamefont {Yamamoto}, \citenamefont {Shintomi}, \citenamefont
  {Numata}, \citenamefont {Waseda}, \citenamefont {Ito},\ and\ \citenamefont
  {Watanabe}}]{yam06}%
  \BibitemOpen
  \bibfield  {author} {\bibinfo {author} {\bibfnamefont {K.}~\bibnamefont
  {Yamamoto}}, \bibinfo {author} {\bibfnamefont {S.}~\bibnamefont {Miyoki}},
  \bibinfo {author} {\bibfnamefont {T.}~\bibnamefont {Uchiyama}}, \bibinfo
  {author} {\bibfnamefont {H.}~\bibnamefont {Ishitsuka}}, \bibinfo {author}
  {\bibfnamefont {M.}~\bibnamefont {Ohashi}}, \bibinfo {author} {\bibfnamefont
  {K.}~\bibnamefont {Kuroda}}, \bibinfo {author} {\bibfnamefont
  {T.}~\bibnamefont {Tomaru}}, \bibinfo {author} {\bibfnamefont
  {N.}~\bibnamefont {Sato}}, \bibinfo {author} {\bibfnamefont {T.}~\bibnamefont
  {Suzuki}}, \bibinfo {author} {\bibfnamefont {T.}~\bibnamefont {Haruyama}},
  \bibinfo {author} {\bibfnamefont {A.}~\bibnamefont {Yamamoto}}, \bibinfo
  {author} {\bibfnamefont {T.}~\bibnamefont {Shintomi}}, \bibinfo {author}
  {\bibfnamefont {K.}~\bibnamefont {Numata}}, \bibinfo {author} {\bibfnamefont
  {K.}~\bibnamefont {Waseda}}, \bibinfo {author} {\bibfnamefont
  {K.}~\bibnamefont {Ito}},\ and\ \bibinfo {author} {\bibfnamefont
  {K.}~\bibnamefont {Watanabe}},\ }\bibfield  {title} {\emph {\bibinfo {title}
  {Measurement of the mechanical loss of a cooled reflective coating for
  gravitational wave detection}},\ }\href
  {https://doi.org/10.1103/PhysRevD.74.022002} {\bibfield  {journal} {\bibinfo
  {journal} {Phys. Rev. D}\ }\textbf {\bibinfo {volume} {74}},\ \bibinfo
  {pages} {022002} (\bibinfo {year} {2006})}\BibitemShut {NoStop}%
\bibitem [{\citenamefont {Robinson}\ \emph {et~al.}(2021)\citenamefont
  {Robinson}, \citenamefont {Oelker}, \citenamefont {Milner}, \citenamefont
  {Kedar}, \citenamefont {Zhang}, \citenamefont {Legero}, \citenamefont
  {Matei}, \citenamefont {H\"{a}fner}, \citenamefont {Riehle}, \citenamefont
  {Sterr},\ and\ \citenamefont {Ye}}]{rob21}%
  \BibitemOpen
  \bibfield  {author} {\bibinfo {author} {\bibfnamefont {J.~M.}\ \bibnamefont
  {Robinson}}, \bibinfo {author} {\bibfnamefont {E.}~\bibnamefont {Oelker}},
  \bibinfo {author} {\bibfnamefont {W.~R.}\ \bibnamefont {Milner}}, \bibinfo
  {author} {\bibfnamefont {D.}~\bibnamefont {Kedar}}, \bibinfo {author}
  {\bibfnamefont {W.}~\bibnamefont {Zhang}}, \bibinfo {author} {\bibfnamefont
  {T.}~\bibnamefont {Legero}}, \bibinfo {author} {\bibfnamefont {D.~G.}\
  \bibnamefont {Matei}}, \bibinfo {author} {\bibfnamefont {S.}~\bibnamefont
  {H\"{a}fner}}, \bibinfo {author} {\bibfnamefont {F.}~\bibnamefont {Riehle}},
  \bibinfo {author} {\bibfnamefont {U.}~\bibnamefont {Sterr}},\ and\ \bibinfo
  {author} {\bibfnamefont {J.}~\bibnamefont {Ye}},\ }\bibfield  {title} {\emph
  {\bibinfo {title} {Thermal noise and mechanical loss of
  {SiO$_2$}/{Ta$_2$O$_5$} optical coatings at cryogenic temperatures}},\ }\href
  {https://doi.org/10.1364/OL.413758} {\bibfield  {journal} {\bibinfo
  {journal} {Opt. Lett.}\ }\textbf {\bibinfo {volume} {46}},\ \bibinfo {pages}
  {592--595} (\bibinfo {year} {2021})}\BibitemShut {NoStop}%
\bibitem [{\citenamefont {H{\"a}fner}\ \emph {et~al.}(2015)\citenamefont
  {H{\"a}fner}, \citenamefont {Falke}, \citenamefont {Grebing}, \citenamefont
  {Vogt}, \citenamefont {Legero}, \citenamefont {Merimaa}, \citenamefont
  {Lisdat},\ and\ \citenamefont {Sterr}}]{hae15a}%
  \BibitemOpen
  \bibfield  {author} {\bibinfo {author} {\bibfnamefont {S.}~\bibnamefont
  {H{\"a}fner}}, \bibinfo {author} {\bibfnamefont {S.}~\bibnamefont {Falke}},
  \bibinfo {author} {\bibfnamefont {C.}~\bibnamefont {Grebing}}, \bibinfo
  {author} {\bibfnamefont {S.}~\bibnamefont {Vogt}}, \bibinfo {author}
  {\bibfnamefont {T.}~\bibnamefont {Legero}}, \bibinfo {author} {\bibfnamefont
  {M.}~\bibnamefont {Merimaa}}, \bibinfo {author} {\bibfnamefont
  {C.}~\bibnamefont {Lisdat}},\ and\ \bibinfo {author} {\bibfnamefont
  {U.}~\bibnamefont {Sterr}},\ }\bibfield  {title} {\emph {\bibinfo {title} {$8
  \times 10^{-17}$ fractional laser frequency instability with a long
  room-temperature cavity}},\ }\href {https://doi.org/10.1364/OL.40.002112}
  {\bibfield  {journal} {\bibinfo  {journal} {Opt. Lett.}\ }\textbf {\bibinfo
  {volume} {40}},\ \bibinfo {pages} {2112--2115} (\bibinfo {year}
  {2015})}\BibitemShut {NoStop}%
\bibitem [{\citenamefont {Legero}\ \emph {et~al.}(2010)\citenamefont {Legero},
  \citenamefont {Kessler},\ and\ \citenamefont {Sterr}}]{leg10}%
  \BibitemOpen
  \bibfield  {author} {\bibinfo {author} {\bibfnamefont {T.}~\bibnamefont
  {Legero}}, \bibinfo {author} {\bibfnamefont {T.}~\bibnamefont {Kessler}},\
  and\ \bibinfo {author} {\bibfnamefont {U.}~\bibnamefont {Sterr}},\ }\bibfield
   {title} {\emph {\bibinfo {title} {Tuning the thermal expansion properties of
  optical reference cavities with fused silica mirrors}},\ }\href
  {https://doi.org/10.1364/JOSAB.27.000914} {\bibfield  {journal} {\bibinfo
  {journal} {J. Opt. Soc. Am. B}\ }\textbf {\bibinfo {volume} {27}},\ \bibinfo
  {pages} {914--919} (\bibinfo {year} {2010})}\BibitemShut {NoStop}%
\bibitem [{\citenamefont {Farsi}\ \emph {et~al.}(2012)\citenamefont {Farsi},
  \citenamefont {Siciliani~de Cumis}, \citenamefont {Marino},\ and\
  \citenamefont {Marin}}]{far12}%
  \BibitemOpen
  \bibfield  {author} {\bibinfo {author} {\bibfnamefont {A.}~\bibnamefont
  {Farsi}}, \bibinfo {author} {\bibfnamefont {M.}~\bibnamefont {Siciliani~de
  Cumis}}, \bibinfo {author} {\bibfnamefont {F.}~\bibnamefont {Marino}},\ and\
  \bibinfo {author} {\bibfnamefont {F.}~\bibnamefont {Marin}},\ }\bibfield
  {title} {\emph {\bibinfo {title} {Photothermal and thermo-refractive effects
  in high reflectivity mirrors at room and cryogenic temperature}},\ }\href
  {https://doi.org/10.1063/1.3684626} {\bibfield  {journal} {\bibinfo
  {journal} {J. Appl. Phys.}\ }\textbf {\bibinfo {volume} {111}},\ \bibinfo
  {pages} {043101} (\bibinfo {year} {2012})}\BibitemShut {NoStop}%
\bibitem [{\citenamefont {Ma}\ \emph {et~al.}(2024)\citenamefont {Ma},
  \citenamefont {Yu}, \citenamefont {Legero}, \citenamefont {Herbers},
  \citenamefont {Nicolodi}, \citenamefont {Kempkes}, \citenamefont {Riehle},
  \citenamefont {Kedar}, \citenamefont {Robinson}, \citenamefont {Ye},\ and\
  \citenamefont {Sterr}}]{ma24a}%
  \BibitemOpen
  \bibfield  {author} {\bibinfo {author} {\bibfnamefont {C.~Y.}\ \bibnamefont
  {Ma}}, \bibinfo {author} {\bibfnamefont {J.}~\bibnamefont {Yu}}, \bibinfo
  {author} {\bibfnamefont {T.}~\bibnamefont {Legero}}, \bibinfo {author}
  {\bibfnamefont {S.}~\bibnamefont {Herbers}}, \bibinfo {author} {\bibfnamefont
  {D.}~\bibnamefont {Nicolodi}}, \bibinfo {author} {\bibfnamefont
  {M.}~\bibnamefont {Kempkes}}, \bibinfo {author} {\bibfnamefont
  {F.}~\bibnamefont {Riehle}}, \bibinfo {author} {\bibfnamefont
  {D.}~\bibnamefont {Kedar}}, \bibinfo {author} {\bibfnamefont {J.~M.}\
  \bibnamefont {Robinson}}, \bibinfo {author} {\bibfnamefont {J.}~\bibnamefont
  {Ye}},\ and\ \bibinfo {author} {\bibfnamefont {U.}~\bibnamefont {Sterr}},\
  }\bibfield  {title} {\emph {\bibinfo {title} {Ultrastable lasers:
  investigations of crystalline mirrors and closed cycle cooling at 124~{K}}},\
  }\href {https://doi.org/10.1088/1742-6596/2889/1/012055} {\bibfield
  {journal} {\bibinfo  {journal} {J. Phys.: Conf. Ser.}\ }\textbf {\bibinfo
  {volume} {2889}},\ \bibinfo {pages} {012055} (\bibinfo {year}
  {2024})}\BibitemShut {NoStop}%
\bibitem [{\citenamefont {Zhu}\ \emph {et~al.}(2023)\citenamefont {Zhu},
  \citenamefont {Cui}, \citenamefont {Kong}, \citenamefont {Yu}, \citenamefont
  {Jiang}, \citenamefont {Xu}, \citenamefont {Dai}, \citenamefont {Chen},\ and\
  \citenamefont {Pan}}]{zhu23}%
  \BibitemOpen
  \bibfield  {author} {\bibinfo {author} {\bibfnamefont {X.-Q.}\ \bibnamefont
  {Zhu}}, \bibinfo {author} {\bibfnamefont {X.-Y.}\ \bibnamefont {Cui}},
  \bibinfo {author} {\bibfnamefont {D.-Q.}\ \bibnamefont {Kong}}, \bibinfo
  {author} {\bibfnamefont {H.-W.}\ \bibnamefont {Yu}}, \bibinfo {author}
  {\bibfnamefont {X.}~\bibnamefont {Jiang}}, \bibinfo {author} {\bibfnamefont
  {P.}~\bibnamefont {Xu}}, \bibinfo {author} {\bibfnamefont {H.-N.}\
  \bibnamefont {Dai}}, \bibinfo {author} {\bibfnamefont {Y.-A.}\ \bibnamefont
  {Chen}},\ and\ \bibinfo {author} {\bibfnamefont {J.-W.}\ \bibnamefont
  {Pan}},\ }\bibfield  {title} {\emph {\bibinfo {title} {Photo-birefringent
  effects of crystalline coatings in ultra-stable cavities}},\ }in\ \href
  {https://doi.org/10.1117/12.3007859} {\emph {\bibinfo {booktitle} {Fourteenth
  International Conference on Information Optics and Photonics (CIOP 2023)}}},\
  Vol.\ \bibinfo {volume} {12935},\ \bibinfo {editor} {edited by\ \bibinfo
  {editor} {\bibfnamefont {Y.}~\bibnamefont {Yang}}},\ \bibinfo {organization}
  {International Society for Optics and Photonics}\ (\bibinfo  {publisher}
  {SPIE},\ \bibinfo {year} {2023})\ p.\ \bibinfo {pages} {1293541}\BibitemShut
  {NoStop}%
\bibitem [{\citenamefont {Kraus}(2023)}]{kra23a}%
  \BibitemOpen
  \bibfield  {author} {\bibinfo {author} {\bibfnamefont {B.}~\bibnamefont
  {Kraus}},\ }\emph {\bibinfo {title} {A highly stable {UV} clock laser}},\
  \href {https://doi.org/10.15488/15360} {Ph.D. thesis},\ \bibinfo  {school}
  {QUEST-Leibniz-Forschungsschule der Gottfried Wilhelm Leibniz Universit\"at
  Hannover} (\bibinfo {year} {2023})\BibitemShut {NoStop}%
\bibitem [{\citenamefont {Cole}\ \emph {et~al.}(2016)\citenamefont {Cole},
  \citenamefont {Zhang}, \citenamefont {Bjork}, \citenamefont {Follman},
  \citenamefont {Heu}, \citenamefont {Deutsch}, \citenamefont {Sonderhouse},
  \citenamefont {Robinson}, \citenamefont {Franz}, \citenamefont
  {Alexandrovski}, \citenamefont {Notcutt}, \citenamefont {Heckl},
  \citenamefont {Ye},\ and\ \citenamefont {Aspelmeyer}}]{col16}%
  \BibitemOpen
  \bibfield  {author} {\bibinfo {author} {\bibfnamefont {G.~D.}\ \bibnamefont
  {Cole}}, \bibinfo {author} {\bibfnamefont {W.}~\bibnamefont {Zhang}},
  \bibinfo {author} {\bibfnamefont {B.~J.}\ \bibnamefont {Bjork}}, \bibinfo
  {author} {\bibfnamefont {D.}~\bibnamefont {Follman}}, \bibinfo {author}
  {\bibfnamefont {P.}~\bibnamefont {Heu}}, \bibinfo {author} {\bibfnamefont
  {C.}~\bibnamefont {Deutsch}}, \bibinfo {author} {\bibfnamefont
  {L.}~\bibnamefont {Sonderhouse}}, \bibinfo {author} {\bibfnamefont
  {J.}~\bibnamefont {Robinson}}, \bibinfo {author} {\bibfnamefont
  {C.}~\bibnamefont {Franz}}, \bibinfo {author} {\bibfnamefont
  {A.}~\bibnamefont {Alexandrovski}}, \bibinfo {author} {\bibfnamefont
  {M.}~\bibnamefont {Notcutt}}, \bibinfo {author} {\bibfnamefont {O.~H.}\
  \bibnamefont {Heckl}}, \bibinfo {author} {\bibfnamefont {J.}~\bibnamefont
  {Ye}},\ and\ \bibinfo {author} {\bibfnamefont {M.}~\bibnamefont
  {Aspelmeyer}},\ }\bibfield  {title} {\emph {\bibinfo {title}
  {High-performance near- and mid-infrared crystalline coatings}},\ }\href
  {https://doi.org/10.1364/OPTICA.3.000647} {\bibfield  {journal} {\bibinfo
  {journal} {Optica}\ }\textbf {\bibinfo {volume} {3}},\ \bibinfo {pages}
  {647--656} (\bibinfo {year} {2016})}\BibitemShut {NoStop}%
\bibitem [{\citenamefont {Perner}\ \emph {et~al.}(2024)\citenamefont {Perner},
  \citenamefont {Winkler}, \citenamefont {Truong}, \citenamefont {Zhao},
  \citenamefont {Bachmann}, \citenamefont {Mayer}, \citenamefont {Fellinger},
  \citenamefont {Follman}, \citenamefont {Heu}, \citenamefont {Deutsch},
  \citenamefont {Bailey}, \citenamefont {Peelaers}, \citenamefont {Puchegger},
  \citenamefont {Fleisher}, \citenamefont {Cole},\ and\ \citenamefont
  {Heckl}}]{per24}%
  \BibitemOpen
  \bibfield  {author} {\bibinfo {author} {\bibfnamefont {L.~W.}\ \bibnamefont
  {Perner}}, \bibinfo {author} {\bibfnamefont {G.}~\bibnamefont {Winkler}},
  \bibinfo {author} {\bibfnamefont {G.-W.}\ \bibnamefont {Truong}}, \bibinfo
  {author} {\bibfnamefont {G.}~\bibnamefont {Zhao}}, \bibinfo {author}
  {\bibfnamefont {D.}~\bibnamefont {Bachmann}}, \bibinfo {author}
  {\bibfnamefont {A.~S.}\ \bibnamefont {Mayer}}, \bibinfo {author}
  {\bibfnamefont {J.}~\bibnamefont {Fellinger}}, \bibinfo {author}
  {\bibfnamefont {D.}~\bibnamefont {Follman}}, \bibinfo {author} {\bibfnamefont
  {P.}~\bibnamefont {Heu}}, \bibinfo {author} {\bibfnamefont {C.}~\bibnamefont
  {Deutsch}}, \bibinfo {author} {\bibfnamefont {D.~M.}\ \bibnamefont {Bailey}},
  \bibinfo {author} {\bibfnamefont {H.}~\bibnamefont {Peelaers}}, \bibinfo
  {author} {\bibfnamefont {S.}~\bibnamefont {Puchegger}}, \bibinfo {author}
  {\bibfnamefont {A.~J.}\ \bibnamefont {Fleisher}}, \bibinfo {author}
  {\bibfnamefont {G.~D.}\ \bibnamefont {Cole}},\ and\ \bibinfo {author}
  {\bibfnamefont {O.~H.}\ \bibnamefont {Heckl}},\ }\bibfield  {title} {\emph
  {\bibinfo {title} {Mid-infrared interference coatings with excess optical
  loss below 10~ppm: erratum}},\ }\href {https://doi.org/10.1364/OPTICA.520398}
  {\bibfield  {journal} {\bibinfo  {journal} {Optica}\ }\textbf {\bibinfo
  {volume} {11}},\ \bibinfo {pages} {619--620} (\bibinfo {year}
  {2024})}\BibitemShut {NoStop}%
\bibitem [{\citenamefont {Monemar}\ \emph {et~al.}(1976)\citenamefont
  {Monemar}, \citenamefont {Shih},\ and\ \citenamefont {Pettit}}]{mon76}%
  \BibitemOpen
  \bibfield  {author} {\bibinfo {author} {\bibfnamefont {B.}~\bibnamefont
  {Monemar}}, \bibinfo {author} {\bibfnamefont {K.~K.}\ \bibnamefont {Shih}},\
  and\ \bibinfo {author} {\bibfnamefont {G.~D.}\ \bibnamefont {Pettit}},\
  }\bibfield  {title} {\emph {\bibinfo {title} {Some optical properties of the
  {Al$_x$Ga$_{1-x}$As} alloys system}},\ }\href
  {https://doi.org/10.1063/1.322979} {\bibfield  {journal} {\bibinfo  {journal}
  {J. Appl. Phys.}\ }\textbf {\bibinfo {volume} {47}},\ \bibinfo {pages}
  {2604--2613} (\bibinfo {year} {1976})}\BibitemShut {NoStop}%
\bibitem [{\citenamefont {Sturge}(1962)}]{stu62}%
  \BibitemOpen
  \bibfield  {author} {\bibinfo {author} {\bibfnamefont {M.~D.}\ \bibnamefont
  {Sturge}},\ }\bibfield  {title} {\emph {\bibinfo {title} {Optical absorption
  of gallium arsenide between 0.6 and 2.75 {eV}}},\ }\href
  {https://doi.org/10.1103/PhysRev.127.768} {\bibfield  {journal} {\bibinfo
  {journal} {Phys. Rev.}\ }\textbf {\bibinfo {volume} {127}},\ \bibinfo {pages}
  {768--773} (\bibinfo {year} {1962})}\BibitemShut {NoStop}%
\bibitem [{\citenamefont {Hutchby}\ and\ \citenamefont
  {Fudurich}(2008)}]{hut08}%
  \BibitemOpen
  \bibfield  {author} {\bibinfo {author} {\bibfnamefont {J.~A.}\ \bibnamefont
  {Hutchby}}\ and\ \bibinfo {author} {\bibfnamefont {R.~L.}\ \bibnamefont
  {Fudurich}},\ }\bibfield  {title} {\emph {\bibinfo {title} {Theoretical
  analysis of {Al$_x$Ga$_{1-x}$As}-{GaAs} graded band-gap solar cell}},\ }\href
  {https://doi.org/10.1063/1.323108} {\bibfield  {journal} {\bibinfo  {journal}
  {J. Appl. Phys.}\ }\textbf {\bibinfo {volume} {47}},\ \bibinfo {pages}
  {3140--3151} (\bibinfo {year} {2008})}\BibitemShut {NoStop}%
\bibitem [{\citenamefont {Dickmann}\ \emph {et~al.}(2018)\citenamefont
  {Dickmann}, \citenamefont {{Rojas Hurtado}}, \citenamefont {Nawrodt},\ and\
  \citenamefont {Kroker}}]{DICKMANN20182275}%
  \BibitemOpen
  \bibfield  {author} {\bibinfo {author} {\bibfnamefont {J.}~\bibnamefont
  {Dickmann}}, \bibinfo {author} {\bibfnamefont {C.}~\bibnamefont {{Rojas
  Hurtado}}}, \bibinfo {author} {\bibfnamefont {R.}~\bibnamefont {Nawrodt}},\
  and\ \bibinfo {author} {\bibfnamefont {S.}~\bibnamefont {Kroker}},\
  }\bibfield  {title} {\emph {\bibinfo {title} {Influence of polarization and
  material on {Brownian} thermal noise of binary grating reflectors}},\ }\href
  {https://doi.org/10.1016/j.physleta.2017.07.006} {\bibfield  {journal}
  {\bibinfo  {journal} {Phys. Lett. A}\ }\textbf {\bibinfo {volume} {382}},\
  \bibinfo {pages} {2275--2281} (\bibinfo {year} {2018})}\BibitemShut {NoStop}%
\bibitem [{\citenamefont {Br\"uckner}\ \emph {et~al.}(2010)\citenamefont
  {Br\"uckner}, \citenamefont {Friedrich}, \citenamefont {Clausnitzer},
  \citenamefont {Britzger}, \citenamefont {Burmeister}, \citenamefont
  {Danzmann}, \citenamefont {Kley}, \citenamefont {T\"unnermann},\ and\
  \citenamefont {Schnabel}}]{Brueckner2010}%
  \BibitemOpen
  \bibfield  {author} {\bibinfo {author} {\bibfnamefont {F.}~\bibnamefont
  {Br\"uckner}}, \bibinfo {author} {\bibfnamefont {D.}~\bibnamefont
  {Friedrich}}, \bibinfo {author} {\bibfnamefont {T.}~\bibnamefont
  {Clausnitzer}}, \bibinfo {author} {\bibfnamefont {M.}~\bibnamefont
  {Britzger}}, \bibinfo {author} {\bibfnamefont {O.}~\bibnamefont
  {Burmeister}}, \bibinfo {author} {\bibfnamefont {K.}~\bibnamefont
  {Danzmann}}, \bibinfo {author} {\bibfnamefont {E.-B.}\ \bibnamefont {Kley}},
  \bibinfo {author} {\bibfnamefont {A.}~\bibnamefont {T\"unnermann}},\ and\
  \bibinfo {author} {\bibfnamefont {R.}~\bibnamefont {Schnabel}},\ }\bibfield
  {title} {\emph {\bibinfo {title} {Realization of a monolithic
  high-reflectivity cavity mirror from a single silicon crystal}},\ }\href
  {https://doi.org/10.1103/PhysRevLett.104.163903} {\bibfield  {journal}
  {\bibinfo  {journal} {Phys. Rev. Lett.}\ }\textbf {\bibinfo {volume} {104}},\
  \bibinfo {pages} {163903} (\bibinfo {year} {2010})}\BibitemShut {NoStop}%
\bibitem [{\citenamefont {{Krosaki Harima Corporation, Ceramics
  Division}}(2023)}]{KrosakiHarimaWebsite}%
  \BibitemOpen
  \bibfield  {author} {\bibinfo {author} {\bibnamefont {{Krosaki Harima
  Corporation, Ceramics Division}}},\ }\href
  {https://krosaki-fc.com/en/ceramics/nexcera.html} {\bibinfo {title}
  {{NEXCERA} ultra low thermal expansion ceramics}} (\bibinfo {year}
  {2023})\BibitemShut {NoStop}%
\bibitem [{\citenamefont {Cesarini}\ \emph {et~al.}(2009)\citenamefont
  {Cesarini}, \citenamefont {Lorenzini}, \citenamefont {Campagna},
  \citenamefont {Martelli}, \citenamefont {Piergiovanni}, \citenamefont
  {Vetrano}, \citenamefont {Losurdo},\ and\ \citenamefont
  {Cagnoli}}]{Cesarini2009}%
  \BibitemOpen
  \bibfield  {author} {\bibinfo {author} {\bibfnamefont {E.}~\bibnamefont
  {Cesarini}}, \bibinfo {author} {\bibfnamefont {M.}~\bibnamefont {Lorenzini}},
  \bibinfo {author} {\bibfnamefont {E.}~\bibnamefont {Campagna}}, \bibinfo
  {author} {\bibfnamefont {F.}~\bibnamefont {Martelli}}, \bibinfo {author}
  {\bibfnamefont {F.}~\bibnamefont {Piergiovanni}}, \bibinfo {author}
  {\bibfnamefont {F.}~\bibnamefont {Vetrano}}, \bibinfo {author} {\bibfnamefont
  {G.}~\bibnamefont {Losurdo}},\ and\ \bibinfo {author} {\bibfnamefont
  {G.}~\bibnamefont {Cagnoli}},\ }\bibfield  {title} {\emph {\bibinfo {title}
  {A “gentle” nodal suspension for measurements of the acoustic attenuation
  in materials}},\ }\href {https://doi.org/10.1063/1.3124800} {\bibfield
  {journal} {\bibinfo  {journal} {Rev. Sci. Instrum.}\ }\textbf {\bibinfo
  {volume} {80}},\ \bibinfo {pages} {053904} (\bibinfo {year}
  {2009})}\BibitemShut {NoStop}%
\bibitem [{\citenamefont {Wagner}\ \emph
  {et~al.}(2024{\natexlab{a}})\citenamefont {Wagner}, \citenamefont
  {Narożnik}, \citenamefont {Bober}, \citenamefont {Sauer}, \citenamefont
  {Zawada},\ and\ \citenamefont {Kroker}}]{Wagner2024_nexcera}%
  \BibitemOpen
  \bibfield  {author} {\bibinfo {author} {\bibfnamefont {N.}~\bibnamefont
  {Wagner}}, \bibinfo {author} {\bibfnamefont {M.}~\bibnamefont {Narożnik}},
  \bibinfo {author} {\bibfnamefont {M.}~\bibnamefont {Bober}}, \bibinfo
  {author} {\bibfnamefont {S.}~\bibnamefont {Sauer}}, \bibinfo {author}
  {\bibfnamefont {M.}~\bibnamefont {Zawada}},\ and\ \bibinfo {author}
  {\bibfnamefont {S.}~\bibnamefont {Kroker}},\ }\href
  {https://doi.org/10.48550/arXiv.2409.14120} {\bibinfo {title} {Mechanical
  losses and stability performance of {NEXCERA} in ultra-stable laser
  cavities}},\ \bibinfo {howpublished} {arXiv:2409.14120 [physics.optics]}
  (\bibinfo {year} {2024}{\natexlab{a}})\BibitemShut {NoStop}%
\bibitem [{\citenamefont {Numata}\ \emph {et~al.}(2004)\citenamefont {Numata},
  \citenamefont {Kemery},\ and\ \citenamefont {Camp}}]{Numata2004}%
  \BibitemOpen
  \bibfield  {author} {\bibinfo {author} {\bibfnamefont {K.}~\bibnamefont
  {Numata}}, \bibinfo {author} {\bibfnamefont {A.}~\bibnamefont {Kemery}},\
  and\ \bibinfo {author} {\bibfnamefont {J.}~\bibnamefont {Camp}},\ }\bibfield
  {title} {\emph {\bibinfo {title} {Thermal-noise limit in the frequency
  stabilization of lasers with rigid cavities}},\ }\href
  {https://doi.org/10.1103/PhysRevLett.93.250602} {\bibfield  {journal}
  {\bibinfo  {journal} {Phys. Rev. Lett.}\ }\textbf {\bibinfo {volume} {93}},\
  \bibinfo {pages} {250602} (\bibinfo {year} {2004})}\BibitemShut {NoStop}%
\bibitem [{\citenamefont {Schenkel}\ \emph {et~al.}(2024)\citenamefont
  {Schenkel}, \citenamefont {Vogt},\ and\ \citenamefont
  {Schiller}}]{Schenkel2024}%
  \BibitemOpen
  \bibfield  {author} {\bibinfo {author} {\bibfnamefont {M.~R.}\ \bibnamefont
  {Schenkel}}, \bibinfo {author} {\bibfnamefont {V.~A.}\ \bibnamefont {Vogt}},\
  and\ \bibinfo {author} {\bibfnamefont {S.}~\bibnamefont {Schiller}},\
  }\bibfield  {title} {\emph {\bibinfo {title} {Metrology-grade spectroscopy
  source based on an optical parametric oscillator}},\ }\href
  {https://doi.org/10.1364/OE.538442} {\bibfield  {journal} {\bibinfo
  {journal} {Opt. Express}\ }\textbf {\bibinfo {volume} {32}},\ \bibinfo
  {pages} {43350--43365} (\bibinfo {year} {2024})}\BibitemShut {NoStop}%
\bibitem [{\citenamefont {Kwong}\ \emph {et~al.}(2018)\citenamefont {Kwong},
  \citenamefont {Hansen}, \citenamefont {Sugawara},\ and\ \citenamefont
  {Schiller}}]{Kwong_2018}%
  \BibitemOpen
  \bibfield  {author} {\bibinfo {author} {\bibfnamefont {C.~J.}\ \bibnamefont
  {Kwong}}, \bibinfo {author} {\bibfnamefont {M.~G.}\ \bibnamefont {Hansen}},
  \bibinfo {author} {\bibfnamefont {J.}~\bibnamefont {Sugawara}},\ and\
  \bibinfo {author} {\bibfnamefont {S.}~\bibnamefont {Schiller}},\ }\bibfield
  {title} {\emph {\bibinfo {title} {Characterization of the long-term
  dimensional stability of a {NEXCERA} block using the optical resonator
  technique}},\ }\href {https://doi.org/10.1088/1361-6501/aac3b0} {\bibfield
  {journal} {\bibinfo  {journal} {Meas. Sci. Technol.}\ }\textbf {\bibinfo
  {volume} {29}},\ \bibinfo {pages} {075011} (\bibinfo {year}
  {2018})}\BibitemShut {NoStop}%
\bibitem [{\citenamefont {Zhang}\ \emph {et~al.}(2017)\citenamefont {Zhang},
  \citenamefont {Robinson}, \citenamefont {Sonderhouse}, \citenamefont
  {Oelker}, \citenamefont {Benko}, \citenamefont {Hall}, \citenamefont
  {Legero}, \citenamefont {Matei}, \citenamefont {Riehle}, \citenamefont
  {Sterr},\ and\ \citenamefont {Ye}}]{Zhang2017}%
  \BibitemOpen
  \bibfield  {author} {\bibinfo {author} {\bibfnamefont {W.}~\bibnamefont
  {Zhang}}, \bibinfo {author} {\bibfnamefont {J.~M.}\ \bibnamefont {Robinson}},
  \bibinfo {author} {\bibfnamefont {L.}~\bibnamefont {Sonderhouse}}, \bibinfo
  {author} {\bibfnamefont {E.}~\bibnamefont {Oelker}}, \bibinfo {author}
  {\bibfnamefont {C.}~\bibnamefont {Benko}}, \bibinfo {author} {\bibfnamefont
  {J.~L.}\ \bibnamefont {Hall}}, \bibinfo {author} {\bibfnamefont
  {T.}~\bibnamefont {Legero}}, \bibinfo {author} {\bibfnamefont {D.~G.}\
  \bibnamefont {Matei}}, \bibinfo {author} {\bibfnamefont {F.}~\bibnamefont
  {Riehle}}, \bibinfo {author} {\bibfnamefont {U.}~\bibnamefont {Sterr}},\ and\
  \bibinfo {author} {\bibfnamefont {J.}~\bibnamefont {Ye}},\ }\bibfield
  {title} {\emph {\bibinfo {title} {Ultrastable silicon cavity in a
  continuously operating closed-cycle cryostat at 4~{K}}},\ }\href
  {https://doi.org/10.1103/PhysRevLett.119.243601} {\bibfield  {journal}
  {\bibinfo  {journal} {Phys. Rev. Lett.}\ }\textbf {\bibinfo {volume} {119}},\
  \bibinfo {pages} {243601} (\bibinfo {year} {2017})}\BibitemShut {NoStop}%
\bibitem [{\citenamefont {Schiller}\ \emph {et~al.}(2004)\citenamefont
  {Schiller}, \citenamefont {L\"ammerzahl}, \citenamefont {M\"uller},
  \citenamefont {Braxmaier}, \citenamefont {Herrmann},\ and\ \citenamefont
  {Peters}}]{Schiller2004}%
  \BibitemOpen
  \bibfield  {author} {\bibinfo {author} {\bibfnamefont {S.}~\bibnamefont
  {Schiller}}, \bibinfo {author} {\bibfnamefont {C.}~\bibnamefont
  {L\"ammerzahl}}, \bibinfo {author} {\bibfnamefont {H.}~\bibnamefont
  {M\"uller}}, \bibinfo {author} {\bibfnamefont {C.}~\bibnamefont {Braxmaier}},
  \bibinfo {author} {\bibfnamefont {S.}~\bibnamefont {Herrmann}},\ and\
  \bibinfo {author} {\bibfnamefont {A.}~\bibnamefont {Peters}},\ }\bibfield
  {title} {\emph {\bibinfo {title} {Experimental limits for low-frequency
  space-time fluctuations from ultrastable optical resonators}},\ }\href
  {https://doi.org/10.1103/PhysRevD.69.027504} {\bibfield  {journal} {\bibinfo
  {journal} {Phys. Rev. D}\ }\textbf {\bibinfo {volume} {69}},\ \bibinfo
  {pages} {027504} (\bibinfo {year} {2004})}\BibitemShut {NoStop}%
\bibitem [{\citenamefont {Wang}\ \emph {et~al.}(2018)\citenamefont {Wang},
  \citenamefont {Dovale-\'Alvarez}, \citenamefont {Collins}, \citenamefont
  {Brown}, \citenamefont {Wang}, \citenamefont {Mow-Lowry}, \citenamefont
  {Han},\ and\ \citenamefont {Freise}}]{Wang2018}%
  \BibitemOpen
  \bibfield  {author} {\bibinfo {author} {\bibfnamefont {H.}~\bibnamefont
  {Wang}}, \bibinfo {author} {\bibfnamefont {M.}~\bibnamefont
  {Dovale-\'Alvarez}}, \bibinfo {author} {\bibfnamefont {C.}~\bibnamefont
  {Collins}}, \bibinfo {author} {\bibfnamefont {D.~D.}\ \bibnamefont {Brown}},
  \bibinfo {author} {\bibfnamefont {M.}~\bibnamefont {Wang}}, \bibinfo {author}
  {\bibfnamefont {C.~M.}\ \bibnamefont {Mow-Lowry}}, \bibinfo {author}
  {\bibfnamefont {S.}~\bibnamefont {Han}},\ and\ \bibinfo {author}
  {\bibfnamefont {A.}~\bibnamefont {Freise}},\ }\bibfield  {title} {\emph
  {\bibinfo {title} {Feasibility of near-unstable cavities for future
  gravitational wave detectors}},\ }\href
  {https://doi.org/10.1103/PhysRevD.97.022001} {\bibfield  {journal} {\bibinfo
  {journal} {Phys. Rev. D}\ }\textbf {\bibinfo {volume} {97}},\ \bibinfo
  {pages} {022001} (\bibinfo {year} {2018})}\BibitemShut {NoStop}%
\bibitem [{\citenamefont {Hauck}\ \emph {et~al.}(1980)\citenamefont {Hauck},
  \citenamefont {Kortz},\ and\ \citenamefont {Weber}}]{Hauck1980}%
  \BibitemOpen
  \bibfield  {author} {\bibinfo {author} {\bibfnamefont {R.}~\bibnamefont
  {Hauck}}, \bibinfo {author} {\bibfnamefont {H.}~\bibnamefont {Kortz}},\ and\
  \bibinfo {author} {\bibfnamefont {H.}~\bibnamefont {Weber}},\ }\bibfield
  {title} {\emph {\bibinfo {title} {Misalignment sensitivity of optical
  resonators}},\ }\href {https://doi.org/10.1364/AO.19.000598} {\bibfield
  {journal} {\bibinfo  {journal} {Appl. Opt.}\ }\textbf {\bibinfo {volume}
  {19}},\ \bibinfo {pages} {598--601} (\bibinfo {year} {1980})}\BibitemShut
  {NoStop}%
\bibitem [{\citenamefont {Siegman}(1986)}]{Siegman1986}%
  \BibitemOpen
  \bibfield  {author} {\bibinfo {author} {\bibfnamefont {A.~E.}\ \bibnamefont
  {Siegman}},\ }\href@noop {} {\emph {\bibinfo {title} {Lasers}}}\ (\bibinfo
  {publisher} {University science books},\ \bibinfo {year} {1986})\BibitemShut
  {NoStop}%
\bibitem [{\citenamefont {{\'A}lvarez}(2019)}]{Alvarez2019}%
  \BibitemOpen
  \bibfield  {author} {\bibinfo {author} {\bibfnamefont {M.~D.}\ \bibnamefont
  {{\'A}lvarez}},\ }\href {https://doi.org/10.1007/978-3-030-20863-9} {\emph
  {\bibinfo {title} {Optical cavities for optical atomic clocks, atom
  interferometry and gravitational-wave detection}}}\ (\bibinfo  {publisher}
  {Springer},\ \bibinfo {year} {2019})\BibitemShut {NoStop}%
\bibitem [{\citenamefont {Amairi}\ \emph {et~al.}(2013)\citenamefont {Amairi},
  \citenamefont {Legero}, \citenamefont {Kessler}, \citenamefont {Sterr},
  \citenamefont {W{\"u}bbena}, \citenamefont {Mandel},\ and\ \citenamefont
  {Schmidt}}]{Amairi2013}%
  \BibitemOpen
  \bibfield  {author} {\bibinfo {author} {\bibfnamefont {S.}~\bibnamefont
  {Amairi}}, \bibinfo {author} {\bibfnamefont {T.}~\bibnamefont {Legero}},
  \bibinfo {author} {\bibfnamefont {T.}~\bibnamefont {Kessler}}, \bibinfo
  {author} {\bibfnamefont {U.}~\bibnamefont {Sterr}}, \bibinfo {author}
  {\bibfnamefont {J.~B.}\ \bibnamefont {W{\"u}bbena}}, \bibinfo {author}
  {\bibfnamefont {O.}~\bibnamefont {Mandel}},\ and\ \bibinfo {author}
  {\bibfnamefont {P.~O.}\ \bibnamefont {Schmidt}},\ }\bibfield  {title} {\emph
  {\bibinfo {title} {Reducing the effect of thermal noise in optical
  cavities}},\ }\href {https://doi.org/10.1007/s00340-013-5464-8} {\bibfield
  {journal} {\bibinfo  {journal} {Applied Physics B}\ }\textbf {\bibinfo
  {volume} {113}},\ \bibinfo {pages} {233--242} (\bibinfo {year}
  {2013})}\BibitemShut {NoStop}%
\bibitem [{\citenamefont {Thorpe}\ \emph {et~al.}(2011)\citenamefont {Thorpe},
  \citenamefont {Rippe}, \citenamefont {Fortier}, \citenamefont {Kirchner},\
  and\ \citenamefont {Rosenband}}]{thorpe2011frequency}%
  \BibitemOpen
  \bibfield  {author} {\bibinfo {author} {\bibfnamefont {M.~J.}\ \bibnamefont
  {Thorpe}}, \bibinfo {author} {\bibfnamefont {L.}~\bibnamefont {Rippe}},
  \bibinfo {author} {\bibfnamefont {T.~M.}\ \bibnamefont {Fortier}}, \bibinfo
  {author} {\bibfnamefont {M.~S.}\ \bibnamefont {Kirchner}},\ and\ \bibinfo
  {author} {\bibfnamefont {T.}~\bibnamefont {Rosenband}},\ }\bibfield  {title}
  {\emph {\bibinfo {title} {Frequency stabilization to $\rm{6\times10^{-16}}$
  via spectral-hole burning}},\ }\href
  {https://doi.org/10.1038/nphoton.2011.215} {\bibfield  {journal} {\bibinfo
  {journal} {Nature Photonics}\ }\textbf {\bibinfo {volume} {5}},\ \bibinfo
  {pages} {688--693} (\bibinfo {year} {2011})}\BibitemShut {NoStop}%
\bibitem [{\citenamefont {Wagner}\ \emph
  {et~al.}(2024{\natexlab{b}})\citenamefont {Wagner}, \citenamefont {Dickmann},
  \citenamefont {Fang}, \citenamefont {Hartman},\ and\ \citenamefont
  {Kroker}}]{Wagner2024_yso}%
  \BibitemOpen
  \bibfield  {author} {\bibinfo {author} {\bibfnamefont {N.}~\bibnamefont
  {Wagner}}, \bibinfo {author} {\bibfnamefont {J.}~\bibnamefont {Dickmann}},
  \bibinfo {author} {\bibfnamefont {B.}~\bibnamefont {Fang}}, \bibinfo {author}
  {\bibfnamefont {M.~T.}\ \bibnamefont {Hartman}},\ and\ \bibinfo {author}
  {\bibfnamefont {S.}~\bibnamefont {Kroker}},\ }\href
  {https://doi.org/10.48550/arXiv.2409.14126} {\bibinfo {title}
  {Temperature-dependent mechanical losses of {Eu$^{3+}$:Y$_{2}$SiO$_{5}$} for
  spectral hole burning laser stabilization}},\ \bibinfo {howpublished}
  {arXiv:2409.14126 [physics.optics]} (\bibinfo {year}
  {2024}{\natexlab{b}})\BibitemShut {NoStop}%
\bibitem [{\citenamefont {Nawrodt}\ \emph {et~al.}(2013)\citenamefont
  {Nawrodt}, \citenamefont {Schwarz}, \citenamefont {Kroker}, \citenamefont
  {Martin}, \citenamefont {Bassiri}, \citenamefont {Br\"uckner}, \citenamefont
  {Cunningham}, \citenamefont {Hammond}, \citenamefont {Heinert}, \citenamefont
  {Hough}, \citenamefont {K\"asebier}, \citenamefont {Kley}, \citenamefont
  {Neubert}, \citenamefont {Reid}, \citenamefont {Rowan}, \citenamefont
  {Seidel},\ and\ \citenamefont {T\"unnermann}}]{Nawrodt_2013}%
  \BibitemOpen
  \bibfield  {author} {\bibinfo {author} {\bibfnamefont {R.}~\bibnamefont
  {Nawrodt}}, \bibinfo {author} {\bibfnamefont {C.}~\bibnamefont {Schwarz}},
  \bibinfo {author} {\bibfnamefont {S.}~\bibnamefont {Kroker}}, \bibinfo
  {author} {\bibfnamefont {I.~W.}\ \bibnamefont {Martin}}, \bibinfo {author}
  {\bibfnamefont {R.}~\bibnamefont {Bassiri}}, \bibinfo {author} {\bibfnamefont
  {F.}~\bibnamefont {Br\"uckner}}, \bibinfo {author} {\bibfnamefont
  {L.}~\bibnamefont {Cunningham}}, \bibinfo {author} {\bibfnamefont {G.~D.}\
  \bibnamefont {Hammond}}, \bibinfo {author} {\bibfnamefont {D.}~\bibnamefont
  {Heinert}}, \bibinfo {author} {\bibfnamefont {J.}~\bibnamefont {Hough}},
  \bibinfo {author} {\bibfnamefont {T.}~\bibnamefont {K\"asebier}}, \bibinfo
  {author} {\bibfnamefont {E.-B.}\ \bibnamefont {Kley}}, \bibinfo {author}
  {\bibfnamefont {R.}~\bibnamefont {Neubert}}, \bibinfo {author} {\bibfnamefont
  {S.}~\bibnamefont {Reid}}, \bibinfo {author} {\bibfnamefont {S.}~\bibnamefont
  {Rowan}}, \bibinfo {author} {\bibfnamefont {P.}~\bibnamefont {Seidel}},\ and\
  \bibinfo {author} {\bibfnamefont {A.}~\bibnamefont {T\"unnermann}},\
  }\bibfield  {title} {\emph {\bibinfo {title} {Investigation of mechanical
  losses of thin silicon flexures at low temperatures}},\ }\href
  {https://doi.org/10.1088/0264-9381/30/11/115008} {\bibfield  {journal}
  {\bibinfo  {journal} {Class. Quantum Grav.}\ }\textbf {\bibinfo {volume}
  {30}},\ \bibinfo {pages} {115008} (\bibinfo {year} {2013})}\BibitemShut
  {NoStop}%
\bibitem [{\citenamefont {Hartman}\ \emph {et~al.}(2024)\citenamefont
  {Hartman}, \citenamefont {Wagner}, \citenamefont {Seidelin},\ and\
  \citenamefont {Fang}}]{har24}%
  \BibitemOpen
  \bibfield  {author} {\bibinfo {author} {\bibfnamefont {M.~T.}\ \bibnamefont
  {Hartman}}, \bibinfo {author} {\bibfnamefont {N.}~\bibnamefont {Wagner}},
  \bibinfo {author} {\bibfnamefont {S.}~\bibnamefont {Seidelin}},\ and\
  \bibinfo {author} {\bibfnamefont {B.}~\bibnamefont {Fang}},\ }\href
  {https://doi.org/10.48550/arXiv.2412.08665} {\bibinfo {title} {Thermal-noise
  limits to the frequency stability of burned spectral holes}},\ \bibinfo
  {howpublished} {arXiv:2412.08665 [cond-mat.mtrl-sci]} (\bibinfo {year}
  {2024})\BibitemShut {NoStop}%
\bibitem [{\citenamefont {Galland}\ \emph
  {et~al.}(2020{\natexlab{a}})\citenamefont {Galland}, \citenamefont
  {Lu{\v{c}}i{\'c}}, \citenamefont {Fang}, \citenamefont {Zhang}, \citenamefont
  {Le~Targat}, \citenamefont {Ferrier}, \citenamefont {Goldner}, \citenamefont
  {Seidelin},\ and\ \citenamefont {Le~Coq}}]{galland2020mechanical}%
  \BibitemOpen
  \bibfield  {author} {\bibinfo {author} {\bibfnamefont {N.}~\bibnamefont
  {Galland}}, \bibinfo {author} {\bibfnamefont {N.}~\bibnamefont
  {Lu{\v{c}}i{\'c}}}, \bibinfo {author} {\bibfnamefont {B.}~\bibnamefont
  {Fang}}, \bibinfo {author} {\bibfnamefont {S.}~\bibnamefont {Zhang}},
  \bibinfo {author} {\bibfnamefont {R.}~\bibnamefont {Le~Targat}}, \bibinfo
  {author} {\bibfnamefont {A.}~\bibnamefont {Ferrier}}, \bibinfo {author}
  {\bibfnamefont {P.}~\bibnamefont {Goldner}}, \bibinfo {author} {\bibfnamefont
  {S.}~\bibnamefont {Seidelin}},\ and\ \bibinfo {author} {\bibfnamefont
  {Y.}~\bibnamefont {Le~Coq}},\ }\bibfield  {title} {\emph {\bibinfo {title}
  {Mechanical tunability of an ultranarrow spectral feature of a
  rare-earth-doped crystal via uniaxial stress}},\ }\href
  {https://doi.org/10.1103/PhysRevApplied.13.044022} {\bibfield  {journal}
  {\bibinfo  {journal} {Phys. Rev. Appl.}\ }\textbf {\bibinfo {volume} {13}},\
  \bibinfo {pages} {044022} (\bibinfo {year} {2020}{\natexlab{a}})}\BibitemShut
  {NoStop}%
\bibitem [{\citenamefont {Callen}\ and\ \citenamefont
  {Greene}(1952)}]{Callen1985}%
  \BibitemOpen
  \bibfield  {author} {\bibinfo {author} {\bibfnamefont {H.~B.}\ \bibnamefont
  {Callen}}\ and\ \bibinfo {author} {\bibfnamefont {R.~F.}\ \bibnamefont
  {Greene}},\ }\bibfield  {title} {\emph {\bibinfo {title} {On a theorem of
  irreversible thermodynamics}},\ }\href
  {https://doi.org/10.1103/PhysRev.86.702} {\bibfield  {journal} {\bibinfo
  {journal} {Phys. Rev.}\ }\textbf {\bibinfo {volume} {86}},\ \bibinfo {pages}
  {702--710} (\bibinfo {year} {1952})}\BibitemShut {NoStop}%
\bibitem [{\citenamefont {Lin}\ \emph {et~al.}(2023)\citenamefont {Lin},
  \citenamefont {Hartman}, \citenamefont {Zhang}, \citenamefont {Seidelin},
  \citenamefont {Fang},\ and\ \citenamefont {Le~Coq}}]{lin2023multi-mode}%
  \BibitemOpen
  \bibfield  {author} {\bibinfo {author} {\bibfnamefont {X.}~\bibnamefont
  {Lin}}, \bibinfo {author} {\bibfnamefont {M.}~\bibnamefont {Hartman}},
  \bibinfo {author} {\bibfnamefont {S.}~\bibnamefont {Zhang}}, \bibinfo
  {author} {\bibfnamefont {S.}~\bibnamefont {Seidelin}}, \bibinfo {author}
  {\bibfnamefont {B.}~\bibnamefont {Fang}},\ and\ \bibinfo {author}
  {\bibfnamefont {Y.}~\bibnamefont {Le~Coq}},\ }\bibfield  {title} {\emph
  {\bibinfo {title} {Multi-mode heterodyne laser interferometry realized via
  software defined radio}},\ }\href {https://doi.org/10.1364/OE.500077}
  {\bibfield  {journal} {\bibinfo  {journal} {Opt. Express}\ }\textbf {\bibinfo
  {volume} {31}},\ \bibinfo {pages} {38475--38493} (\bibinfo {year}
  {2023})}\BibitemShut {NoStop}%
\bibitem [{\citenamefont {Zhang}\ \emph
  {et~al.}(2020{\natexlab{a}})\citenamefont {Zhang}, \citenamefont {Lučić},
  \citenamefont {Galland}, \citenamefont {Le~Targat}, \citenamefont {Goldner},
  \citenamefont {Fang}, \citenamefont {Seidelin},\ and\ \citenamefont
  {Le~Coq}}]{zhang2020precision}%
  \BibitemOpen
  \bibfield  {author} {\bibinfo {author} {\bibfnamefont {S.}~\bibnamefont
  {Zhang}}, \bibinfo {author} {\bibfnamefont {N.}~\bibnamefont {Lučić}},
  \bibinfo {author} {\bibfnamefont {N.}~\bibnamefont {Galland}}, \bibinfo
  {author} {\bibfnamefont {R.}~\bibnamefont {Le~Targat}}, \bibinfo {author}
  {\bibfnamefont {P.}~\bibnamefont {Goldner}}, \bibinfo {author} {\bibfnamefont
  {B.}~\bibnamefont {Fang}}, \bibinfo {author} {\bibfnamefont {S.}~\bibnamefont
  {Seidelin}},\ and\ \bibinfo {author} {\bibfnamefont {Y.}~\bibnamefont
  {Le~Coq}},\ }\bibfield  {title} {\emph {\bibinfo {title} {Precision
  measurements of electric-field-induced frequency displacements of an
  ultranarrow optical transition in ions in a solid}},\ }\href
  {https://doi.org/10.1063/5.0025356} {\bibfield  {journal} {\bibinfo
  {journal} {Appl. Phys. Lett.}\ }\textbf {\bibinfo {volume} {117}},\ \bibinfo
  {pages} {221102} (\bibinfo {year} {2020}{\natexlab{a}})}\BibitemShut
  {NoStop}%
\bibitem [{\citenamefont {Thorpe}\ \emph {et~al.}(2013)\citenamefont {Thorpe},
  \citenamefont {Leibrandt},\ and\ \citenamefont {Rosenband}}]{tho13}%
  \BibitemOpen
  \bibfield  {author} {\bibinfo {author} {\bibfnamefont {M.~J.}\ \bibnamefont
  {Thorpe}}, \bibinfo {author} {\bibfnamefont {D.~R.}\ \bibnamefont
  {Leibrandt}},\ and\ \bibinfo {author} {\bibfnamefont {T.}~\bibnamefont
  {Rosenband}},\ }\bibfield  {title} {\emph {\bibinfo {title} {Shifts of
  optical frequency references based on spectral-hole burning in
  {Eu$^{3+}$:Y$_2$SiO$_5$}}},\ }\href
  {https://doi.org/10.1088/1367-2630/15/3/033006} {\bibfield  {journal}
  {\bibinfo  {journal} {New J. Phys.}\ }\textbf {\bibinfo {volume} {15}},\
  \bibinfo {pages} {033006} (\bibinfo {year} {2013})}\BibitemShut {NoStop}%
\bibitem [{\citenamefont {K{\"o}nz}\ \emph {et~al.}(2003)\citenamefont
  {K{\"o}nz}, \citenamefont {Sun}, \citenamefont {Thiel}, \citenamefont {Cone},
  \citenamefont {Equall}, \citenamefont {Hutcheson},\ and\ \citenamefont
  {Macfarlane}}]{konz2003temperature}%
  \BibitemOpen
  \bibfield  {author} {\bibinfo {author} {\bibfnamefont {F.}~\bibnamefont
  {K{\"o}nz}}, \bibinfo {author} {\bibfnamefont {Y.}~\bibnamefont {Sun}},
  \bibinfo {author} {\bibfnamefont {C.}~\bibnamefont {Thiel}}, \bibinfo
  {author} {\bibfnamefont {R.}~\bibnamefont {Cone}}, \bibinfo {author}
  {\bibfnamefont {R.}~\bibnamefont {Equall}}, \bibinfo {author} {\bibfnamefont
  {R.}~\bibnamefont {Hutcheson}},\ and\ \bibinfo {author} {\bibfnamefont
  {R.}~\bibnamefont {Macfarlane}},\ }\bibfield  {title} {\emph {\bibinfo
  {title} {Temperature and concentration dependence of optical dephasing,
  spectral-hole lifetime, and anisotropic absorption in
  {Eu$^{3+}$:Y$_{2}$SiO$_{5}$}}},\ }\href
  {https://doi.org/10.1103/PhysRevB.68.085109} {\bibfield  {journal} {\bibinfo
  {journal} {Phys. Rev. B}\ }\textbf {\bibinfo {volume} {68}},\ \bibinfo
  {pages} {085109} (\bibinfo {year} {2003})}\BibitemShut {NoStop}%
\bibitem [{\citenamefont {Oswald}\ \emph {et~al.}(2018)\citenamefont {Oswald},
  \citenamefont {Hansen}, \citenamefont {Wiens}, \citenamefont {Nevsky},\ and\
  \citenamefont {Schiller}}]{Oswald2018}%
  \BibitemOpen
  \bibfield  {author} {\bibinfo {author} {\bibfnamefont {R.}~\bibnamefont
  {Oswald}}, \bibinfo {author} {\bibfnamefont {M.~G.}\ \bibnamefont {Hansen}},
  \bibinfo {author} {\bibfnamefont {E.}~\bibnamefont {Wiens}}, \bibinfo
  {author} {\bibfnamefont {A.~Y.}\ \bibnamefont {Nevsky}},\ and\ \bibinfo
  {author} {\bibfnamefont {S.}~\bibnamefont {Schiller}},\ }\bibfield  {title}
  {\emph {\bibinfo {title} {Characteristics of long-lived persistent spectral
  holes in {Eu$^{3+}$:Y$_2$SiO$_5$} at 1.2~{K}}},\ }\href
  {https://doi.org/10.1103/PhysRevA.98.062516} {\bibfield  {journal} {\bibinfo
  {journal} {Phys. Rev. A}\ }\textbf {\bibinfo {volume} {98}},\ \bibinfo
  {pages} {1--8} (\bibinfo {year} {2018})}\BibitemShut {NoStop}%
\bibitem [{\citenamefont {Galland}\ \emph
  {et~al.}(2020{\natexlab{b}})\citenamefont {Galland}, \citenamefont
  {Lu{\v{c}}i{\'c}}, \citenamefont {Zhang}, \citenamefont {Alvarez-Martinez},
  \citenamefont {Le~Targat}, \citenamefont {Ferrier}, \citenamefont {Goldner},
  \citenamefont {Fang}, \citenamefont {Seidelin},\ and\ \citenamefont
  {Le~Coq}}]{galland2020double}%
  \BibitemOpen
  \bibfield  {author} {\bibinfo {author} {\bibfnamefont {N.}~\bibnamefont
  {Galland}}, \bibinfo {author} {\bibfnamefont {N.}~\bibnamefont
  {Lu{\v{c}}i{\'c}}}, \bibinfo {author} {\bibfnamefont {S.}~\bibnamefont
  {Zhang}}, \bibinfo {author} {\bibfnamefont {H.}~\bibnamefont
  {Alvarez-Martinez}}, \bibinfo {author} {\bibfnamefont {R.}~\bibnamefont
  {Le~Targat}}, \bibinfo {author} {\bibfnamefont {A.}~\bibnamefont {Ferrier}},
  \bibinfo {author} {\bibfnamefont {P.}~\bibnamefont {Goldner}}, \bibinfo
  {author} {\bibfnamefont {B.}~\bibnamefont {Fang}}, \bibinfo {author}
  {\bibfnamefont {S.}~\bibnamefont {Seidelin}},\ and\ \bibinfo {author}
  {\bibfnamefont {Y.}~\bibnamefont {Le~Coq}},\ }\bibfield  {title} {\emph
  {\bibinfo {title} {Double-heterodyne probing for an ultra-stable laser based
  on spectral hole burning in a rare-earth-doped crystal}},\ }\href
  {https://doi.org/10.1364/OL.389833} {\bibfield  {journal} {\bibinfo
  {journal} {Opt. Lett.}\ }\textbf {\bibinfo {volume} {45}},\ \bibinfo {pages}
  {1930--1933} (\bibinfo {year} {2020}{\natexlab{b}})}\BibitemShut {NoStop}%
\bibitem [{\citenamefont {Zhang}\ \emph {et~al.}(2023)\citenamefont {Zhang},
  \citenamefont {Seidelin}, \citenamefont {Le~Targat}, \citenamefont {Goldner},
  \citenamefont {Fang},\ and\ \citenamefont {Le~Coq}}]{zhang2023first}%
  \BibitemOpen
  \bibfield  {author} {\bibinfo {author} {\bibfnamefont {S.}~\bibnamefont
  {Zhang}}, \bibinfo {author} {\bibfnamefont {S.}~\bibnamefont {Seidelin}},
  \bibinfo {author} {\bibfnamefont {R.}~\bibnamefont {Le~Targat}}, \bibinfo
  {author} {\bibfnamefont {P.}~\bibnamefont {Goldner}}, \bibinfo {author}
  {\bibfnamefont {B.}~\bibnamefont {Fang}},\ and\ \bibinfo {author}
  {\bibfnamefont {Y.}~\bibnamefont {Le~Coq}},\ }\bibfield  {title} {\emph
  {\bibinfo {title} {First-order thermal insensitivity of the frequency of a
  narrow spectral hole in a crystal}},\ }\href
  {https://doi.org/10.1103/PhysRevA.107.013518} {\bibfield  {journal} {\bibinfo
   {journal} {Phys. Rev. A}\ }\textbf {\bibinfo {volume} {107}},\ \bibinfo
  {pages} {013518} (\bibinfo {year} {2023})}\BibitemShut {NoStop}%
\bibitem [{\citenamefont {Zhang}\ \emph
  {et~al.}(2020{\natexlab{b}})\citenamefont {Zhang}, \citenamefont {Galland},
  \citenamefont {Lu\ifmmode \check{c}\else \v{c}\fi{}i\ifmmode~\acute{c}\else
  \'{c}\fi{}}, \citenamefont {Le~Targat}, \citenamefont {Ferrier},
  \citenamefont {Goldner}, \citenamefont {Fang}, \citenamefont {{Le Coq}},\
  and\ \citenamefont {Seidelin}}]{zhang2020inhomogeneous}%
  \BibitemOpen
  \bibfield  {author} {\bibinfo {author} {\bibfnamefont {S.}~\bibnamefont
  {Zhang}}, \bibinfo {author} {\bibfnamefont {N.}~\bibnamefont {Galland}},
  \bibinfo {author} {\bibfnamefont {N.}~\bibnamefont {Lu\ifmmode \check{c}\else
  \v{c}\fi{}i\ifmmode~\acute{c}\else \'{c}\fi{}}}, \bibinfo {author}
  {\bibfnamefont {R.}~\bibnamefont {Le~Targat}}, \bibinfo {author}
  {\bibfnamefont {A.}~\bibnamefont {Ferrier}}, \bibinfo {author} {\bibfnamefont
  {P.}~\bibnamefont {Goldner}}, \bibinfo {author} {\bibfnamefont
  {B.}~\bibnamefont {Fang}}, \bibinfo {author} {\bibfnamefont {Y.}~\bibnamefont
  {{Le Coq}}},\ and\ \bibinfo {author} {\bibfnamefont {S.}~\bibnamefont
  {Seidelin}},\ }\bibfield  {title} {\emph {\bibinfo {title} {Inhomogeneous
  response of an ion ensemble from mechanical stress}},\ }\href
  {https://doi.org/10.1103/PhysRevResearch.2.013306} {\bibfield  {journal}
  {\bibinfo  {journal} {Phys. Rev. Res.}\ }\textbf {\bibinfo {volume} {2}},\
  \bibinfo {pages} {013306} (\bibinfo {year} {2020}{\natexlab{b}})}\BibitemShut
  {NoStop}%
\bibitem [{\citenamefont {Oswald}\ \emph {et~al.}(2021)\citenamefont {Oswald},
  \citenamefont {Nevsky},\ and\ \citenamefont
  {Schiller}}]{PhysRevA.104.063111}%
  \BibitemOpen
  \bibfield  {author} {\bibinfo {author} {\bibfnamefont {R.}~\bibnamefont
  {Oswald}}, \bibinfo {author} {\bibfnamefont {A.~Y.}\ \bibnamefont {Nevsky}},\
  and\ \bibinfo {author} {\bibfnamefont {S.}~\bibnamefont {Schiller}},\
  }\bibfield  {title} {\emph {\bibinfo {title} {Burning and reading ensembles
  of spectral holes by optical frequency combs: Demonstration in
  rare-earth-doped solids and application to laser frequency stabilization}},\
  }\href {https://doi.org/10.1103/PhysRevA.104.063111} {\bibfield  {journal}
  {\bibinfo  {journal} {Phys. Rev. A}\ }\textbf {\bibinfo {volume} {104}},\
  \bibinfo {pages} {063111} (\bibinfo {year} {2021})}\BibitemShut {NoStop}%
\bibitem [{\citenamefont {Gobron}\ \emph {et~al.}(2017)\citenamefont {Gobron},
  \citenamefont {Jung}, \citenamefont {Galland}, \citenamefont {Predehl},
  \citenamefont {Le~Targat}, \citenamefont {Ferrier}, \citenamefont {Goldner},
  \citenamefont {Seidelin},\ and\ \citenamefont {Le~Coq}}]{gob17}%
  \BibitemOpen
  \bibfield  {author} {\bibinfo {author} {\bibfnamefont {O.}~\bibnamefont
  {Gobron}}, \bibinfo {author} {\bibfnamefont {K.}~\bibnamefont {Jung}},
  \bibinfo {author} {\bibfnamefont {N.}~\bibnamefont {Galland}}, \bibinfo
  {author} {\bibfnamefont {K.}~\bibnamefont {Predehl}}, \bibinfo {author}
  {\bibfnamefont {R.}~\bibnamefont {Le~Targat}}, \bibinfo {author}
  {\bibfnamefont {A.}~\bibnamefont {Ferrier}}, \bibinfo {author} {\bibfnamefont
  {P.}~\bibnamefont {Goldner}}, \bibinfo {author} {\bibfnamefont
  {S.}~\bibnamefont {Seidelin}},\ and\ \bibinfo {author} {\bibfnamefont
  {Y.}~\bibnamefont {Le~Coq}},\ }\bibfield  {title} {\emph {\bibinfo {title}
  {Dispersive heterodyne probing method for laser frequency stabilization based
  on spectral hole burning in rare-earth doped crystals.}},\ }\href
  {https://doi.org/10.1364/oe.25.015539} {\bibfield  {journal} {\bibinfo
  {journal} {Opt. Express}\ }\textbf {\bibinfo {volume} {25}},\ \bibinfo
  {pages} {15539--15548} (\bibinfo {year} {2017})}\BibitemShut {NoStop}%
\bibitem [{\citenamefont {Gustavsson}(2024)}]{gus24a}%
  \BibitemOpen
  \bibfield  {author} {\bibinfo {author} {\bibfnamefont {D.}~\bibnamefont
  {Gustavsson}},\ }\emph {\bibinfo {title} {Laser frequency stabilization using
  a slow light cavity}},\ \href
  {https://portal.research.lu.se/en/publications/laser-frequency-stabilization-using-a-slow-light-cavity}
  {Ph.D. thesis},\ \bibinfo  {school} {Department of Physics, Lund University}
  (\bibinfo {year} {2024})\BibitemShut {NoStop}%
\bibitem [{\citenamefont {Horvath}\ \emph {et~al.}(2022)\citenamefont
  {Horvath}, \citenamefont {Shi}, \citenamefont {Gustavsson}, \citenamefont
  {Walther}, \citenamefont {Kinos}, \citenamefont {Kr{\"{o}}ll},\ and\
  \citenamefont {Rippe}}]{Horvath2022}%
  \BibitemOpen
  \bibfield  {author} {\bibinfo {author} {\bibfnamefont {S.~P.}\ \bibnamefont
  {Horvath}}, \bibinfo {author} {\bibfnamefont {C.}~\bibnamefont {Shi}},
  \bibinfo {author} {\bibfnamefont {D.}~\bibnamefont {Gustavsson}}, \bibinfo
  {author} {\bibfnamefont {A.}~\bibnamefont {Walther}}, \bibinfo {author}
  {\bibfnamefont {A.}~\bibnamefont {Kinos}}, \bibinfo {author} {\bibfnamefont
  {S.}~\bibnamefont {Kr{\"{o}}ll}},\ and\ \bibinfo {author} {\bibfnamefont
  {L.}~\bibnamefont {Rippe}},\ }\bibfield  {title} {\emph {\bibinfo {title}
  {Slow light frequency reference cavities - proof of concept for reducing the
  frequency sensitivity due to length fluctuations}},\ }\href
  {https://doi.org/10.1088/1367-2630/ac5932} {\bibfield  {journal} {\bibinfo
  {journal} {New J. Phys.}\ }\textbf {\bibinfo {volume} {24}},\ \bibinfo
  {pages} {033034} (\bibinfo {year} {2022})}\BibitemShut {NoStop}%
\bibitem [{\citenamefont {Gustavsson}\ \emph {et~al.}(tion)\citenamefont
  {Gustavsson}, \citenamefont {Lindén}, \citenamefont {Walther}, \citenamefont
  {Kinos}, \citenamefont {Kr{\"{o}}ll},\ and\ \citenamefont
  {Rippe}}]{Gustavsson}%
  \BibitemOpen
  \bibfield  {author} {\bibinfo {author} {\bibfnamefont {D.}~\bibnamefont
  {Gustavsson}}, \bibinfo {author} {\bibfnamefont {M.}~\bibnamefont {Lindén}},
  \bibinfo {author} {\bibfnamefont {A.}~\bibnamefont {Walther}}, \bibinfo
  {author} {\bibfnamefont {A.}~\bibnamefont {Kinos}}, \bibinfo {author}
  {\bibfnamefont {S.}~\bibnamefont {Kr{\"{o}}ll}},\ and\ \bibinfo {author}
  {\bibfnamefont {L.}~\bibnamefont {Rippe}},\ }\href@noop {} {\emph {\bibinfo
  {title} {Laser frequency stabilization using the slow light effect in
  Eu:YSO}}}\ (\bibinfo {year} {Manuscript in preparation})\BibitemShut
  {NoStop}%
\bibitem [{\citenamefont {Lindén}\ \emph {et~al.}(tion)\citenamefont
  {Lindén}, \citenamefont {Gustavsson}, \citenamefont {Walther}, \citenamefont
  {Kinos}, \citenamefont {Kr{\"{o}}ll},\ and\ \citenamefont
  {Rippe}}]{LindenDesign}%
  \BibitemOpen
  \bibfield  {author} {\bibinfo {author} {\bibfnamefont {M.}~\bibnamefont
  {Lindén}}, \bibinfo {author} {\bibfnamefont {D.}~\bibnamefont {Gustavsson}},
  \bibinfo {author} {\bibfnamefont {A.}~\bibnamefont {Walther}}, \bibinfo
  {author} {\bibfnamefont {A.}~\bibnamefont {Kinos}}, \bibinfo {author}
  {\bibfnamefont {S.}~\bibnamefont {Kr{\"{o}}ll}},\ and\ \bibinfo {author}
  {\bibfnamefont {L.}~\bibnamefont {Rippe}},\ }\href@noop {} {\emph {\bibinfo
  {title} {Highly tuneable in-situ cryogenic switch bank resonator for magnetic
  field generation at radio-frequencies}}}\ (\bibinfo {year} {Manuscript in
  preparation})\BibitemShut {NoStop}%
\bibitem [{\citenamefont {Lind{\'{e}}n}\ \emph {et~al.}(tion)\citenamefont
  {Lind{\'{e}}n}, \citenamefont {Gustavsson}, \citenamefont {Walther},
  \citenamefont {Kinos}, \citenamefont {Kr{\"{o}}ll},\ and\ \citenamefont
  {Rippe}}]{LindenAtomic}%
  \BibitemOpen
  \bibfield  {author} {\bibinfo {author} {\bibfnamefont {M.}~\bibnamefont
  {Lind{\'{e}}n}}, \bibinfo {author} {\bibfnamefont {D.}~\bibnamefont
  {Gustavsson}}, \bibinfo {author} {\bibfnamefont {A.}~\bibnamefont {Walther}},
  \bibinfo {author} {\bibfnamefont {A.}~\bibnamefont {Kinos}}, \bibinfo
  {author} {\bibfnamefont {S.}~\bibnamefont {Kr{\"{o}}ll}},\ and\ \bibinfo
  {author} {\bibfnamefont {L.}~\bibnamefont {Rippe}},\ }\href@noop {} {\emph
  {\bibinfo {title} {Erasing spectral hole features in {Eu:YSO} using a highly
  tunable {RF} resonance circuit}}}\ (\bibinfo {year} {Manuscript in
  preparation})\BibitemShut {NoStop}%
\bibitem [{\citenamefont {Cook}\ \emph {et~al.}(2015)\citenamefont {Cook},
  \citenamefont {Rosenband},\ and\ \citenamefont {Leibrandt}}]{coo15}%
  \BibitemOpen
  \bibfield  {author} {\bibinfo {author} {\bibfnamefont {S.}~\bibnamefont
  {Cook}}, \bibinfo {author} {\bibfnamefont {T.}~\bibnamefont {Rosenband}},\
  and\ \bibinfo {author} {\bibfnamefont {D.~R.}\ \bibnamefont {Leibrandt}},\
  }\bibfield  {title} {\emph {\bibinfo {title} {Laser frequency stabilization
  based on steady-state spectral-hole burning in {Eu$^{3+}$}:{Y$_2$SiO$_5$}}},\
  }\href {https://doi.org/10.1103/PhysRevLett.114.253902} {\bibfield  {journal}
  {\bibinfo  {journal} {Phys. Rev. Lett.}\ }\textbf {\bibinfo {volume} {114}},\
  \bibinfo {pages} {253902} (\bibinfo {year} {2015})}\BibitemShut {NoStop}%
\bibitem [{\citenamefont {Peterson}(1993)}]{pet93}%
  \BibitemOpen
  \bibfield  {author} {\bibinfo {author} {\bibfnamefont {J.}~\bibnamefont
  {Peterson}},\ }\href
  {https://www.mttmllr.com/ADS/DATA/peterson_usgs_seismic_noise_ofr93-322.pdf}
  {\bibinfo {title} {Observations and modeling of seismic background noise}},\
  \bibinfo {howpublished} {Open File. Report 93-322, U. S. Geological Survey,
  Albuquerque, NM} (\bibinfo {year} {1993})\BibitemShut {NoStop}%
\bibitem [{\citenamefont {Strollo}\ \emph {et~al.}(2008)\citenamefont
  {Strollo}, \citenamefont {Bindi}, \citenamefont {Parolai},\ and\
  \citenamefont {J\"ackel}}]{str08}%
  \BibitemOpen
  \bibfield  {author} {\bibinfo {author} {\bibfnamefont {A.}~\bibnamefont
  {Strollo}}, \bibinfo {author} {\bibfnamefont {D.}~\bibnamefont {Bindi}},
  \bibinfo {author} {\bibfnamefont {S.}~\bibnamefont {Parolai}},\ and\ \bibinfo
  {author} {\bibfnamefont {K.-H.}\ \bibnamefont {J\"ackel}},\ }\bibfield
  {title} {\emph {\bibinfo {title} {On the suitability of 1~s geophone for
  ambient noise measurements in the {0.1--20~Hz} frequency range: experimental
  outcomes}},\ }\href {https://doi.org/10.1007/s10518-008-9061-x} {\bibfield
  {journal} {\bibinfo  {journal} {Bull. Earthquake Eng.}\ }\textbf {\bibinfo
  {volume} {6}},\ \bibinfo {pages} {141--147} (\bibinfo {year}
  {2008})}\BibitemShut {NoStop}%
\bibitem [{\citenamefont {van Kann}\ and\ \citenamefont
  {Winterflood}(2005)}]{kan05a}%
  \BibitemOpen
  \bibfield  {author} {\bibinfo {author} {\bibfnamefont {F.}~\bibnamefont {van
  Kann}}\ and\ \bibinfo {author} {\bibfnamefont {J.}~\bibnamefont
  {Winterflood}},\ }\bibfield  {title} {\emph {\bibinfo {title} {Simple method
  for absolute calibration of geophones, seismometers, and other inertial
  vibration sensors}},\ }\href {https://doi.org/10.1063/1.1867432} {\bibfield
  {journal} {\bibinfo  {journal} {Rev. Sci. Instrum.}\ }\textbf {\bibinfo
  {volume} {76}},\ \bibinfo {pages} {034501} (\bibinfo {year}
  {2005})}\BibitemShut {NoStop}%
\bibitem [{\citenamefont {Yu}\ \emph {et~al.}(2022)\citenamefont {Yu},
  \citenamefont {Legero}, \citenamefont {Riehle}, \citenamefont {Ma},
  \citenamefont {Herbers}, \citenamefont {Nicolodi}, \citenamefont {Kedar},
  \citenamefont {Oelker}, \citenamefont {Ye},\ and\ \citenamefont
  {Sterr}}]{yu22}%
  \BibitemOpen
  \bibfield  {author} {\bibinfo {author} {\bibfnamefont {J.}~\bibnamefont
  {Yu}}, \bibinfo {author} {\bibfnamefont {T.}~\bibnamefont {Legero}}, \bibinfo
  {author} {\bibfnamefont {F.}~\bibnamefont {Riehle}}, \bibinfo {author}
  {\bibfnamefont {C.~Y.}\ \bibnamefont {Ma}}, \bibinfo {author} {\bibfnamefont
  {S.}~\bibnamefont {Herbers}}, \bibinfo {author} {\bibfnamefont
  {D.}~\bibnamefont {Nicolodi}}, \bibinfo {author} {\bibfnamefont
  {D.}~\bibnamefont {Kedar}}, \bibinfo {author} {\bibfnamefont
  {E.}~\bibnamefont {Oelker}}, \bibinfo {author} {\bibfnamefont
  {J.}~\bibnamefont {Ye}},\ and\ \bibinfo {author} {\bibfnamefont
  {U.}~\bibnamefont {Sterr}},\ }\bibfield  {title} {\emph {\bibinfo {title}
  {Novel noise contributions in crystalline mirror coatings}},\ }in\ \href
  {https://doi.org/10.1109/EFTF/IFCS54560.2022.9850553} {\emph {\bibinfo
  {booktitle} {2022 Joint Conference of the European Frequency and Time Forum
  and IEEE International Frequency Control Symposium (EFTF/IFCS)}}}\ (\bibinfo
  {year} {2022})\ pp.\ \bibinfo {pages} {1--3}\BibitemShut {NoStop}%
\bibitem [{\citenamefont {Yu}(2023)}]{yu23}%
  \BibitemOpen
  \bibfield  {author} {\bibinfo {author} {\bibfnamefont {J.}~\bibnamefont
  {Yu}},\ }\emph {\bibinfo {title} {Cryogenic silicon {Fabry}-{Perot} resonator
  with {Al$_{0.92}$Ga$_{0.08}$As/GaAs} mirror coatings.}},\ \href
  {https://doi.org/10.15488/13416} {Ph.D. thesis},\ \bibinfo  {school}
  {QUEST-Leibniz-Forschungsschule der Gottfried Wilhelm Leibniz Universit\"at
  Hannover} (\bibinfo {year} {2023})\BibitemShut {NoStop}%
\bibitem [{\citenamefont {Hagemann}(2013)}]{hag13a}%
  \BibitemOpen
  \bibfield  {author} {\bibinfo {author} {\bibfnamefont {C.}~\bibnamefont
  {Hagemann}},\ }\emph {\bibinfo {title} {Ultra-stable laser based on a
  cryogenic single-crystal silicon cavity}},\ \href
  {https://doi.org/10.15488/8071} {Ph.D. thesis},\ \bibinfo  {school}
  {Fakult\"at f\"ur Mathematik und Physik der Gottfried Wilhelm Leibniz
  Universit\"at Hannover} (\bibinfo {year} {2013})\BibitemShut {NoStop}%
\bibitem [{\citenamefont {Oon}\ and\ \citenamefont {Dumke}(2022)}]{oon22}%
  \BibitemOpen
  \bibfield  {author} {\bibinfo {author} {\bibfnamefont {F.~E.}\ \bibnamefont
  {Oon}}\ and\ \bibinfo {author} {\bibfnamefont {R.}~\bibnamefont {Dumke}},\
  }\bibfield  {title} {\emph {\bibinfo {title} {Compact active vibration
  isolation and tilt stabilization for a portable high-precision atomic
  gravimeter}},\ }\href {https://doi.org/10.1103/PhysRevApplied.18.044037}
  {\bibfield  {journal} {\bibinfo  {journal} {Phys. Rev. Applied}\ }\textbf
  {\bibinfo {volume} {18}},\ \bibinfo {pages} {044037} (\bibinfo {year}
  {2022})}\BibitemShut {NoStop}%
\bibitem [{\citenamefont {Kawohl}(2021)}]{kaw21}%
  \BibitemOpen
  \bibfield  {author} {\bibinfo {author} {\bibfnamefont {J.}~\bibnamefont
  {Kawohl}},\ }\emph {\bibinfo {title} {Entwicklung eines digitalen
  {R}eglersystems zur aktiven {S}chwingungsisolation optischer
  {R}esonatoren}},\ \href@noop {} {Master's thesis},\ \bibinfo  {school}
  {Technische Universit\"at Braunschweig}, \bibinfo {address}
  {Pockelsstra{\ss}e 38, Braunschweig} (\bibinfo {year} {2021})\BibitemShut
  {NoStop}%
\bibitem [{Red()}]{RedPitaya}%
  \BibitemOpen
  \href {https://www.redpitaya.com.} {\bibinfo {title} {{Instrumentation
  Technologies, LLC, Red Pitaya}}},\ \bibinfo {howpublished}
  {https://www.redpitaya.com.}\BibitemShut {Stop}%
\bibitem [{\citenamefont {Anders}(2023)}]{and23}%
  \BibitemOpen
  \bibfield  {author} {\bibinfo {author} {\bibfnamefont {L.}~\bibnamefont
  {Anders}},\ }\emph {\bibinfo {title} {Entwicklung und {C}harakterisierung
  einer {F}eedforward-{K}orrektur von vibrationsinduzierten {S}t\"orungen einer
  ultrastabilen {L}aserfrequenz}},\ \href@noop {} {Master's thesis},\ \bibinfo
  {school} {Technische Universit\"at Braunschweig}, \bibinfo {address}
  {Pockelsstra{\ss}e 38, Braunschweig} (\bibinfo {year} {2023})\BibitemShut
  {NoStop}%
\bibitem [{\citenamefont {Barbarat}\ \emph {et~al.}(2024)\citenamefont
  {Barbarat}, \citenamefont {Gillot}, \citenamefont {Millo}, \citenamefont
  {Lacro\^ute}, \citenamefont {Legero}, \citenamefont {Giordano},\ and\
  \citenamefont {Kersal\'e}}]{bar24}%
  \BibitemOpen
  \bibfield  {author} {\bibinfo {author} {\bibfnamefont {J.}~\bibnamefont
  {Barbarat}}, \bibinfo {author} {\bibfnamefont {J.}~\bibnamefont {Gillot}},
  \bibinfo {author} {\bibfnamefont {J.}~\bibnamefont {Millo}}, \bibinfo
  {author} {\bibfnamefont {C.}~\bibnamefont {Lacro\^ute}}, \bibinfo {author}
  {\bibfnamefont {T.}~\bibnamefont {Legero}}, \bibinfo {author} {\bibfnamefont
  {V.}~\bibnamefont {Giordano}},\ and\ \bibinfo {author} {\bibfnamefont
  {Y.}~\bibnamefont {Kersal\'e}},\ }\bibfield  {title} {\emph {\bibinfo {title}
  {Towards a sub-kelvin cryogenic {Fabry}-{Perot} silicon cavity}},\ }\href
  {https://doi.org/10.1088/1742-6596/2889/1/012056} {\bibfield  {journal}
  {\bibinfo  {journal} {J. Phys.: Conf. Ser.}\ }\textbf {\bibinfo {volume}
  {2889}},\ \bibinfo {pages} {012056} (\bibinfo {year} {2024})}\BibitemShut
  {NoStop}%
\bibitem [{\citenamefont {Millo}\ \emph {et~al.}(2009)\citenamefont {Millo},
  \citenamefont {Magalh\~aes}, \citenamefont {Mandache}, \citenamefont
  {Le~Coq}, \citenamefont {English}, \citenamefont {Westergaard}, \citenamefont
  {Lodewyck}, \citenamefont {Bize}, \citenamefont {Lemonde},\ and\
  \citenamefont {Santarelli}}]{jak2009}%
  \BibitemOpen
  \bibfield  {author} {\bibinfo {author} {\bibfnamefont {J.}~\bibnamefont
  {Millo}}, \bibinfo {author} {\bibfnamefont {D.~V.}\ \bibnamefont
  {Magalh\~aes}}, \bibinfo {author} {\bibfnamefont {C.}~\bibnamefont
  {Mandache}}, \bibinfo {author} {\bibfnamefont {Y.}~\bibnamefont {Le~Coq}},
  \bibinfo {author} {\bibfnamefont {E.~M.~L.}\ \bibnamefont {English}},
  \bibinfo {author} {\bibfnamefont {P.~G.}\ \bibnamefont {Westergaard}},
  \bibinfo {author} {\bibfnamefont {J.}~\bibnamefont {Lodewyck}}, \bibinfo
  {author} {\bibfnamefont {S.}~\bibnamefont {Bize}}, \bibinfo {author}
  {\bibfnamefont {P.}~\bibnamefont {Lemonde}},\ and\ \bibinfo {author}
  {\bibfnamefont {G.}~\bibnamefont {Santarelli}},\ }\bibfield  {title} {\emph
  {\bibinfo {title} {Ultrastable lasers based on vibration insensitive
  cavities}},\ }\href {https://doi.org/10.1103/PhysRevA.79.053829} {\bibfield
  {journal} {\bibinfo  {journal} {Phys. Rev. A}\ }\textbf {\bibinfo {volume}
  {79}},\ \bibinfo {pages} {053829} (\bibinfo {year} {2009})}\BibitemShut
  {NoStop}%
\bibitem [{\citenamefont {Caparrelli}\ \emph {et~al.}(2006)\citenamefont
  {Caparrelli}, \citenamefont {Majorana}, \citenamefont {Moscatelli},
  \citenamefont {Pascucci}, \citenamefont {Perciballi}, \citenamefont {Puppo},
  \citenamefont {Rapagnani},\ and\ \citenamefont {Ricci}}]{cap06}%
  \BibitemOpen
  \bibfield  {author} {\bibinfo {author} {\bibfnamefont {S.}~\bibnamefont
  {Caparrelli}}, \bibinfo {author} {\bibfnamefont {E.}~\bibnamefont
  {Majorana}}, \bibinfo {author} {\bibfnamefont {V.}~\bibnamefont
  {Moscatelli}}, \bibinfo {author} {\bibfnamefont {E.}~\bibnamefont
  {Pascucci}}, \bibinfo {author} {\bibfnamefont {M.}~\bibnamefont
  {Perciballi}}, \bibinfo {author} {\bibfnamefont {P.}~\bibnamefont {Puppo}},
  \bibinfo {author} {\bibfnamefont {P.}~\bibnamefont {Rapagnani}},\ and\
  \bibinfo {author} {\bibfnamefont {F.}~\bibnamefont {Ricci}},\ }\bibfield
  {title} {\emph {\bibinfo {title} {Vibration-free cryostat for low-noise
  applications of a pulse tube cryocooler}},\ }\href
  {https://doi.org/10.1063/1.2349609} {\bibfield  {journal} {\bibinfo
  {journal} {Rev. Sci. Instrum.}\ }\textbf {\bibinfo {volume} {77}},\ \bibinfo
  {pages} {095102} (\bibinfo {year} {2006})}\BibitemShut {NoStop}%
\bibitem [{\citenamefont {Wang}\ \emph {et~al.}(2023)\citenamefont {Wang},
  \citenamefont {Chen}, \citenamefont {Zhang}, \citenamefont {Du},
  \citenamefont {Wu}, \citenamefont {Qiao}, \citenamefont {Kuang},\ and\
  \citenamefont {Zhang}}]{wan23}%
  \BibitemOpen
  \bibfield  {author} {\bibinfo {author} {\bibfnamefont {W.-W.}\ \bibnamefont
  {Wang}}, \bibinfo {author} {\bibfnamefont {Z.-A.}\ \bibnamefont {Chen}},
  \bibinfo {author} {\bibfnamefont {H.}~\bibnamefont {Zhang}}, \bibinfo
  {author} {\bibfnamefont {S.}~\bibnamefont {Du}}, \bibinfo {author}
  {\bibfnamefont {R.}~\bibnamefont {Wu}}, \bibinfo {author} {\bibfnamefont
  {C.}~\bibnamefont {Qiao}}, \bibinfo {author} {\bibfnamefont {S.}~\bibnamefont
  {Kuang}},\ and\ \bibinfo {author} {\bibfnamefont {X.}~\bibnamefont {Zhang}},\
  }\bibfield  {title} {\emph {\bibinfo {title} {Design and realization of a
  {3-K} cryostat for a 10-cm ultrastable silicon cavity}},\ }\href
  {https://doi.org/10.3389/fphy.2023.1176783} {\bibfield  {journal} {\bibinfo
  {journal} {Frontiers in Physics}\ }\textbf {\bibinfo {volume} {11}},\
  \bibinfo {pages} {1176783} (\bibinfo {year} {2023})}\BibitemShut {NoStop}%
\bibitem [{\citenamefont {Valencia}\ \emph {et~al.}(2024)\citenamefont
  {Valencia}, \citenamefont {Iskander}, \citenamefont {Nardelli}, \citenamefont
  {Leibrandt},\ and\ \citenamefont {Hume}}]{val24}%
  \BibitemOpen
  \bibfield  {author} {\bibinfo {author} {\bibfnamefont {J.}~\bibnamefont
  {Valencia}}, \bibinfo {author} {\bibfnamefont {G.}~\bibnamefont {Iskander}},
  \bibinfo {author} {\bibfnamefont {N.~V.}\ \bibnamefont {Nardelli}}, \bibinfo
  {author} {\bibfnamefont {D.~R.}\ \bibnamefont {Leibrandt}},\ and\ \bibinfo
  {author} {\bibfnamefont {D.~B.}\ \bibnamefont {Hume}},\ }\href
  {https://doi.org/10.48550/arXiv.2404.14310} {\bibinfo {title} {Cryogenic
  sapphire optical reference cavity with crystalline coatings at $\mathrm{1
  \times 10^{-16}}$ fractional instability}},\ \bibinfo {howpublished}
  {arXiv:2404.14310 [physics.optics]} (\bibinfo {year} {2024})\BibitemShut
  {NoStop}%
\bibitem [{\citenamefont {Wang}\ and\ \citenamefont {Hartnett}(2010)}]{wan10b}%
  \BibitemOpen
  \bibfield  {author} {\bibinfo {author} {\bibfnamefont {C.}~\bibnamefont
  {Wang}}\ and\ \bibinfo {author} {\bibfnamefont {J.~G.}\ \bibnamefont
  {Hartnett}},\ }\bibfield  {title} {\emph {\bibinfo {title} {A vibration free
  cryostat using pulse tube cryocooler}},\ }\href
  {https://doi.org/10.1016/j.cryogenics.2010.01.003} {\bibfield  {journal}
  {\bibinfo  {journal} {Cryogenics}\ }\textbf {\bibinfo {volume} {50}},\
  \bibinfo {pages} {336 -- 341} (\bibinfo {year} {2010})}\BibitemShut {NoStop}%
\bibitem [{\citenamefont {{Stirling Cryogenics}}()}]{cryofan}%
  \BibitemOpen
  \bibfield  {author} {\bibinfo {author} {\bibnamefont {{Stirling
  Cryogenics}}},\ }\href
  {https://stirlingcryogenics.com/products/cryogenic-fans/} {\bibinfo {title}
  {{CryoFans}}},\ \bibinfo {note}
  {https://stirlingcryogenics.com/products/cryogenic-fans/}\BibitemShut
  {NoStop}%
\bibitem [{\citenamefont {Bluefors}(2024{\natexlab{a}})}]{cryomech}%
  \BibitemOpen
  \bibfield  {author} {\bibinfo {author} {\bibnamefont {Bluefors}},\ }\href
  {https://bluefors.com/products/cryomech-products/} {\bibinfo {title}
  {{Cryomec Pulse Tube Coolers}}} (\bibinfo {year} {2024}{\natexlab{a}}),\
  \bibinfo {note}
  {https://bluefors.com/products/cryomech-products/}\BibitemShut {NoStop}%
\bibitem [{\citenamefont {Hartnett}\ \emph {et~al.}(2010)\citenamefont
  {Hartnett}, \citenamefont {Nand}, \citenamefont {Wang},\ and\ \citenamefont
  {{Le Floch}}}]{har10a}%
  \BibitemOpen
  \bibfield  {author} {\bibinfo {author} {\bibfnamefont {J.~G.}\ \bibnamefont
  {Hartnett}}, \bibinfo {author} {\bibfnamefont {N.~R.}\ \bibnamefont {Nand}},
  \bibinfo {author} {\bibfnamefont {C.}~\bibnamefont {Wang}},\ and\ \bibinfo
  {author} {\bibfnamefont {J.-M.}\ \bibnamefont {{Le Floch}}},\ }\bibfield
  {title} {\emph {\bibinfo {title} {Cryogenic sapphire oscillator using a
  low-vibration design pulse-tube cryocooler: {F}irst results.}},\ }\href
  {https://doi.org/10.1109/TUFFC.2010.1515} {\bibfield  {journal} {\bibinfo
  {journal} {IEEE Trans. Ultrason. Ferroelectr. Freq. Control}\ }\textbf
  {\bibinfo {volume} {57}},\ \bibinfo {pages} {1034--1038} (\bibinfo {year}
  {2010})}\BibitemShut {NoStop}%
\bibitem [{\citenamefont {Hartnett}\ \emph {et~al.}(2012)\citenamefont
  {Hartnett}, \citenamefont {Nand},\ and\ \citenamefont {Lu}}]{har12}%
  \BibitemOpen
  \bibfield  {author} {\bibinfo {author} {\bibfnamefont {J.~G.}\ \bibnamefont
  {Hartnett}}, \bibinfo {author} {\bibfnamefont {N.~R.}\ \bibnamefont {Nand}},\
  and\ \bibinfo {author} {\bibfnamefont {C.}~\bibnamefont {Lu}},\ }\bibfield
  {title} {\emph {\bibinfo {title} {Ultra-low-phase-noise cryocooled microwave
  dielectric-sapphire-resonator oscillators}},\ }\href
  {https://doi.org/10.1063/1.4709479} {\bibfield  {journal} {\bibinfo
  {journal} {Appl. Phys. Lett.}\ }\textbf {\bibinfo {volume} {100}},\ \bibinfo
  {pages} {183501} (\bibinfo {year} {2012})}\BibitemShut {NoStop}%
\bibitem [{\citenamefont {Bluefors}(2024{\natexlab{b}})}]{bluefors}%
  \BibitemOpen
  \bibfield  {author} {\bibinfo {author} {\bibnamefont {Bluefors}},\ }\href
  {https://bluefors.com/products/cryomech-products/} {\bibinfo {title}
  {{Cryogen-Free Dilution Refrigerator System, BF-XLD-SERIES}}} (\bibinfo
  {year} {2024}{\natexlab{b}}),\ \bibinfo {note}
  {https://bluefors.com/}\BibitemShut {NoStop}%
\bibitem [{\citenamefont {Kessler}\ \emph {et~al.}(2012)\citenamefont
  {Kessler}, \citenamefont {Hagemann}, \citenamefont {Grebing}, \citenamefont
  {Legero}, \citenamefont {Sterr}, \citenamefont {Riehle}, \citenamefont
  {Martin}, \citenamefont {Chen},\ and\ \citenamefont {Ye}}]{kes12a}%
  \BibitemOpen
  \bibfield  {author} {\bibinfo {author} {\bibfnamefont {T.}~\bibnamefont
  {Kessler}}, \bibinfo {author} {\bibfnamefont {C.}~\bibnamefont {Hagemann}},
  \bibinfo {author} {\bibfnamefont {C.}~\bibnamefont {Grebing}}, \bibinfo
  {author} {\bibfnamefont {T.}~\bibnamefont {Legero}}, \bibinfo {author}
  {\bibfnamefont {U.}~\bibnamefont {Sterr}}, \bibinfo {author} {\bibfnamefont
  {F.}~\bibnamefont {Riehle}}, \bibinfo {author} {\bibfnamefont {M.~J.}\
  \bibnamefont {Martin}}, \bibinfo {author} {\bibfnamefont {L.}~\bibnamefont
  {Chen}},\ and\ \bibinfo {author} {\bibfnamefont {J.}~\bibnamefont {Ye}},\
  }\bibfield  {title} {\emph {\bibinfo {title} {A sub-40-{mHz}-linewidth laser
  based on a silicon single-crystal optical cavity}},\ }\href
  {https://doi.org/10.1038/NPHOTON.2012.217} {\bibfield  {journal} {\bibinfo
  {journal} {Nature Photonics}\ }\textbf {\bibinfo {volume} {6}},\ \bibinfo
  {pages} {687--692} (\bibinfo {year} {2012})}\BibitemShut {NoStop}%
\bibitem [{\citenamefont {Lin}\ \emph {et~al.}(2024{\natexlab{a}})\citenamefont
  {Lin}, \citenamefont {Hartman}, \citenamefont {Goldner}, \citenamefont
  {Fang}, \citenamefont {Le~Coq},\ and\ \citenamefont {Seidelin}}]{lin24a}%
  \BibitemOpen
  \bibfield  {author} {\bibinfo {author} {\bibfnamefont {X.}~\bibnamefont
  {Lin}}, \bibinfo {author} {\bibfnamefont {M.~T.}\ \bibnamefont {Hartman}},
  \bibinfo {author} {\bibfnamefont {P.}~\bibnamefont {Goldner}}, \bibinfo
  {author} {\bibfnamefont {B.}~\bibnamefont {Fang}}, \bibinfo {author}
  {\bibfnamefont {Y.}~\bibnamefont {Le~Coq}},\ and\ \bibinfo {author}
  {\bibfnamefont {S.}~\bibnamefont {Seidelin}},\ }\href
  {https://doi.org/10.48550/arXiv.2412.10403} {\bibinfo {title} {Homogeneous
  linewidth behaviour of narrow optical emitters at sub-kelvin temperatures}},\
  \bibinfo {howpublished} {arXiv:2412.10403 [cond-mat.mes-hall]} (\bibinfo
  {year} {2024}{\natexlab{a}})\BibitemShut {NoStop}%
\bibitem [{\citenamefont {White}\ and\ \citenamefont
  {Minges}(1997)}]{white1997}%
  \BibitemOpen
  \bibfield  {author} {\bibinfo {author} {\bibfnamefont {G.}~\bibnamefont
  {White}}\ and\ \bibinfo {author} {\bibfnamefont {M.}~\bibnamefont {Minges}},\
  }\bibfield  {title} {\emph {\bibinfo {title} {Thermophysical properties of
  some key solids: an update}},\ }\href {https://doi.org/10.1007/BF02575261}
  {\bibfield  {journal} {\bibinfo  {journal} {Int. J. Thermophys.}\ }\textbf
  {\bibinfo {volume} {18}},\ \bibinfo {pages} {1269--1327} (\bibinfo {year}
  {1997})}\BibitemShut {NoStop}%
\bibitem [{\citenamefont {Middelmann}\ \emph {et~al.}(2015)\citenamefont
  {Middelmann}, \citenamefont {Walkov}, \citenamefont {Bartl},\ and\
  \citenamefont {Sch\"odel}}]{mid15}%
  \BibitemOpen
  \bibfield  {author} {\bibinfo {author} {\bibfnamefont {T.}~\bibnamefont
  {Middelmann}}, \bibinfo {author} {\bibfnamefont {A.}~\bibnamefont {Walkov}},
  \bibinfo {author} {\bibfnamefont {G.}~\bibnamefont {Bartl}},\ and\ \bibinfo
  {author} {\bibfnamefont {R.}~\bibnamefont {Sch\"odel}},\ }\bibfield  {title}
  {\emph {\bibinfo {title} {Thermal expansion coefficient of single-crystal
  silicon from 7~{K} to 293~{K}}},\ }\href
  {https://doi.org/10.1103/PhysRevB.92.174113} {\bibfield  {journal} {\bibinfo
  {journal} {Phys. Rev. B}\ }\textbf {\bibinfo {volume} {92}},\ \bibinfo
  {pages} {174113} (\bibinfo {year} {2015})}\BibitemShut {NoStop}%
\bibitem [{\citenamefont {Wiens}\ \emph {et~al.}(2014)\citenamefont {Wiens},
  \citenamefont {Chen}, \citenamefont {Ernsting}, \citenamefont {Luckmann},
  \citenamefont {Rosowski}, \citenamefont {Nevsky},\ and\ \citenamefont
  {Schiller}}]{wiens2014}%
  \BibitemOpen
  \bibfield  {author} {\bibinfo {author} {\bibfnamefont {E.}~\bibnamefont
  {Wiens}}, \bibinfo {author} {\bibfnamefont {Q.-F.}\ \bibnamefont {Chen}},
  \bibinfo {author} {\bibfnamefont {I.}~\bibnamefont {Ernsting}}, \bibinfo
  {author} {\bibfnamefont {H.}~\bibnamefont {Luckmann}}, \bibinfo {author}
  {\bibfnamefont {U.}~\bibnamefont {Rosowski}}, \bibinfo {author}
  {\bibfnamefont {A.}~\bibnamefont {Nevsky}},\ and\ \bibinfo {author}
  {\bibfnamefont {S.}~\bibnamefont {Schiller}},\ }\bibfield  {title} {\emph
  {\bibinfo {title} {Silicon single-crystal cryogenic optical resonator}},\
  }\href {https://doi.org/10.1364/OL.39.003242} {\bibfield  {journal} {\bibinfo
   {journal} {Opt. Lett.}\ }\textbf {\bibinfo {volume} {39}},\ \bibinfo {pages}
  {3242--3245} (\bibinfo {year} {2014})}\BibitemShut {NoStop}%
\bibitem [{\citenamefont {Wiens}\ \emph {et~al.}(2020)\citenamefont {Wiens},
  \citenamefont {Kwong}, \citenamefont {M{\"u}ller},\ and\ \citenamefont
  {Schiller}}]{wiens2020}%
  \BibitemOpen
  \bibfield  {author} {\bibinfo {author} {\bibfnamefont {E.}~\bibnamefont
  {Wiens}}, \bibinfo {author} {\bibfnamefont {C.~J.}\ \bibnamefont {Kwong}},
  \bibinfo {author} {\bibfnamefont {T.}~\bibnamefont {M{\"u}ller}},\ and\
  \bibinfo {author} {\bibfnamefont {S.}~\bibnamefont {Schiller}},\ }\bibfield
  {title} {\emph {\bibinfo {title} {A simplified cryogenic optical resonator
  apparatus providing ultra-low frequency drift}},\ }\href
  {https://doi.org/10.1063/1.5140321} {\bibfield  {journal} {\bibinfo
  {journal} {Rev. Sci. Instr.}\ }\textbf {\bibinfo {volume} {91}},\ \bibinfo
  {pages} {045112} (\bibinfo {year} {2020})}\BibitemShut {NoStop}%
\bibitem [{\citenamefont {Wiens}\ \emph {et~al.}(2023)\citenamefont {Wiens},
  \citenamefont {Kwong}, \citenamefont {M\"{u}ller}, \citenamefont {Bongs},
  \citenamefont {Singh},\ and\ \citenamefont {Schiller}}]{wiens2023}%
  \BibitemOpen
  \bibfield  {author} {\bibinfo {author} {\bibfnamefont {E.}~\bibnamefont
  {Wiens}}, \bibinfo {author} {\bibfnamefont {C.~J.}\ \bibnamefont {Kwong}},
  \bibinfo {author} {\bibfnamefont {T.}~\bibnamefont {M\"{u}ller}}, \bibinfo
  {author} {\bibfnamefont {K.}~\bibnamefont {Bongs}}, \bibinfo {author}
  {\bibfnamefont {Y.}~\bibnamefont {Singh}},\ and\ \bibinfo {author}
  {\bibfnamefont {S.}~\bibnamefont {Schiller}},\ }\bibfield  {title} {\emph
  {\bibinfo {title} {Optical frequency reference based on a cryogenic silicon
  resonator}},\ }\href {https://doi.org/10.1364/OE.497365} {\bibfield
  {journal} {\bibinfo  {journal} {Opt. Express}\ }\textbf {\bibinfo {volume}
  {31}},\ \bibinfo {pages} {42059--42076} (\bibinfo {year} {2023})}\BibitemShut
  {NoStop}%
\bibitem [{\citenamefont {Lin}\ \emph {et~al.}(2024{\natexlab{b}})\citenamefont
  {Lin}, \citenamefont {Hartman}, \citenamefont {Pointard}, \citenamefont
  {Le~Targat}, \citenamefont {Goldner}, \citenamefont {Seidelin}, \citenamefont
  {Fang},\ and\ \citenamefont {Le~Coq}}]{lin24}%
  \BibitemOpen
  \bibfield  {author} {\bibinfo {author} {\bibfnamefont {X.}~\bibnamefont
  {Lin}}, \bibinfo {author} {\bibfnamefont {M.~T.}\ \bibnamefont {Hartman}},
  \bibinfo {author} {\bibfnamefont {B.}~\bibnamefont {Pointard}}, \bibinfo
  {author} {\bibfnamefont {R.}~\bibnamefont {Le~Targat}}, \bibinfo {author}
  {\bibfnamefont {P.}~\bibnamefont {Goldner}}, \bibinfo {author} {\bibfnamefont
  {S.}~\bibnamefont {Seidelin}}, \bibinfo {author} {\bibfnamefont
  {B.}~\bibnamefont {Fang}},\ and\ \bibinfo {author} {\bibfnamefont
  {Y.}~\bibnamefont {Le~Coq}},\ }\bibfield  {title} {\emph {\bibinfo {title}
  {Anomalous subkelvin thermal frequency shifts of ultranarrow linewidth solid
  state emitters}},\ }\href {https://doi.org/10.1103/PhysRevLett.133.183803}
  {\bibfield  {journal} {\bibinfo  {journal} {Phys. Rev. Lett.}\ }\textbf
  {\bibinfo {volume} {133}},\ \bibinfo {pages} {183803} (\bibinfo {year}
  {2024}{\natexlab{b}})}\BibitemShut {NoStop}%
\bibitem [{\citenamefont {Lyon}\ \emph {et~al.}(1977)\citenamefont {Lyon},
  \citenamefont {Salinger}, \citenamefont {Swenson},\ and\ \citenamefont
  {White}}]{lyon2008}%
  \BibitemOpen
  \bibfield  {author} {\bibinfo {author} {\bibfnamefont {K.~G.}\ \bibnamefont
  {Lyon}}, \bibinfo {author} {\bibfnamefont {G.~L.}\ \bibnamefont {Salinger}},
  \bibinfo {author} {\bibfnamefont {C.~A.}\ \bibnamefont {Swenson}},\ and\
  \bibinfo {author} {\bibfnamefont {G.~K.}\ \bibnamefont {White}},\ }\bibfield
  {title} {\emph {\bibinfo {title} {Linear thermal expansion measurements on
  silicon from 6 to 340 {{K}}}},\ }\href {https://doi.org/10.1063/1.323747}
  {\bibfield  {journal} {\bibinfo  {journal} {J. Appl. Phys.}\ }\textbf
  {\bibinfo {volume} {48}},\ \bibinfo {pages} {865--868} (\bibinfo {year}
  {1977})}\BibitemShut {NoStop}%
\bibitem [{\citenamefont {Telle}\ \emph {et~al.}(2002)\citenamefont {Telle},
  \citenamefont {Lipphardt},\ and\ \citenamefont {Stenger}}]{telle}%
  \BibitemOpen
  \bibfield  {author} {\bibinfo {author} {\bibfnamefont {H.~R.}\ \bibnamefont
  {Telle}}, \bibinfo {author} {\bibfnamefont {B.}~\bibnamefont {Lipphardt}},\
  and\ \bibinfo {author} {\bibfnamefont {J.}~\bibnamefont {Stenger}},\
  }\bibfield  {title} {\emph {\bibinfo {title} {Kerr-lens, mode-locked lasers
  as transfer oscillators for optical frequency measurements}},\ }\href
  {https://doi.org/10.1007/s003400100735} {\bibfield  {journal} {\bibinfo
  {journal} {Appl. Phys. B}\ }\textbf {\bibinfo {volume} {74}},\ \bibinfo
  {pages} {1--6} (\bibinfo {year} {2002})}\BibitemShut {NoStop}%
\bibitem [{\citenamefont {Diddams}\ \emph {et~al.}(2020)\citenamefont
  {Diddams}, \citenamefont {Vahala},\ and\ \citenamefont {Udem}}]{diddams}%
  \BibitemOpen
  \bibfield  {author} {\bibinfo {author} {\bibfnamefont {S.~A.}\ \bibnamefont
  {Diddams}}, \bibinfo {author} {\bibfnamefont {K.}~\bibnamefont {Vahala}},\
  and\ \bibinfo {author} {\bibfnamefont {T.}~\bibnamefont {Udem}},\ }\bibfield
  {title} {\emph {\bibinfo {title} {Optical frequency combs: {C}oherently
  uniting the electromagnetic spectrum}},\ }\href
  {https://doi.org/10.1126/science.aay3676} {\bibfield  {journal} {\bibinfo
  {journal} {Science}\ }\textbf {\bibinfo {volume} {369}},\ \bibinfo {pages}
  {eaay3676} (\bibinfo {year} {2020})}\BibitemShut {NoStop}%
\bibitem [{\citenamefont {Johnson}\ \emph {et~al.}(2015)\citenamefont
  {Johnson}, \citenamefont {Gill},\ and\ \citenamefont {Margolis}}]{johnson}%
  \BibitemOpen
  \bibfield  {author} {\bibinfo {author} {\bibfnamefont {L.~A.~M.}\
  \bibnamefont {Johnson}}, \bibinfo {author} {\bibfnamefont {P.}~\bibnamefont
  {Gill}},\ and\ \bibinfo {author} {\bibfnamefont {H.~S.}\ \bibnamefont
  {Margolis}},\ }\bibfield  {title} {\emph {\bibinfo {title} {Evaluating the
  performance of the {NPL} femtosecond frequency combs: agreement at the
  $10^{-21}$ level}},\ }\href {https://doi.org/10.1088/0026-1394/52/1/62}
  {\bibfield  {journal} {\bibinfo  {journal} {Metrologia}\ }\textbf {\bibinfo
  {volume} {52}},\ \bibinfo {pages} {62} (\bibinfo {year} {2015})}\BibitemShut
  {NoStop}%
\bibitem [{\citenamefont {Williams}\ \emph {et~al.}(2008)\citenamefont
  {Williams}, \citenamefont {Swann},\ and\ \citenamefont
  {Newbury}}]{williams2008}%
  \BibitemOpen
  \bibfield  {author} {\bibinfo {author} {\bibfnamefont {P.~A.}\ \bibnamefont
  {Williams}}, \bibinfo {author} {\bibfnamefont {W.~C.}\ \bibnamefont
  {Swann}},\ and\ \bibinfo {author} {\bibfnamefont {N.~R.}\ \bibnamefont
  {Newbury}},\ }\bibfield  {title} {\emph {\bibinfo {title} {High-stability
  transfer of an optical frequency over long fiber-optic links}},\ }\href
  {https://doi.org/10.1364/JOSAB.25.001284} {\bibfield  {journal} {\bibinfo
  {journal} {J. Opt. Soc. Am. B}\ }\textbf {\bibinfo {volume} {25}},\ \bibinfo
  {pages} {1284--1293} (\bibinfo {year} {2008})}\BibitemShut {NoStop}%
\bibitem [{\citenamefont {Rauf}\ \emph {et~al.}(2018)\citenamefont {Rauf},
  \citenamefont {Vélez~López}, \citenamefont {Thoumany}, \citenamefont
  {Pizzocaro},\ and\ \citenamefont {Calonico}}]{rauf2018}%
  \BibitemOpen
  \bibfield  {author} {\bibinfo {author} {\bibfnamefont {B.}~\bibnamefont
  {Rauf}}, \bibinfo {author} {\bibfnamefont {M.~C.}\ \bibnamefont
  {Vélez~López}}, \bibinfo {author} {\bibfnamefont {P.}~\bibnamefont
  {Thoumany}}, \bibinfo {author} {\bibfnamefont {M.}~\bibnamefont
  {Pizzocaro}},\ and\ \bibinfo {author} {\bibfnamefont {D.}~\bibnamefont
  {Calonico}},\ }\bibfield  {title} {\emph {\bibinfo {title} {{Phase noise
  cancellation in polarisation-maintaining fibre links}}},\ }\href
  {https://doi.org/10.1063/1.5016514} {\bibfield  {journal} {\bibinfo
  {journal} {Rev. Sci. Instrum.}\ }\textbf {\bibinfo {volume} {89}},\ \bibinfo
  {pages} {033103} (\bibinfo {year} {2018})}\BibitemShut {NoStop}%
\bibitem [{\citenamefont {Clivati}\ \emph {et~al.}(2020)\citenamefont
  {Clivati}, \citenamefont {Savio}, \citenamefont {Abrate}, \citenamefont
  {Curri}, \citenamefont {Gaudino}, \citenamefont {Pizzocaro},\ and\
  \citenamefont {Calonico}}]{clivati2020}%
  \BibitemOpen
  \bibfield  {author} {\bibinfo {author} {\bibfnamefont {C.}~\bibnamefont
  {Clivati}}, \bibinfo {author} {\bibfnamefont {P.}~\bibnamefont {Savio}},
  \bibinfo {author} {\bibfnamefont {S.}~\bibnamefont {Abrate}}, \bibinfo
  {author} {\bibfnamefont {V.}~\bibnamefont {Curri}}, \bibinfo {author}
  {\bibfnamefont {R.}~\bibnamefont {Gaudino}}, \bibinfo {author} {\bibfnamefont
  {M.}~\bibnamefont {Pizzocaro}},\ and\ \bibinfo {author} {\bibfnamefont
  {D.}~\bibnamefont {Calonico}},\ }\bibfield  {title} {\emph {\bibinfo {title}
  {Robust optical frequency dissemination with a dual-polarization coherent
  receiver}},\ }\href {https://doi.org/10.1364/OE.378602} {\bibfield  {journal}
  {\bibinfo  {journal} {Opt. Express}\ }\textbf {\bibinfo {volume} {28}},\
  \bibinfo {pages} {8494--8511} (\bibinfo {year} {2020})}\BibitemShut {NoStop}%
\bibitem [{\citenamefont {Barbieri}\ \emph {et~al.}(2019)\citenamefont
  {Barbieri}, \citenamefont {Clivati}, \citenamefont {Pizzocaro}, \citenamefont
  {Levi},\ and\ \citenamefont {Calonico}}]{barbieri2019}%
  \BibitemOpen
  \bibfield  {author} {\bibinfo {author} {\bibfnamefont {P.}~\bibnamefont
  {Barbieri}}, \bibinfo {author} {\bibfnamefont {C.}~\bibnamefont {Clivati}},
  \bibinfo {author} {\bibfnamefont {M.}~\bibnamefont {Pizzocaro}}, \bibinfo
  {author} {\bibfnamefont {F.}~\bibnamefont {Levi}},\ and\ \bibinfo {author}
  {\bibfnamefont {D.}~\bibnamefont {Calonico}},\ }\bibfield  {title} {\emph
  {\bibinfo {title} {Spectral purity transfer with $5\times10^{-17}$
  instability at 1 s using a multibranch {Er}:fiber frequency comb}},\ }\href
  {https://doi.org/10.1088/1681-7575/ab2b0f} {\bibfield  {journal} {\bibinfo
  {journal} {Metrologia}\ }\textbf {\bibinfo {volume} {56}},\ \bibinfo {pages}
  {045008} (\bibinfo {year} {2019})}\BibitemShut {NoStop}%
\bibitem [{\citenamefont {Rolland}\ \emph {et~al.}(2018)\citenamefont
  {Rolland}, \citenamefont {Li}, \citenamefont {Kuse}, \citenamefont {Jiang},
  \citenamefont {Cassinerio}, \citenamefont {Langrock},\ and\ \citenamefont
  {Fermann}}]{rolland2018}%
  \BibitemOpen
  \bibfield  {author} {\bibinfo {author} {\bibfnamefont {A.}~\bibnamefont
  {Rolland}}, \bibinfo {author} {\bibfnamefont {P.}~\bibnamefont {Li}},
  \bibinfo {author} {\bibfnamefont {N.}~\bibnamefont {Kuse}}, \bibinfo {author}
  {\bibfnamefont {J.}~\bibnamefont {Jiang}}, \bibinfo {author} {\bibfnamefont
  {M.}~\bibnamefont {Cassinerio}}, \bibinfo {author} {\bibfnamefont
  {C.}~\bibnamefont {Langrock}},\ and\ \bibinfo {author} {\bibfnamefont
  {M.~E.}\ \bibnamefont {Fermann}},\ }\bibfield  {title} {\emph {\bibinfo
  {title} {Ultra-broadband dual-branch optical frequency comb with $10^{-18}$
  instability}},\ }\href {https://doi.org/10.1364/OPTICA.5.001070} {\bibfield
  {journal} {\bibinfo  {journal} {Optica}\ }\textbf {\bibinfo {volume} {5}},\
  \bibinfo {pages} {1070--1077} (\bibinfo {year} {2018})}\BibitemShut {NoStop}%
\bibitem [{\citenamefont {Giunta}\ \emph {et~al.}(2019)\citenamefont {Giunta},
  \citenamefont {H\"ansel}, \citenamefont {Fischer}, \citenamefont {Lezius},\
  and\ \citenamefont {Holzwarth}}]{giunta2019}%
  \BibitemOpen
  \bibfield  {author} {\bibinfo {author} {\bibfnamefont {M.}~\bibnamefont
  {Giunta}}, \bibinfo {author} {\bibfnamefont {W.}~\bibnamefont {H\"ansel}},
  \bibinfo {author} {\bibfnamefont {M.}~\bibnamefont {Fischer}}, \bibinfo
  {author} {\bibfnamefont {T.}~\bibnamefont {Lezius}, \bibfnamefont
  {Matthias~Udem}},\ and\ \bibinfo {author} {\bibfnamefont {R.}~\bibnamefont
  {Holzwarth}},\ }\bibfield  {title} {\emph {\bibinfo {title} {Real-time phase
  tracking for wide-band optical frequency measurements at the 20th decimal
  place}},\ }\href {https://doi.org/10.1038/s41566-019-0520-5} {\bibfield
  {journal} {\bibinfo  {journal} {Nature Photonics}\ }\textbf {\bibinfo
  {volume} {14}},\ \bibinfo {pages} {44--49} (\bibinfo {year}
  {2019})}\BibitemShut {NoStop}%
\bibitem [{\citenamefont {Yeaton-Massey}\ and\ \citenamefont
  {Adhikari}(2012)}]{yeaton2012}%
  \BibitemOpen
  \bibfield  {author} {\bibinfo {author} {\bibfnamefont {D.}~\bibnamefont
  {Yeaton-Massey}}\ and\ \bibinfo {author} {\bibfnamefont {R.~X.}\ \bibnamefont
  {Adhikari}},\ }\bibfield  {title} {\emph {\bibinfo {title} {A new bound on
  excess frequency noise in second harmonic generation in {PPKTP} at the
  $10^{-19}$ level}},\ }\href {https://doi.org/10.1364/OE.20.021019} {\bibfield
   {journal} {\bibinfo  {journal} {Opt. Express}\ }\textbf {\bibinfo {volume}
  {20}},\ \bibinfo {pages} {21019--21024} (\bibinfo {year} {2012})}\BibitemShut
  {NoStop}%
\bibitem [{\citenamefont {Leopardi}\ \emph {et~al.}(2017)\citenamefont
  {Leopardi}, \citenamefont {Davila-Rodriguez}, \citenamefont {Quinlan},
  \citenamefont {Olson}, \citenamefont {Sherman}, \citenamefont {Diddams},\
  and\ \citenamefont {Fortier}}]{leopardi2017}%
  \BibitemOpen
  \bibfield  {author} {\bibinfo {author} {\bibfnamefont {H.}~\bibnamefont
  {Leopardi}}, \bibinfo {author} {\bibfnamefont {J.}~\bibnamefont
  {Davila-Rodriguez}}, \bibinfo {author} {\bibfnamefont {F.}~\bibnamefont
  {Quinlan}}, \bibinfo {author} {\bibfnamefont {J.}~\bibnamefont {Olson}},
  \bibinfo {author} {\bibfnamefont {J.~A.}\ \bibnamefont {Sherman}}, \bibinfo
  {author} {\bibfnamefont {S.~A.}\ \bibnamefont {Diddams}},\ and\ \bibinfo
  {author} {\bibfnamefont {T.~M.}\ \bibnamefont {Fortier}},\ }\bibfield
  {title} {\emph {\bibinfo {title} {Single-branch {Er}:fiber frequency comb for
  precision optical metrology with $10^{-18}$ fractional instability}},\ }\href
  {https://doi.org/10.1364/OPTICA.4.000879} {\bibfield  {journal} {\bibinfo
  {journal} {Optica}\ }\textbf {\bibinfo {volume} {4}},\ \bibinfo {pages}
  {879--885} (\bibinfo {year} {2017})}\BibitemShut {NoStop}%
\bibitem [{\citenamefont {Herbers}\ \emph {et~al.}(2019)\citenamefont
  {Herbers}, \citenamefont {D\"{o}rscher}, \citenamefont {Benkler},\ and\
  \citenamefont {Lisdat}}]{herbers2019}%
  \BibitemOpen
  \bibfield  {author} {\bibinfo {author} {\bibfnamefont {S.}~\bibnamefont
  {Herbers}}, \bibinfo {author} {\bibfnamefont {S.}~\bibnamefont
  {D\"{o}rscher}}, \bibinfo {author} {\bibfnamefont {E.}~\bibnamefont
  {Benkler}},\ and\ \bibinfo {author} {\bibfnamefont {C.}~\bibnamefont
  {Lisdat}},\ }\bibfield  {title} {\emph {\bibinfo {title} {Phase noise of
  frequency doublers in optical clock lasers}},\ }\href
  {https://doi.org/10.1364/OE.27.023262} {\bibfield  {journal} {\bibinfo
  {journal} {Opt. Express}\ }\textbf {\bibinfo {volume} {27}},\ \bibinfo
  {pages} {23262--23273} (\bibinfo {year} {2019})}\BibitemShut {NoStop}%
\bibitem [{\citenamefont {Risaro}\ \emph {et~al.}(2022)\citenamefont {Risaro},
  \citenamefont {Savio}, \citenamefont {Pizzocaro}, \citenamefont {Levi},
  \citenamefont {Calonico},\ and\ \citenamefont {Clivati}}]{risaro2022}%
  \BibitemOpen
  \bibfield  {author} {\bibinfo {author} {\bibfnamefont {M.}~\bibnamefont
  {Risaro}}, \bibinfo {author} {\bibfnamefont {P.}~\bibnamefont {Savio}},
  \bibinfo {author} {\bibfnamefont {M.}~\bibnamefont {Pizzocaro}}, \bibinfo
  {author} {\bibfnamefont {F.}~\bibnamefont {Levi}}, \bibinfo {author}
  {\bibfnamefont {D.}~\bibnamefont {Calonico}},\ and\ \bibinfo {author}
  {\bibfnamefont {C.}~\bibnamefont {Clivati}},\ }\bibfield  {title} {\emph
  {\bibinfo {title} {Improving the resolution of comb-based frequency
  measurements using a track-and-hold amplifier}},\ }\href
  {https://doi.org/10.1103/PhysRevApplied.18.064010} {\bibfield  {journal}
  {\bibinfo  {journal} {Phys. Rev. Appl.}\ }\textbf {\bibinfo {volume} {18}},\
  \bibinfo {pages} {064010} (\bibinfo {year} {2022})}\BibitemShut {NoStop}%
\bibitem [{\citenamefont {Sinclair}\ \emph {et~al.}(2015)\citenamefont
  {Sinclair}, \citenamefont {Deschênes}, \citenamefont {Sonderhouse},
  \citenamefont {Swann}, \citenamefont {Khader}, \citenamefont {Baumann},
  \citenamefont {Newbury},\ and\ \citenamefont {Coddington}}]{sinclair2015}%
  \BibitemOpen
  \bibfield  {author} {\bibinfo {author} {\bibfnamefont {L.~C.}\ \bibnamefont
  {Sinclair}}, \bibinfo {author} {\bibfnamefont {J.-D.}\ \bibnamefont
  {Deschênes}}, \bibinfo {author} {\bibfnamefont {L.}~\bibnamefont
  {Sonderhouse}}, \bibinfo {author} {\bibfnamefont {W.~C.}\ \bibnamefont
  {Swann}}, \bibinfo {author} {\bibfnamefont {I.~H.}\ \bibnamefont {Khader}},
  \bibinfo {author} {\bibfnamefont {E.}~\bibnamefont {Baumann}}, \bibinfo
  {author} {\bibfnamefont {N.~R.}\ \bibnamefont {Newbury}},\ and\ \bibinfo
  {author} {\bibfnamefont {I.}~\bibnamefont {Coddington}},\ }\bibfield  {title}
  {\emph {\bibinfo {title} {Invited article: {A} compact optically coherent
  fiber frequency comb}},\ }\href {https://doi.org/10.1063/1.4928163}
  {\bibfield  {journal} {\bibinfo  {journal} {Rev. Sci. Instrum.}\ }\textbf
  {\bibinfo {volume} {86}},\ \bibinfo {pages} {081301} (\bibinfo {year}
  {2015})}\BibitemShut {NoStop}%
\bibitem [{\citenamefont {Desch\^{e}nes}\ and\ \citenamefont
  {Genest}(2013)}]{deschenes2013}%
  \BibitemOpen
  \bibfield  {author} {\bibinfo {author} {\bibfnamefont {J.-D.}\ \bibnamefont
  {Desch\^{e}nes}}\ and\ \bibinfo {author} {\bibfnamefont {J.}~\bibnamefont
  {Genest}},\ }\bibfield  {title} {\emph {\bibinfo {title} {Heterodyne beats
  between a continuous-wave laser and a frequency comb beyond the shot-noise
  limit of a single comb mode}},\ }\href
  {https://doi.org/10.1103/PhysRevA.87.023802} {\bibfield  {journal} {\bibinfo
  {journal} {Phys. Rev. A}\ }\textbf {\bibinfo {volume} {87}},\ \bibinfo
  {pages} {023802} (\bibinfo {year} {2013})}\BibitemShut {NoStop}%
\bibitem [{\citenamefont {Tourigny-Plante}\ \emph {et~al.}(2018)\citenamefont
  {Tourigny-Plante}, \citenamefont {Michaud-Belleau}, \citenamefont
  {Bourbeau~H\'ebert}, \citenamefont {Bergeron}, \citenamefont {Genest},\ and\
  \citenamefont {Desch\^enes}}]{tourigny2018}%
  \BibitemOpen
  \bibfield  {author} {\bibinfo {author} {\bibfnamefont {A.}~\bibnamefont
  {Tourigny-Plante}}, \bibinfo {author} {\bibfnamefont {V.}~\bibnamefont
  {Michaud-Belleau}}, \bibinfo {author} {\bibfnamefont {N.}~\bibnamefont
  {Bourbeau~H\'ebert}}, \bibinfo {author} {\bibfnamefont {H.}~\bibnamefont
  {Bergeron}}, \bibinfo {author} {\bibfnamefont {J.}~\bibnamefont {Genest}},\
  and\ \bibinfo {author} {\bibfnamefont {J.-D.}\ \bibnamefont {Desch\^enes}},\
  }\bibfield  {title} {\emph {\bibinfo {title} {An open and flexible digital
  phase-locked loop for optical metrology}},\ }\href
  {https://doi.org/10.1063/1.5039344} {\bibfield  {journal} {\bibinfo
  {journal} {Rev. Sci. Inst.}\ }\textbf {\bibinfo {volume} {89}},\ \bibinfo
  {pages} {093103} (\bibinfo {year} {2018})}\BibitemShut {NoStop}%
\bibitem [{\citenamefont {Sherman}\ and\ \citenamefont
  {Jördens}(2016)}]{sherman2014}%
  \BibitemOpen
  \bibfield  {author} {\bibinfo {author} {\bibfnamefont {J.~A.}\ \bibnamefont
  {Sherman}}\ and\ \bibinfo {author} {\bibfnamefont {R.}~\bibnamefont
  {Jördens}},\ }\bibfield  {title} {\emph {\bibinfo {title} {{Oscillator
  metrology with software defined radio}}},\ }\href
  {https://doi.org/10.1063/1.4950898} {\bibfield  {journal} {\bibinfo
  {journal} {Rev. Sci. Instrum.}\ }\textbf {\bibinfo {volume} {87}},\ \bibinfo
  {pages} {054711} (\bibinfo {year} {2016})}\BibitemShut {NoStop}%
\bibitem [{\citenamefont {Donadello}\ \emph {et~al.}(2023)\citenamefont
  {Donadello}, \citenamefont {Bertacco}, \citenamefont {Calonico},\ and\
  \citenamefont {Clivati}}]{donadello2023}%
  \BibitemOpen
  \bibfield  {author} {\bibinfo {author} {\bibfnamefont {S.}~\bibnamefont
  {Donadello}}, \bibinfo {author} {\bibfnamefont {E.~K.}\ \bibnamefont
  {Bertacco}}, \bibinfo {author} {\bibfnamefont {D.}~\bibnamefont {Calonico}},\
  and\ \bibinfo {author} {\bibfnamefont {C.}~\bibnamefont {Clivati}},\
  }\bibfield  {title} {\emph {\bibinfo {title} {Embedded digital phase noise
  analyzer for optical frequency metrology}},\ }\href
  {https://doi.org/10.1109/TIM.2023.3288255} {\bibfield  {journal} {\bibinfo
  {journal} {IEEE Trans. Instrum. Meas.}\ }\textbf {\bibinfo {volume} {72}},\
  \bibinfo {pages} {1--12} (\bibinfo {year} {2023})}\BibitemShut {NoStop}%
\bibitem [{\citenamefont {Savio}\ \emph {et~al.}(2025)\citenamefont {Savio},
  \citenamefont {Goti}, \citenamefont {Pizzocaro}, \citenamefont {Levi},
  \citenamefont {Calonico},\ and\ \citenamefont {Clivati}}]{sav25}%
  \BibitemOpen
  \bibfield  {author} {\bibinfo {author} {\bibfnamefont {P.}~\bibnamefont
  {Savio}}, \bibinfo {author} {\bibfnamefont {I.}~\bibnamefont {Goti}},
  \bibinfo {author} {\bibfnamefont {M.}~\bibnamefont {Pizzocaro}}, \bibinfo
  {author} {\bibfnamefont {F.}~\bibnamefont {Levi}}, \bibinfo {author}
  {\bibfnamefont {D.}~\bibnamefont {Calonico}},\ and\ \bibinfo {author}
  {\bibfnamefont {C.}~\bibnamefont {Clivati}},\ }\bibfield  {title} {\emph
  {\bibinfo {title} {Optical-comb-based frequency stability transfer across the
  spectrum with a multi-channel {FPGA}}},\ }\href
  {https://doi.org/10.1109/TUFFC.2025.3526761} {\bibfield  {journal} {\bibinfo
  {journal} {IEEE Trans. Ultrason. Ferroelectr. Freq. Control}\ }\textbf
  {\bibinfo {volume} {72}},\ \bibinfo {pages} {397--406} (\bibinfo {year}
  {2025})}\BibitemShut {NoStop}%
\bibitem [{\citenamefont {Santarelli}\ \emph {et~al.}(1998)\citenamefont
  {Santarelli}, \citenamefont {Audoin}, \citenamefont {Makdissi}, \citenamefont
  {Laurent}, \citenamefont {Dick},\ and\ \citenamefont {Clairon}}]{santarelli}%
  \BibitemOpen
  \bibfield  {author} {\bibinfo {author} {\bibfnamefont {G.}~\bibnamefont
  {Santarelli}}, \bibinfo {author} {\bibfnamefont {C.}~\bibnamefont {Audoin}},
  \bibinfo {author} {\bibfnamefont {A.}~\bibnamefont {Makdissi}}, \bibinfo
  {author} {\bibfnamefont {P.}~\bibnamefont {Laurent}}, \bibinfo {author}
  {\bibfnamefont {G.}~\bibnamefont {Dick}},\ and\ \bibinfo {author}
  {\bibfnamefont {A.}~\bibnamefont {Clairon}},\ }\bibfield  {title} {\emph
  {\bibinfo {title} {Frequency stability degradation of an oscillator slaved to
  a periodically interrogated atomic resonator}},\ }\href
  {https://doi.org/10.1109/58.710548} {\bibfield  {journal} {\bibinfo
  {journal} {IEEE Trans. Ultrason. Ferroelectr. Freq. Control}\ }\textbf
  {\bibinfo {volume} {45}},\ \bibinfo {pages} {887--894} (\bibinfo {year}
  {1998})}\BibitemShut {NoStop}%
\bibitem [{\citenamefont {Dick}(1987)}]{dick}%
  \BibitemOpen
  \bibfield  {author} {\bibinfo {author} {\bibfnamefont {G.~J.}\ \bibnamefont
  {Dick}},\ }\bibfield  {title} {\emph {\bibinfo {title} {Local oscillator
  induced instabilities in trapped ion frequency standards}},\ }in\ \href
  {https://apps.dtic.mil/sti/citations/ADA502386} {\emph {\bibinfo {booktitle}
  {Proc. 19th Annual Precise Time and Time Interval (PTTI) Applications and
  Planning Meeting.}}}\ (\bibinfo  {publisher} {Ed. by R. L. Sydnor. US Naval
  Observatory/Goddard Space Flight Center},\ \bibinfo {year} {1987})\ pp.\
  \bibinfo {pages} {133--147}\BibitemShut {NoStop}%
\bibitem [{\citenamefont {Benkler}\ \emph {et~al.}(2020)\citenamefont
  {Benkler}, \citenamefont {Lipphardt}, \citenamefont {Puppe}, \citenamefont
  {Wilk}, \citenamefont {Rohde},\ and\ \citenamefont {Sterr}}]{ben20}%
  \BibitemOpen
  \bibfield  {author} {\bibinfo {author} {\bibfnamefont {E.}~\bibnamefont
  {Benkler}}, \bibinfo {author} {\bibfnamefont {B.}~\bibnamefont {Lipphardt}},
  \bibinfo {author} {\bibfnamefont {T.}~\bibnamefont {Puppe}}, \bibinfo
  {author} {\bibfnamefont {R.}~\bibnamefont {Wilk}}, \bibinfo {author}
  {\bibfnamefont {F.}~\bibnamefont {Rohde}},\ and\ \bibinfo {author}
  {\bibfnamefont {U.}~\bibnamefont {Sterr}},\ }\bibfield  {title} {\emph
  {\bibinfo {title} {End-to-end topology for fiber comb based optical frequency
  transfer at the $10^{-21}$ level: erratum}},\ }\href
  {https://doi.org/10.1364/OE.27.036886} {\bibfield  {journal} {\bibinfo
  {journal} {Opt. Express}\ }\textbf {\bibinfo {volume} {28}},\ \bibinfo
  {pages} {15023--15024} (\bibinfo {year} {2020})}\BibitemShut {NoStop}%
\end{thebibliography}%

\end{document}